\documentclass[twocolumn]{aastex631}

\usepackage{style}

\NewPageAfterKeywords

\begin{document}

\title{The NANOGrav 15-year Data Set: Search for Signals from New Physics}
\author[0000-0003-4439-5342]{Adeela Afzal}
\affiliation{Institute for Theoretical Physics, University of M\"{u}nster, D-48149 M\"{u}nster, Germany}
\affiliation{Department of Physics, Quaid-i-Azam University, Islamabad, 45320, Pakistan}
\author[0000-0001-5134-3925]{Gabriella Agazie}
\affiliation{Center for Gravitation, Cosmology and Astrophysics, Department of Physics, University of Wisconsin-Milwaukee,\\ P.O. Box 413, Milwaukee, WI 53201, USA}
\author[0000-0002-8935-9882]{Akash Anumarlapudi}
\affiliation{Center for Gravitation, Cosmology and Astrophysics, Department of Physics, University of Wisconsin-Milwaukee,\\ P.O. Box 413, Milwaukee, WI 53201, USA}
\author[0000-0003-0638-3340]{Anne M. Archibald}
\affiliation{Newcastle University, NE1 7RU, UK}
\author{Zaven Arzoumanian}
\affiliation{X-Ray Astrophysics Laboratory, NASA Goddard Space Flight Center, Code 662, Greenbelt, MD 20771, USA}
\author[0000-0003-2745-753X]{Paul T. Baker}
\affiliation{Department of Physics and Astronomy, Widener University, One University Place, Chester, PA 19013, USA}
\author[0000-0003-0909-5563]{Bence B\'{e}csy}
\affiliation{Department of Physics, Oregon State University, Corvallis, OR 97331, USA}
\author[0000-0003-2260-9047]{Jose Juan Blanco-Pillado}
\affiliation{Department of Physics, University of Basque Country, UPV/EHU, E-48080, Bilbao, Spain}
\affiliation{EHU Quantum Center, University of Basque Country, UPV/EHU, Bilbao, Spain}
\affiliation{IKERBASQUE, Basque Foundation for Science, E-48011, Bilbao, Spain}
\author[0000-0002-2183-1087]{Laura Blecha}
\affiliation{Physics Department, University of Florida, Gainesville, FL 32611, USA}
\author[0000-0003-1928-4667]{Kimberly K. Boddy}
\affiliation{Department of Physics, The University of Texas at Austin, Austin, TX 78712, USA}
% ...
\shorttitle{NANOGrav 15-year New-Physics Signals}
\shortauthors{the NANOGrav Collaboration}

\correspondingauthor{The NANOGrav Collaboration}
\email{comments@nanograv.org}

\begin{abstract}
The 15-year pulsar timing data set collected by the North American Nanohertz Observatory for Gravitational Waves (NANOGrav) shows positive evidence for the presence of a low-frequency gravitational-wave (GW) background. In this paper, we investigate potential cosmological interpretations of this signal, specifically cosmic inflation, scalar-induced GWs, first-order phase transitions, cosmic strings, and domain walls. We find that, with the exception of stable cosmic strings of field theory origin, all these models can reproduce the observed signal. When compared to the standard interpretation in terms of inspiraling supermassive black hole binaries (SMBHBs), many cosmological models seem to provide a better fit resulting in Bayes factors in the range from 10 to 100. However, these results strongly depend on modeling assumptions about the cosmic SMBHB population and, at this stage, should not be regarded as evidence for new physics. Furthermore, we identify excluded parameter regions where the predicted GW signal from cosmological sources significantly exceeds the NANOGrav signal. These parameter constraints are independent of the origin of the NANOGrav signal and illustrate how pulsar timing data provide a new way to constrain the parameter space of these models. Finally, we search for deterministic signals produced by models of ultralight dark matter (ULDM) and dark matter substructures in the Milky Way. We find no evidence for either of these signals and thus report updated constraints on these models. In the case of ULDM, these constraints outperform torsion balance and atomic clock constraints for ULDM coupled to electrons, muons, or gluons.
\end{abstract}

\keywords{
Gravitational waves --
Cosmology:~early universe --
Methods:~data analysis
}

\tableofcontents

\section{Introduction} 
\label{sec:introduction}

The Standard Model (SM) of particle physics currently provides our best description of the laws governing the universe at subatomic scales. However, it fails to explain several observed properties of our universe, such as the origin of the matter--antimatter asymmetry, the nature of dark matter (DM) and dark energy, and the origin of neutrino masses. These shortcomings have motivated the development of several theories for physics beyond the SM, or BSM theories for short, accompanied by a rich experimental program trying to test them. 
The generation of gravitational waves (GWs) is a ubiquitous feature of many BSM theories~\citep{Maggiore:1999vm,Caprini:2018mtu,Christensen:2018iqi}. These GWs form a stochastic background and propagate essentially unimpeded over cosmic distances to be detected today, whereas electromagnetic radiation does not start free streaming until after recombination. Thus, detecting a stochastic GW background (GWB) of cosmological origin would offer a unique and direct glimpse into the very early universe and herald a new era for using GWs to study fundamental physics.

Cosmological GWBs can be produced by a number of particle physics models of the early universe. Notably, cosmic inflation generically produces GWs~\citep{Guzzetti:2016mkm}, which may be observable at nanohertz frequencies if their energy density spectrum is sufficiently blue-tilted. Similarly, an enhanced spectrum of short-wavelength scalar perturbations produced during inflation can source so-called scalar-induced GWs~\citep[SIGWs;][]{Domenech:2021ztg,Yuan:2021qgz}. Another potential source of GWs are cosmological first-order phase transitions~\citep{Caprini:2015zlo,Caprini:2019egz,Hindmarsh:2020hop}, which proceed through bubble nucleation; bubble collisions and bubble interactions with the primordial plasma giving rise to sound waves contribute to GW production. Finally, topological defects left behind by cosmological phase transitions, such as cosmic strings and domain walls~\citep{Vilenkin:1984ib,Hindmarsh:1994re,Saikawa:2017hiv}, can radiate GWs and hence contribute to the GWB.

The North American Nanohertz Observatory for Gravitational Waves~\citep[NANOGrav;][]{McLaughlin:2013ira} has recently found the first convincing evidence for a stochastic GWB signal, as detailed in~\citet[hereafter \citetalias{aaa+23_gwb}]{aaa+23_gwb}. Analyzing 15-year of pulsar timing observations, NANOGrav has detected a red-noise process whose spectral properties are common among all pulsars and that is spatially correlated among pulsar pairs in a manner consistent with an isotropic GWB. In the following, we will refer to this observation as ``the NANOGrav signal,'' ``the GWB signal,'' or simply ``the signal,'' keeping in mind the level of statistical significance at which the GW nature of the signal has been demonstrated in~\citetalias{aaa+23_gwb}. While the GWB is primarily expected to arise from a population of inspiraling supermassive black hole binaries~\citep[SMBHBs;][]{Rahagopal+Romani-1995,Jaffe:2002rt,Wyithe+Loeb-2003,Sesana:2004sp,Burke-Spolaor:2018bvk}, cosmological sources may also contribute to it.

The SMBHB interpretation of the signal is considered in \citet[hereafter \citetalias{aaa+23_smbh}]{aaa+23_smbh}. In this paper, we analyze the NANOGrav 15-year data set~\cite[hereafter \citetalias{aaa+23_timing}]{aaa+23_timing} to investigate the possibility that the observed signal is cosmological in nature or that it arises from a combination of SMBHBs and a cosmological source. In particular, we consider phenomenological models of cosmic inflation, SIGWs, first-order phase transitions, cosmic strings (stable, metastable, and superstrings), and domain walls. We find that all of these models, except for stable cosmic strings of field theory origin, are consistent with the observed GWB signal. Many models provide in fact a better fit of the NANOGrav data than the baseline SMBHB model, which is reflected in the outcome of a comprehensive Bayesian model comparison analysis that we perform: several new-physics models result in Bayes factors between 10 and 100. We also consider composite models where the GWB spectrum receives contributions from new physics and SMBHBs. Comparing these composite models to the SMBHB reference model leads to comparable results, again with many Bayes factors falling into the range from 10 to 100. Cosmic superstrings, as predicted by string theory, are among the models that provide a good fit of the data, while stable cosmic strings of field theory origin only result in Bayes factors in the range from 0.1 to 1.

The reason that some of the Bayes factors reach large values is that the SMBHB signal expected from the theoretical model used in this analysis agrees somewhat poorly (only at the level of $95\%$ regions) with the observed data, leaving room for improvement by adding additional sources or better noise modeling. It is perhaps an intriguing idea that this disagreement may point to the presence of a cosmological source, but the present evidence is quite weak. We stress that Bayes factors for additional models beyond the SMBHB interpretation are highly dependent on the range of priors with which these models are introduced. Thus, one should not assign too much meaning to the exact numerical values of the Bayes factors reported in this work.

In many models, there are ranges of parameter values that would produce signals in conflict with the NG15 data. In those cases, we show the excluded regions and give numerical upper limits for individual parameters. We do so in terms of a new statistical test, introducing what we call the $K$ ratio. These parameter constraints are independent of the origin of the signal in the NG15 data and a testament to the constraining power of PTA data in the search for new physics. In our parameter plots, we label the $K$-ratio constraints by NG15, and where applicable, we compare them to other existing bounds. In many cases, the NG15 bounds are complementary to existing bounds, highlighting the fact that new-physics searches at the PTA frontier venture into previously unexplored regions of parameter space.

Aside from cosmological GWBs, signals of new physics can appear in GW detectors in a deterministic manner. Although pulsar timing arrays (PTAs) are primarily used to search for a GWB, we can also leverage their remarkable sensitivity to search for these deterministic signals. Specifically, DM substructures within the Milky Way can produce a Doppler effect by accelerating the Earth or a pulsar~\citep{Seto:2007kj}, or a Shapiro delay of the photons' arrival times by perturbing the metric along the photon geodesic~\citep{Siegel:2007fz}. PTAs can also probe models of ultralight DM (ULDM), which can cause shifts in the observed pulse timing via metric fluctuations \citep{Khmelnitsky:2013lxt,Porayko2014} or via couplings between ULDM and SM particles~\citep{Graham:2015ifn,Kaplan2022}. We search for both of these deterministic signals, and after finding no evidence for either of them, we derive new bounds on both these models.

This paper is organized as follows. We describe the NG15 data set in Section~\ref{sec:pta_data} and our general analysis methods in Section~\ref{sec:data_analysis}. In Section~\ref{sec:astro_signal}, we discuss the GWB expected from SMBHBs. We present the analysis and results for new-physics models that generate a cosmological GWB in Section~\ref{sec:gwb_new_physics} and for models that produce deterministic signals in Section~\ref{sec:det_new_physics}. We conclude in Section~\ref{sec:results}. Additionally, we include a list of parameters for each model, the prior ranges we use in our analysis, and the corresponding recovered posterior ranges in Appendix~\ref{app:parameters}. We present median GW spectra for all cosmological models based on our recovered posterior distributions in Appendix~\ref{app:mediangw}, and we provide supplementary material for specific models in Appendix~\ref{app:supplementary} .

\section{PTA data}
\label{sec:pta_data}

The NANOGrav 15-year (NG15) data set consists of observations of 68 millisecond pulsars made between 2004 July and 2020 August. This updated data set adds 21 pulsars and 3 yr of observations to the previous 12.5\;yr data set~\citep{NANOGrav:2020gpb}. One pulsar, J0614--3329, was observed for less than 3 yr, which is why it is not included in our analysis. The remaining 67 pulsars were all observed for more than 3 yr with an approximate cadence of 1 month (with the exception of six pulsars that were observed weekly as part of a high-cadence campaign, which started in 2013 at the Green Bank Telescope and in 2015 at the Arecibo Observatory). 

The pulse times of arrival (TOAs) were generated from the raw data following the procedure discussed in~\cite{NANOGrav:2015qfw, NANOGrav:2017wvv} and~\cite{NANOGrav:2020gpb}. The resulting cleaned TOAs were fit to a timing model that accounts for the pulsar's period and spin period derivative, sky location, proper motion, and parallax. For pulsars in a binary system, we included in the timing model five Keplerian binary parameters and an additional non-Keplerian parameter if they improved the fit as determined by an $F$-test. Pulse dispersion was modeled as a piecewise constant with the inclusion of DMX parameters~\citep{NANOGrav:2015qfw, Jones:2016fkk}. The timing model fits were performed using the TT(BIPM2019) timescale and the JPL Solar System Ephemeris model DE440~\citep{2021AJ....161..105P}. Additional detail about the data set and the processing of the TOAs can be found in~\citetalias{aaa+23_timing} and~\citet[hereafter \citetalias{aaa+23_detchar}]{aaa+23_detchar}.

\section{Data analysis methods}
\label{sec:data_analysis}

The statistical tools needed to describe noise sources, GWBs, and deterministic signals in pulsar timing data have already been extensively discussed in the literature~\citep[see, e.g.,][]{NANOGrav:2015aud, NANOGRAV:2018hou}. In the following brief overview, we focus on the implementation of new-physics signals within this framework. 

\subsection{Likelihood}
\label{subsec:likelihood}

Our search for a new-physics signal utilizes the pulsars' timing residuals, $\boldsymbol{\delta t}$. These timing residuals measure the discrepancy between the observed TOAs and the ones predicted by the pulsar timing model described in~\citetalias{aaa+23_timing} and briefly summarized in Section~\ref{sec:pta_data}. There are three main contributions to these timing residuals: white noise, time-correlated stochastic processes (also known as red noise), and small errors in the fit to the timing-ephemeris parameters~\citep{NANOGrav:2020tig}. Specifically, we can model the timing residuals as
\begin{equation}
\label{eq:res_model}
\boldsymbol{\delta t}=\boldsymbol{n}+\mat{F}\,\boldsymbol{a}+\mat{M}\,\boldsymbol{\epsilon}\,.
\end{equation}
In the remainder of this section, we will define and discuss each of these three terms and define the PTA likelihood.

The first term on the right-hand side of Eq.~\eqref{eq:res_model}, $\boldsymbol{n}$, describes the white noise that is assumed to be left in each of the $N_{\rm TOA}$ timing residuals after subtracting all known systematics. White noise is assumed to be a zero mean normal random variable, fully characterized by its covariance. For the receiver/back-end combination $I$, the white-noise covariance matrix reads
\begin{equation}
\langle n_i n_j\rangle=\mathcal{F}_{ I}^2\left[\sigma_{i \;{ \rm S/N}}^2+\mathcal{Q}_{ I}^2\right]\; \delta_{ij}+ \mathcal{J}_{ I}^2\;\mathcal{U}_{ij}\,,
\end{equation}
where $i$ and $j$ index the TOAs, $\sigma_{i \;{\rm S/N}}$ is the TOA uncertainty for the $i$th observation, $\mathcal{F}_{ I}$ is the {\it Extra FACtor} (EFAC) parameter, $\mathcal{Q}_{ I}$ is the {\it Extra QUADrature} (EQUAD) parameter, and $\mathcal{J}_{ I}$ is the ECORR parameter. ECORR is modeled using a block diagonal matrix, $\mathcal{U}$, with values of 1 for TOAs from the same observing epoch and zeros for all other entries. Following the approach of previous works~\citep{NANOGrav:2015aud, NANOGRAV:2018hou}, we fix all white-noise parameters to their values at the maxima in the posterior probability distributions recovered from single pulsar noise studies in order to increase computational efficiency~\citepalias{aaa+23_detchar}.

Time-correlated stochastic processes, like pulsar-intrinsic red noise and GWB signals, are modeled using a Fourier basis of frequencies $i/T_{\rm obs}$, where $i$ indexes the harmonics of the basis and $T_{\rm obs}$ is the timing baseline, extending from the first to the last recorded TOA in the full PTA data set. Since we are generally interested in processes that exhibit long-timescale correlations, the expansion is truncated after $N_f$ frequency bins. In this paper, we use $N_f=30$ for pulsar-intrinsic red noise and $N_f=14$ for GWBs. The latter choice stems from the observation that most of the evidence for a GWB comes from the first 14 frequency bins. More specifically, fitting a common-spectrum uncorrelated red-noise process with a broken power-law spectral shape to the NG15 data, the posterior distribution for the break frequency reaches it maximum around the 14th frequency bin~\citepalias{aaa+23_gwb}. This set of 2$N_f$ sine--cosine pairs evaluated at the different observation times is contained in the Fourier design matrix, $\mat{F}$. The Fourier coefficients of this expansion, $\boldsymbol{a}$, are assumed to be normally distributed random variables with zero mean and covariance matrix, $\langle\boldsymbol{a}\boldsymbol{a}^{\rm T}\rangle=\mat{\phi}$, given by 
\begin{equation}
\label{eq:red_cov}
[\phi]_{(ak)(bj)}=\delta_{ij}\left(\Gamma_{ab}\Phi_{i}+\delta_{ab}\varphi_{a,i}\right)
\end{equation}
where $a$ and $b$ index the pulsars, $i$ and $j$ index the frequency harmonics, and $\Gamma_{ab}$ is the GWB overlap reduction function, which describes average correlations between pulsars $a$ and $b$ as a function of their angular separation in the sky. For an isotropic and unpolarized GWB, $\Gamma_{ab}$ is given by the Hellings \& Downs correlation~\citep{1983ApJ...265L..39H}, also known as ``quadrupolar'' or ``HD'' correlation.

The first term on the right-hand side of Eq.~\eqref{eq:red_cov} parameterizes the contribution to the timing residuals induced by a GWB in terms of the model-dependent coefficients $\Phi_i$. In this work, we consider two kinds of GWB sources: one of astrophysical origin, namely a population of inspiraling SMBHBs (discussed in section \ref{sec:astro_signal}), and one of cosmological origin, induced by one of the exotic new-physics models under consideration (discussed in section \ref{sec:gwb_new_physics}). The last term in Eq.~\eqref{eq:red_cov} models pulsar-intrinsic red-noise in terms of the coefficients $\varphi_{a,i}$, where
\begin{equation}
\label{eq:kappaa}
\varphi_a(f) = \frac{A_a^2}{12\pi^2}\frac{1}{T_{\rm obs}}\left(\frac{f}{1 \, {\rm yr}^{-1}}\right)^{-\gamma_a}\,{\rm yr}^3 
\end{equation}
and $\varphi_{a,i} = \varphi_a(i/T_{\rm obs})$ for all $N_f$ frequencies. The priors for the red noise parameters are reported in Table \ref{tab:base_priors}.

Finally, deviations from the initial best-fit values of the $m$ timing-ephemeris parameters are accounted for by the term $\mat{M}\boldsymbol{\epsilon}$. The {\it design matrix}, $\mat{M}$, is an $N_{ \rm TOA}\times m$ matrix containing the partial derivatives of the TOAs with respect to each timing-ephemeris parameter (evaluated at the initial best-fit value), and $\boldsymbol{\epsilon}$ is a vector containing the linear offset from these best-fit parameters. 

Since in this analysis we are not interested in the specific realization of the noise but only in its statistical properties, we can analytically marginalize over all the possible noise realizations (i.e., integrate over all the possible values of $\boldsymbol{a}$ and $\boldsymbol{\epsilon}$). This leaves us with a marginalized likelihood that depends only on the (unknown) parameters describing the red-noise covariance matrix \citep[i.e., $A_a$, $\gamma_a$, plus any other parameters describing $\Phi_i$;][]{van_Haasteren_2012,Lentati_2013}: 
\begin{equation}
\label{eq:likelihood}
p(\boldsymbol{\delta t}|\mat{\phi})=\frac{\exp\left(-\frac{1}{2}\boldsymbol{\delta t}^{ T}\mat{C}^{-1}\boldsymbol{\delta t}\right)}{\sqrt{{\rm det}(2\pi\mat{C})}}\,,
\end{equation}
where $\mat{C}=\mat{N}+\mat{TBT}^{ T}$. Here $\mat{N}$ is the covariance matrix of white noise, $\mat{T}=[\mat{M},\mat{F}]$, and $\mat{B}={\rm diag}(\mat{\infty},\mat{\phi})$, where $\mat{\infty}$ is a diagonal matrix of infinities, which effectively means that we assume flat priors for the parameters in $\boldsymbol{\epsilon}$. Since in our calculations we always deal with the inverse of $\mat{B}$, all these infinities reduce to zeros.

Eq.~\eqref{eq:likelihood} can be easily generalized to take into account deterministic signals (like the ones that will be discussed in Sections~\ref{subsec:uldm} and \ref{subsec:dm_substructures}). In the presence of a deterministic signal, $\boldsymbol{h}(\boldsymbol{\theta})$, which depends on a set of parameters $\boldsymbol{\theta}$, we just need to shift the residuals, $\boldsymbol{\delta t}\to\boldsymbol{\delta t}-\boldsymbol{h}(\boldsymbol{\theta})$.

Finally, we relate our characterization of the GWB given in Eq.~\eqref{eq:red_cov} in terms of $\Phi_i$ to the commonly adopted spectral representation in terms of the GWB energy density per logarithmic frequency interval, $d\rho_{\scriptscriptstyle\rm GW}/d\ln f$, as a fraction of the closure density, i.e., the total energy density of our universe, $\rho_c$~\citep{Allen:1997ad}
\begin{equation}
\label{eq:OGW}
\Omega_{\scriptscriptstyle\rm GW}(f) \equiv \frac{1}{\rho_c}\frac{d\rho_{\scriptscriptstyle\rm GW}(f)}{d\ln f}=\frac{8 \pi^4 f^5}{H_0^2}\,\frac{\Phi(f)}{\Delta f}\,.
\end{equation}
Here $H_0$ is the present-day value of the Hubble rate, $\Delta f=1/T_{\rm obs}$ is the separation between the $N_f$ frequency bins, and $\Phi(f)$ determines the coefficients $\Phi_i$ in Eq.~\eqref{eq:red_cov}, i.e., $\Phi_i = \Phi\left(i/T_{\rm obs}\right)$. Note that $\Phi(f)$ is identical to the timing residual power spectral density (PSD), $S(f) = \Phi(f)/\Delta f$, up to the constant factor of $1/\Delta f$. In the remainder of this paper, we will often work with $h^2\Omega_{\scriptscriptstyle\rm GW}$ instead of $\Omega_{\scriptscriptstyle\rm GW}$, where $h$ is the dimensionless Hubble constant, $H_0 = h\;\times\;100\,\textrm{km}\textrm{s}^{-1}\textrm{Mpc}^{-1}$, such that the explicit value of $H_0$ cancels in the product $h^2\Omega_{\scriptscriptstyle\rm GW}$. 

\subsection{Bayesian analysis}
\label{subsec:bayes_analysis}

The goal of this work is to investigate a series of cosmological interpretations of the GWB signal in our data. Specifically, we would like to answer two questions. First, what is the region in the parameter space of the new-physics models that could produce the observed GWB? And second, is there any preference between the astrophysical and cosmological interpretations of the signal?

To answer these questions, we make use of Bayesian inference. Bayesian inference is a statistical method in which Bayes' rule of conditional probabilities is used to update one's knowledge as observations are acquired. Given a model $\mathcal{H}$, a set of parameters $\Theta$, and data $\mathcal{D}$, we can use Bayes' rule to write 
\begin{equation}\label{eq:posterior}
    P(\Theta|\mathcal{D}, \mathcal{H})=\frac{P(\mathcal{D}|\Theta,\mathcal{H})P(\Theta|\mathcal{H})}{P(\mathcal{D}|\mathcal{H})}\,,
\end{equation}
where $P(\Theta|\mathcal{D},\mathcal{H})$ is the posterior probability distribution for the model parameters, $P(\mathcal{D}|\Theta,\mathcal{H})$ is the likelihood, $P(\Theta|\mathcal{H})$ is the prior probability distribution, and
\begin{equation}\label{eq:evidence}
    \mathcal{Z}\equiv P(\mathcal{D}|\mathcal{H})=\int \textrm{d}\Theta\: P(\mathcal{D}|\Theta,\mathcal{H})P(\Theta|\mathcal{H})
\end{equation}
is the marginalized likelihood, or evidence. In the context of this work, $\mathcal{H}$ is the timing residual model given in Eq.~\eqref{eq:res_model}, $\Theta$ contains the parameter describing the covariance matrix $\mat{\phi}$, and the data are the timing residuals $\boldsymbol{\delta t}$. The likelihood function for our analysis is given by Eq.~\eqref{eq:likelihood} and implemented using the \enterprise~\citep{2019ascl.soft12015E} and \texttt{ ENTEPRISE\_EXTENSIONS}~\citep{enterprise} packages. Our prior choices are summarized in Tables \ref{tab:base_priors} and \ref{tab:np_priors}.

The posterior distribution on the left-hand side of Eq.~\eqref{eq:posterior} is the central result of the Bayesian analysis and contains all the information needed to answer our two original questions. Indeed, integrating over all the model parameters except one (two) allows us to derive marginalized distributions that can be used to obtain 1D (2D) credible intervals. At the same time, given two models $\mathcal{H}_0$ and $\mathcal{H}_1$, we can perform model selection by calculating the Bayes factor defined as
\begin{equation}
\mathcal{B}_{10}(\mathcal{D}) = \frac{\mathcal{Z}_1}{\mathcal{Z}_0} = \frac{P(\mathcal{D}|\mathcal{H}_1)}{P(\mathcal{D}|\mathcal{H}_0)} \,.
\end{equation}
The numerical value of the Bayes factor for a given model comparison can then be interpreted as evidence against or in favor of model hypothesis $\mathcal{H}_1$ according to the Jeffreys scale~\citep{Jeffreys:1961}: $\mathcal{B}_{10} < 1$ means that $\mathcal{H}_1$ is disfavored, while $\mathcal{B}_{10}$ values in the ranges $[10^{0.0},10^{0.5}]$, $[10^{0.5},10^{1.0}]$, $[10^{1.0},10^{1.5}]$, $[10^{1.5},10^{2.0}]$, $[10^{2.0},\infty)$ are interpreted as negligibly small, substantial, strong, very strong, and decisive evidence in favor of $\mathcal{H}_1$, respectively.

Given the large number of parameters, the integration required to derive marginalized distributions and Bayes factors needs to be performed through Monte Carlo sampling. Specifically, we use the Markov Chain Monte Carlo (MCMC) tools implemented in the \texttt{PTMCMCSampler} package~\citep{justin_ellis_2017_1037579} to sample from the posterior distributions. The marginalized posterior densities shown in our plots are then derived by applying kernel density estimates to the MCMC samples via the methods implemented in the \texttt{GetDist} package~\citep{Lewis:2019xzd}. 

In order to compute the Bayes factor between two models, we use product space methods~\citep{10.2307/2346151, 10.2307/1391010, 10.1093/mnras/stv2217}, instead of calculating the evidence $\mathcal{Z}$ for each model separately. This procedure recasts model selection as a parameter estimation problem, introducing a model indexing variable that is sampled along with the parameters of the competing models and controls which model likelihood is active at each MCMC iteration. The ratio of samples spent in each bin of the model indexing variable returns the posterior odds ratio between models, which coincides with the Bayes factor for equal model priors, $P(\mathcal{H}_1) = P(\mathcal{H}_0)$. The Monte Carlo sampling uncertainties associated with this derivation of the Bayes factors can be estimated through statistical bootstrapping~\citep{Efron:1986hys}. Bootstrapping creates new sets of Monte Carlo draws by resampling (with replacement) the original set of draws. These sets of draws act as independent realizations of the sampling procedure and allow us to obtain a distribution for the Bayes factors from which we derive point values and uncertainties on our Bayes factors corresponding to mean and standard deviation. Specifically, the central values and corresponding errors quoted in the following for the Bayes factors were derived by creating $5\times 10^4$ realizations of our Monte Carlo draws. 

From Eq.~\eqref{eq:evidence}, it is evident that models' evidence and, therefore, Bayes factors depend on the prior choice. In our analysis, we will often restrict priors to the region of parameter space for which cosmological models produce an observable signal in the PTA frequency band. However, a more appropriate prior choice would cover the entire allowed region of parameter space. Nonetheless, when working with flat priors, it is easy to rescale the Bayes factors to account for wider prior ranges. Specifically, if the priors are extended to a region of parameter space for which the likelihood $P\left(\mathcal{D}|\Theta, \mathcal{H}\right)$ is approximately zero, the Bayes factors decrease by a factor proportional to the increase in prior volume.

\begin{figure}
    \centering
    \includegraphics[width=1\linewidth]{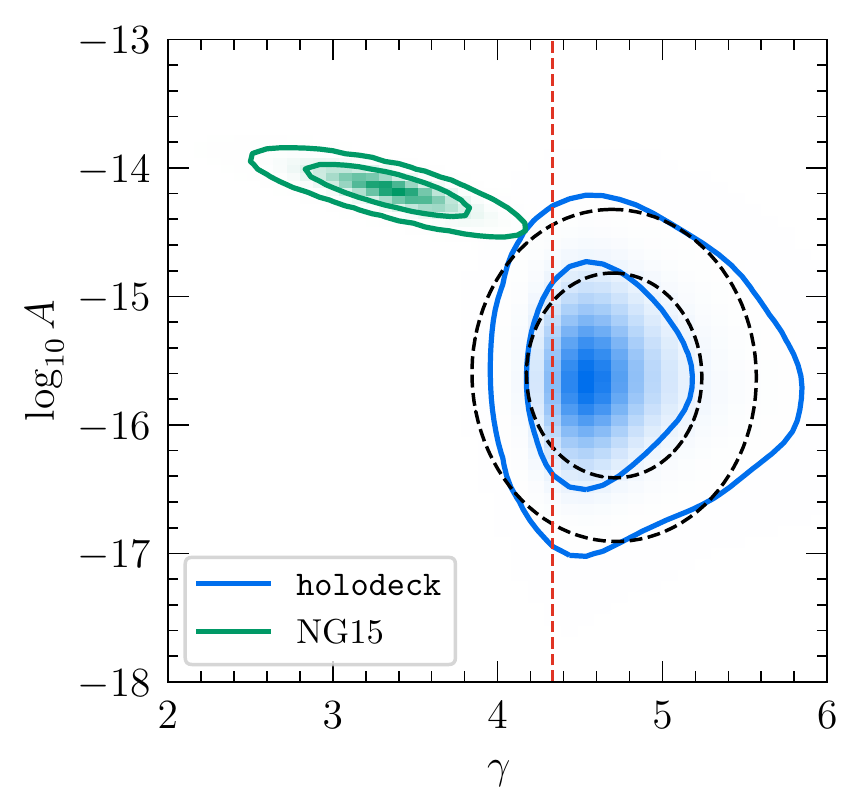}
    \caption{Comparison of the $68\%$ and $95\%$ probability regions for the amplitude and slope of a power-law fit to the observed GWB signal (green contours) and predicted for purely GW-driven SMBHB populations with circular orbits~\citepalias[blue contours;][]{aaa+23_smbh}. The black dashed lines represent a 2D Gaussian fit of the blue contours. The vertical red line indicates $\gamma=13/3$, the naive expectation for a GWB produced by a GW-driven SMBHB population~\citep{Phinney-2001}. \label{fig:A_gamma_comparison}}
\end{figure}

\begin{figure*}
    \centering
    \includegraphics[width=\textwidth]{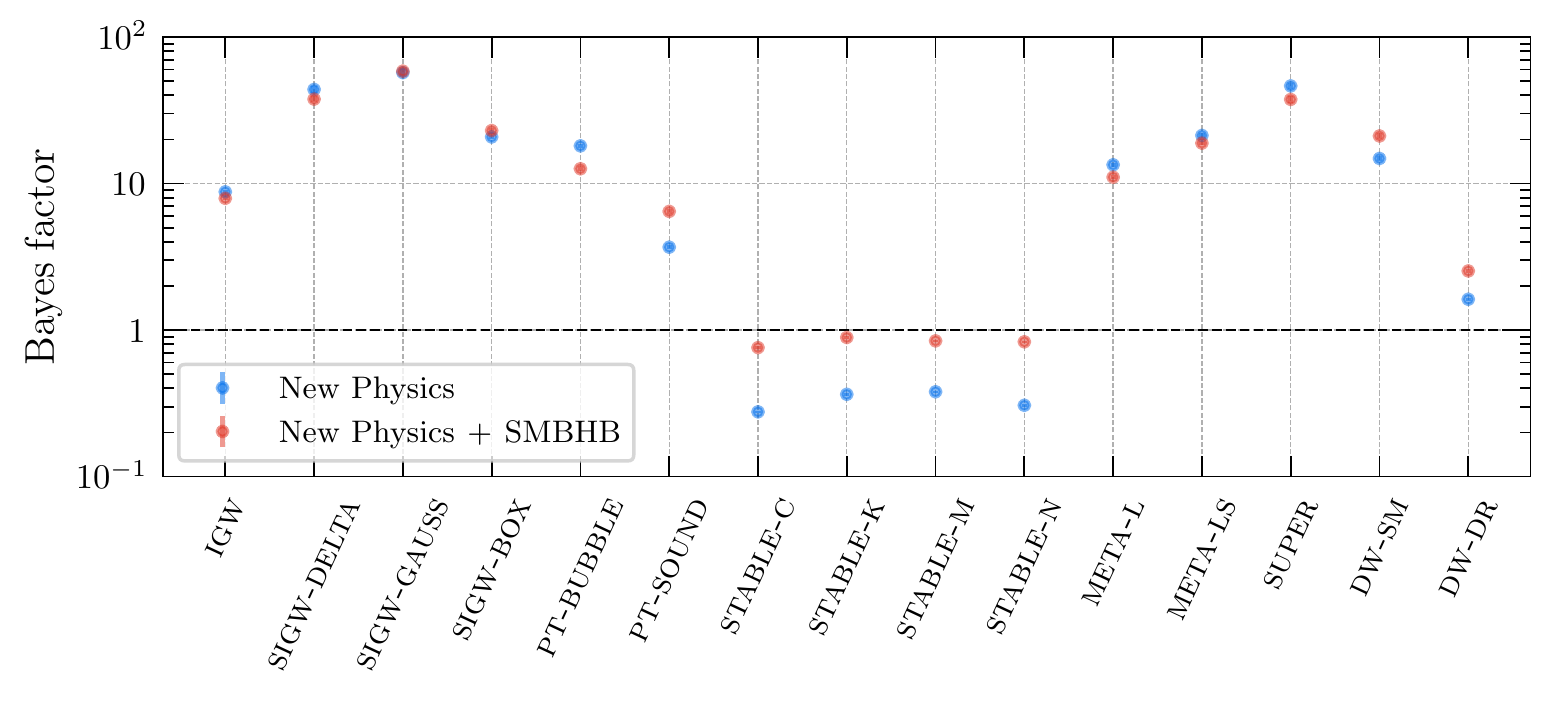}
    \caption{\label{fig:bf_table} Bayes factors for the model comparisons between the new-physics interpretations of the signal considered in this work and the interpretation in terms of SMBHBs alone.  Blue points are for the new physics alone, and red points are for the new physics in combination with the SMBHB signal. We also plot the error bars of all Bayes factors, which we obtain following the bootstrapping method outlined in Section~\ref{subsec:bayes_analysis}. In most cases, however, these error bars are small and not visible.}
\end{figure*}

For each model $\mathcal{H}$ considered in our analysis, we use the reconstructed posterior distribution, $P\left(\Theta|\mathcal{D},\mathcal{H}\right)$, to identify relevant parameter ranges and set upper limits. Specifically, we identify 68\% (95\%) Bayesian credible intervals~\citep{Bernardo:2000} by integrating the posterior over the regions of highest density until the integral covers $68\%$ ($95\%$) of the posterior probability. Moreover, we give upper limits above which the additional model is ``strongly disfavored'' according to the Jeffreys scale~\citep{Jeffreys:1961}. For instance, to place a bound on a single parameter $\theta$, we first marginalize over all other model parameters and then determine the parameter value at which the likelihood ratio
\begin{equation}
\label{eq:Kratio}
K(\theta) = \frac{P(\mathcal{D}|\theta,\mathcal{H})}{P(\mathcal{D}|\theta_0,\mathcal{H})} \,,
\end{equation}
has dropped to $K=\sfrac{1}{10}$. Here $\theta_0$ refers to the parameter limit in which the new-physics contribution to the total signal becomes negligible and $P(\mathcal{D}|\theta,\mathcal{H})$ no longer depends on the exact value of $\theta$. Graphing $P(\mathcal{D}|\theta,\mathcal{H})$ as a function of $\theta$, this parameter region appears as a plateau, with $P(\mathcal{D}|\theta_0,\mathcal{H})$ denoting the height of this plateau. Assuming a flat prior on $\theta$, the ratio in Eq.~\eqref{eq:Kratio} is identical to the corresponding ratio of marginalized posteriors. Furthermore, multiplying and dividing by the prior on~$\theta$, 
\begin{equation}
K(\theta) = \frac{P(\theta|\mathcal{H})}{P(\theta_0|\mathcal{D}, \mathcal{H})}\frac{P(\theta|\mathcal{D}, \mathcal{H})}{P(\theta|\mathcal{H})} \,.
\end{equation}
The first factor is the Savage--Dickey density ratio and can hence be identified as the Bayes factor $\mathcal{B} = P(\mathcal{D}|\mathcal{H})/P(\mathcal{D}|\mathcal{H}_0)$, where $\mathcal{H}_0$ is the model that results from model $\mathcal{H}$ when omitting the signal contribution controlled by the parameter $\theta$. The $K$ ratio can thus be written as the product of the global Bayes factor and the local posterior-to-prior ratio for the parameter $\theta$,
\begin{equation}
\label{eq:KratioB}
K(\theta) = \mathcal{B}\,\frac{P(\theta|\mathcal{D}, \mathcal{H})}{P(\theta|\mathcal{H})} \,.
\end{equation}
Once $\mathcal{B}$ is known, it is straightforward to evaluate Eq.~\eqref{eq:KratioB} and determine the $K$-ratio bound on $\theta$. Eq.~\eqref{eq:KratioB} is useful for numerically evaluating $K$, as it automatically encodes the height of the plateau in the marginalized posterior, $P(\theta_0|\mathcal{D}, \mathcal{H}) = P(\theta|\mathcal{H}) / \mathcal{B}$, which we would otherwise have to obtain from a fit to our MCMC data. However, we stress that $K$ is defined as a likelihood ratio, which renders it immune to  prior effects~\citep[prior choice, range, etc.;][]{Azzalini:1996}. For more than one parameter dimension, we proceed analogously and derive bounds based on the criterion $K(\Theta) > \sfrac{1}{10}$.

All Bayesian inference analyses discussed in this work were implemented into \enterprise\ via a newly developed wrapper that we call \texttt{PTArcade}~\citep{andrea_mitridate_2023_7876429, NG15-ptarcade}. This wrapper is intended to allow easy implementation of new-physics searches in PTA data. We make this wrapper publicly available at \url{https://doi.org/10.5281/zenodo.7876429}. Similarly, all MCMC chains analyzed in this work can be downloaded at \url{https://zenodo.org/record/8010909}.

\section{GWB signal from \texorpdfstring{SMBHB\MakeLowercase{s}}{SMBHBs}}
\label{sec:astro_signal}

Most galaxies are expected to host a supermassive black hole (SMBH) at their center~\citep{Kormendy:2013dxa,EventHorizonTelescope:2019dse}. During the hierarchical merging of galaxies taking place in the course of structure formation~\citep{10.1093/mnras/183.3.341}, these black holes are expected to sink to the center of the merger remnants, eventually forming binary systems~\citep{Begelman1980}. The gravitational radiation emitted by this population of inspiraling SMBHBs forms a GWB in the PTA band~\citep{Rahagopal+Romani-1995, Jaffe:2002rt, Wyithe+Loeb-2003} and is a natural candidate for the source of the signal observed in our data. 

\begin{figure*}[htpb]
	\centering
        \includegraphics[width=\textwidth]{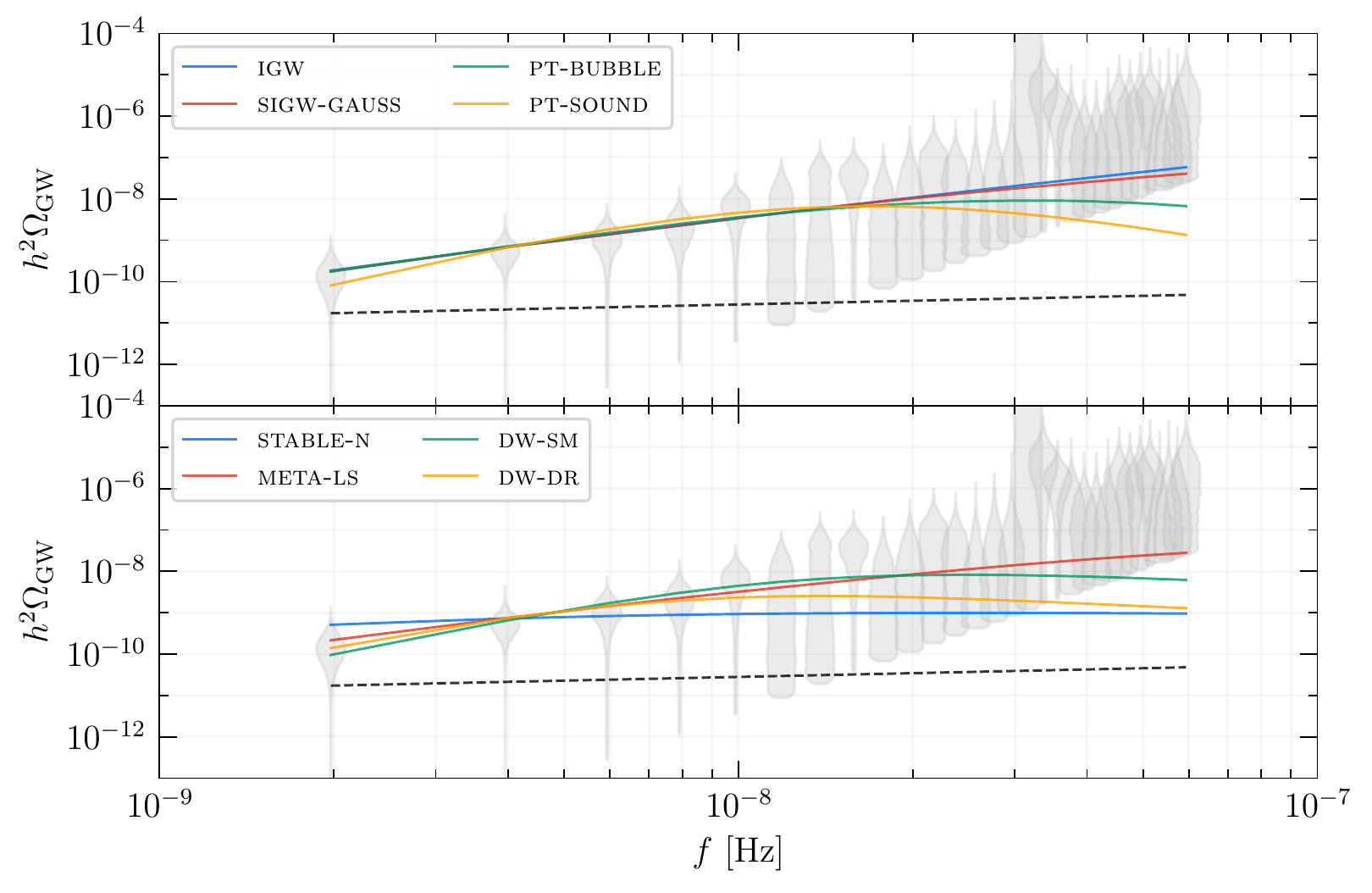}
	\caption{\label{fig:mean_spectra}
       Median GWB spectra produced by a subset of the new-physics models, which we construct by mapping our model parameter posterior distributions to $h^2\Omega_{\scriptscriptstyle \rm GW}$ distributions at every frequency $f$ (see Appendix~\ref{app:mediangw} for more details and Figs.~\ref{fig:mean_spectra_all_1} and \ref{fig:mean_spectra_all_2} for the models not included here). We also show the periodogram for an HD-correlated free spectral process (gray violins) and the GWB spectrum produced by an astrophysical population of inspiraling SMBHBs with the parameters $\abhb$ and $\gbhb$ fixed at the central values $\boldsymbol{\mu}_{\scriptscriptstyle \rm BHB}$ of the 2D Gaussian prior distribution specified in Eq.~\eqref{eq:musigma} (black dashed line).}
\end{figure*}

The shape and normalization of this GWB depend on the properties of the SMBHB population and on its dynamical evolution~\citep{Enoki+Nagashima-2007, Sesana:2008mz, Kocsis+Sesana-2011, Kelley:2017lek}. As discussed in~\citetalias{aaa+23_smbh}, the normalization is primarily controlled by the typical masses and abundance of SMBHBs, while the shape of the spectrum is determined by subparsec-scale binary evolution, which is currently unconstrained by observations. For a population of binaries whose orbital evolution is driven purely by GW emission, the resulting timing residual PSD is a power law with a spectral index (defined below in Eq.~\eqref{eq:BHB_signal}) of $-\gbhb = -13/3$~\citep{Phinney-2001}, produced by the increasing rate of inspiral and decreasing number of binaries emitting over each frequency interval. However, as GW emission alone is typically insufficient to merge SMBHBs within a Hubble time, the number of binaries emitting in the PTA band depends on interactions between binaries and their local galactic environment to extract orbital energy and drive systems toward merger~\citep{Begelman1980}. If these environmental effects extend into the PTA band, or if binary orbits are substantially eccentric, then the GWB spectrum can flatten at low frequencies~\citep[typically expected at $f \ll 1 \, {\rm yr}^{-1}$;][]{Kocsis+Sesana-2011}.  At high frequencies, once the expected number of binaries dominating the GWB  approaches unity, the spectrum steepens below $13/3$ \citep[typically expected at $f \gg 1 \, {\rm yr}^{-1}$;][]{Sesana:2008mz}.

Unfortunately, current observations and numerical simulations provide only weak constraints on the spectral amplitude or the specific locations and strengths of power-law deviations. Despite these uncertainties, the sensitivity range of PTAs is sufficiently narrowband that it is reasonable, to first approximation, to model the signal by a power law in this frequency range:
\begin{equation}\label{eq:BHB_signal}
	\Phi_{\scriptscriptstyle\rm BHB}(f)=\frac{\abhb^2}{12\pi^2}\frac{1}{T_{\rm obs}}\left(\frac{f}{{\rm yr}^{-1}}\right)^{-\gbhb}\,{\rm yr}^3\,,
\end{equation}
where $\Phi_{\scriptscriptstyle\rm BHB}/\Delta f$ is the timing residual PSD (see Eq.~\eqref{eq:OGW}). 

\begin{figure*}[htpb]
	\centering
        \includegraphics[width=\textwidth]{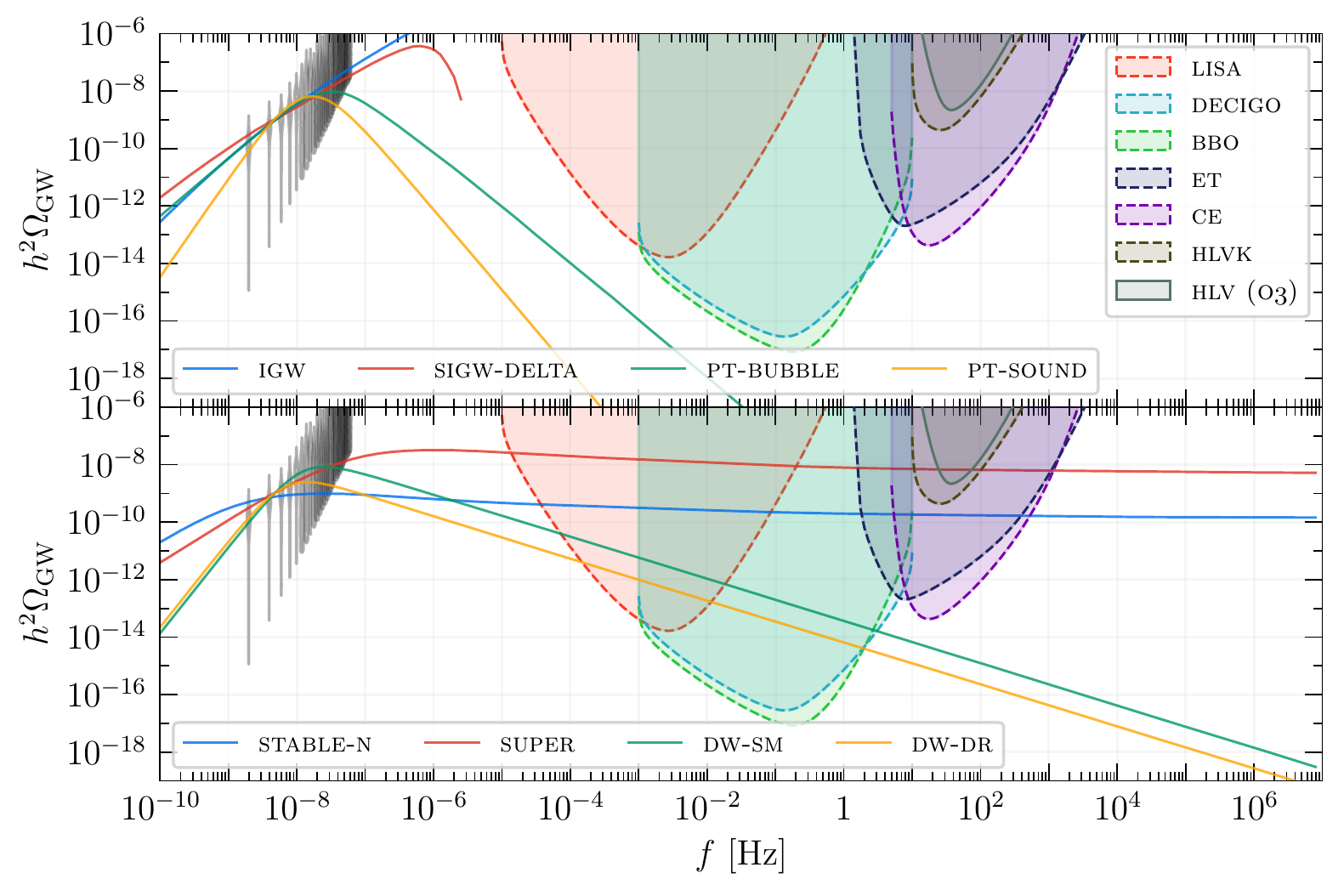}
	\caption{Same as Fig.~\ref{fig:mean_spectra} but for a different selection of models and showing a larger frequency range. The solid lines represent median GWB spectra for a subset of new-physics models (see Appendix~\ref{app:mediangw} for more details); the gray violins correspond to the posteriors of an HD-correlated free spectral reconstruction of the NANOGrav signal; and the shaded regions indicate the power-law-integrated sensitivity~\citep{Thrane:2013oya} of various existing and planned GW interferometer experiments: LISA~\citep{LISA:2017pwj}, DECIGO~\citep{Kawamura:2011zz}, BBO~\citep{Crowder:2005nr}, Einstein Telescope~\citep[ET;][]{Punturo:2010zz}, Cosmic Explorer \citep[CE;][]{Reitze:2019iox}, the HLVK detector network (consisting of aLIGO in Hanford and Livingston~\citep{LIGOScientific:2014pky}, aVirgo~\citep{VIRGO:2014yos}, and KAGRA~\citep{KAGRA:2018plz}) at design sensitivity, and the HLV detector network during the third observing run (O3).  All sensitivity curves are normalized to a signal-to-noise ratio of unity and, for planned experiments, an observing time of one year. For the HLV detector network, we use the O3 observing time. Different signal-to-noise thresholds $\rho_{\rm thr}$ and observing times $t_{\rm obs}$ can be easily implemented by rescaling the sensitivity curves by a factor of $\rho_{\rm thr}/\sqrt{t_{\rm obs}}$. More details on the construction of the sensitivity curves can be found in \cite{Schmitz:2020syl}. We emphasize that models whose median GWB spectrum exceeds the sensitivity of existing experiments are not automatically ruled out. This applies, e.g., to cosmic superstrings (\textsc{super}) and the O3 sensitivity of the HLV detector network. Typically, no single GWB spectrum in a given model will coincide with the median GWB spectrum, which is constructed from distributions of $h^2\Omega_{\scriptscriptstyle\rm GW}$ values at any given frequency. Therefore, if the median GWB spectrum is in conflict with existing bounds, typically only some regions in the model parameter space will be ruled out, while others remain viable (see, e.g., Fig.~\ref{fig:string_super_corner} for the \textsc{super} model). Finally, note that any primordial GWB signal is subject to the upper limit on the amount of dark radiation in Eq.~\eqref{eq:igwneff}, which requires the total integrated GW energy density to remain smaller than $\mathcal{O}(10^{-(5\cdots6)})$ (see Section~\ref{subsec:inflation}).\label{fig:landscape_spectra}}
\end{figure*}

Following~\cite{Middleton:2020asl}, we can gain some insight into the allowed range of values for the amplitude, $\abhb$, and slope, $\gbhb$, of this power law by simulating a large number of SMBHB populations covering the entire range of allowed astrophysical parameters. Specifically, we consider the SMBHB populations contained in the \texttt{GWOnly-Ext} library generated as part of the~\citetalias{aaa+23_smbh} analysis (and discussed in additional detail there).  This library was constructed with the \texttt{holodeck} package \citep{holodeck} using semianalytic models of SMBHB mergers.  These models use simple, parameterized forms of galaxy stellar mass functions, pair fractions, merger rates, and SMBH-mass~versus~galaxy-mass relations to produce binary populations and derived GWB spectra.  While some parameters in these models are fairly well known (e.g., concerning the galaxy stellar mass function), others are almost entirely unconstrained---particularly those governing the dynamical evolution of SMBHBs on subparsec scales~\citep{Begelman1980}. The \texttt{GWOnly-Ext} library assumes purely GW-driven binary evolution and uses relatively narrow distributions of model parameters based on literature constraints from galaxy-merger observations \cite[e.g.,][]{Tomczak:2014} in addition to more detailed numerical studies of SMBHB evolution \cite[e.g.,][]{Sesana-2013}.

For each population contained in the \texttt{GWOnly-Ext} library, we perform a power-law fit of the corresponding GWB spectrum across the first 14 frequency bins that we use in our analysis. The distribution for $\abhb$ and $\gbhb$ obtained in this way is reported in Fig.~\ref{fig:A_gamma_comparison} (blue contours) and compared to the results of a simple power-law fit to the GWB signal in the NG15 data set (green contours). The $95\%$ regions of the two distributions barely overlap, signaling a mild tension between the astrophysical prediction and the reconstructed spectral shape of the GWB. In view of this observation, we stress again that while these simulated populations are consistent with systematic investigations of the GWB spectrum~\citep[e.g.,][]{Sesana-2013}, they assume circular orbits and GW-only driven evolution. Adopting models that include either significant coupling between binaries and their local environments or very high eccentricities could serve to flatten the spectral shape and lead to SMBHB signals that better align with the observed data (see~\citetalias{aaa+23_smbh} for an extended discussion). Neither of these effects, however, is expected to significantly impact the amplitudes of the predicted spectra that, for expected values of astrophysical parameters, remain in mild tension with observed data. 
As discussed in \citetalias{aaa+23_smbh}, in order to reproduce the observed amplitude, SMBHB models require one or more of the astrophysical parameters describing the binaries' population to differ from expected values. For the present analysis, the spectra derived from the \texttt{GWOnly-Ext} library thus represent a convenient benchmark that is simple, well defined, and easy to use. By using theory-motivated priors, our reference model constitutes an important step toward a more realistic modeling of the GWB spectrum from inspiraling SMBHBs that goes beyond a power-law parameterization with spectral index $\gbhb = 13/3$, which has been the standard reference model in much of the PTA literature over the past decades.

The black dashed contours in Fig.~\ref{fig:A_gamma_comparison} show the results of a 2D Gaussian fit to the distribution of $\abhb$ and $\gbhb$ values derived from the simulated SMBHB populations (see Eq.~\eqref{eq:musigma} in Appendix~\ref{app:parameters} for the parameters of this Gaussian distribution). This fitted distribution is what we adopt as a prior distribution for $\abhb$ and $\gbhb$ in all parts of the analysis described in this paper.

\section{GWB signals from new physics}
\label{sec:gwb_new_physics}

In this section, we discuss the GWB produced by various new-physics models and investigate each model alone and in combination with the SMBHB signal as a possible explanation of the observed GWB signal. For each model, we give a brief review of the mechanism behind the GWB production and discuss the parametrization of its signal prediction. We report the reconstructed posterior distributions of the model parameters and compute the Bayes factors against the baseline SMBHB interpretation. In Fig.~\ref{fig:bf_table}, we show a summary of these Bayes factors; in Fig.~\ref{fig:mean_spectra}, we present median reconstructed GWB spectra in the PTA band for a number of select new-physics models; and in Fig.~\ref{fig:landscape_spectra}, we show similar median reconstructed GWB spectra in the broader landscape of present and future GW experiments.

As discussed in Section~\ref{sec:astro_signal} and in more detail in~\citetalias{aaa+23_smbh}, there is a mild tension between the NG15 data and the predictions of SMBHB models. The models generally prefer a weaker and less blue-tilted $h^2\Omega_{\scriptscriptstyle\rm GW}$ spectrum than the data. This discrepancy presents an opportunity for new-physics models to fit the data better than the conventional SMBHB signal. Eventually, this tension may grow to the point of giving strong evidence for new physics, or it may be resolved with better modeling and more data. Specifically, models of SMBHB evolution with a significant coupling between binaries and their local environment could lead to a signal that better aligns with the data and reduce the evidence for new physics. For all these reasons, we caution against over-interpreting the observed evidence in favor of some of the new-physics models discussed in the following sections. 

\subsection{Cosmic inflation}
\label{subsec:inflation}

\subsubsection*{Model description}

Cosmic inflation denotes a stage of exponential expansion in the early universe that provides an explanation for the initial conditions of Big Bang cosmology~\citep{Liddle:2000cg}. At the level of the background expansion, inflation accounts for the size, homogeneity, isotropy, and flatness of the observable universe on cosmological scales; at the level of perturbations, it  provides the seeds for structure formation in the form of primordial density fluctuations. In the standard scenario of inflation, these primordial perturbations are sourced by scalar quantum vacuum fluctuations of the spacetime metric and inflaton field, which are first stretched to superhorizon scales during inflation and then reenter the horizon in the form of classical density perturbations after inflation. In addition to scalar perturbations, inflation also leads to the amplification of tensor perturbations of the metric, which reenter the horizon in the form of stochastic GWs after inflation. These primordial or inflationary gravitational waves (IGWs)~\citep{Grishchuk:1974ny,Starobinsky:1979ty,Rubakov:1982df,Fabbri:1983us,Abbott:1984fp} represent a prime GW signal from the early universe. For earlier work on the IGW interpretation of the signal in recent PTA data sets, see~\cite{Vagnozzi:2020gtf}, \cite{Kuroyanagi:2020sfw}, and \cite{Benetti:2021uea}.

IGWs leave an imprint in the temperature and polarization anisotropies of the cosmic microwave background (CMB) whose relative strength compared to the contributions from scalar perturbations is quantified in terms of the tensor-to-scalar ratio, $r$. For the simplest type of inflation---standard single-field slow-roll inflation---the $h^2\Omega_{\scriptscriptstyle\rm GW}$ spectrum is red-tilted at CMB scales, with the tensor spectral index $n_t$ being given by the so-called consistency relation, $n_t = - r/8 < 0$. Meanwhile, $r$ is bounded from above by current CMB observations, $r \leq 0.036$ at $95\%\,\textrm{C.L.}$~\citep{BICEP:2021xfz}. A vanishing tensor spectral index, $n_t \approx 0$, would imply an upper bound on the GW energy density spectrum of $\Omega_{\scriptscriptstyle\rm GW}h^2 \sim 10^{-16}$ at PTA frequencies, rendering any detection of an IGW signal in PTA observations hopeless. This conclusion, however, only applies to the standard case of single-field slow-roll inflation. Nonminimal scenarios may have significantly better detection prospects.

We remain agnostic about the microphysics of inflation and restrict ourselves to a model-independent analysis, in which we parameterize the IGW signal in terms of four parameters: the tensor-to-scalar ratio $r$ and tensor spectral index $n_t$ at the CMB pivot scale, $f_{\scriptscriptstyle\rm CMB} = 0.05\,\textrm{Mpc}^{-1}/(2\pi a_0) \simeq 7.73\times10^{-17}\,\textrm{Hz}$, which quantify the efficiency and scale dependence of GW production during inflation, and the reheating temperature $T_{\rm rh}$ and the number of $e$-folds during reheating $N_{\rm rh}$, which describe the reheating process after inflation. Here the factor $a_0$ in the expression for $f_{\scriptscriptstyle\rm CMB}$ denotes the present value of the cosmological scale factor $a(t)$ in the Robertson--Walker metric; in our convention, $a_0 = 1$.

We do not impose the standard consistency relation between $r$ and $n_t$; instead, we allow both parameters to vary independently across large prior ranges, $\log_{10} r \in [-40,0]$ and $n_t \in [0,6]$. We note that blue values of the tensor spectral index can be generated, e.g., from axion--vector dynamics during inflation~\citep{Anber:2012du,Cook:2011hg,Namba:2015gja,Dimastrogiovanni:2016fuu,Caldwell:2017chz} or in other non-minimal inflation models (see~\cite{Piao:2004tq,Satoh:2008ck,Kobayashi:2010cm,Endlich:2012pz,Fujita:2018ehq} for an incomplete list).

Similarly, we allow for more flexibility for $T_{\rm rh}$ and $N_{\rm rh}$ than in the standard treatment of single-field slow-roll inflation. To illustrate this point, note that the number of $e$-folds during reheating, $N_{\rm rh}$, can be written as~\citep{Liddle:2000cg}
\begin{equation}
\label{eq:Nrh}
N_{\rm rh} = \frac{1}{3\left(1+w_{\rm rh}\right)} \,\ln\left(\frac{3H_{\rm end}^2M_{\scriptscriptstyle \rm Pl}^2}{\pi^2/30\,g_*^{\rm rh}\,T_{\rm rh}^4}\right) \,,
\end{equation}
where $H_{\rm end}$ is the Hubble rate at the end of inflation, $w_{\rm rh}$ is the equation-of-state parameter during reheating, $M_{\scriptscriptstyle \rm Pl} \simeq 2.44 \times 10^{18}\,\textrm{GeV}$ is the reduced Planck mass, and $g_*^{\rm rh}$ is defined below. We assume for definiteness that reheating is dominated by the coherent oscillations of the inflaton field, such that the equation of state is equivalent to the one of pressureless dust (i.e., matter), $w_{\rm rh} = 0$. In typical models of single-field slow-roll inflation, one can often approximate $H_{\rm end}$ by the Hubble rate at the time of CMB horizon exit during inflation, such that
\begin{equation}
\label{eq:Nnaive}
N_{\rm rh}^{\rm naive} \approx \frac{1}{3}\ln\left(\frac{3H_{\rm naive}^2M_{\scriptscriptstyle \rm Pl}^2}{\pi^2/30\,g_*^{\rm rh}\,T_{\rm rh}^4}\right) \,,
\end{equation}
where $H_{\rm naive}$ is fixed by the tensor-to-scalar ratio $r$ and the amplitude of the primordial scalar power spectrum, $A_s \simeq 2.10\times 10^{-9}$~\citep{Planck:2018vyg},
\begin{equation}
\label{eq:Hnaive}
H_{\rm end} \approx H_{\rm naive} = \left(\frac{\pi^2}{2} r A_s\right)^{1/2} M_{\scriptscriptstyle \rm Pl} \,.
\end{equation}
However, we already assume nonminimal dynamics in order to motivate a strongly blue-tilted $h^2\Omega_{\scriptscriptstyle\rm GW}$ spectrum, so there is no reason why we should make use of this approximation. In our analysis, we therefore treat $H_{\rm end}$ and correspondingly $N_{\rm rh}$ as independent parameters and do not fix them in terms of $r$ and $T_{\rm rh}$ as in Eqs.~\eqref{eq:Nnaive} and \eqref{eq:Hnaive}. This flexibility provides us with more parametric freedom that we can use in order to ensure that the IGW signal does not violate constraints on the amplitude of the stochastic GWB set by the LIGO--Virgo--KAGRA (LVK) Collaboration~\citep{KAGRA:2021kbb} and on the amount of dark radiation, i.e., the effective number of neutrino species, $N_{\rm eff}$, inferred from Big Bang nucleosynthesis (BBN) and the CMB~\citep{Pisanti:2020efz,Yeh:2020mgl}. As for the latter constraint, we specifically work with $\Delta N_{\rm eff} = \rho_{\scriptscriptstyle\rm DR}/\rho_\nu$, where $\rho_{\scriptscriptstyle\rm DR}$ is the energy density of dark radiation (i.e., the integrated GW energy density in the context of the \textsc{igw} model) and $\rho_\nu$ denotes the energy density of a single neutrino species. $\Delta N_{\rm eff}$ characterizes the excess energy in radiation beyond the SM expectation (i.e., dark radiation) after neutrino decoupling and $e^+e^-$ annihilation, $N_{\rm eff} = N_{\rm eff}^{\scriptscriptstyle\rm SM} + \Delta N_{\rm eff}$, where $N_{\rm eff}^{\scriptscriptstyle\rm SM} \simeq 3.0440$~\citep{Bennett:2020zkv}.

Under the assumptions outlined above, we are able to model the IGW spectrum at PTA frequencies as
\begin{equation}
\label{eq:Ogwinf}
\Omega_{\scriptscriptstyle\rm GW}^{\rm inf}\left(f\right) = \frac{\Omega_{\rm r}}{24} \left(\frac{g_*\left(f\right)}{g_*^0}\right)\bigg(\frac{g_{*,s}^0}{g_{*,s}\left(f\right)}\bigg)^{4/3} \hspace{-0.25em}\mathcal{P}_t\left(f\right) \mathcal{T}\left(f\right) \,.
\end{equation}
Here $\Omega_{\rm r}/g_*^0 \simeq 2.72 \times 10^{-5}$ is the current radiation energy density per relativistic degree of freedom, in units of the critical (closure) density, $g_{*,s}^0 \simeq 3.93$ counts the effective number of relativistic degrees of freedom contributing to the radiation entropy today, and $g_*(f)$ and $g_{*,s}(f)$ denote the effective numbers of relativistic degrees of freedom in the early universe when GWs with comoving wavenumber $k = 2\pi a_0 f$ reentered the Hubble horizon after inflation. In order to evaluate $g_*(f)$ and $g_{*,s}(f)$, we use the numbers of relativistic degrees of freedom as functions of temperature tabulated in~\cite{Saikawa:2020swg}, $g_*(T)$ and $g_{*,s}(T)$, in combination with the standard temperature--frequency relation in $\Lambda$CDM (i.e., the cosmological Lambda cold dark matter standard model) that follows from the condition $k=a(T)H(T)$ at the time of horizon reentry. In the remainder of this paper, whenever we need $g_*$ or $g_{*,s}$ in a different part of our analysis, we will use the same functions $g_*(f)$, $g_{*,s}(f)$, $g_*(T)$, and $g_{*,s}(T)$.

The inflationary dynamics give rise to the primordial tensor power spectrum $\mathcal{P}_t$, while the transfer function $\mathcal{T}$ accounts for the redshifting behavior of GWs after horizon reentry. In our analysis, we assume a constant tensor spectral tilt (i.e., zero running of $n_t$) from CMB to PTA frequencies, such that
\begin{equation}
\label{eq:Ptensor}
\mathcal{P}_t\left(f\right) = r\,A_s \left(\frac{f}{f_{\scriptscriptstyle\rm CMB}}\right)^{n_t} \,.
\end{equation}
Meanwhile, the only relevant contribution to $\mathcal{T}$ in the PTA band corresponds to the transfer function that connects the radiation-dominated era to reheating,
\begin{equation}
\label{eq:transfer}
\mathcal{T}\left(f\right) \approx \frac{\Theta\left(f_{\rm end} - f\right)}{1-0.22\left(f/f_{\rm rh}\right)^{1.5} + 0.65\left(f/f_{\rm rh}\right)^2} \,.
\end{equation}
Here the fit function in the denominator of this expression is taken from~\cite{Kuroyanagi:2014nba,Kuroyanagi:2020sfw} and describes the spectral turnover, $f^{n_t} \rightarrow f^{n_t-2}$, at frequencies around $f \sim f_{\rm rh}$, which marks the end of the reheating period,
\begin{equation}
\label{eq:frh}
f_{\rm rh} = \frac{1}{2\pi}\: \bigg(\frac{g_{*,s}^0}{g_{*,s}^{\rm rh}}\bigg)^{1/3}\left(\frac{\pi^2 g_*^{\rm rh}}{90}\right)^{1/2} \frac{T_{\rm rh}T_0}{M_{\scriptscriptstyle \rm Pl}} \,,
\end{equation}
with $g_*^{\rm rh} = g_*(T_{\rm rh})$ and $g_{*,s}^{\rm rh} = g_{*,s}(T_{\rm rh})$ and the present-day CMB temperature $T_0 \simeq 2.73\,\textrm{K}$~\citep{Fixsen:2009ug}.
The Heaviside theta function in Eq.~\eqref{eq:transfer} denotes the endpoint of the IGW spectrum at $f = f_{\rm end}$, which marks the end of inflation and hence the onset of reheating,
\begin{equation}
\label{eq:fend}
f_{\rm end} = \frac{1}{2\pi}\: \bigg(\frac{g_{*,s}^0}{g_{*,s}^{\rm rh}}\bigg)^{1/3}\left(\frac{\pi^2 g_*^{\rm rh}}{90}\right)^{1/3} \frac{T_{\rm rh}^{1/3} H_{\rm end}^{1/3}\,T_0}{M_{\scriptscriptstyle \rm Pl}^{2/3}} \,.
\end{equation}

For fixed $T_{\rm rh}$, the frequency $f_{\rm end}$ is solely controlled by $H_{\rm end}$, which follows from our choice of $N_{\rm rh}$ according to Eq.~\eqref{eq:Nrh} (recall that we set $w_{\rm rh} = 0$).
In our MCMC analysis, we do not sample over $f_{\rm end}$, since its precise value does not affect the shape of the IGW spectrum in the PTA band and hence the quality of our fit. 
Instead, we constrain its maximally allowed value (i.e., $N_{\rm rh}$) after identifying the viable region in the $r$\,--\,$n_t$\,--\,$T_{\rm rh}$ parameter space, in order to make sure that the IGW spectrum does not violate the $N_{\rm eff}$ and LVK bounds.
The frequency $f_{\rm rh}$, on the other hand, can easily fall into the PTA band: from Eq.~\eqref{eq:frh}, we have $f_{\rm rh} \sim 30\,\textrm{nHz}\left(T_{\rm rh}/1\,\textrm{GeV}\right)$.
Therefore, we sample the reheating temperature in a symmetric interval around $T_{\rm rh} = 1\,\textrm{GeV}$ extending down to temperatures relevant for BBN, $T_{\scriptscriptstyle\rm BBN} \sim 1\,\textrm{MeV}$.
That is, we work with the log-uniform prior $\log_{10}\left(T_{\rm rh}/1\,\textrm{GeV}\right) \in \left[-3,+3\right]$.

\subsubsection*{Results and discussion}

We now discuss the outcome of our Bayesian fit analysis. First, we note that the \textsc{igw} model fits the NG15 data slightly better than the baseline \textsc{smbhb} model. This is evident from the Bayes factor that we find for the \textsc{igw} versus \textsc{smbhb} model comparison, $\mathcal{B} = 8.8 \pm 0.3$ (mean value and one standard deviation), and simply follows from the larger parametric freedom of the \textsc{igw} model. Both the \textsc{igw} and \textsc{smbhb} models basically correspond to a power-law approximation of the GW spectrum. However, in the case of the \textsc{igw} model, the parameters controlling this power law are drawn from prior distributions that allow for a larger amplitude and a steeper slope of the spectrum, which improves the quality of the fit. Meanwhile, the combined GW spectrum from inflation with an additional SMBHB signal on top compared to the \textsc{smbhb} model alone results in a Bayes factor of $\mathcal{B} = 7.9 \pm 0.3$. We thus observe a slight decrease in the Bayes factor, which accounts for the fact that adding the SMBHB signal on top of the IGW signal does not improve the quality of fit but merely increases the prior volume compared to the \textsc{igw} model.

\begin{figure}
\centering
\includegraphics[width=0.48\textwidth]{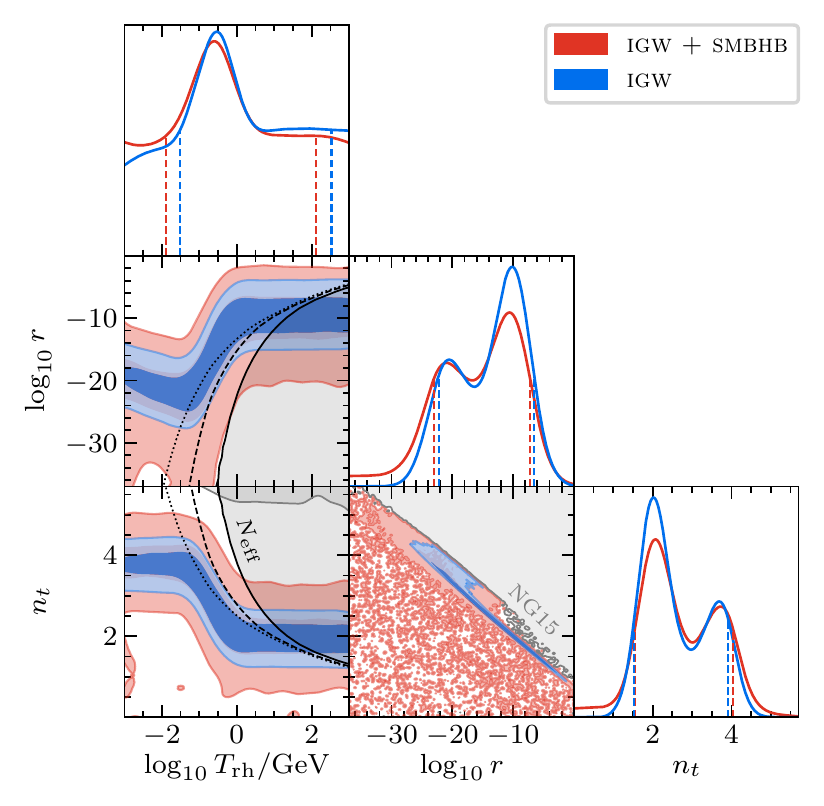}
\caption{Reconstructed posterior distributions for the parameters of the \textsc{igw} (blue) and \textsc{igw$+$smbhb} (red) models. On the diagonal of the corner plot, we report the 1D marginalized distributions together with the $68\%$ Bayesian credible intervals (vertical lines), while the off-diagonal panels show the $68\%$ (darker) and $95\%$ (lighter) Bayesian credible regions in the 2D posterior distributions. We construct all credible intervals and regions by integrating over the regions of highest posterior density. The gray-shaded regions are strongly disfavored by the NG15 data as they result in a $K$ ratio of less than $\sfrac{1}{10}$ [see Eq.~\eqref{eq:Kratio}]. The black-shaded region results in a violation of the $N_{\rm eff}$ bound in Eq.~\eqref{eq:igwneff} (see Appendix~\ref{app:inflation} and Fig.~\ref{fig:nrhmax}), assuming $N_{\rm rh} = 0$ (solid line), $N_{\rm rh} = 5$ (dashed line), $N_{\rm rh} = 10$ (dotted line). Fig.~\ref{fig:igw_ext_corner} in Appendix~\ref{app:inflation} shows an extended version of this plot that includes the SMBHB parameters $\abhb$ and $\gbhb$.}
\label{fig:igw_corner}
\end{figure}

The reconstructed posterior distributions for the parameters of the \textsc{igw} model and its \textsc{igw$+$smbhb} extension are shown in Figs.~\ref{fig:igw_corner}.\footnote{The noise in the 95\% credible interval for the $n_t$-$\log_{10}r$ posterior distribution of the \textsc{igw+smbhb} model is due to the presence of an extended plateau region in the posterior distribution, which renders the kernel density reconstruction sensitive to Poisson fluctuations in the binned MCMC data.} For both models, we find a strong covariance between the spectral index $n_t$ and the tensor-to-scalar ratio $r$, which is approximately fit by 
\begin{equation}
\label{eq:ntr}
n_t = -0.14 \log_{10}r + 0.58 \,,
\end{equation}
and which can be explained as follows: the \textsc{igw} interpretation of the PTA signal requires the primordial tensor power spectrum $\mathcal{P}_t$ to take values of $\mathcal{O}\left(10^{-4}\right)$ at nHz frequencies. This requirement fixes the parameter combination $r \left(f_{\scriptscriptstyle\rm PTA}/f_{\scriptscriptstyle\rm CMB}\right)^{n_t}$ in Eq.~\eqref{eq:Ptensor} and thus allows us to estimate the coefficients in Eq.~\eqref{eq:ntr} as $1/\log_{10}\left(f_{\scriptscriptstyle\rm PTA}/f_{\scriptscriptstyle\rm CMB}\right) \approx 0.14$ and $\log_{10}\left(\mathcal{P}_t/A_s\right)/\log_{10}\left(f_{\scriptscriptstyle\rm PTA}/f_{\scriptscriptstyle\rm CMB}\right) \approx 0.58$, respectively, where we used $f_{\scriptscriptstyle\rm PTA} = 1\,\textrm{nHz}$ and $\mathcal{P}_t = 0.3 \times 10^{-4}$.

In addition to the strong covariance between $n_t$ and $r$, we note that the posterior probabilities of both parameters exhibit a bimodal distribution for both \textsc{igw} and \textsc{igw$+$smbhb}. In the 2D distributions of the parameter pairs $\left(T_{\rm rh}, n_t\right)$ and $\left(T_{\rm rh}, r\right)$, this bimodality is accompanied by an approximate reflection symmetry with respect to the points $\left(\log_{10}T_{\rm rh}/\textrm{GeV}, n_t\right) \sim \left(-0.5,2.75\right)$ and $\left(\log_{10}T_{\rm rh}/\textrm{GeV}, \log_{10} r\right) \sim \left(-0.5,-15\right)$, respectively. These features of the corner plot in Fig.~\ref{fig:igw_corner} indicate that the \textsc{igw} model can operate in two regimes: for $T_{\rm rh} \gg 1\,\textrm{GeV}$, the reference frequency $f_{\rm rh}$ is larger than the frequencies in the PTA band, and the GW spectrum seen by NANOGrav is composed of tensor modes that re-entered the horizon during the radiation-dominated era. For $T_{\rm rh} \ll 1\,\textrm{GeV}$, on the other hand, $f_{\rm rh}$ can be pushed below PTA frequencies, and the GW spectrum in the PTA band is composed of tensor modes that re-entered the horizon during reheating after inflation. In the first case, the tilt of the spectrum is directly given by $n_t$; in the second case, it corresponds to $n_t-2$. Clearly, the mirror symmetry in the 2D distributions of $\left(T_{\rm rh}, n_t\right)$ and $\left(T_{\rm rh}, r\right)$ is not exact. At the level of the GW spectrum, it is explicitly broken by the frequency dependence of $g_*$ and $g_{*,s}$ as well as by the nontrivial shape of the transfer function in Eq.~\eqref{eq:transfer}.

At small or large values of the reheating temperature, the posterior distributions develop approximately flat directions along the $T_{\rm rh}$ axis at $\left(n_t,\log_{10}r\right) \sim \left(2,-10\right)$ in the large-$T_{\rm rh}$ regime and at $\left(n_t,\log_{10}r\right) \sim \left(4,-20\right)$ in the small-$T_{\rm rh}$ regime. This behavior is broadly consistent with the linear fit in Eq.~\eqref{eq:ntr}, the reflection symmetry discussed above, and the fact that, past a certain point, raising or lowering $T_{\rm rh}$ no longer influences the shape of the GW signal in the PTA band. A tensor index of $n_t = 2$ at large $T_{\rm rh}$, moreover, corresponds to an index $\gamma = 3$ in the timing-residual PSD, which is among the best-fitting values---see the 2D posterior for $A$ and $\gamma$ in Fig.~\ref{fig:A_gamma_comparison}. The same conclusion holds at low $T_{\rm rh}$, where $\gamma = 3$ is mapped onto $n_t - 2 = 2$.

Finally, we derive the $N_{\rm eff}$ and LVK bounds on the \textsc{igw} parameter space. For the $N_{\rm eff}$ bound, the GW spectrum, integrated from BBN scales to the cutoff frequency $f_{\rm end}$, must not exceed a certain upper limit that is set by the allowed amount of extra relativistic degrees of freedom at the time of BBN and recombination,
\begin{equation}
\label{eq:igwneff}
\int_{f_{\scriptscriptstyle\rm BBN}}^{f_{\rm end}} \frac{\textrm{d}f}{f} \: h^2 \Omega_{\scriptscriptstyle\rm GW}^{\rm inf}\left(f\right) \lesssim 5.6 \times 10^{-6}\,\Delta N_{\rm eff}^{\rm max} \,.
\end{equation}
Here, $f_{\scriptscriptstyle\rm BBN} \sim 10^{-12}\,\textrm{Hz}$ refers to tensor modes that re-entered the Hubble horizon around the onset of BBN at $T \sim 10^{-4}\,\textrm{GeV}$~\citep{Caprini:2018mtu}, and $\Delta N_{\rm eff}^{\rm max} \sim \textrm{few} \times 0.1$ denotes the upper limit on the amount of dark radiation. For definiteness, we set $f_{\scriptscriptstyle\rm BBN} = 10^{-12}\,\textrm{Hz}$ and $\Delta N_{\rm eff}^{\rm max} = 0.5$ in our analysis; the precise $\Delta N_{\rm eff}^{\rm max}$ value at $95\%$ confidence level varies across different combinations of data sets~\citep{Pisanti:2020efz,Yeh:2020mgl}. For given values of the parameters $T_{\rm rh}$, $r$, and $n_t$, we can then ask if there is a cutoff frequency $f_{\rm end}^{\rm max}$ that leads to the saturation of the inequality in Eq.~\eqref{eq:igwneff}. If this is the case, $f_{\rm end}^{\rm max}$ yields an upper bound on the allowed number of $e$-folds during reheating, $N_{\rm rh}^{\rm max}$, according to Eqs.~\eqref{eq:Nrh} and \eqref{eq:fend}. A second constraint on $N_{\rm rh}$ follows from the LVK bound on the amplitude of the GWB~\citep{KAGRA:2021kbb},
\begin{equation}
\label{eq:LVKbound}
\Omega_{\scriptscriptstyle\rm GW} \leq 1.7 \times 10^{-8}  \quad \textrm{at} \quad  f_{\rm lvk} = 25\,\textrm{Hz} \,,
\end{equation}
assuming a flat GW spectrum. For a power-law spectrum, $\Omega_{\scriptscriptstyle\rm GW} = \Omega_{{\scriptscriptstyle\rm GW}, \alpha} \left(f/f_{\rm lvk}\right)^\alpha$ with $\alpha = n_t - 2$ at $f \gg f_{\rm rh}$, this bound can be generalized following~\cite{Kuroyanagi:2014nba,Kuroyanagi:2020sfw},
\begin{equation}
\label{eq:LVKboundalpha}
\Omega_{\scriptscriptstyle\rm GW} \leq 1.7 \times 10^{-8} \left(\frac{5-2\alpha}{5}\right)^{1/2} \left(\frac{20\,\textrm{Hz}}{f_{\rm lvk}}\right)^{-\alpha} \,.
\end{equation}
which approximately holds if $\alpha \ll 5/2$. For given values of $T_{\rm rh}$, $r$, and $n_t$, we can then evaluate $\Omega_{\scriptscriptstyle\rm GW}^{\rm inf}$ at $f_{\rm lvk}$ and check whether it exceeds the LVK bound. If so, we place an upper bound on $N_{\rm rh}$ by demanding that $f_{\rm end}^{\rm max} = 20\,\textrm{Hz}$ (the lower end of the LVK frequency band).

The outcome of our analysis is shown in Fig.~\ref{fig:nrhmax} in Appendix~\ref{app:inflation}. In both plots of this figure, the dependence on the tensor-to-scalar ratio $r$ is eliminated by means of the linear relation in Eq.~\eqref{eq:ntr}. In the left panel of Fig.~\ref{fig:nrhmax}, we present contour lines of $N_{\rm rh}^{\rm max}$ derived from the $N_{\rm eff}$ bound and the LVK bound, respectively. In a realization of the \textsc{igw} model with a given duration of reheating, these contour lines can be thought of as bounds on the $T_{\rm rh}$\,--\,$n_t$ parameter plane---for fixed $N_{\rm rh}$, the regions with $N_{\rm rh}^{\rm max} < N_{\rm rh}$ are excluded. We find that the $N_{\rm eff}$ bound rules out most values of the spectral index $n_t$ in the large-$T_{\rm rh}$ regime. At the same time, large regions of parameter space remain viable as long as $N_{\rm rh} < N_{\rm rh}^{\rm max}$. In fact, away from the region where $N_{\rm rh}^{\rm max}$ turns negative, our upper bound is typically rather large, $N_{\rm rh}^{\rm max} \sim \mathcal{O}\left(10\cdots100\right)$, and hence easy to satisfy in realistic models of reheating, where $N_{\rm rh}\sim \mathcal{O}\left(1\cdots10\right)$. In the right panel of Fig.~\ref{fig:nrhmax}, we compare our result to the naive expectation $N_{\rm rh}^{\rm naive}$ in single-field slow-roll inflation with a nearly constant Hubble rate [see Eq.~\eqref{eq:Nnaive}]. Across large parts of parameter space, we find that $N_{\rm rh}^{\rm naive}$ assumes unrealistically large values, $N_{\rm rh}^{\rm naive} \gg 10$.

In Fig.~\ref{fig:igw_corner}, we highlight the constraints on $T_{\rm rh}$ and $n_t$ (as well as $T_{\rm rh}$ and $r$) that we deduce from the $N_{\rm eff}$ bound assuming $N_{\rm rh}$ values of $N_{\rm rh} = 0$, $5$, and $10$. Here, the constraints for $N_{\rm rh} = 0$ correspond to the assumption of instantaneous reheating after inflation and hence represent the most conservative bound on the $T_{\rm rh}$\,--\,$n_t$ parameter plane. A longer duration of reheating results in tighter constraints on $T_{\rm rh}$ and $n_t$, as illustrated by the contours for $N_{\rm rh} = 5$ and $10$. For an even larger number of $e$-folds during reheating, see Fig.~\ref{fig:nrhmax} in Appendix~\ref{app:inflation}. 

In view of Figs.~\ref{fig:igw_corner} and \ref{fig:nrhmax}, we conclude that the \textsc{igw} model is indeed capable of fitting the NANOGrav signal across large regions in parameter space. An interesting viable scenario consists, e.g., of a large tensor spectral index, $n_t \sim 3\cdots 4$, a tiny tensor-to-scalar ratio, $r \sim 10^{-(23\cdots16)}$, a low reheating temperature, $T_{\rm rh} \sim 10^{-(3\cdots0)}\,\textrm{GeV}$, and a moderate number of $e$-folds during reheating, $N_{\rm rh} \lesssim 20$. It remains to be seen whether it is possible to identify explicit microscopic models that realize inflation in this parametric regime.

\subsection{Scalar-induced gravitational waves}
\label{subsec:sigw}

\subsubsection*{Model description}

The amplitude of the primordial scalar power spectrum is well measured by CMB observations, $A_s \simeq 2.10 \times 10^{-9}$ at the CMB pivot scale $k_{\scriptscriptstyle\rm CMB} = 0.05\,\textrm{Mpc}^{-1}$~\citep{Planck:2018vyg}. If we naively extrapolate this value down to smaller scales, assuming a fixed and slightly red-tilted $h^2\Omega_{\scriptscriptstyle\rm GW}$ spectrum with index $n_s \sim 0.96$, we are led to conclude that there must be increasingly less power in scalar perturbations on shorter scales. This conclusion can, however, be easily avoided in models that deviate from the standard picture of single-field slow-roll inflation giving rise to a nearly scale-invariant spectrum of scalar perturbations. A prominent example, among many other mechanisms, consists in a stage of inflation close to an inflection point in the scalar potential, which readily amplifies the scalar perturbations leaving the horizon (see, e.g.,~\cite{Garcia-Bellido:2017mdw,Ezquiaga:2017fvi,Ballesteros:2017fsr}). An enhanced scalar power spectrum at small scales is, therefore, a viable possibility. Moreover, it promises a rich phenomenology with regard to the production of GWs and potentially the origin of primordial black holes (PBHs)~\citep{Carr:2016drx,Garcia-Bellido:2016dkw,Inomata:2016rbd,Inomata:2018epa,Wang:2019kaf,Escriva:2022duf}. The possibility of having PBH formation in models of single-field inflation is the subject of ongoing debate~\citep{Kristiano:2022maq,Riotto:2023hoz,Choudhury:2023vuj,Choudhury:2023jlt,Kristiano:2023scm,Riotto:2023gpm,Choudhury:2023rks,Firouzjahi:2023ahg}. Below, we comment on the implications of this debate for our PBH-related parameter bounds.

In cosmological perturbation theory, scalar and tensor perturbations evolve independently at linear order. Starting at second order, however, they are coupled, which means that large first-order scalar perturbations can act as a source of second-order tensor perturbations. We refer to these tensor perturbations, which are produced as soon as the enhanced scalar perturbations reenter the Hubble horizon after inflation, as scalar-induced gravitational waves (SIGWs)~\citep{Matarrese:1992rp,Matarrese:1993zf,Matarrese:1997ay,Mollerach:2003nq,Ananda:2006af,Baumann:2007zm} (see~\cite{Domenech:2021ztg} for a review and more details on the history of SIGWs). At the same time, large overdensities in the tail of the distribution of scalar perturbations can collapse into PBHs upon horizon reentry. This PBH production mechanism thus results in a PBH population whose properties are strongly correlated with the spectral shape of the SIGW signal, as both phenomena are controlled by the scalar spectrum generated during inflation~\citep{Yuan:2021qgz}.
For earlier works on the PBH/SIGW interpretation of the signal in recent PTA data sets, see~\cite{Vaskonen:2020lbd,DeLuca:2020agl,Kohri:2020qqd}. For earlier Bayesian searches for an SIGW signal in PTA data sets, see~\cite{Chen:2019xse,Bian:2020urb,Zhao:2022kvz,Yi:2022ymw,Dandoy:2023jot}, and for a search in LVK data, see~\cite{Romero-Rodriguez:2021aws}.

In our analysis, we consider SIGWs in the PTA band and model the associated GW spectrum as follows:
\begin{equation}
\Omega_{\scriptscriptstyle\rm GW}^{\rm ind}\left(f\right) = \Omega_{\rm r} \left(\frac{g_*\left(f\right)}{g_*^0}\right)\bigg(\frac{g_{*,s}^0}{g_{*,s}\left(f\right)}\bigg)^{4/3} \bar{\Omega}_{\scriptscriptstyle\rm GW}^{\rm ind}\left(f\right) \,,
\end{equation}
where the first three factors account for the cosmological redshift as in Eq.~\eqref{eq:Ogwinf}, and where the last factor denotes the GW spectrum at the time of production, which we assume to be during the radiation-dominated era,
\begin{equation}
\label{eq:Osigw}
\bar{\Omega}_{\scriptscriptstyle\rm GW}^{\rm ind}\left(f\right) = \int_0^\infty \textrm{d}v \int_{\left|1-v\right|}^{1+v} \textrm{d}u\: \mathcal{K}\left(u,v\right) \mathcal{P}_\mathcal{\scriptscriptstyle R}\left(uk\right) \mathcal{P}_\mathcal{\scriptscriptstyle R}\left(vk\right) \,.
\end{equation}
The present-day GW frequency is related to the comoving wavenumber by $f = k/\left(2\pi a_0\right)$, $\mathcal{P}_{\mathcal{\scriptscriptstyle R}}$ denotes the primordial spectrum of the comoving curvature perturbation $\mathcal{R}$, and the integration kernel $\mathcal{K}$ is given by~\citep{Espinosa:2018eve,Kohri:2018awv,Pi:2020otn,Gong:2019mui}
\begin{widetext}
\begin{equation}
\mathcal{K}\left(u,v\right) = \frac{3\left(4v^2-(1+v^2-u^2)^2\right)^2\left(u^2+v^2-3\right)^4}{1024\,u^8 v^8}\left[\left(\ln\left|\frac{3-(u+v)^2}{3-(u-v)^2}\right| - \frac{4uv}{u^2+v^2-3}\right)^2 + \pi^2 \Theta(u+v-\sqrt{3}) \right] \,.
\end{equation}
\end{widetext}
The expression in Eq.~\eqref{eq:Osigw} illustrates the dependence of the SIGW signal on the scalar input spectrum; in particular, it shows that $\Omega_{\scriptscriptstyle\rm GW}^{\rm ind}$ scales as $\Omega_{\scriptscriptstyle\rm GW}^{\rm ind} \propto \mathcal{P}_{\mathcal{\scriptscriptstyle R}}^2$. We stress that the expression in Eq.~\eqref{eq:Osigw} assumes Gaussian perturbations, whose statistics are fully described by the power spectrum $\mathcal{P}_{\mathcal{\scriptscriptstyle R}}$. We do not consider any non-Gaussian contributions to the primordial scalar power spectrum in our analysis. The impact of non-Gaussianities on the SIGW signal was studied in~\cite{Nakama:2016gzw,Cai:2018dig,Unal:2018yaa,Atal:2019cdz,Yuan:2020iwf,Atal:2021jyo,Adshead:2021hnm,Ferrante:2022mui}.

To remain as model-independent as possible, we refrain from choosing a particular inflation model capable of generating an enhanced spectrum $\mathcal{P}_{\mathcal{\scriptscriptstyle R}}$. Instead, we ignore the microphysics of inflation and work with three characteristic templates for $\mathcal{P}_{\mathcal{\scriptscriptstyle R}}$ that reflect the range of possibilities that one may expect in realistic models. Specifically, we consider the following templates:\medskip\\
\textsc{sigw-delta}: Sharp spectral feature in $\mathcal{P}_{\mathcal{\scriptscriptstyle R}}$ modeled by a Dirac \textit{delta} function in logarithmic $k$ space,
\begin{equation}
\label{eq:PRdelta}
\mathcal{P}_{\mathcal{\scriptscriptstyle R}}\left(k\right) = A\,\delta\left(\ln k-\ln k_*\right) \,.
\end{equation}
\textsc{sigw-gauss}: Broad spectral feature in $\mathcal{P}_{\mathcal{\scriptscriptstyle R}}$ modeled by a \textit{Gaussian} peak in logarithmic $k$ space,
\begin{equation}
\label{eq:PRgauss}
\mathcal{P}_{\mathcal{\scriptscriptstyle R}}\left(k\right) = \frac{A}{\sqrt{2\pi}\,\Delta}\,\exp\left[-\frac{1}{2}\left(\frac{\ln k -\ln k_*}{\Delta}\right)^2\right] \,.
\end{equation}
\textsc{sigw-box}: Flat and continuous spectral feature in $\mathcal{P}_{\mathcal{\scriptscriptstyle R}}$ modeled by a \textit{box} function in logarithmic $k$ space,
\begin{equation}
\label{eq:PRbox}
\mathcal{P}_{\mathcal{\scriptscriptstyle R}}\left(k\right) = A\,\Theta\left(\ln k_{\rm max}- \ln k\right)\Theta\left(\ln k- \ln k_{\rm min}\right) \,.
\end{equation}

Note that the Gaussian power spectrum in logarithmic $k$ space that we assume in the \textsc{sigw-gauss} model corresponds to a lognormal power spectrum in linear $k$ space. As evident from the above expressions, \textsc{sigw-delta} represents a two-parameter model, while \textsc{sigw-gauss} and \textsc{sigw-box} are three-parameter models. Our prior choices for the respective parameters are listed in Table~\ref{tab:np_priors} in Appendix~\ref{app:parameters}, where we use again $f = k/\left(2\pi a_0\right)$ to convert from wavenumber to frequency. For a given set of parameter values, we are then able to use the scalar power spectrum in Eq.~\eqref{eq:PRdelta}, Eq.~\eqref{eq:PRgauss}, or Eq.~\eqref{eq:PRbox} to evaluate the integrals in Eq.~\eqref{eq:Osigw} and compute the GW spectrum. For \textsc{sigw-gauss} and \textsc{sigw-box}, the integration needs to be carried out numerically; for \textsc{sigw-delta}, we can resort to the exact analytical expression provided in~\cite{Wang:2019kaf,Yuan:2021qgz}.

\subsubsection*{Results and discussion}

We now turn to the outcome of our Bayesian fit analysis. Compared to the \textsc{igw} model discussed in Section~\ref{subsec:inflation}, we obtain even larger Bayes factors, indicating that SIGWs tend to provide an even better fit of the NG15 data than IGWs. Specifically, we obtain $\mathcal{B} = 44 \pm 2$, $\mathcal{B} = 57 \pm 3$, and $\mathcal{B} = 21 \pm 1$ for \textsc{sigw-delta}, \textsc{sigw-gauss}, and \textsc{sigw-box}, respectively, and $\mathcal{B} = 38 \pm 2$, $\mathcal{B} = 59 \pm 3$, and $\mathcal{B} = 23 \pm 1$ for \textsc{sigw-delta$+$smbhb}, \textsc{sigw-gauss$+$smbhb}, and \textsc{sigw-box$+$smbhb}, respectively (see Fig.~\ref{fig:bf_table}). This improvement in the quality of the fit reflects the fact that the SIGW spectra deviate from a pure power law and thus manage to provide a better fit across the whole frequency range probed by NANOGrav (see Fig.~\ref{fig:mean_spectra}). For \textsc{sigw-delta}, we observe again that adding the SMBHB signal does not improve the quality of the fit. The larger prior volume of \textsc{sigw-delta$+$smbhb} compared to \textsc{sigw-delta} therefore results in a slight decrease of the Bayes factor. For the other two SIGW models, the Bayes factors remain more or less the same, within the statistical uncertaintity of our bootstrapping analysis.

\begin{figure}
\centering
\includegraphics[width=0.48\textwidth]{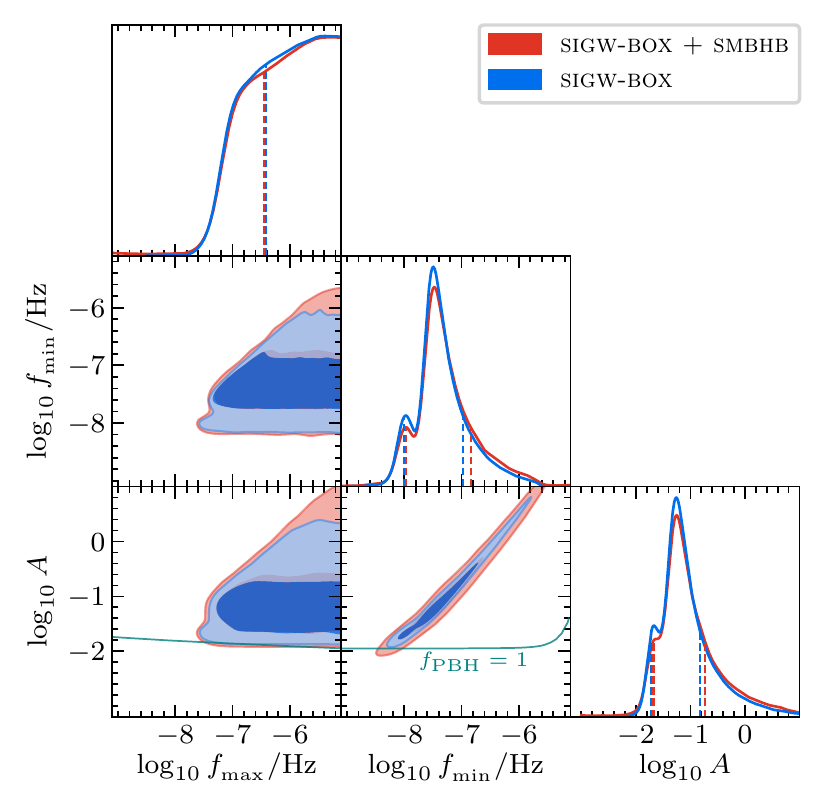}
\caption{Same as in Fig.~\ref{fig:igw_corner} but for the \textsc{sigw-box} model. The regions \textit{above} the teal contour lines labeled $f_{\scriptscriptstyle \rm PBH} = 1$ lead to the overproduction of PBHs, according to our analysis in Appendix~\ref{app:sigw}; however, see text for more discussion. Fig.~\ref{fig:sigw_box_ext_corner} in Appendix~\ref{app:sigw} shows an extended version of this plot that includes the SMBHB parameters $\abhb$ and $\gbhb$.
\label{fig:sigw_box_corner}}
\end{figure}

\begin{figure*}
\centering
\includegraphics[width=0.48\textwidth]{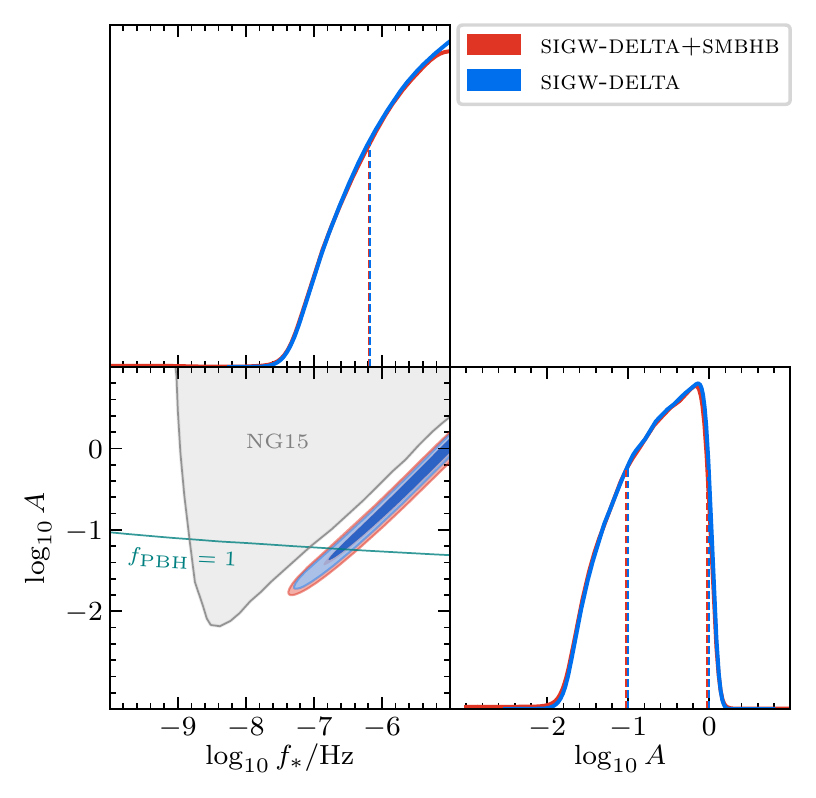}
\includegraphics[width=0.48\textwidth]{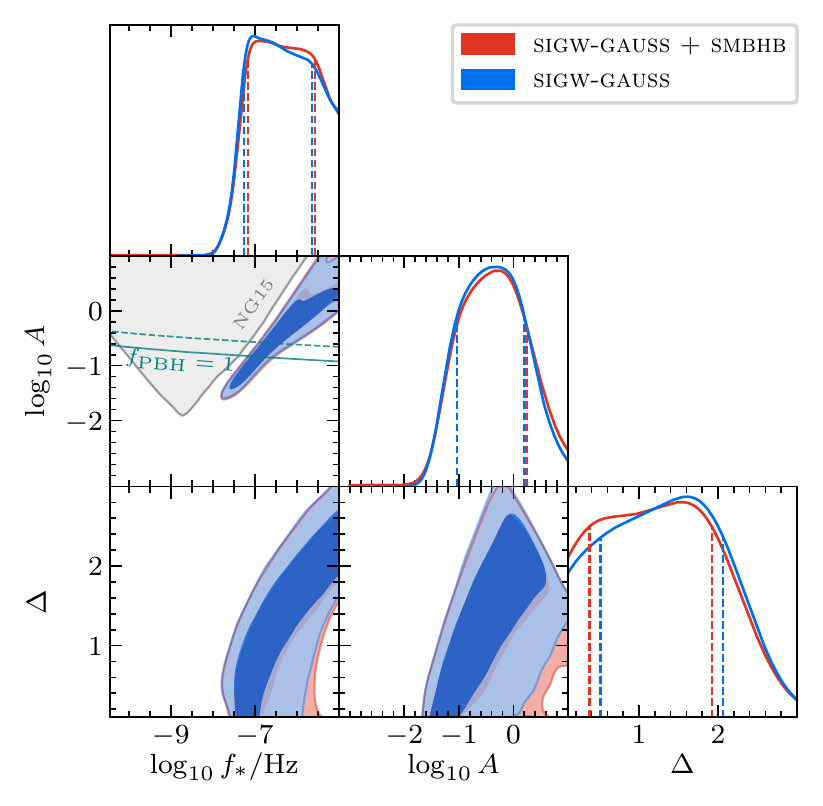}
\caption{Same as in Fig.~\ref{fig:sigw_box_corner} but for the \textsc{sigw-delta} (left panel) and \textsc{sigw-gauss} (right panel) models. The solid and dashed teal contour lines labeled $f_{\scriptscriptstyle \rm PBH} = 1$ in the right panel indicate the PBH bound for $\Delta =1$ and $\Delta = 2$, respectively. Fig.~\ref{fig:sigw_ext_corner} in Appendix~\ref{app:sigw} shows extended versions of the two plots that include the SMBHB parameters $\abhb$ and $\gbhb$.} 
 \label{fig:sigw_corner}
\end{figure*}

The reconstructed posterior distributions for all SIGW models under consideration are shown in Figs.~\ref{fig:sigw_box_corner} and \ref{fig:sigw_corner}. Our first conclusion in view of these results is that a successful explanation of the NANOGrav signal in terms of SIGWs requires the primordial scalar power spectrum to have a large amplitude $A$. We can quantify this statement in terms of the lower limits of the $95\%$ Bayesian credible intervals for $A$ that we obtain from integrating the corresponding 1D marginalized posteriors over the regions of highest posterior density: for \textsc{sigw-delta}, \textsc{sigw-gauss}, and \textsc{sigw-box}, we respectively find $\log_{10}A \gtrsim -1.55$, $\log_{10}A \gtrsim -1.43$, and $\log_{10}A \gtrsim -1.90$. At the same time, the enhanced power in scalar perturbations must be localized at the right scales, so that the resulting SIGW signal falls into the PTA band. This requirement leads to bounds on the frequencies $f_{\rm min}$, $f_{\rm max}$, and $f_*$ that can be read off from Figs.~\ref{fig:sigw_box_corner} and \ref{fig:sigw_corner} and that are summarized in Table~\ref{tab:list}.

A notable feature in this context is that the posterior distributions for $f_{\rm min}$, $f_{\rm max}$, and $f_*$ all extend to large frequencies, much beyond the upper limit of the NANOGrav band. The reason for this is that the NG15 data are best fit by the low-frequency tail of the SIGW spectrum (see Fig.~\ref{fig:mean_spectra_all_1}). Beyond the NANOGrav band, the SIGW spectrum keeps increasing until it reaches its maximal value at $f \gg 1\,\textrm{nHz}$. This observation also explains the flat directions in the 2D posterior distribution for $A$ and $f_{\rm min}$ in Fig.~\ref{fig:sigw_box_corner} and the 2D posterior distributions for $A$ and $f_*$ in Fig.~\ref{fig:sigw_corner}. A simultaneous increase in $A$ and $f_{\rm min}$ or $f_*$ along these flat directions moves the peak in the GW spectrum to higher frequencies, but it preserves the shape of the low-frequency tail in the PTA band and hence does not affect the quality of the fit.

We also observe that the 2D posterior distributions for $A$ and $f_*$ for \textsc{sigw-delta} and \textsc{sigw-gauss} are roughly similar to each other, with the distribution for the \textsc{sigw-gauss} model being slightly broader. This result is consistent with the fact that \textsc{sigw-delta} and \textsc{sigw-gauss} are nested models; \textsc{sigw-delta} can be obtained from \textsc{sigw-gauss} in the limit $\Delta \rightarrow 0$. The slightly broader posterior distribution for $A$ and $f_*$ in the right panel of Fig.~\ref{fig:sigw_corner} thus reflects the extra dimension in the parameter space of the \textsc{sigw-gauss} model and the additional parametric freedom that comes with it.

Finally, we comment on the bounds on the parameter space of the three SIGW models. As in the case of the \textsc{igw} model, we identify regions of the parameter space where the $K$ ratio in Eq.~\eqref{eq:Kratio} falls below $\sfrac{1}{10}$. In these regions, which are shaded in gray in Fig.~\ref{fig:sigw_corner} and labeled NG15, adding the SIGW contribution to the GW signal leads to much worse agreement with the data than in the case of no SIGW contribution at all. In fact, parameter points in these regions lead to an SIGW signal that exceeds the observed signal---in other words, the gray shaded regions are ruled out because they result in too strong of a SIGW signal. For \textsc{sigw-box}, we are not able to place a $K$-ratio bound on the 2D parameter space spanned by $A$ and $f_{\rm min}$, because we additionally marginalize over $f_{\rm max}$. That is, for any pair of values of $A$ and $f_{\rm min}$, the SIGW signal can be made arbitrarily small if we choose $f_{\rm max}$ close enough to $f_{\rm min}$. 

These bounds are an important result of our analysis that is independent of the origin of the NANOGrav signal. They provide valid constraints on the parameter space for both the \textsc{sigw-delta} and \textsc{sigw-gauss} models, regardless of whether these models contribute to the explanation of the observed GWB. In addition, they are complementary to other existing bounds, in particular, the requirement that the mass density of PBHs produced alongside SIGWs must not exceed the energy density of DM,
\begin{equation}
\label{eq:fpbh}
f_{\scriptscriptstyle\rm PBH} \leq 1 \,,
\end{equation}
where $f_{\scriptscriptstyle \rm PBH} = \Omega_{\scriptscriptstyle \rm PBH}/\Omega_{\scriptscriptstyle \rm DM}$ denotes the PBH DM fraction integrated over the entire PBH mass spectrum. We provide more details on how we compute $f_{\scriptscriptstyle \rm PBH}$ in Appendix~\ref{app:sigw}; here we simply report our final results in terms of the teal contour lines labeled $f_{\scriptscriptstyle \rm PBH} = 1$ in Figs.~\ref{fig:sigw_box_corner} and \ref{fig:sigw_corner}. For \textsc{sigw-box}, the PBH bound in the $A$\,--\,$f_{\rm max}$ plane is computed for $f_{\rm min} = 10^{-11}\,\textrm{Hz}$ (the lower end of our prior range), and the PBH bound in the $A$\,--\,$f_{\rm min}$ plane is computed for $f_{\rm max} = 10^{-5}\,\textrm{Hz}$ (the upper end of our prior range). For \textsc{sigw-gauss}, we show the PBH bound in the $A$\,--\,$f_*$ plane for $\Delta = 1$ (solid contour line) and $\Delta = 2$ (dashed contour line).

In all three cases, we find that the PBH bound is very restrictive, ruling out most of the parameter space favored by the NG15 data. If taken at face value, the PBH bound therefore renders the SIGW interpretation of the NANOGrav signal less likely. However, we stress that the computation of $f_{\scriptscriptstyle \rm PBH}$ is very sensitive to various assumptions and numerical steps in the analysis. Slight changes in the computational strategy may result in very different results for $f_{\scriptscriptstyle \rm PBH}$, which is why the results reported in  Figs.~\ref{fig:sigw_box_corner} and \ref{fig:sigw_corner} need to be treated with caution. In view of the large conceptional uncertainty in the computation of $f_{\scriptscriptstyle \rm PBH}$, one needs to be careful not to draw any premature conclusions. At the same time, the PBH bounds in Figs.~\ref{fig:sigw_box_corner} and \ref{fig:sigw_corner} illustrate that small regions of parameter space do remain viable. In fact, for \textsc{sigw-delta} and \textsc{sigw-gauss}, it is even possible to realize $f_{\scriptscriptstyle \rm PBH} = 1$ inside the $68\%$ credible regions. That is, along the $f_{\scriptscriptstyle \rm PBH} = 1$ contour lines and inside the $68\%$ credible regions, we find scenarios where SIGWs manage to provide a good fit to the NG15 data and PBHs account for the entire DM in our universe. 

In closing, we remark that the above conclusions are endangered by the recent claim of a no-go theorem for PBH formation from single-field inflation~\citep{Kristiano:2022maq,Kristiano:2023scm}. \cite{Kristiano:2022maq,Kristiano:2023scm} argue that enhanced scalar perturbations at small scales lead to unacceptably large one-loop corrections to the scalar power spectrum at large scales. In terms of the model parameters discussed in this section, this means that the amplitude $A$ must be small, $\log_{10} A \ll -2$; otherwise, the loop corrections to the scalar power spectrum will exceed the measured amplitude at CMB scales, $A_s \simeq 2.10 \times 10^{-9}$. At present, this claim is subject to an ongoing debate; it is notably challenged by~\cite{Riotto:2023hoz,Riotto:2023gpm,Firouzjahi:2023ahg}. However, if it should prove to be valid, the requirement of $\log_{10} A \ll -2$ would clash with the lower bounds on $A$ listed above. In this case, one would then have to either give up on the SIGW interpretation of the NANOGrav signal or seek to construct inflation models that evade the arguments of~\cite{Kristiano:2022maq,Kristiano:2023scm} and still lead to a large SIGW signal. 

\subsection{Cosmological phase transitions}
\label{subsec:phase_transitions}

\subsubsection*{Model description}

In the cosmological context, first-order phase transitions take place when a potential barrier separates the true and false minima of scalar field potential.\footnote{The scalar field can either be an elementary field of a weakly coupled theory or correspond to the vacuum condensate of a strongly coupled theory. Scenarios with several scalars are also possible.} The phase transition is triggered by quantum or thermal fluctuations that cause the scalar field to tunnel through or fluctuate over the barrier, which results in the nucleation of bubbles within which the scalar field is in the true vacuum configuration. If sufficiently large, these bubbles expand in the surrounding plasma where the scalar field still resides in the false vacuum. The expansion and collision of these bubbles~\citep{Kosowsky:1991ua,Kosowsky:1992rz,astro-ph/9211004, astro-ph/9310044, Caprini:2007xq, Huber:2008hg}, together with sound waves generated in the plasma~\citep{Hindmarsh:2013xza,Giblin:2013kea,Giblin:2014qia,Hindmarsh:2015qta}, can source a primordial GWB (see \cite{Witten:1984rs,Hogan:1986qda} for seminal work).\footnote{Turbulent motion of the plasma can also source GWs; however, its contribution is usually subleading compared to the two other contributions (see the discussion in~\cite{Caprini:2019egz}). Therefore, we ignore GWs sourced by turbulence in our analysis.} For earlier Bayesian searches for a phase transition signal in PTA data, see~\cite{NANOGrav:2021flc,Xue:2021gyq}.

Generally, the GWB produced during the phase transition is a superposition of the bubble and sound-wave contributions. However, if the bubble walls interact with the surrounding plasma, most of the energy released in the phase transition is expected to be converted to plasma motion, causing the sound-wave contribution to dominate the resulting GWB. An exception to this scenario is provided by models in which there are no (or only very feeble) interactions between the bubble walls and the plasma, or by models where the energy released in the phase transition is large enough that the friction exerted by the plasma is not enough to stop the walls from accelerating ({\it runaway scenario}). However, determining whether or not the runaway regime is reached is either model dependent or affected by large theoretical uncertainties (see, e.g.,~\cite{LiLi:2023dlc,Ai:2023see,Krajewski:2023clt} for recent work on this topic). Therefore, in this work, we perform two separate analyses: a {\it sound-wave-only analysis} (\textsc{pt-sound}), where we assume that the runaway regime is not reached and sound waves dominate the GW spectrum, and a {\it bubble-collisions-only analysis} (\textsc{pt-bubble}), where we assume that the runaway regime is reached and bubble collisions dominate the GW spectrum.

We parameterize the GWB produced by both sound waves and bubble collisions in a model-independent way in terms of the following phase transition parameters:

\medskip\noindent\textbullet~$T_*$, the percolation temperature, i.e., the temperature of the universe when $\sim34\%$ of its volume has been converted to the true vacuum~\citep{Ellis:2018mja}. For weak transitions, this temperature coincides with the temperature at the time of bubble nucleation, $T_n \sim T_*$. Conversely, for supercooled transitions, we typically have $T_n \ll T_*$. Barring the case of extremely strong transitions, $\alpha_* \ggg 1$ (see below), which we do not consider in this work, $T_*$ also determines the reheating temperature after percolation, $T_{\rm rh} \sim T_*$~\citep{Ellis:2018mja}.

\medskip\noindent\textbullet~$\alpha_*$, the strength of the transition, i.e., the ratio of the change in the trace of the energy--momentum tensor across the transition and the radiation energy density at percolation~\citep{Caprini:2019egz,Ellis:2019oqb}. 

\medskip\noindent\textbullet~$H_*R_* = R_* / H_*^{-1}$, the average bubble separation in units of the Hubble radius at percolation, $H_*^{-1}$. For relativistic bubble velocities, which is what we consider in the following, $R_*$ is related to the bubble nucleation rate, $\beta$, by the relation $H_*R_*=(8\pi)^{1/3}H_*/\beta$.

\begin{figure*}
	\centering
        \includegraphics[width=0.48\textwidth]{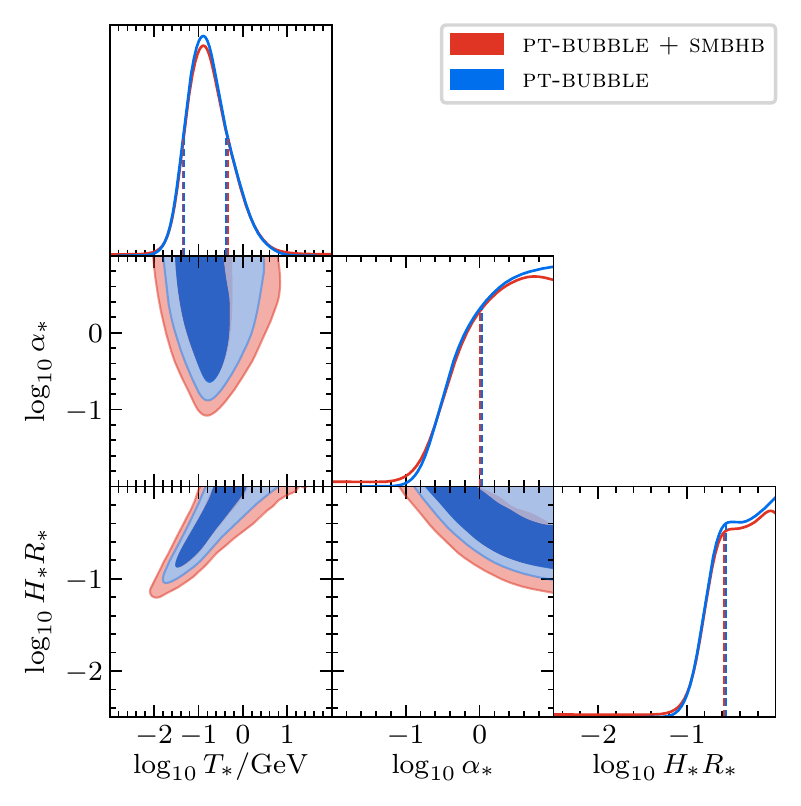}
        \includegraphics[width=0.48\textwidth]{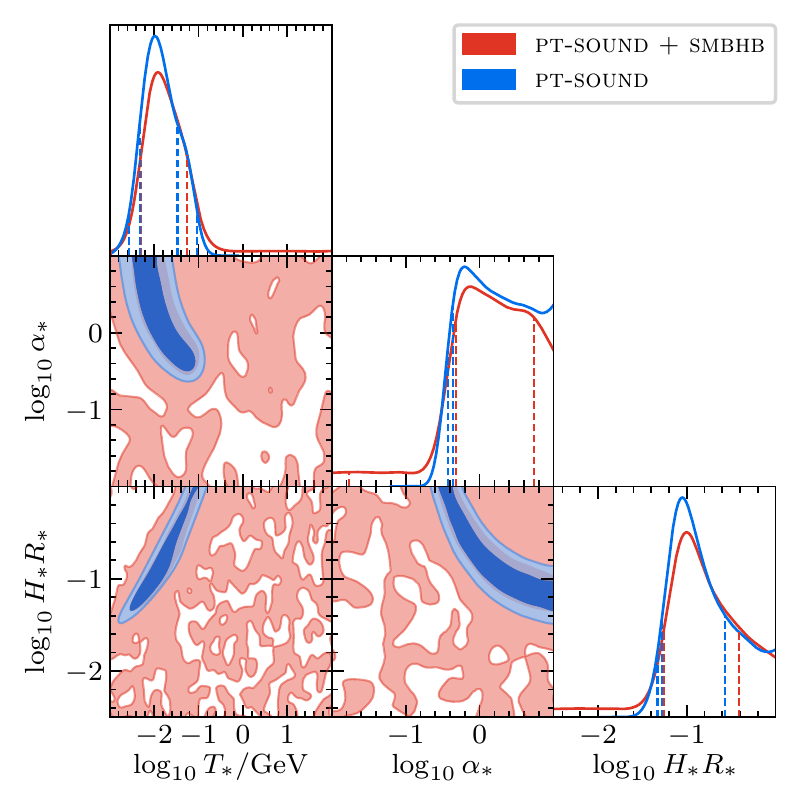}
	\caption{\label{fig:pt_corner}
        Same as in Fig.~\ref{fig:igw_corner} but for the \textsc{pt-bubble} (left panel) and \textsc{pt-sound} model (right panel). Fig.~\ref{fig:pt_corner_fix} in Appendix~\ref{app:phase_transitions} shows the same plots but with the parameter $a$ fixed by causality, $a = 3$. Figs.~\ref{fig:pt_bubble_corner_full} and \ref{fig:pt_sound_corner_full} in Appendix~\ref{app:phase_transitions} show extended versions of the two plots that include the spectral shape parameters $a$, $b$, $c$ and the SMBHB parameters $\abhb$ and $\gbhb$.}
\end{figure*}

\medskip
In addition to the parameters $T_*$, $\alpha_*$, and $H_*R_*$, the GWB produced by a phase transition also depends on the velocity of the expanding bubble walls, $v_w$. However, deriving the precise value of this quantity is an open theoretical problem, which depends on model-dependent quantities, such as the strength of the interactions between the bubble walls and the SM plasma. Therefore, in our analysis, we fix the bubble velocity to unity (i.e., the speed of light in natural units). This assumption is well justified for strong phase transitions~\citep{Bodeker:2017cim}, which, realistically, are the only ones that could lead to a detectable signal in our current data. In particular, we fix $v_w = 1$ for both phase transition scenarios that we are interested in, \textsc{pt-sound} and \textsc{pt-bubble}. In the latter case, $v_w \rightarrow 1$ is automatically implied by the runaway behavior of the phase transition; in the former case, one actually expects a subluminal terminal velocity, $v_w < 1$. In this sense, our decision to fix $v_w = 1$ amounts to the optimistic assumption that this terminal velocity is numerically close to $v_w = 1$. A similar approach is followed by the authors of the LISA review paper \cite{Caprini:2019egz} who work with $v_w = 0.95$ throughout most of their analysis in the absence of more detailed microphysical calculations. Finally, we point out that the parametrization of the GWB signal in terms of $H_*R_* = (8\pi)^{1/3}\,v_w\,H_*/\beta$ already absorbs a large part of the dependence on the bubble wall velocity. The remaining $v_w$ dependence is mostly contained in the efficiency factor $\kappa_s$ (see below). However, in the regime of large $\alpha_*$ values, $\alpha_* \sim 0.3 \cdots 10$, which turn out to be preferred by the NG15 data (see Fig.~\ref{fig:pt_corner}), this dependence is rather weak (see Fig.~13 in \cite{Espinosa:2010hh}), which justifies again to keeping $v_w$ fixed. 

The GWB spectrum sourced by bubbles and sound waves can be written in terms of these parameters as
\begin{align}
    \Omega_b(f)&=\mathcal{D}\,\tilde{\Omega}_b\left(\frac{\alpha_*}{1+\alpha_*}\right)^2\left(H_*R_*\right)^2\mathcal{S}(f/f_{b})\label{eq:omega_pt_bubble}\\
    \Omega_s(f)&=\mathcal{D}\,\tilde{\Omega}_s\Upsilon(\tau_{\rm sw})\left(\frac{\kappa_s\,\alpha_*}{1+\alpha_*}\right)^2\left(H_*R_*\right)\mathcal{S}(f/f_{s})\label{eq:omega_pt_sound} \,.
\end{align}
Here $\tilde{\Omega}_b=0.0049$~\citep{Jinno:2016vai} and $\tilde{\Omega}_s=0.036$~\citep{Hindmarsh:2017gnf}; the efficiency factor $\kappa_s=\alpha_*/(0.73+0.083\sqrt{\alpha_*}+\alpha_*)$~\citep{Espinosa:2010hh} gives the fraction of the released energy that is transferred to plasma motion in the form of sound waves, and $\mathcal{D}$ accounts for the redshift of the GW energy density,
\begin{equation}\label{eq:dilution}
    \mathcal{D} = \frac{\pi^2}{90}\frac{T_0^4}{M_{\scriptscriptstyle \rm Pl}^2H_0^2}\,g_{*}\left(\frac{g_{*,s}^{\rm eq}}{g_{*,s}}\right)^{4/3}\simeq 1.67\times 10^{-5} \,.
\end{equation}
We recall that $T_0$ and $H_0$ denote the photon temperature and Hubble rate today. The degrees of freedom $g_*$ and $g_{*,s}$ in Eq.~\eqref{eq:dilution} are evaluated at $T = T_*$, and $g_{*,s}^{\rm eq}$ is the number of degrees of freedom contributing to the radiation entropy at the time of matter--radiation equality.
The production of GWs from sound waves stops after a period $\tau_{\rm sw}$, when the plasma motion turns turbulent~\citep{Ellis:2018mja, Ellis:2019oqb, Ellis:2020awk, Guo:2020grp}. In Eq.~\eqref{eq:omega_pt_sound}, this effect is taken into account by the suppression factor 
\begin{equation}
\Upsilon(\tau_{\rm sw}) = 1 - (1+2\tau_{\rm sw}H_*)^{-1/2},
\end{equation}
where the shock formation time scale, $\tau_{\rm sw}$, can be written in terms of the phase transition parameters as $\tau_{\rm sw}\approx R_*/\bar U_f$, with $\bar U^2_f\approx3\kappa_s\alpha_*/[4(1+\alpha_*)]$~\citep{1705.01783}.

The functions $\mathcal{S}_{b,s}$ describe the spectral shape of the GWB and are expected to peak at the frequencies 
\begin{equation}\label{eq:pt_peak_f}
    f_{b,s}\simeq 48.5\,{\rm nHz}\; g_*^{1/2}\bigg(\frac{g_{*,s}^{\rm eq}}{g_{*,s}}\bigg)^{1/3}\bigg(\frac{T_*}{1\,\rm GeV}\bigg)\:\frac{f_{b,s}^*R_*}{H_*R_*} \,,
\end{equation}
where the values of the peak frequencies at the time of GW emission are given by $f_b^*=0.58/R_*$~\citep{Jinno:2016vai} and $f_s^*=1.58/R_*$~\citep{Hindmarsh:2017gnf}. In passing, we mention that the numerical factors in these estimates may still change in the future, as our understanding of cosmological phase transitions improves. However, at the level of our Bayesian fit analysis, changes in these prefactors can be absorbed in the temperature scale $T_*$, which in its role as an independent fit parameter only controls the peak frequencies in Eq.~\eqref{eq:pt_peak_f}. A similar argument applies to the numerical factors in Eqs.~\eqref{eq:omega_pt_bubble} and \eqref{eq:omega_pt_sound}: changes in these prefactors can always be absorbed in a rescaled version of the fit parameter $\alpha_*$, which only appears in the expressions for the peak amplitudes of the GWB signal.

The shape of the spectral functions can be usually approximated with a broken power law of the form
\begin{equation}\label{eq:pt_spec_shape}
    \mathcal{S}(x)=\frac{1}{\mathcal{N}}\frac{(a+b)^c}{(bx^{-a/c}+ax^{b/c})^c} \,.
\end{equation}
Here $a$ and $b$ are two real and positive numbers that give the slope of the spectrum at low and high frequencies, respectively; $c$ parametrizes the width of the peak. The normalization constant, $\mathcal{N}$, ensures that the logarithmic integral of $\mathcal{S}$ is normalized to unity and is given by
\begin{equation}
    \mathcal{N}=\left(\frac{b}{a}\right)^{a/n}\bigg(\frac{nc}{b}\bigg)^c \:\frac{\Gamma\left(a/n\right)\,\Gamma\left(b/n\right)}{n\,\Gamma(c)} \,,
\end{equation}
where $n=(a+b)/c$ and $\Gamma$ denotes the gamma function. While the values of the coefficients $a$, $b$, and $c$ can in principle be estimated from numerical simulations, we allow them to float within the prior ranges given in Table~\ref{tab:np_priors}. These prior ranges were chosen to roughly reflect the typical uncertainties of numerical simulations and any possible model dependency of these coefficients (see, e.g.,~\cite{Hindmarsh:2017gnf,Hindmarsh:2020hop,Lewicki:2020jiv, Lewicki:2020azd, Cutting:2018tjt, Cutting:2020nla}).\footnote{Causality fixes the spectral index of the phase transition GWB signal to $a=3$ in the low-frequency limit. However, given the simple power-law parametrization adopted in this work, double-peak features observed in the results of numerical simulations~\citep{Hindmarsh:2017gnf, Hindmarsh:2019phv} can appear as a deviation from this behavior near the peak frequency. Nonetheless, in Appendix \ref{app:phase_transitions}, we also report the results of an analysis in which the low-frequency slope is fixed to $a=3$.}

\subsubsection*{Results and discussion}
The reconstructed posterior distributions for the parameters $\alpha_*$, $T_*$ and $H_*R_*$ of the \textsc{pt-sound} and \textsc{pt-bubble} models are reported in Fig.~\ref{fig:pt_corner}, both for the case where the phase transition is assumed to be the only source of GWs (blue contours) and for the scenario where we consider the superposition of the phase transition and SMBHB signals (red contours).\footnote{The noise in the 95\% credible regions of the posterior distributions of the \textsc{pt-sound+smbhb } model is due to the presence of an extended plateau region in the posterior distribution, which renders the kernel density reconstruction sensitive to Poisson fluctuations in the binned MCMC data.} Corner plots including the posterior distributions for the spectral shape parameters $a$, $b$, $c$ and SMBHB parameters $\abhb$ and $\gbhb$ are reported in Figs.~\ref{fig:pt_bubble_corner_full} and~\ref{fig:pt_sound_corner_full} in Appendix~\ref{app:phase_transitions}. 

In all analyses, the data prefer a relatively strong and slow phase transition. Specifically, for \textsc{pt-bubble}, we find $\alpha_*>1.1$ ($0.29$) and $H_*R_* > 0.28$ ($0.14$) at the $68\%$ ($95\%$) credible level. When the SMBHB signal is added on top of the GWB predicted by \textsc{pt-bubble}, we find $\alpha_*>1.0$ ($0.23$) and $H_*R_*>0.26$ ($0.11$) at the $68\%$ ($95\%$) credible level. For the \textsc{pt-sound} model, we find $\alpha_* > 0.42$ ($0.37$) and $H_*R_*\in[0.053,0.27]$ ($[0.046,0.89]$) at the $68\%$ ($95\%$) credible level. Including the SMBHB signal on top of the one predicted by \textsc{pt-sound}, we find $\alpha_*\in [0.46,5.4]$ ($>0.16$) and $H_*R_*\in[0.054,0.35]$ ($>0.0015$) at the $68\%$ ($95\%$) credible level.

As can be seen in Fig.~\ref{fig:mean_spectra}, for both phase transition models, the reconstructed GWB spectrum tends to peak around the higher frequency bins and fit the signal in the lower frequency bins with the left tail of the spectrum. Specifically, for the \textsc{pt-bubble} model we find $T_*\in [0.047,0.41]$ $([0.023,1.75])$ GeV at the $68\%$ ($95\%$) credible level, whereas for the \textsc{pt-sound} model we get $T_*\in[4.7,33]$ $([2.7,93])$ MeV at the $68\%$ ($95\%$) credible level. The shift between these $T_*$ intervals is partially explained by the different numerical factors in the frequencies $f_s^*$ and $f_b^*$ (see Eq.~\eqref{eq:pt_peak_f}). As explained below Eq.~\eqref{eq:pt_peak_f}, any change in these numerical factors can be reabsorbed in a redefinition of the fit parameter $T_*$.

The inclusion of the SMBHB signal, by adding power to the lowest frequency bins, allows the $T_*$ posterior for the \textsc{pt-sound} model to extend to higher values. In this case, we find that $T_*\in [4.9,50]$ $([0.8,2 \times 10^6])$ MeV at the $68\%$ ($95\%$) credible level. Here the increase in the $68\%$ upper limit is reflected in the slight shift between the red and blue dashed vertical lines in the 1D marginalized posterior distribution for $T_*$ in the right panel of Fig.~\ref{fig:pt_corner}. The drastic increase in the $95\%$ upper limit, on the other hand, is related to the fact that adding the SMBHB signal to the GWB results in a flat plateau region in the posterior distribution of the \textsc{pt-sound} model parameters where the NANOGrav signal is mostly explained by the SMBHB contribution to the GWB. The $95\%$ credible regions for the \textsc{pt-sound$+$smbhb} model cover much of this plateau, which explains their large extent and noisy appearance in Fig.~\ref{fig:pt_corner}. For the \textsc{pt-bubble} model, the inclusion of the SMBHB signal is less significant, and we find $T_*\in[0.046,0.46]$ $([0.017,3.27])$ GeV at the $68\%$ ($95\%$) credible level.

The larger phase transition temperatures observed for the \textsc{pt-bubble} model are a consequence of the smaller value of the peak frequency at the time of emission, $f_b^*$, but also of the lower prior range for the low-frequency spectral index adopted for the \textsc{pt-bubble} model. Indeed, a shallower low-frequency tail allows spectra with a higher peak frequency to still produce a sizable signal in the lowest frequency bins. In Appendix~\ref{app:phase_transitions}, we report the results of the analysis in which the low-frequency slope is set to the value predicted by causality ($a=3$). In this case, as expected, the reconstructed phase transition temperatures for the two phase transition models are closer to each other. 

The corner plots in Fig.~\ref{fig:pt_corner} also illustrate that, as expected from the expression for the peak frequency in Eq.~\eqref{eq:pt_peak_f}, there is an approximately linear correlation between $\log_{10} T_*$ and $\log_{10} H_*R_*$. For $\alpha_*\lesssim 1$, we instead find $\alpha_*\sim 1/(H_*R_*)$ for the \textsc{pt-bubble} model and $\alpha_*^2\sim 1/(H_*R_*)$ for the \textsc{pt-sound} model as expected from the factors in Eq.~\eqref{eq:omega_pt_bubble} and Eq.~\eqref{eq:omega_pt_sound}. 

We also notice that, for both models, the posterior distribution for $T_*$ is peaked at significantly larger values compared to what was derived in the 12.5 yr analysis~\citep{NANOGrav:2021flc}. This shift results from the reconstructed $h^2\Omega_{\scriptscriptstyle \rm GW}$ spectrum being bluer than the one derived for the common process observed in the 12.5 yr data set. As a result, the lowest frequency bins, which were fit by the high-frequency tail of the phase transition spectrum in the 12.5 yr analysis, are now fit by the low-frequency tail of the spectrum. This then translates into a higher peak frequency and therefore a higher transition temperature. Incidentally, the larger reconstructed value for the transition temperature allows the phase transition signal to safely evade bounds from BBN and CMB observations~\citep{Bai:2021ibt,Deng:2023seh} for both the \textsc{pt-bubble} and \textsc{pt-sound} models, which constrain the phase transition parameter space at temperatures around $T_* \sim 1\,\textrm{MeV}$.

Instead, we conclude that the reconstructed posterior distribution of $T_*$ is compatible with phase transition scenarios that have been discussed in the literature as a possible source of GWs in the PTA band: (i) BSM models in which the chiral-symmetry-breaking phase transition in quantum chromodynamics (QCD) is a strong first-order phase transition (see, e.g., \cite{Neronov:2020qrl,Li:2021qer}) and (ii) strong first-order phase transitions in a dark sector composed of new BSM degrees of freedom (see, e.g., \cite{Nakai:2020oit,Ratzinger:2020koh}). In view of the NG15 data, both of these options for the particle physics origin of the phase transition signal remain viable. A third option may consist in a strongly supercooled first-order electroweak phase transition~\citep{Kobakhidze:2017mru}.

Finally, we report that that, like the models studied in Sections~\ref{subsec:inflation} and \ref{subsec:sigw}, the phase transition models provide a better fit of the NG15 data than the base \textsc{smbhb} model. The Bayes factors for \textsc{pt-bubble} and \textsc{pt-sound} versus \textsc{smbhb} are $\mathcal{B}=18.1\pm0.6$ and $\mathcal{B}=3.7\pm0.1$, respectively, while the Bayes factors for \textsc{pt-bubble$+$smbhb} and \textsc{pt-sound$+$smbhb} versus \textsc{smbhb} are $\mathcal{B}=12.6\pm0.5$ and $\mathcal{B}=6.5\pm0.3$, respectively. An interesting observation in view of these results is that adding the SMBHB contribution to the GWB signal does not help to improve the quality of the fit for \textsc{pt-bubble}---in this case, we find again a decrease in the Bayes factor going from \textsc{pt-bubble} to \textsc{pt-bubble$+$smbhb} because of the larger prior volume---but it does lead to a better fit for \textsc{pt-sound}. This model benefits from the additional SMBHB contribution because it can add power to the low frequency bins in the GW spectrum that the \textsc{pt-sound} model alone struggles to fit well on its own (see Fig.~\ref{fig:mean_spectra}). The reason for this, in turn, is the prior range for the spectral index at low frequencies, $a$, which can as be as low as $a=1$ for \textsc{pt-bubble}, but which we require to be at least $a=3$ for \textsc{pt-sound} (see Table~\ref{tab:np_priors}). Another consequence of this interplay between the phase transition and SMBHB signals is that the NANOGrav signal may in fact be dominated by SMBHBs. This possibility is realized when the \textsc{pt-sound} model parameters fall into the plateau region in Fig.~\ref{fig:pt_corner} (i.e., the red $95\%$ credible regions in the right panel) and the SMBHB parameters are close to $\log_{10}\abhb \sim -(15 \cdots 14)$ and $\gbhb \sim 3\cdots4$ (see Fig.~\ref{fig:pt_sound_corner_full}).

\subsection{Cosmic strings}
\label{subsec:strings}

\subsubsection*{Model description}

Cosmic strings are effectively 1D topological defects that can form in the early universe as a consequence of a cosmological phase transition~\citep{Kibble:1976sj}. Rigorously speaking, the criterion for the formation of a cosmic-string network is that the underlying phase transition must entail the spontaneous breaking of a local or global symmetry and end on a vacuum manifold with a nontrivial first homotopy group. In practice, the most relevant scenario satisfying this criterion is the cosmological breaking of a $U(1)$ symmetry. The breaking of global $U(1)$ symmetries results in the formation of global-string networks, which happens, e.g., in axion models. The GW signal in this case is suppressed because global strings lose most of their energy by emitting light pseudo-Nambu--Goldstone bosons (i.e., ``axions''). In the following, we therefore focus on local strings from the spontaneous breaking of a local $U(1)$ symmetry as occurs in many particle physics models of the early universe, such as grand unified theories (GUTs), where intermediate symmetry breaking stages can be realized in the form of string-producing phase transitions. Cosmic strings are thus well motivated by ideas about particle physics at very high energies, which prompts us to consider them as yet another possible source of exotic GWs~\citep{Vilenkin:1981bx,Hogan:1984is,Vachaspati:1984gt,Accetta:1988bg,Bennett:1990ry,Caldwell:1991jj}. For earlier work on the cosmic strings interpretation of the signal in recent PTA data sets, see, e.g.,~\cite{Blasi:2020mfx}, \cite{Ellis:2020ena}, and \cite{Blanco-Pillado:2021ygr}.

In our analysis, we study ordinary stable cosmic strings as well as metastable strings, which are unstable against the nucleation of GUT monopoles. In passing, we also comment on cosmic superstrings, which do not have a particle-physics origin but are present in some string-theoretic models of the early universe. Our starting point for describing all these scenarios is the Nambu--Goto action, which treats cosmic strings as featureless 1D objects that can be characterized by a single parameter: their tension, i.e., energy per unit length, $\mu$. In the case of metastable strings, there is a second parameter, $\kappa$, which controls the lifetime of the network. For superstrings, the relevant parameters are $\mu$ and the intercommutation probability $P$ (see below).  Before we move on, we note that, as an alternative approach to the Nambu--Goto approximation, it is possible to describe cosmic strings in a field-theoretical language, e.g., in terms of the Abelian-Higgs model. The comparison of these two approaches, i.e., the relation between Nambu--Goto strings and Abelian--Higgs strings, is the subject of ongoing research~\citep{Blanco-Pillado:2023sap}, and we refer the reader to Section~3.4 of~\cite{Auclair:2019wcv} for an extended discussion. Furthermore, we only consider the GWB produced by cosmic-string loops in our analysis and disregard the subdominant contribution from long (infinite) strings (see a discussion of this point in Section~4.4 of~\cite{Auclair:2019wcv}).

Shortly after their formation, cosmic strings enter a scaling regime where the total energy stored in the network, $\rho_{\rm cs}$, remains a constant fraction of the critical energy density, $\Omega_{\rm cs} = \rho_{\rm cs} / \rho_c \approx \textrm{const}$~\citep{Hindmarsh:1994re}. This behavior is possible because long strings, stretching over superhorizon distances, frequently intercommute with each other, thereby producing an abundance of closed string loops on subhorizon scales that radiate energy in the form of GWs. These GWs are sourced by the oscillations of the loops under their own tension, as well as by localized features (``cusps'' and ``kinks'') propagating along the loops. The superposition of the GWs emitted by the individual loops in the network thus results in a stochastic GWB,
\begin{equation}
\label{eq:Ogwcs}
\Omega_{\scriptscriptstyle\rm GW}^{\rm cs}\left(f\right) = \frac{8\pi}{3H_0^2}\left(G\mu\right)^2\sum_{k=1}^{k_{\rm max}} P_k\,\mathcal{I}_k\left(f\right) \,.
\end{equation}
Here the dimensionless factor $G\mu$ denotes the cosmic-string tension in units of Newton's constant, for which we choose a log-uniform prior in our numerical analysis, $\log_{10}G\mu \in [-14,-6]$ for stable strings and superstrings and $\log_{10}G\mu \in [-14,-1.5]$ for metastable strings. The sum in Eq.~\eqref{eq:Ogwcs} runs over the harmonic excitations of the closed string loops that, given a loop of length $\ell$, correspond to oscillations at frequency $f = 2k/\ell$. We evaluate the sum starting at the fundamental oscillation mode, $k=1$, and including terms up to $k_{\rm max} = 10^5$, which ensures good convergence of the GW spectrum.

The dimensionless factor $P_k$ inside the sum describes the GW power, in units of $G\mu^2$, that is emitted by a loop in its $k$th excitation. For large $k_{\rm max}$, we can write
\begin{equation}
\label{eq:Pk}
P_k = \frac{\Gamma}{\zeta\left(q\right)} \frac{1}{k^q} \,,
\end{equation}
where the prefactor $\Gamma/\zeta\left(q\right)$ ensures that the total power emitted in GWs amounts to $\sum_k P_k = \Gamma$ and where the power-law index $q$ depends on the predominant source of GWs on the loops. In our analysis, we set $\Gamma = 50$, as suggested by numerical simulations~\citep{Blanco-Pillado:2017oxo}, and discuss different possibilities for $q$. The actual average GW spectrum from non-self-intersecting loops is still uncertain. We therefore choose several different models that give an idea of the range of possibilities. Specifically, we consider four different models of stable cosmic strings:\medskip\\
\textsc{stable-c}: \textit{Stable} strings emitting GWs only in the form of GW bursts from \textit{cusps} on closed loops, $q=4/3$ \citep{Vachaspati:1984gt}.\smallskip\\
\textsc{stable-k}: \textit{Stable} strings emitting GWs only in the form of GW bursts from \textit{kinks} on closed loops, $q=5/3$ \citep{Damour:2001bk}.\smallskip\\
\textsc{stable-m}: \textit{Stable} strings emitting \textit{monochromatic} GWs at the fundamental frequency $f = 2/\ell$ of closed loops.\smallskip\\
\textsc{stable-n:} \textit{Stable} strings described by \textit{numerical} simulations including GWs from cusps and kinks~\citep{Blanco-Pillado:2011egf,Blanco-Pillado:2015ana}.\medskip\\
For \textsc{stable-n}, $P_k$ is dominated by cusp emission at large $k$, i.e., $q\approx 4/3$ for $k \gtrsim 100$, while at smaller $k$ it deviates from the pure cusp spectrum, reaching $q$ values of up to $q \sim 1.5$ around $k\sim 10$. Meanwhile, Eq.~\eqref{eq:Pk} is irrelevant for \textsc{stable-m}; for this model, we simply set $P_k = \Gamma$ if $k=1$ and $P_k = 0$ otherwise. More details on our four stable-string models can be found in~\cite{Blanco-Pillado:2021ygr}.

Finally, the frequency-dependent factor $\mathcal{I}_k$ in Eq.~\eqref{eq:Ogwcs} represents an integral of the number density of closed string loops, $n_l(\ell,t)$, over all possible GW emission times,
\begin{equation}
\label{eq:Ikcs}
\mathcal{I}_k\left(f\right) = \frac{2k}{f} \int_{t_{\rm ini}}^{t_0} \textrm{d}t \left(\frac{a\left(t\right)}{a(t_0)}\right)^5 n_l\left(\frac{2k}{f}\frac{a\left(t\right)}{a(t_0)},t\right) \,.
\end{equation}
Here $t_0$ is the present time and $t_{\rm ini}$ is the time when the network reaches the scaling attractor solution. The precise value of $t_{\rm ini}$ only affects the high-frequency part of the GW spectrum and plays no role in our analysis. The loop number density $n_l$ can be estimated based on the velocity-dependent one-scale (VOS) model,
\begin{equation}
\label{eq:nl-stable}
n_l\left(\ell,t\right) = \mathcal{F}\:\frac{C_*\,\Theta\left(t-t_*\right)\Theta\left(t_*-t_{\rm ini}\right)}{\alpha_*\left(\alpha_* + \Gamma G\mu + \dot{\alpha}_* t_*\right)t_*^4} \left(\frac{a_*}{a\left(t\right)}\right)^3 \,,
\end{equation}
where $\mathcal{F} = 0.1$ is an efficiency factor~\citep{Blanco-Pillado:2013qja,Auclair:2019wcv} and an asterisk indicates that the corresponding quantity needs to be evaluated at the time of loop formation, which follows from solving the following relation for $t_*$:
\begin{equation}
t_* = \frac{\ell + \Gamma G\mu\,t}{\alpha_* + \Gamma G \mu} \,, \qquad \alpha_* = \alpha\left(t_*\right) \,.
\end{equation}

The time-dependent functions $C$ and $\alpha$ characterize the efficiency of loop formation from the network and the typical loop size at the time of production, respectively,
\begin{equation}
C\left(t\right) = \frac{\tilde{c}}{\sqrt{2}}\frac{\bar{v}\left(t\right)}{\xi^3\left(t\right)} \,, \qquad \alpha\left(t\right) = \alpha_L\,\xi\left(t\right) \,.
\end{equation}
Here $\tilde{c}$ is the loop-chopping parameter and $\alpha_{\scriptscriptstyle L}$ controls the loop length at the time of production in units of the correlation length of the long-string network, $L = \xi\,t$. We set $\tilde{c} \simeq 0.23$ and $\alpha_{\scriptscriptstyle L} \simeq 0.37$, in agreement with numerical simulations~\citep{Martins:2000cs,Blanco-Pillado:2011egf,Blanco-Pillado:2013qja}. With $\tilde{c}$ fixed, we are able to solve the VOS equations for the dimensionless correlation length $\xi$ and the root-mean-square velocity of the long strings, $\bar{v}$, and hence determine the time dependence of $C$ and $\alpha$. In doing so, we account for the exact evolution of the scale factor, $a$, in $\Lambda$CDM, including the temperature-dependent variation in the number of relativistic degrees of freedom. At very high temperatures, this analysis returns $\xi_r \simeq 0.27$ and $\bar{v}_r \simeq 0.66$, such that $\alpha \simeq 0.10$ deep in the radiation-dominated era. The loop number density $n_l$ obtained in this way agrees very well with the result of numerical simulations~\citep{Blanco-Pillado:2013qja} in the limit of constant $g_*$ and $g_{*,s}$.

In the case of metastable strings, the loop number density in Eq.~\eqref{eq:nl-stable} receives two correction factors,
\begin{equation}
\label{eq:nl-meta}
n_l^{\rm meta}\left(\ell,t\right) =  \Theta\left(t_s - t_*\right) E\left(\ell,t\right) n_l\left(\ell,t\right) \,.
\end{equation}
The Heaviside function accounts for the fact that we expect loop formation to cease when monopole nucleation becomes efficient and the network transitions from the scaling regime to the decay regime at times around
\begin{equation}
t_s = \frac{1}{\Gamma_d^{1/2}} \,,
\end{equation}
with the decay rate, $\Gamma_d$, counting the number of monopole nucleation events per time and per unit string length,
\begin{equation}
\label{eq:Gammad}
\Gamma_d = \frac{\mu}{2\pi}\,e^{-\pi\kappa} \,.
\end{equation}
Here $\kappa$ is a measure for the ratio of the GUT and $U(1)$ symmetry breaking scales in the underlying GUT model. Specifically, $\sqrt{\kappa}$ describes the ratio of the GUT monopole mass and the square root of the $U(1)$ string tension,
\begin{equation}
\sqrt{\kappa} = \frac{m_{\scriptscriptstyle \rm GUT}}{\mu^{1/2}} \sim \frac{\Lambda_{\scriptscriptstyle \rm\textsc{GUT}}}{\Lambda_{\scriptscriptstyle U(1)}} \,.
\end{equation}
Meanwhile, the second new factor in Eq.~\eqref{eq:nl-meta} represents an exponential suppression term, reflecting the depletion of the loop number density in the decaying network,
\begin{equation}
\label{eq:Elt}
E\left(\ell,t\right) = e^{-\Gamma_d\left[\ell\left(t-t_*\right) + \sfrac{1}{2}\,\Gamma G\mu\left(t-t_*\right)^2\right]} \,.
\end{equation}
As evident from Eqs.~\eqref{eq:Gammad} and \eqref{eq:Elt}, the loop number density of a metastable network is exponentially sensitive to the decay parameter $\kappa$. For $\sqrt{\kappa} \gg 10$, the lifetime of the network exceeds the age of the universe, such that the resulting GW signal is indistinguishable from the signal of a stable-string network. For $\sqrt{\kappa} \sim 1$, on the other hand, the network decays very fast in the early universe, such that no GW signal at PTA frequencies is generated. For these reasons, we choose a uniform prior on $\sqrt{\kappa}$ in a rather narrow range, $\sqrt{\kappa} \in [7.0,9.5]$, which is enough to capture the nontrivial aspects of the metastable scenario.

Monopole nucleation in a metastable network results in string segments, dumbbell-like shaped pieces of string with monopoles attached on either end. In many GUT models, these monopoles still carry unconfined flux, such that we actually need to distinguish between monopoles and antimonopoles. Monopoles and antimonopoles with unconfined flux occur, e.g., in GUT models involving, schematically, the symmetry breaking pattern
\begin{equation}
SU(2)_1 \times U(1)_2 \rightarrow U(1)_1 \times U(1)_2 \rightarrow U(1) \,.
\end{equation}
In this case, we expect the string segments to dissipate most of their energy via the emission of massless vector bosons soon after their formation. Alternatively, monopoles may carry no unconfined flux, in which case they are able to lose energy only via the emission of GWs. In our analysis, we cover both possibilities:\medskip\\
\textsc{meta-l}: \textit{Metastable} strings; monopoles carry unconfined flux; GW emission from \textit{loops} only.\smallskip\\
\textsc{meta-ls}: \textit{Metastable} strings; monopoles carry no unconfined flux; GW emission from \textit{loops and segments}.\medskip\\
For \textsc{meta-ls}, we also require the number density of segments that result from the decaying network; the relevant expressions are summarized in Appendix~\ref{app:strings}. Moreover, we need to specify the power spectrum of GWs emitted by string segments. Following~\cite{Buchmuller:2021mbb}, we use again Eq.~\eqref{eq:Pk} and set $q = 1$, which can be analytically derived in the straight-string approximation~\citep{Martin:1996cp}. At the same time, we assume a pure cusp spectrum ($q=4/3$) for the GW emission from loops in both \textsc{meta-l} and \textsc{meta-ls}.

\begin{figure*}
	\centering
        \includegraphics[width=0.48\textwidth]{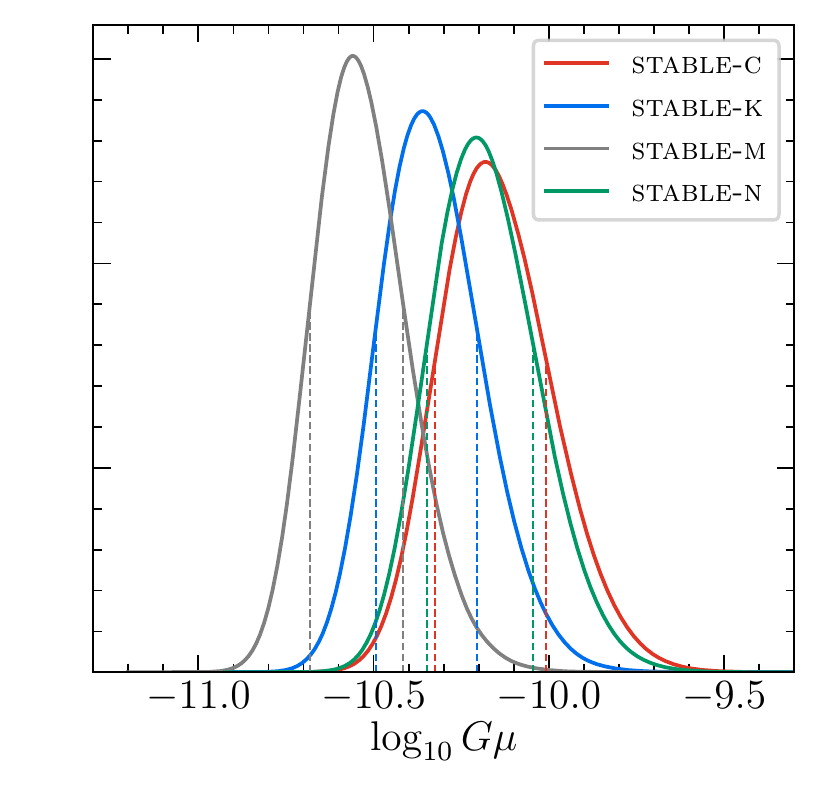}
        \includegraphics[width=0.48\textwidth]{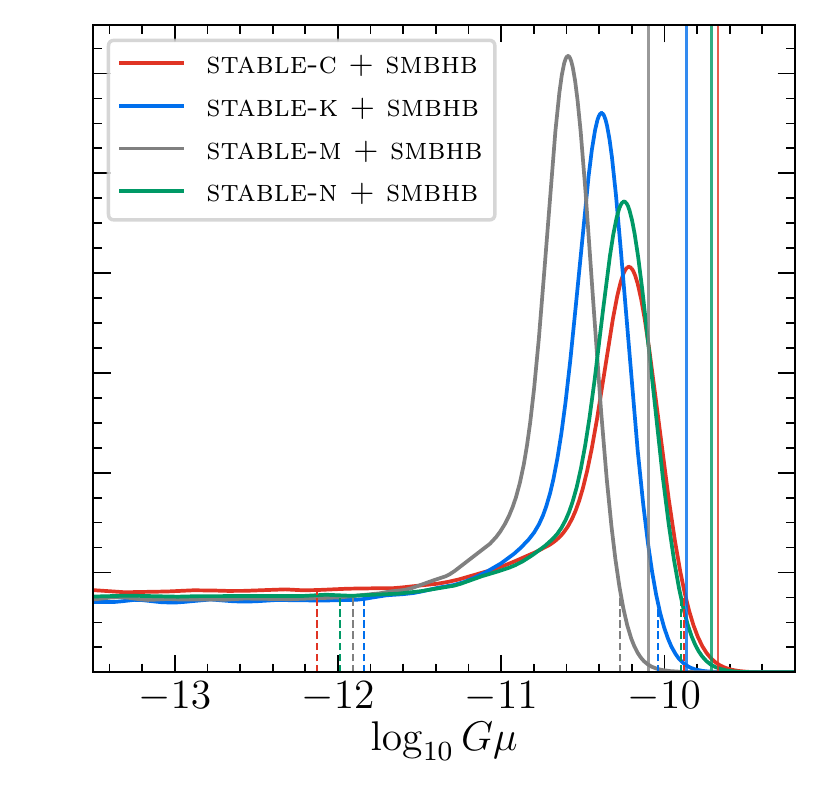}
	\caption{Reconstructed 1D marginalized posterior distributions for the dimensionless cosmic-string tension parameter $G\mu$ in the four stable-string models without SMBHBs (\textsc{stable-c}, \textsc{stable-k}, \textsc{stable-m}, and \textsc{stable-n}) in the left panel and including SMBHBs (\textsc{stable-c$+$smbhb}, \textsc{stable-k$+$smbhb}, \textsc{stable-m$+$smbhb}, and \textsc{stable-n$+$smbhb}) in the right panel. The dashed vertical lines show the $68\%$ Bayesian credible intervals, which we construct as described in the text, and the solid vertical lines mark the $G\mu$ values where the $K$ ratio in Eq.~\eqref{eq:Kratio} reaches a value of $\sfrac{1}{10}$. Fig.~\ref{fig:string_stable_corner} in Appendix~\ref{app:strings} shows an extended version of this plot that includes the SMBHB parameters $\abhb$ and $\gbhb$.\label{fig:string_stable_post}}
\end{figure*}

Finally, we also consider cosmic superstrings:\medskip\\
\textsc{super}: \textit{Cosmic superstrings}; suppressed intercommutation probability $P$; GWs from cusps on loops, $q=4/3$.\medskip\\
Cosmic superstrings are characterized by a reduced intercommutation probability $P$ as a consequence of their quantum-mechanical nature. In superstring networks, intercommutation events are, therefore, less frequent, which demands longer numerical simulations (potentially up to a factor $10^3$ longer). Such simulations have not yet been carried out, which makes the prediction of the GW signal from cosmic superstrings rather uncertain at present. In the absence of a more sophisticated treatment, we follow the standard practice in the literature and simply estimate the GW spectrum from cosmic superstrings by rescaling the spectrum in Eq.~\eqref{eq:Ogwcs},
\begin{equation}
\label{eq:OmegacsP}
\Omega_{\scriptscriptstyle\rm GW}^{\rm cs}\left(f\right) \quad\rightarrow\quad \frac{1}{P}\,\Omega_{\scriptscriptstyle\rm GW}^{\rm cs}\left(f\right) \,,
\end{equation}
where we allow for $P$ values drawn from a log-uniform prior, $\log_{10}P \in \left[-3,0\right]$, which covers the theoretically expected range for cosmic $F$- and $D$-superstrings~\citep{Jackson:2004zg}.

\subsubsection*{Results and discussion}

We now turn to the outcome of our Bayesian fit analysis, discussing first our results for stable strings. As evident from Figs.~\ref{fig:bf_table} and \ref{fig:mean_spectra}, we find that stable strings are not able to provide a better fit of the PTA data than the benchmark \textsc{smbhb} model. In fact, among all exotic GWB sources considered in this work, stable cosmic strings represent the only case that consistently yields a Bayes factor in favor of the \textsc{smbhb} interpretation. Comparing the \textsc{stable-c}, \textsc{stable-k}, \textsc{stable-m}, and \textsc{stable-n} models to the \textsc{smbhb} model, we obtain $\mathcal{B}=0.277\pm 0.006$, $\mathcal{B}=0.364 \pm 0.008$, $\mathcal{B}=0.379 \pm0.008$, and $\mathcal{B}=0.307 \pm 0.006$, respectively; comparing the \textsc{stable-c$+$smbhb}, \textsc{stable-k$+$smbhb}, \textsc{stable-m$+$smbhb}, and \textsc{stable-n$+$smbhb} models to the \textsc{smbhb} model, we obtain $\mathcal{B}=0.76 \pm 0.01$, $\mathcal{B}=0.89 \pm0.02$, $\mathcal{B}=0.84 \pm 0.02$, and $\mathcal{B}=0.83 \pm 0.01$, respectively. These Bayes factors are close to unity, which means that adding a cosmic-string contribution to the GWB signal on top of the SMBHB signal does not improve the fit. Meanwhile, the larger prior volume pushes the Bayes factors to values slightly smaller than unity.

The reason for the poor performance of the stable-string models is straightforward. In order to explain the relatively large amplitude of the signal, a comparatively large value of the cosmic-string tension $G\mu$ is necessary. Large $G\mu$ values, however, tend to result in a rather flat GW spectrum in the PTA band as seen in Fig.~\ref{fig:mean_spectra_all_2} (see also, e.g., Fig.~1 in~\cite{Blanco-Pillado:2011egf} or Fig.~1 in~\cite{Blasi:2020mfx}), which clashes with the fact that the data seem to prefer a blue-tilted $h^2\Omega_{\scriptscriptstyle\rm GW}$ spectrum. If we approximate the GW signal from stable strings by a simple power law, this observation can be reformulated in the language of the $\gamma$\,--\,$A$ plot in Fig.~\ref{fig:A_gamma_comparison}. In terms of $\gamma$ and $A$, we then conclude that stable strings allow one to obtain $\gamma \lesssim 4$ only for $\log_{10} A \lesssim -15.0$, while larger values in the range $\log_{10} A \sim -\left(14.5\cdots14.0\right)$ can only be achieved for $\gamma \sim 5$. The GW spectrum from stable strings is therefore always either too weak or too flat.

In Fig.~\ref{fig:string_stable_post}, we present the reconstructed posterior distributions for the cosmic-string tension in our four stable-string models. The left panel of Fig.~\ref{fig:string_stable_post} shows our results in the absence of an additional SMBHB contribution to the signal, while the right panel displays the posterior distributions in the combined strings-plus-SMBHBs models. In the former case, we find peaked distributions centered around values of the order of $\log_{10}G\mu \sim -\left(10.5\cdots10.0\right)$. Values of the cosmic-string tension in this range represent a compromise between the two extremes discussed above: at smaller $\log_{10}G\mu$ the GW signal becomes too weak, while at larger $\log_{10}G\mu$ it becomes too flat. Moreover, we note that the order of the peaks in the posterior distributions---\textsc{stable-m}, \textsc{stable-k}, \textsc{stable-n}, \textsc{stable-c} from left to right---agrees with the ordering found in~\cite{Blanco-Pillado:2021ygr} at $\gamma \lesssim 4.6$. The underlying reason is that, for fixed $G\mu$, \textsc{stable-m} predicts the strongest GW signal in the nanohertz frequency band, followed by \textsc{stable-k} and \textsc{stable-n}, while \textsc{stable-c} predicts the weakest signal.

If we extend the stable-string models by including an SMBHB contribution to the GW signal, the posterior distributions feature not only a peak toward the upper end of the $\log_{10}G\mu$ range but also an extended plateau at small $\log_{10}G\mu$ values. This plateau corresponds precisely to the plateau that we alluded to in the definition of the $K$ ratio in Eq.~\eqref{eq:Kratio} and hence allows us to derive the following upper limits on the cosmic-string tension in the \textsc{stable-c$+$smbhb}, \textsc{stable-k$+$smbhb}, \textsc{stable-m$+$smbhb}, and \textsc{stable-n$+$smbhb} models, respectively: $\log_{10}G\mu < -9.67$, $-9.87$, $-10.10$, and $-9.71$. We recall that, for cosmic-string tensions exceeding these values, the likelihood of the strings-plus-SMBHBs model drops below $\sfrac{1}{10}$ of the likelihood of the \textsc{smbhb} benchmark model. In this region of parameter space, adding GWs from stable strings to the signal does more harm than good and is strongly disfavored by the data. For comparison, we also quote the upper limits of the $95\%$ Bayesian credible intervals for $G\mu$, specifically of the $95\%$ highest posterior density intervals (HPDIs). For each model, the HPDI is the narrowest possible range that includes 95\% of the posterior probability. By construction, the boundaries of an HPDI are at points with the same posterior probability density, and each parameter value inside the HPDI has higher posterior probability density than any parameter value outside the HPDI.\footnote{In the case of the four strings-plus-SMBHBs models, the probability density threshold for this procedure may coincide with the height of the plateau at low frequencies. If this happens, sampling fluctuations in the plateau region will lead to a set of disjoint intervals. To avoid these fictitious intervals, we set the upper limit where the posterior probability density falls to the level of the plateau and adjust the lower limit within the plateau so that the intervals contain 95\% of the posterior probability.} 
We find $\log_{10}G\mu < -9.88$, $-10.04$, $-10.26$, and $-9.90$ for \textsc{stable-c$+$smbhb}, \textsc{stable-k$+$smbhb}, \textsc{stable-m$+$smbhb}, and \textsc{stable-n$+$smbhb}, respectively.
\begin{figure*}
\centering
\includegraphics[width=0.48\textwidth]{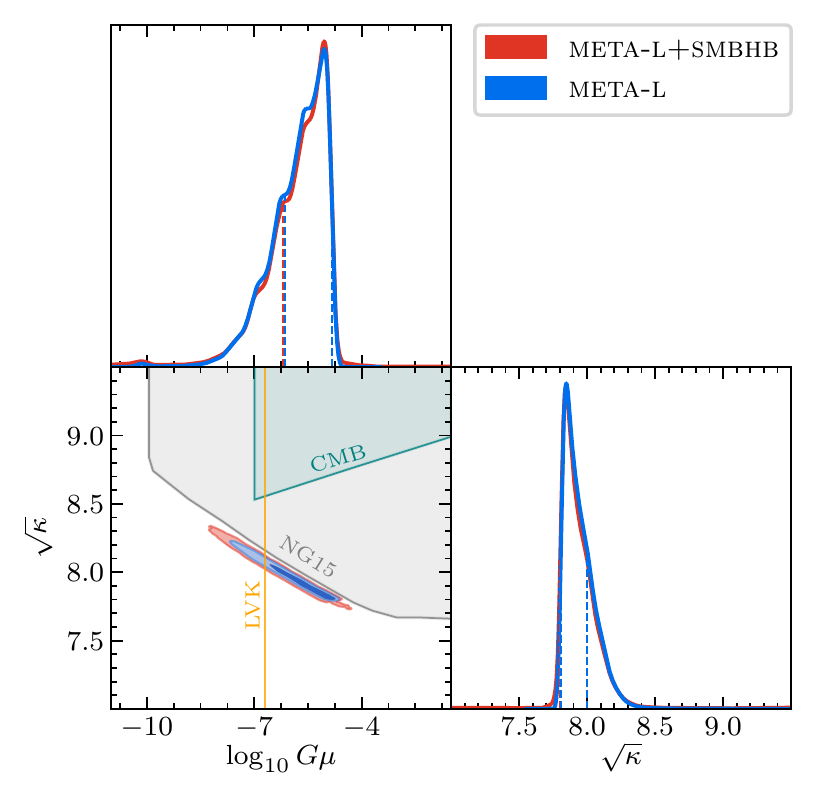}\quad
\includegraphics[width=0.48\textwidth]{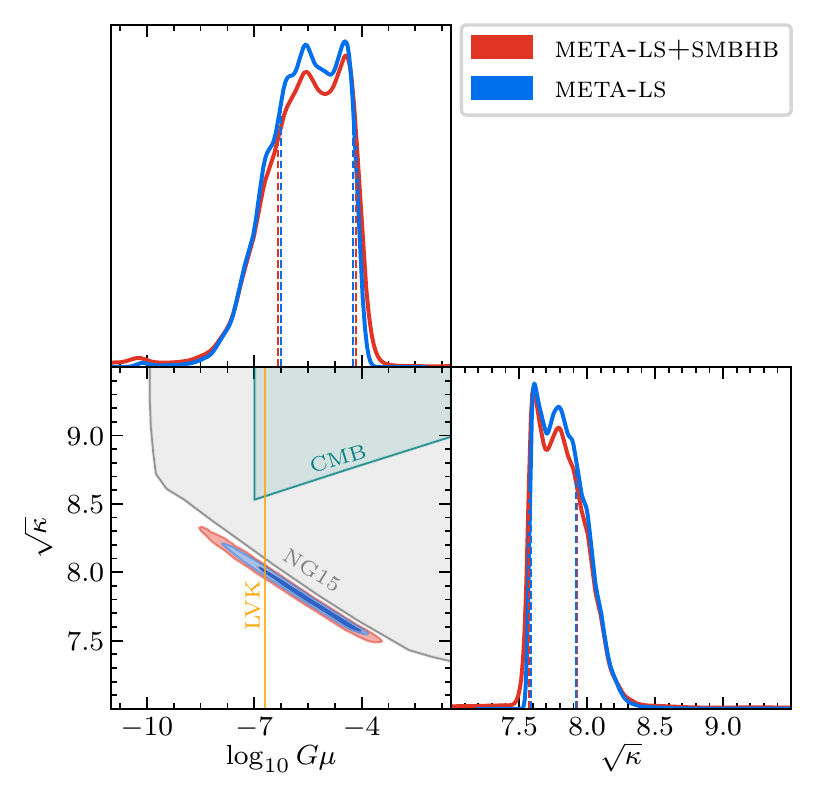}
\caption{Same as in Fig.~\ref{fig:igw_corner} but for the metastable-string models \textsc{meta-l} (GW emissions from loops only) in the left panel and \textsc{meta-ls} (GW emission from loops and segments) in the right panel. The gray shaded regions are strongly disfavored by the NG15 data, as they result in a $K$ ratio of less than $\sfrac{1}{10}$ (see Eq.~\eqref{eq:Kratio}); the teal shaded regions are ruled out by the CMB bound on cosmic strings~\citep{Planck:2015fie,Charnock:2016nzm,Lizarraga:2016onn}; and the regions \textit{to the right} of the yellow contour lines violate the LVK bound in Eq.~\eqref{eq:LVKbound}. Fig.~\ref{fig:string_meta_ext_corner} in Appendix~\ref{app:strings} shows extended versions of the two plots that include the SMBHB parameters $\abhb$ and $\gbhb$.}
\label{fig:string_meta_corner}
\end{figure*}
Next, we turn to metastable strings, which, as can be read off from Fig.~\ref{fig:bf_table}, provide a better fit to the data. In the absence of an SMBHB contribution to the GW signal, the Bayes factors for the \textsc{meta-l} and \textsc{meta-ls} models are $\mathcal{B}=13.4 \pm 0.4$ and $\mathcal{B}=21.3 \pm 0.8$, respectively. Adding an SMBHB contribution to the GW signal, the Bayes factors for the \textsc{meta-l} and \textsc{meta-ls} models are instead $\mathcal{B}=11.1 \pm 0.3$ and $\mathcal{B}=18.9 \pm 0.7$, respectively. Again, the fit does not become better, but the larger prior volume results in a decrease of the Bayes factors. 

The reconstructed 1D and 2D posterior distributions for the two metastable-string models are shown in the corner plots in Fig.~\ref{fig:string_meta_corner}. From these plots, we read off that the NG15 data prefer values of the decay parameter $\sqrt{\kappa}$ of around $\sqrt{\kappa} \sim 8$ in combination with a large cosmic-string tension, $\log_{10}G\mu \sim -\left(7\cdots4\right)$. Here $\sqrt{\kappa} \sim 8$ causes a suppression of the GW signal in the nanohertz band compared to stable strings because of the exponential factor in Eq.~\eqref{eq:Elt}. This suppression results in a decreased signal strength and a larger spectral tilt. The decrease in signal strength can, however, be compensated by a large cosmic-string tension, such that the final spectrum still has the right amplitude but is now more blue-tilted than in the stable-string case. This interplay of the different factors affecting the GW spectrum from metastable strings effectively leads to a better fit.

Let us comment on a few characteristic features of the posterior distributions shown in Fig.~\ref{fig:string_meta_corner}. First, we find that the $\log_{10} G\mu$ posterior in the \textsc{meta-ls} model extends to slightly larger values than in the \textsc{meta-l} model, which demonstrates the effect of the additional contribution to the GW signal from string segments. In fact, in the region of highest posterior density, the segment contribution dominates over the loop contribution in the \textsc{meta-ls} model. Second, we point out that the 1D marginalized posterior distributions for $\log_{10}G\mu$ exhibit small local maxima at $\log_{10}G\mu \sim -\left(11\cdots10\right)$, which correspond to the stable-string limit within the metastable-string models. This limit is realized for large values of the decay parameter, $\sqrt{\kappa} \gtrsim 9$, which pushes the effect of the network decay to frequencies below the PTA band. Next, we observe that for \textsc{meta-l} the $\log_{10} G\mu$ posterior experiences a sharp drop-off at $\log_{10} G\mu \sim -5$, whereas for \textsc{meta-ls}, there is a small dip in the $\log_{10} G\mu$ posterior at $\log_{10} G\mu \sim -5$. Both features can be traced back to the Heaviside theta function in Eq.~\eqref{eq:nl-meta}, which ensures that no more new loops are formed during the decay regime of the network. Because of this Heaviside theta function, the loop contribution to the GW spectrum moves to frequencies above the PTA band if we raise $\log_{10} G\mu$ above $\log_{10} G\mu \sim -5$. In this sense, the drop-off and the dip in the $\log_{10} G\mu$ posteriors should be regarded as being due to our simplified theoretical modeling of the GW spectrum. More work is necessary to improve on the description in terms of a simple Heaviside theta function and obtain a better understanding of the evolution of the decaying network. We expect that a more accurate description of the transition from the scaling to the decay regime would result in smoother $\log_{10} G\mu$ posteriors. Finally, we note that some of the fluctuations in the posteriors in Fig.~\ref{fig:string_meta_corner} can be attributed to the fact that, in the case of the metastable-string models, we have to work with tabulated data for the GW spectrum, based on the numerical evaluation of Eq.~\eqref{eq:Ogwcs}. 

We also use the corner plots in Fig.~\ref{fig:string_meta_corner} to highlight the relevant bounds on the parameter space spanned by $\log_{10}G\mu$ and $\sqrt{\kappa}$. The gray shaded areas notably indicate the $K$-ratio bound, which marks the parameter regions that are ruled out by the NANOGrav data. In these regions, the GW signal from metastable strings exceeds the observed signal and hence is unacceptably large. We find that the NANOGrav bound is stronger than the well-known CMB bound, which demands that a cosmic-string network that has not yet decayed by the time of recombination must not have a cosmic-string tension larger than $\log_{10}G\mu \simeq -7$~\citep{Planck:2015fie,Charnock:2016nzm,Lizarraga:2016onn}. In order to derive the CMB bound, we estimate that a decaying network completely disappears because of GW emission at times around $t_e \simeq \left(2/\left(\Gamma G\mu\right)\right)^{1/2}t_s$, which is the time when the second term in the exponent in Eq.~\eqref{eq:Elt} becomes large. The teal shaded areas in Fig.~\ref{fig:string_meta_corner} indicate where the conditions $\log_{10}G\mu > -7$ and $t_e > t_{\rm rec} \simeq 370,000\,\textrm{yr}$ are satisfied simultaneously. Last but not least, the yellow solid lines represent the LVK bound in Eq.~\eqref{eq:LVKbound}, which translates to the upper limit $G\mu \lesssim 2 \times 10^{-7}$.%
\footnote{In our analysis, we work with the upper limit on the amplitude of a generic isotropic and stochastic GWB that was reported by the LVK Collaboration in~\cite{KAGRA:2021kbb}. Combining this limit with our own GW spectra, we find $G\mu \lesssim 2 \times 10^{-7}$. This bound needs to be compared to the results of~\cite{LIGOScientific:2021nrg}, a dedicated search for GWs from stable cosmic strings by the LVK Collaboration. The analysis in~\cite{LIGOScientific:2021nrg} models the GW spectrum from cosmic strings based on slightly different assumptions. Model A in~\cite{LIGOScientific:2021nrg} is, however, similar to our approach and leads the same bound on $G\mu$ in the limit of a large number of kinks, $N_k \gg 1$ (see the panel for model A in Fig.~3 of~\cite{LIGOScientific:2021nrg} at $N_k \sim 100\cdots200$).}

The LVK bound on the amplitude of the stochastic GWB appears to be in tension with most of the parameter space preferred by the NANOGrav data. In particular, the $68\%$ credible regions in the $\log_{10}G\mu$\,--\,$\sqrt{\kappa}$ parameter plane are mostly ruled out by this bound. However, we stress that the LVK bound in Fig.~\ref{fig:string_meta_corner} relies on an extrapolation of the GW signal across 10 orders of magnitude in frequency, from PTA frequencies, $f \sim \textrm{few} \times 10^{-9}\,\textrm{nHz}$, to LVK frequencies, $f \sim \textrm{few} \times  10\,\textrm{Hz}$. In order to perform this extrapolation, we have to assume a cosmological expansion history across 10 orders of magnitude in temperature, even though not much is known about the equation of state of the universe prior to BBN. Any deviation from the expansion history of standard Big Bang cosmology can therefore affect the extrapolation of the GW spectrum to higher frequencies and potentially render the LVK bound harmless. A simple example is an early stage of matter domination~\citep{Allahverdi:2020bys}, which would suppress the GW signal from metastable strings at high frequencies and thus allow for the possibility of cosmic-string tensions as large as those in Fig.~\ref{fig:string_meta_corner}. On the other hand, if we trust the extrapolation to higher frequencies, we conclude that some parts of the $95\%$ credible regions in Fig.~\ref{fig:string_meta_corner} are in fact not ruled out. In this case, metastable strings with $\sqrt{\kappa} \sim 8$ and $\log_{10}G\mu \sim -7$ are still able to provide a good fit of the data, which represents a particularly interesting scenario for two reasons: a value of $\log_{10}G\mu \sim -7$ would point to a GUT origin of the string network, and ground-based interferometers would be poised to detect the stochastic GWB signal from such a network in the near future. 

Finally, we discuss our results for cosmic superstrings, for which we obtain the largest Bayes factors among all cosmic-string models considered in this work: $\mathcal{B}=46 \pm  2$ for GWs from superstrings alone and $\mathcal{B}=37 \pm 2$ for an SMBHB contribution added to the superstring GW signal, both compared to the \textsc{smbhb} model.

\begin{figure}
	\centering
        \includegraphics[width=0.48\textwidth]{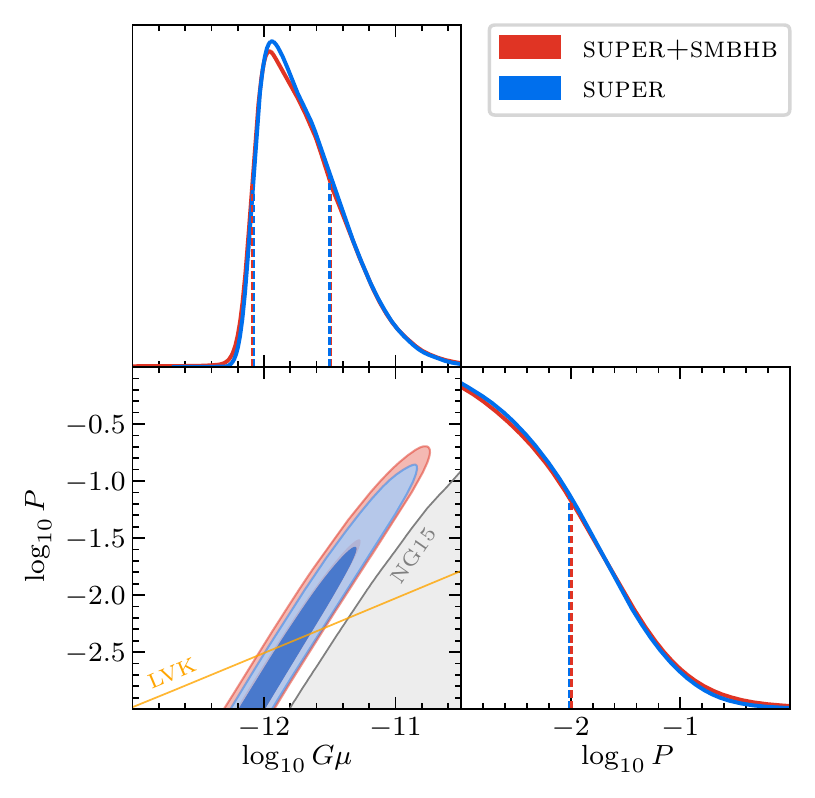}
	\caption{Same as in Fig.~\ref{fig:string_meta_corner}, but for the \textsc{super} model. Fig.~\ref{fig:superstrings_ext_corner} in Appendix~\ref{app:strings} shows an extended version of this plot that includes the SMBHB parameters $\abhb$ and $\gbhb$.\label{fig:string_super_corner}}
\end{figure}

Superstrings result in a good fit of the data because, unlike ordinary stable strings, they permit a GW signal that is neither too weak nor too flat. To see this, note that small cosmic-string tensions, $\log_{10}G\mu \lesssim -11$, yield a blue-tilted $h^2\Omega_{\scriptscriptstyle\rm GW}$ spectrum at nanohertz frequencies (see our discussion above). In the case of ordinary strings, the amplitude of this spectrum would be too weak; however, for superstrings the suppression of the GW signal at $\log_{10}G\mu \lesssim -11$ can be compensated by the $1/P$ enhancement factor in Eq.~\eqref{eq:OmegacsP}. By choosing $P$ appropriately, the amplitude of the GW signal can then be raised until a good fit of the data is reached. This interplay between the two parameters of the \textsc{super} model can also be observed in Fig.~\ref{fig:string_super_corner}. The 2D posterior distribution for $\log_{10} G\mu$ and $P$ in this corner plot displays a strong covariance in line with our heuristic understanding.

The highest posterior density is achieved at small intercommutation probabilities and cosmic-string tensions, $\log_{10} P \sim -3$ and $\log_{10} G\mu \sim -12$, where $\log_{10}P = -3$ corresponds in fact to the edge of the prior range for $P$. This parameter region is in tension with the LVK bound in Eq.~\eqref{eq:LVKbound} (see the yellow solid line in Fig.~\ref{fig:string_super_corner}) but may be viable assuming a nonstandard expansion history at high temperatures. Similarly to what is found in Fig.~\ref{fig:string_meta_corner}, we highlight the region ruled out by the NG15 data because it leads to a $K$ ratio of less than $\sfrac{1}{10}$. As expected, this region is located to the lower right of the $68\%$ and $95\%$ credible regions, where small $\log_{10}P$ values cause the amplitude of the GW spectrum to exceed the measured strength of the signal. The tension of cosmic superstrings can also be constrained by CMB observations~\citep{Charnock:2016nzm}. Existing upper limits are, however, located at larger $G\mu$ values and thus not relevant for our results in Fig.~\ref{fig:string_super_corner}. 

In conclusion, we stress again that the GW spectrum from cosmic superstrings is not well understood at present, and more work is needed to arrive at a reliable prediction. Such an effort will be important to confirm that GWs from cosmic superstrings do indeed represent a realistic and viable interpretation of the PTA signal.

\subsection{Domain walls}
\label{subsec:domain_walls}

\subsubsection*{Model description}
Domain walls are 2D topological defects that form when a cosmological phase transition results in the spontaneous breaking of a discrete symmetry, filling the universe with patches in different degenerate vacua~\citep{Kibble:1976sj}. In the absence of significant interactions with relativistic particles, domain wall networks are expected to reach a scaling regime in which each Hubble volume, $H^{-3}$, contains $\mathcal{A}\sim\mathcal{O}(1)$ domain walls (see, e.g.,~\citet{Press:1989yh, Hindmarsh:1996xv, Hindmarsh:2002bq,Garagounis:2002kt}). In this regime, the energy density stored in domain walls is given by 
\begin{equation}\label{eq:dw_sclaing}
    \rho_{\scriptscriptstyle\rm DW}\sim \mathcal{A}\,\sigma H \,,
\end{equation}
where $\mathcal{A}$ is nearly constant and $\sigma$ is the domain wall tension, which gives the domain walls their surface energy density. During the scaling regime, the energy density stored in domain walls dilutes more slowly than that of relativistic radiation and nonrelativistic matter. Indeed, the total energy density of the background always scales like $\rho_c \propto H^2$, which means that $\Omega_{\scriptscriptstyle\rm DW} = \rho_{\scriptscriptstyle\rm DW}/\rho_c$ grows like the Hubble radius when the domain wall network is in the scaling regime, $\Omega_{\scriptscriptstyle\rm DW} \propto H^{-1}$. Therefore, domain walls eventually overclose the universe and alter the cosmological evolution in a way that is incompatible with CMB observations~\citep{Zeldovich:1974uw}. A possible solution to this problem is to have domain walls decay at some temperature $T_*$ by assuming, e.g., that the global symmetry responsible for domain wall formation is explicitly broken. Indeed, an explicit breaking of the symmetry introduces a bias among the possible low-energy vacua, thus lifting their degeneracy, which eventually leads to the collapse of the domain wall network~\citep{Vilenkin:1981zs, Gelmini:1988sf, Larsson:1996sp}.

\begin{figure*}
        \includegraphics[width=0.48\textwidth]{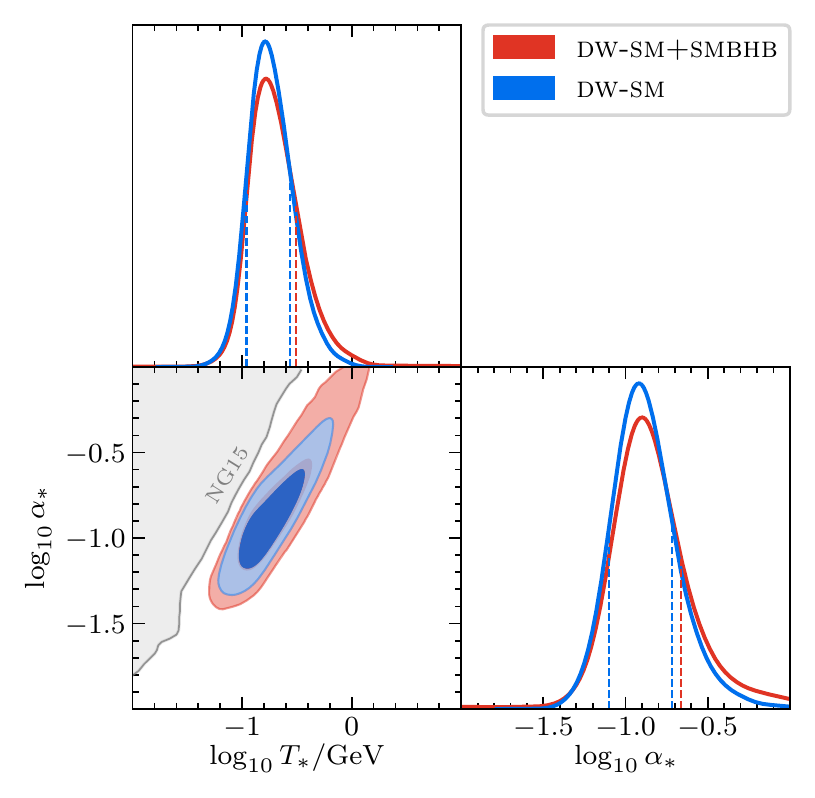}
        \includegraphics[width=0.48\textwidth]{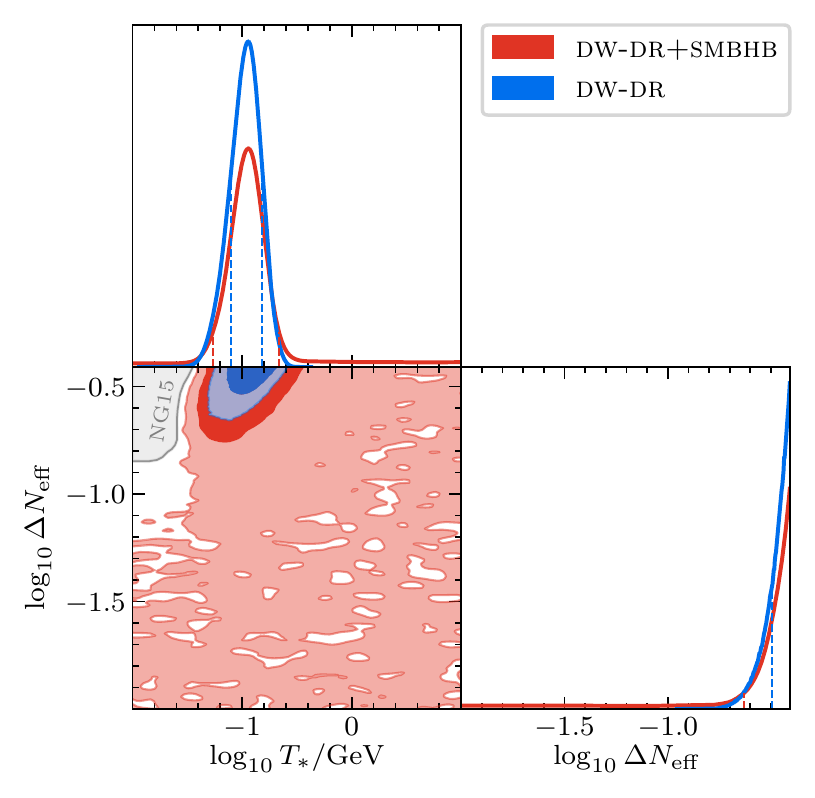}
	\caption{\label{fig:dw_corner}
        Same as in Fig.~\ref{fig:igw_corner}, but for domain walls decaying to SM particles (left panel) and a dark sector (right panel). Fig.~\ref{fig:dw_corner_full} in Appendix~\ref{app:dw} shows extended versions of the two plots that include the spectral shape parameters $b$, $c$ and the SMBHB parameters $\abhb$ and $\gbhb$.}
\end{figure*}

During the scaling regime, domain walls continuously change their configuration and shrink in order to maintain the scaling relation in Eq.~\eqref{eq:dw_sclaing}. While these processes take place, a fraction of the domain wall energy is released in the form of GWs, which produce a GWB with a present-day relic abundance given by~\citep{Vilenkin:1981zs, Preskill:1991kd,  Gleiser:1998na,Chang:1998tb, Hiramatsu:2010yz, Kawasaki:2011vv}
\begin{equation}
    h^2\Omega_{\scriptscriptstyle\rm GW}(f) = \frac{3}{32\pi}\mathcal{D}\,\tilde\epsilon\,\alpha_*^2\, \mathcal{S}(f/f_p) \,,
\end{equation}
where $\mathcal{D}$ is the dilution factor defined in Eq.~\eqref{eq:dilution}, $\tilde\epsilon=0.7$ is an efficiency coefficient derived from numerical simulations~\citep{Hiramatsu:2013qaa}, and $\alpha_*$ is the fraction of the total energy density stored in domain walls at $T_*$,
\begin{equation}
    \alpha_* \equiv \Omega_{\scriptscriptstyle\rm DW} \left(T_*\right) = \left.\frac{\rho_{\scriptscriptstyle\rm DW}}{3H^2M_{\scriptscriptstyle \rm Pl}^2}\right|_{T=T_*}\,.
\end{equation}
The GWB is dominated by the emission taking place just before the decay of the domain wall network. As the GWs are predominantly sourced by horizon-scale structures in the network, the typical frequency of the GWs emitted around this time is $H_*$, i.e., the Hubble scale at $T_*$. Therefore, after redshifting until today, the GWB is expected to exhibit a peak frequency
\begin{equation}
    f_p=1.14\,{\rm nHz}\left(\frac{10.75}{g_{*,s}}\right)^{1/3}\bigg(\frac{g_*}{10.75}\bigg)^{1/2}\left(\frac{T_*}{10\,{\rm MeV}}\right)\,,
\end{equation}
with both $g_*$ and $g_{*,s}$ evaluated at $T_*$. 
For the spectral shape, $\mathcal{S}(x)$, we use a parameterization similar to the one used for the phase transition spectrum in Section \ref{subsec:phase_transitions}:\footnote{Since the efficiency coefficient $\tilde\epsilon$ is derived by matching the value of the domain wall spectrum at the peak frequency, Eq.~\eqref{eq:dw_spec_shape} does not contain the normalization coefficient $\mathcal{N}$ appearing in Eq.~\eqref{eq:pt_spec_shape}.}
\begin{equation}\label{eq:dw_spec_shape}
    \mathcal{S}(x)= \frac{(a+b)^c}{(bx^{-a/c}+ax^{b/c})^c}\,.
\end{equation}
Causality fixes $a=3$, while numerical simulations~\citep{Hiramatsu:2013qaa} suggest $b\simeq c\simeq1$. In our analysis, we therefore fix the low-frequency slope of the domain wall signal to the value predicted by causality, and, following~\cite{Ferreira:2022zzo}, we allow $b$ and $c$ to float within the uniform prior ranges $b\in[0.5,1]$ and $c\in[0.3,3]$.

In our analysis we consider two possible decay channels for the domain wall network: dark radiation (\textsc{dw-dr}) and SM particles (\textsc{dw-sm}). In the \textsc{dw-dr} model, instead of using $\alpha_*$, we parameterize the strength of the domain wall signal in terms of the dark radiation produced in the decay, $\rho_{\scriptscriptstyle\rm DR}$. This quantity is usually expressed in terms of the effective number of extra neutrino species, $\Delta N_{\rm eff}$, which we defined in Section~\ref{subsec:inflation} and which is bounded from above by BBN and CMB observations, $\Delta N_{\rm eff}^{\rm max} \sim \textrm{few} \times 0.1$~\citep{Pisanti:2020efz,Yeh:2020mgl}. In our MCMC analysis of the \textsc{dw-dr} model, we follow \cite{Ferreira:2022zzo} and use $\Delta N_{\rm eff}^{\rm max} = 0.39$ as the upper bound on our prior for the parameter $\Delta N_{\rm eff}$. Assuming that all the domain wall energy is converted to dark radiation at $T_*$, $\Delta N_{\rm eff}$ is related to $\alpha_*$ by
\begin{equation}
    \Delta N_{\rm eff} \simeq 0.62\,\left(\frac{10.75}{g_*}\right)^{1/3}\left(\frac{g_{*,s}}{g_*}\right)\bigg(\frac{\alpha_*}{0.1}\bigg)\,,
\end{equation}
where, as before, both $g_*$ and $g_{*,s}$ are evaluated at $T_*$. 

In the \textsc{dw-sm} model, BBN restricts the possible values of the decay temperature to $T_*\gtrsim2.7\,{\rm MeV}$~\citep{Jedamzik:2006xz, Bai:2021ibt} for any detectable value of $\alpha_*$. Following~\cite{Ferreira:2022zzo}, we also impose $\alpha_*<0.3$ to avoid any possible deviation from radiation domination and to evade bounds from $\Delta N_{\rm eff}$.

\subsubsection*{Results and discussion}
The reconstructed posterior distributions for the parameters $T_*$ and $\alpha_*$ ($T_*$ and $\Delta N_{\rm eff}$) of the \textsc{dw-sm} (\textsc{dw-dr}) model are reported in Fig.~\ref{fig:dw_corner}, for both the case where the domain walls are assumed to be the only source of GWs (blue contours) and the scenario where we consider the superposition of the domain wall and SMBHB signals (red contours). Full corner plots including the posterior distributions of the spectral shape parameters $b$ and $c$ are reported in Fig.~\ref{fig:dw_corner_full} in Appendix~\ref{app:dw}. 

For both the \textsc{dw-sm} and \textsc{dw-dr} models, with and without the inclusion of the SMBHB signal, we find that the GWB produced by domain walls peaks around $10^{-8}$ Hz such that most of the low frequency bins are fit by the low-frequency tail of the spectrum (see Figs.~\ref{fig:mean_spectra} and \ref{fig:mean_spectra_all_1}). Specifically, for the \textsc{dw-sm} (\textsc{dw-dr}) model, we find that $T_*\in [110, 275]$ ($[79, 153]$) MeV at the $68\%$ credible level and $T_*\in[76, 505]$ (${[54, 198]}$) MeV at the $95\%$ credible level. When including the SMBHB contribution to the GWB, we find $T_*\in[108, 309]$ ($[54, 216]$) MeV at the $68\%$ credible level and $T_*\in{[67, 843]}$ (no bound on $T_*$) MeV at the $95\%$ credible level for the \textsc{dw-sm} (\textsc{dw-dr}) model. We notice that, with and without the inclusion of SMBHBs, the recovered transition temperature for the \textsc{dw-sm} model is high enough to evade BBN constraints. 

For the \textsc{dw-dr} model, the posterior distribution for $\Delta N_{\rm eff}$ is peaked near the upper prior boundary, signaling that larger values fit the observed signal better. Specifically, we find $\Delta N_{\rm eff}\gtrsim{0.32}$ at the $68\%$ credible level and $\Delta N_{\rm eff}\gtrsim{0.25}$ at the $95\%$ credible level. Including the contribution from SMBHBs allows the distribution for $\Delta N_{\rm eff}$ to extend to lower values, and we find $\Delta N_{\rm eff}\gtrsim{0.23}$ at the $68\%$ credible level and no bound at the $95\%$ credible level. We thus conclude that the \textsc{dw-dr} model prefers large $\Delta N_{\rm eff}$ values in the vicinity of existing bounds. This means that the most promising parameter regions, i.e., regions that are not yet ruled by $\Delta N_{\rm eff}$ but still manage to fit the NANOGrav signal, point to $\Delta N_{\rm eff}$ values within the reach of upcoming experiments, including CMB-S4~\citep{CMB-S4:2022ght}, which promises to be sensitive to $\Delta N_{\rm eff}$  values as small as $\Delta N_{\rm eff} \simeq 0.06$ at the $95\%$ confidence level.

For the \textsc{dw-sm} model, we find $\alpha_*\in[0.080, 0.19]$ at the $68\%$ credible level and $\alpha_*\in[0.053, 0.35]$ at the $95\%$ credible level. With the inclusion of the SMBHB contribution, smaller values of $\alpha_*$ become allowed, and we find $\alpha_*\in[0.079, 0.22]$ at the $68\%$ credible level and $\alpha_*\in[0.047, 0.61]$ at the $95\%$ credible level. We also notice a partial degeneracy between $T_*$ and $\alpha_*$, due to low frequency bins (which are the ones contributing the most to the evidence for a GWB) being fit by the low-frequency tail of the domain wall spectrum. Therefore, a shift to higher transition temperatures (and therefore to higher peak frequencies) can be partially reabsorbed with a shift to higher $\alpha_*$ values. 

For both \textsc{dw} models, we identify regions of the parameter space where the GW signal from domain walls is too strong to be compatible with NG15 data. These regions, shaded in gray in Fig.~\ref{fig:dw_corner}, illustrate for the first time how PTA data can be used to constrain models of domain walls.

Finally, we state the Bayes factors for the model comparison between the domain wall models and the \textsc{smbhb} reference model. The Bayes factors for \textsc{dw-sm} and \textsc{dw-dr} versus \textsc{smbhb} are $\mathcal{B}={14.8 \pm 0.5}$ and $\mathcal{B}={1.62 \pm 0.05}$, respectively, while the Bayes factors for \textsc{dw-sm$+$smbhb} and \textsc{dw-dr$+$smbhb} versus \textsc{smbhb} are $\mathcal{B}={21.1 \pm 0.9}$ and $\mathcal{B}={2.53 \pm 0.10}$, respectively. In both cases, the extra SMBHB contribution helps to improve the fit of the NG15 data. In the case of \textsc{dw-dr}, we moreover observe the same effect as for \textsc{pt-sound} in Section~\ref{subsec:phase_transitions}: adding the SMBHB contribution to the GWB signal results in a plateau region in the posterior distribution of the \textsc{dw-dr} model parameters that manifests itself as part of the $95\%$ credible region in Fig.~\ref{fig:dw_corner}.

\section{Deterministic signals from new physics}
\label{sec:det_new_physics}
In addition to the GWB signals discussed previously, there are several new-physics theories that can imprint a deterministic signal, described by a time series $\boldsymbol{h}$, in pulsar timing data. In this section, we consider the deterministic signals induced by ultralight dark matter (ULDM) and DM substructures. After finding no statistically significant evidence for such signals in our data, we report upper limits on the allowed strength of these signals.  

\subsection{Ultralight dark matter}
\label{subsec:uldm}
\subsubsection*{Model description}

While DM constitutes roughly 27\%~\citep{ParticleDataGroup:2020ssz} of the energy density of the universe, very little is known about its fundamental properties. Consequently, a wide range of DM models remain consistent with cosmological and astrophysical observations. The lightest possible DM particles are classified as ultralight, or fuzzy, DM. These particles must be bosonic; otherwise, they could not be packed into galaxies owing to Fermi degeneracy pressure~\citep{Tremaine1979,DiPaolo2018, Savchenko2019,Alvey2020,Davoudiasl2021}. These ULDM models also generically suppress structure on small scales, allowing them to be potential solutions to the small-scale structure problems of the standard $\Lambda$CDM paradigm~\citep{Bullock:2017xww}. However, too much suppression on large scales would be in conflict with CMB measurements~\citep{Hlozek2015,Hlozek2017,Hlozek2018}, which sets a lower bound on the DM mass, $10^{-24} \, \text{eV} < m_\phi$. PTAs can probe these miniscule masses, since, as we describe below, the frequency of the ULDM signal, $f$, is generally proportional to the ULDM mass, $2 \pi f \sim m_\phi$. Therefore, the sensitivity window of NANOGrav is expected to be $10^{-23} \, \text{eV} \lesssim m_\phi \lesssim 10^{-20} \, \text{eV}$.

Other astrophysical constraints, such as measurements of the Lyman-$\alpha$ forest~\citep{Armengaud2017a,Irsic2017a,Kobayashi2017,Rogers2021}, galactic subhalo mass functions~\citep{Schutz2020a, Banik2021, DES:2020fxi}, and stellar kinematics~\citep{Dalal2022}, can push this bound further up, ranging from $10^{-21} \, \text{eV}$ to $10^{-19} \, \text{eV}$, and the nonobservation of superradiance at SMBHs~\citep{Arvanitaki:2014wva,Stott:2020gjj,Unal:2020jiy} pushes the bound up to $10^{-21} \cdots 10^{-17}  \, \text{eV}$. PTA searches for ULDM in the $10^{-23} \, \text{eV} \lesssim m_\phi \lesssim 10^{-20} \, \text{eV}$ window can then be viewed as complementary to these astrophysical searches. Signals in PTAs do not depend on the same astrophysical uncertainties, e.g., modeling the nonlinear small-scale matter power spectrum with analytic or numeric methods~\citep{Zhang2018, Zhang2019a} or the evolution of specific density profiles. Moreover, the power of these methods quickly degrades when considering \textit{subcomponents} of DM, or populations of DM that do not compose the whole DM density. The searches discussed here have a weaker dependence on the subcomponent fraction and are therefore still useful in the hunt for DM.

\setlength{\tabcolsep}{10pt}
\begin{table*}
    \centering
    \begin{tabular}{ccccc}
        \toprule\addlinespace
        Effect & $A_{ E}^i(\boldsymbol{x})$ & $A_{ P, I}^i(\boldsymbol{x})$ & $\omega = 2 \pi f$ & Spin ($S_\phi$) \\\addlinespace
        \cmidrule(lr){1-5} \\
        Metric fluctuations & $(2S_\phi+1)\displaystyle\frac{\pi G \rho_\phi}{2 m_\phi^3} \hat{\phi}^2(\boldsymbol{x})$ & $(2S_\phi+1)\displaystyle\frac{\pi G \rho_\phi}{2 m_\phi^3} \hat{\phi}^2(\boldsymbol{x})$ & $\displaystyle 2m_\phi$ & $0, 1$ \\ \addlinespace
        Doppler--$U(1)$ forces & $\displaystyle g_{ E} \frac{Q_{ E}}{M_{ E}} \frac{\sqrt{2\rho_\phi}}{\sqrt{3} m_\phi^2} \left( \hat{\boldsymbol{n}}_{ I} \cdot \bm{\epsilon}_i \right) \hat{\phi}_i(\boldsymbol{x})$ & $\displaystyle g_{ P} \frac{Q_{ P}}{M_{ P}} \frac{\sqrt{2\rho_\phi}}{\sqrt{3} m_\phi^2} \left( \hat{\boldsymbol{n}}_{ I} \cdot \bm{\epsilon}_i \right) \hat{\phi}_i(\boldsymbol{x})$ & $\displaystyle m_\phi$ & $1$  \\\addlinespace
        Pulsar spin fluctuations & 0 & $\displaystyle y_{ P} d_{ P} \frac{\sqrt{2\rho_\phi}}{m_\phi^2 \Lambda} \hat{\phi}(\boldsymbol{x})$ & $m_\phi$ & 0 \\ \addlinespace
        Reference clock shift & $\displaystyle y_{ E} d_{ E} \frac{\sqrt{2\rho_\phi}}{m_\phi^2 \Lambda} \hat{\phi}(\boldsymbol{x})$ & 0 & $m_\phi$ & $0$  \\ \addlinespace\bottomrule
    \end{tabular}
    \caption{Summary of the different effects induced by ULDM that generate a deterministic signal of the form given in Eq.~\eqref{eq:uldm_signal_shape}. ``E'' and ``P'' subscripts denote Earth and pulsar term contributions, respectively. $A^i$ is the signal amplitude for the $i$th DM polarization, $\omega$ is the signal frequency, and the ``Spin" column refers to whether the effect occurs for scalar (spin-0) or vector (spin-1) ULDM candidates. $\rho_\phi$ is the local DM density, taken to be $0.4 \, \text{GeV}/\text{cm}^3$, $m_\phi$ is the ULDM mass, $g$ are gauge couplings, $Q$ is the charge under the corresponding gauge group, $y$ parameterizes the sensitivity to different fundamental constant fluctuations, $d$ are defined in Eq.~\eqref{eq:uldm_L}, and $\Lambda = M_{\scriptscriptstyle \rm Pl}/\sqrt{4 \pi}$. $\hat{\phi}(\boldsymbol{x})$ is a random variable representing the fluctuations in the ULDM field in different correlation patches; the correlated limit corresponds to $\hat{\phi}(\boldsymbol{x}) \rightarrow \hat{\phi}$. $n_{ I}$ is a vector pointing from the Earth to the $I$th pulsar, and $\boldsymbol{\epsilon}_i$ are the ULDM polarization vectors.}
    \label{tab:ULDM_effect_summary}
\end{table*}

While there is a large phenomenology of ULDM signals that can be produced in PTAs, the timing residuals for the $I$th pulsar, $h_{ I}$, are deterministic and can be written in the form
\begin{align}
    h_{ I}(t) = & \sum_i A_{ E, I}^i(\boldsymbol{x}_{ E}) \sin( \omega t + \gamma_{ E}^i ) \nonumber \\
    \quad & + A_{ P, I}^i(\boldsymbol{x}_{ P, I}) \sin( \omega t + \gamma_{ P, I}^i ) \, ,
    \label{eq:uldm_signal_shape}
\end{align}
where $i$ indexes the DM field polarization,\footnote{Here we focus on signals from scalar (spin-0) and vector (spin-1) ULDM. We leave the search for the effects of spin-2 ULDM~\citep{Marzola:2017lbt,Armaleo:2019gil,Armaleo:2020yml,Xia:2023hov} to future work. For scalar DM, $i$ is absent, since the field has only one component. For vector DM, $i \in \{ 1, 2, 3\}$ for each massive vector polarization.} $\omega$ is the signal frequency, and we have split the signal into two terms: an ``Earth term'' with amplitude $A_{ E, I}^i(\boldsymbol{x}_{ E})$ and time-independent phase $\gamma_{ E}^i$ and a ``pulsar term'' with amplitude $A_{ P, I}^i(\boldsymbol{x}_{ P, I})$ and time-independent phase $\gamma_{ P, I}^i$. The phases can be written as
\begin{align}
    \gamma_{ E}^i & = \alpha^i(\boldsymbol{x}_{ E} )\,,\\
    \gamma_{ P,I}^i &= \omega |\boldsymbol{x}_{ E} - \boldsymbol{x}_{ P, I}| + \alpha^i(\boldsymbol{x}_{ P, I}) \,,
\end{align}
where $\alpha^i(\boldsymbol{x})$ is dependent on the underlying ULDM field phase at point $\boldsymbol{x}$, and the additional term in $\gamma_{ P, I}^i$ is due to the light-travel time between the Earth and the pulsar.

An important scale in understanding these signals is given by the correlation length of the ULDM field, $\ell_c$,
\begin{equation}
	\ell_c\simeq\frac{2\pi}{m_\phi v_\phi}\sim0.4\,{\rm kpc}\left(\frac{10^{-22}\,{\rm eV}}{m_\phi}\right)\,,
\end{equation}
where $v_\phi \sim 10^{-3}$ is the ULDM velocity. If the correlation length is much larger than the distance between Earth and the pulsar, $\ell_c \gg |\boldsymbol{x}_{ E} - \boldsymbol{x}_{ P, I}|$, both experience the same DM field. In this ``correlated'' limit of the signals, the amplitudes and phases can be taken to be position independent:
\begin{equation}
    A_{ E, I}^i(\boldsymbol{x}_{ E})  \rightarrow A_{ E}^i, \quad A_{ P}^i(\boldsymbol{x}_{ P, I}) \rightarrow A_{ P, I}^i,\quad    \alpha^i(\boldsymbol{x})  \rightarrow \alpha^i.
\end{equation}
Generally, a correlated analysis can drastically reduce the number of free amplitude parameters in the MCMC. For example, if ULDM couples to all pulsars identically, then there is only one amplitude, $A_{ P}$, instead of one for each pulsar, to account for fluctuations in the ULDM field. However, since $m_\phi |\boldsymbol{x}_{ E} - \boldsymbol{x}_{ P, I}| \gg 1$, $\gamma_{ P, I}$ and $\gamma_{ E}$ must still be taken to be independent. In the uncorrelated limit, $\ell_c \ll |\boldsymbol{x}_{ E} - \boldsymbol{x}_{ P, I}|$, the DM field is no longer correlated, and its amplitude fluctuations in different patches need to be accounted for.

In Table~\ref{tab:ULDM_effect_summary}, we summarize the ULDM signals searched for in our analysis by providing the corresponding expression for the amplitudes and signal frequency appearing in Eq.~\eqref{eq:uldm_signal_shape}. In the following, we give a brief review of each of these signals and refer the reader to the original references for more details:

\medskip\noindent\textbullet~\textit{Metric fluctuations}~\citep{Khmelnitsky:2013lxt, Porayko2014, Porayko:2018sfa, Kato2020,Nomura2020, Unal:2022ooa, Wu2022}: Oscillations in the ULDM field generate fluctuations in the local stress-energy tensor, $T_{\mu \nu}$, which are independent of any direct couplings the ULDM may have with SM fields. These fluctuations in $T_{\mu \nu}$ generate fluctuations in metric perturbations by Einstein's equations. These metric perturbations then affect the photon geodesic on its path from the pulsar and generate a timing residual with pulsar and Earth amplitudes given in Table~\ref{tab:ULDM_effect_summary}. While the amplitudes shown in Table~\ref{tab:ULDM_effect_summary} are specific to scalar (spin-$0$) ULDM, vector (spin-$1$) ULDM also generates this purely gravitational signal. The vector DM scenario is more complicated~\citep{Nomura2020}, as off-diagonal components of the metric fluctuations can generate nontrivial angular correlations between signals seen in different pulsars. In our analysis, we ignore these spatial correlations and set approximate bounds by treating the vector as three scalar components, such that the vector signal is three times larger than the scalar one~\citep{Nomura2020}. 
    
\medskip\noindent\textbullet~\textit{Doppler--$U(1)$ forces}~\citep{Graham:2015ifn, PPTA:2021uzb}: A background of ultralight vector DM can create dark ``electric" fields, which can accelerate pulsars, or the Earth, and Doppler-shift light arrival times. For example, if the DM is the gauge field of a local baryon symmetry, $U(1)_{B}$, then it couples to the Earth and pulsar baryon current density. This coupling generates a force on both Earth and pulsars proportional to their baryon number, $Q_{B} = N_{B} \times q_{B}$, where $N_{B}$ is the number of baryons and $q_{B}$ is the gauge charge. This scenario is straightforwardly generalized to other $U(1)$ models, e.g., models where the difference between baryon and lepton numbers, $B\!-\!L$, is a local symmetry. These forces cause periodic displacements between the Earth and pulsars and generate timing residuals with amplitudes given in Table~\ref{tab:ULDM_effect_summary}. These forces also exist for scalar DM coupling to SM operators (e.g., $\phi \,\bar{n} n$, where $\phi$ is the DM and $n$ is the neutron field). However, in the scalar case, these forces originate from the field gradient $\nabla \phi$~\citep{Graham:2015ifn}, which causes them to be velocity suppressed compared to the vector DM scenario.

\medskip
The next two signals we discuss are specific to scalar ULDM. All the linear couplings of a scalar ULDM field to the SM can be summarized in a single Lagrangian,
\begin{align}
    \mathcal{L} \supset\frac{\phi}{\Lambda}& \Bigg[ \frac{ d_\gamma}{4e^2}F_{\mu\nu}F^{\mu\nu}+\frac{d_g\beta_3}{2g_3}G_{\mu\nu}^{A}G_{A}^{\mu\nu}\nonumber\\ - & \sum_{f=e,\mu} d_{f} m_f\bar f f -\sum_{q=u,d}(d_q+\gamma_q d_g)m_q\bar qq\Bigg] \, ,
    \label{eq:uldm_L}
\end{align}
where $F_{\mu\nu}$ and $G_{\mu\nu}$ are the photon and gluon field strengths, respectively, $\Lambda=M_{ \rm Pl}/\sqrt{4\pi}$, $\beta_3$ is the QCD beta function, $\gamma_q$ are the light quark anomalous dimensions, $m_f$ are the fermion masses, and the $d$ values are dimensionless couplings. These couplings induce periodic oscillations in the values of fundamental constants, i.e., particle masses and couplings, which can affect timing residuals in two ways:

\medskip\noindent\textbullet~\textit{Pulsar spin fluctuations}~\citep{Kaplan2022}: Particle mass fluctuations change the moment of inertia of the pulsar, which, by conservation of angular momentum, leads to pulsar spin fluctuations. The amplitude of the signal is given in Table~\ref{tab:ULDM_effect_summary}. The $y$ parameters are pulsar specific and denote the pulsar sensitivity to a specific coupling, e.g., $y_e$ parameterizes the pulsar sensitivity to changes in the electron mass. $y_{\hat{m}}$ parameterizes the pulsar sensitivity to the mass-weighted combination of quark mass couplings, $d_{\hat{m}} = (m_u d_u + m_d d_d) / (m_u + m_d)$. To be explicit, following~\cite{Kaplan2022}, we employ in our analysis
\begin{alignat}{2}
    y_g  &= -5 \,, \qquad\qquad y_{\hat{m}} &&= -2.4 \times 10^{-1} \,,  \nonumber\\
     y_\mu &= 2 \times 10^{-3} \,,  \qquad  y_e &&= 1.7 \times 10^{-5} \,,
\end{alignat}
for each pulsar, which assumes the simplest possible model for the pulsars' moment of inertia.
    
\medskip\noindent\textbullet~\textit{Reference clock shifts}~\citep{Graham:2015ifn, Kaplan2022}: PTA timing residuals are measured with respect to a collection of mostly cesium~\citep{McCarthy2009} atomic clocks. Therefore, changes to these atomic clock frequencies can appear in PTA data as apparent shifts in timing residuals. These shifts are perfectly correlated among pulsars and have the amplitude reported in Table~\ref{tab:ULDM_effect_summary}. Similar to the pulsar spin fluctuation signal, the $y$ parameters denote the atomic clock sensitivity to specific couplings. To be explicit, following~\cite{Kaplan2022}, we take these parameters to be
\begin{alignat}{2}
        y_g  &= 1\qquad\qquad\qquad y_{\hat{m}} &&= 0.158 \nonumber \\
         y_e &= 2\quad\qquad\qquad\quad  y_\gamma &&= 4.83 \,,
\end{alignat}
assuming cesium clocks to be used as a reference.

\begin{figure}
	\centering
        \includegraphics[width=0.48\textwidth]{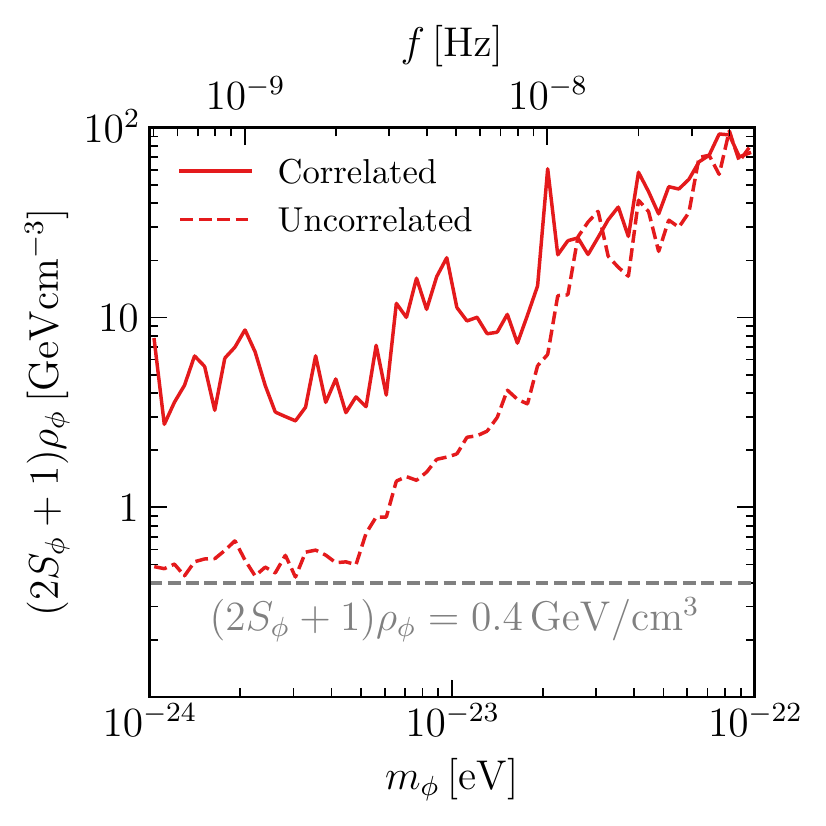}\quad
	\caption{Constraints on the local ULDM density for the correlated (solid line) and uncorrelated (dashed line) signals at the $95\%$ credible level (see the discussion in the text). The gray dashed line indicates the predicted DM abundance.  \label{fig:uldm_gravitational}}
\end{figure}

\begin{figure*}
	\centering
    \includegraphics[width=\textwidth]{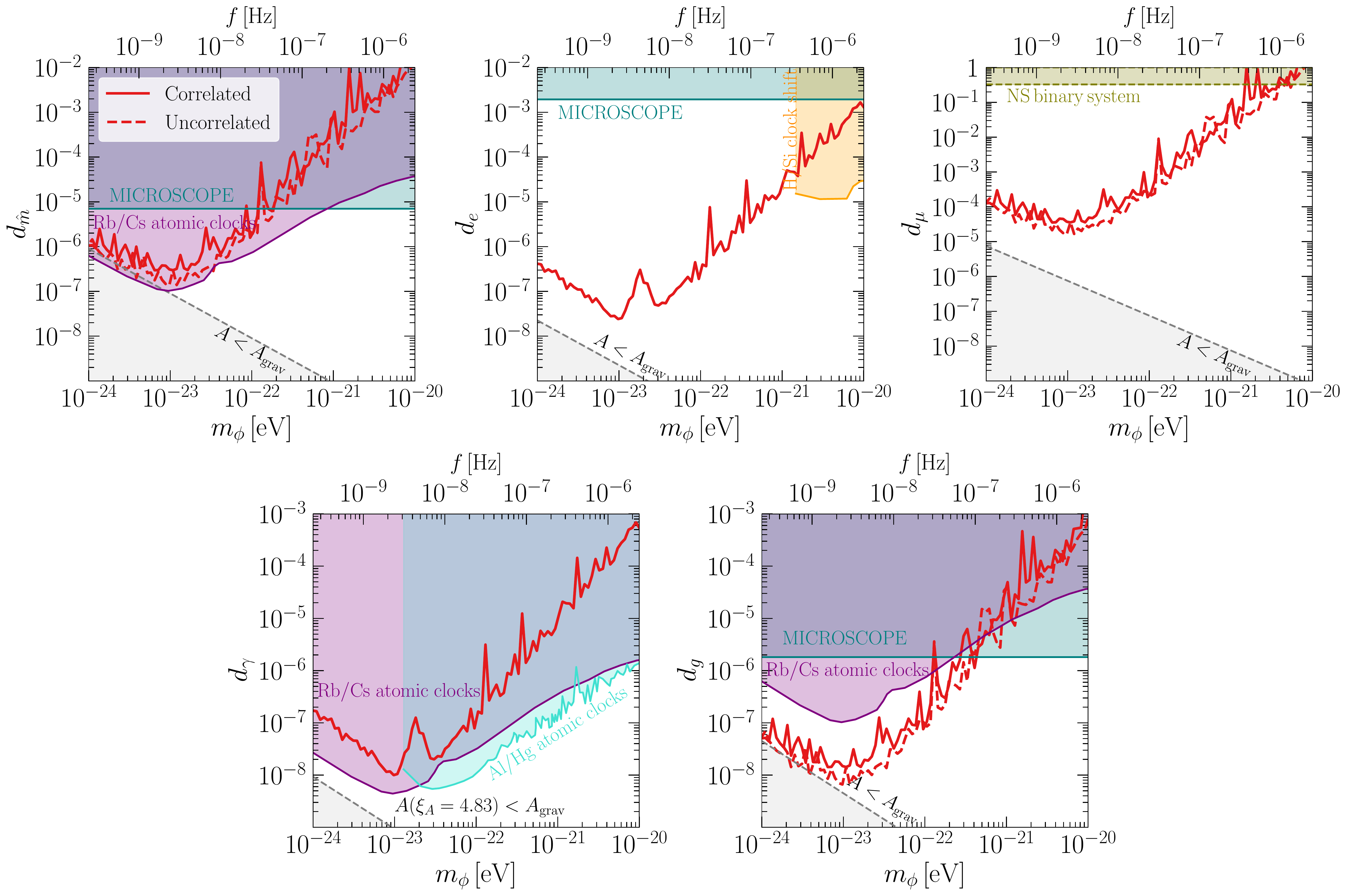}
	\caption{The red solid (dashed) lines show our constraints at the 95\% credible level on the $d$ parameters of the ULDM model, Eq.~\eqref{eq:uldm_L}, derived by searching for an (un)correlated signal. We assume that all reference clocks use the transition between the two hyperfine levels of the $^{133}{\rm Cs}$ ground state. Additionally, for each panel we assume that the $d$ parameters not shown are set to zero. The lower gray shaded regions correspond to regions of parameter space where the signal amplitude is less than the purely gravitational signal. Current constraints ``Rb/Cs atomic clocks" (purple) are from~\cite{Hees2016}, ``Al/Hg atomic clocks" (turquoise) are from~\cite{2020arXiv200514694O}, ``MICROSCOPE" (teal) are from~\cite{Berge:2017ovy}, ``H/Si clock shift" (orange) are from~\cite{2020arXiv200514694O}, and ``NS binary system" are from~\cite{KumarPoddar:2019ceq} and~\cite{ Dror:2019uea}.\label{fig:uldm_direct}}
\end{figure*}

\begin{figure*}
	\centering
        \includegraphics[width=\textwidth]{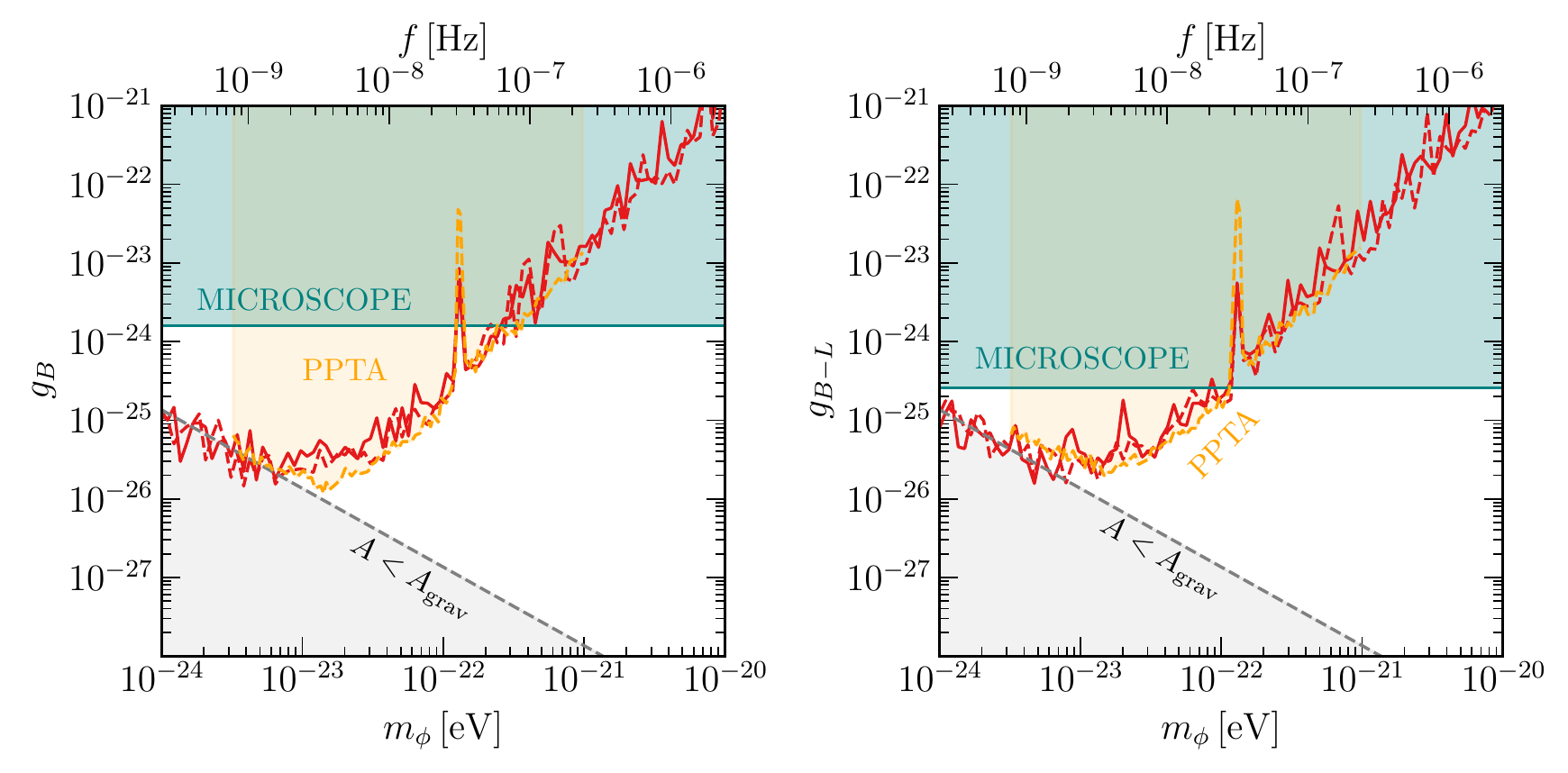}
	\caption{The red solid (dashed) lines show our constraints at the $95\%$ credible level for the two vector ULDM models considered in this analysis, for both the correlated (solid line) and uncorrelated (dashed line) regimes. The lower gray shaded regions correspond to regions of parameter space where the signal amplitude is smaller than the purely gravitational signal. The constraints from tests of the equivalence principle (``MICROSCOPE") are shown in teal~\citep{Berge:2017ovy}, and the constraints previously set by the PPTA Collaboration are reported in yellow~\citep{PPTA:2021uzb}.\label{fig:uldm_vec}}
\end{figure*}

\subsubsection*{Results and discussion}

We search for ULDM signals by performing a variety of Bayesian fit analyses on the NG15 data set. The priors for the ULDM model parameters are summarized in Table~\ref{tab:np_priors}, while the priors for the intrinsic red-noise parameters are reported in Table~\ref{tab:base_priors}. Since in this analysis we want to remain agnostic about the origin of the GWB, we model the GWB as a power law and allow the values of the amplitude and spectral index to float within the following prior ranges: $\log_{10}A_{\scriptscriptstyle\rm GWB}\in[-18,-11]$ and $\gamma_{\scriptscriptstyle\rm GWB}\in[0,7]$.

Over most of the ULDM mass range, $10^{-24} \, \text{eV} \lesssim m_\phi \lesssim 10^{-20} \, \text{eV}$, we find no significant evidence for any of the ULDM signals described in the previous section (with $\mathcal{B}\sim 1$ in favor of models including a ULDM signal on top of the GWB). However, for ULDM models that give rise to an ``Earth term" signal, we find mild evidence for a ULDM signal with frequency $f\simeq4\,{\rm nHz}$. Specifically, restricting the prior range to $f\in[3.05,6.09]\,{\rm nHz}$, we find a Bayes factor of $\mathcal{B}\sim 2$ ($\mathcal{B}\sim 1.5$) in favor of the model including the ULDM signal in the correlated (uncorrelated) regime. This result is expected given the excess of monopolar-correlated power observed in the second frequency bin of the common red-noise process (see the discussion in~\citetalias{aaa+23_gwb} for more details). Unfortunately, an ULDM interpretation of this monopolar signal is difficult, since the corresponding ULDM masses, $m_\phi\sim2\times10^{-23}\,{\rm eV}$, needed to explain this excess are in tension with other astrophysical bounds: Lyman-$\alpha$ forest~\citep{Armengaud2017a,Irsic2017a,Kobayashi2017,Rogers2021}, galactic subhalo mass functions~\citep{Schutz2020a, Banik2021, DES:2020fxi}, and stellar kinematics~\citep{Dalal2022}. 

Without convincing evidence for a signal, we compute constraints on the ULDM model parameters, shown in Figs.~\ref{fig:uldm_gravitational}-\ref{fig:uldm_vec}. All constraints are the 95th percentile of the marginalized posterior distribution for the parameter on the vertical axis. The curves labeled ``(un)correlated" correspond to the analysis done in the (un)correlated limit, discussed in the previous section.

In Fig.~\ref{fig:uldm_gravitational}, we show the constraints on the local ULDM energy density that can be derived assuming only gravitational coupling between the ULDM and SM fields. In Fig.~\ref{fig:uldm_gravitational}, we show the constraints on the local ULDM energy density that can be derived assuming only gravitational coupling between the ULDM and SM fields. The strongest bounds are obtained in the mass range $m_\phi \lesssim 10^{-23} \, \text{eV}$, where we nearly constrain ULDM to be a subcomponent of the total DM abundance. While we show constraints down to $m_\phi=10^{-24}\,{\rm eV}$, it is easy to extrapolate them to lower masses, where we expect them to remain flat down to $m\sim10^{-26}\,{\rm eV}$, where $1/m_\phi$ becomes of the same order of the interpulsar separation and the ULDM signal is additionally suppressed~\citep{Khmelnitsky:2013lxt, Unal:2022ooa}. While future PTA analyses will be able to improve on these constraints, constraining ULDM with $m_\phi\gtrsim 10^{-23}\,{\rm eV}$ to have a local abundance smaller than $\rho_\phi=0.4\,{\rm GeV}\,{\rm cm}^{-3}$ will be challenging, given that the constraints scale as $\rho_\phi^\text{lim} \propto m_\phi^3$ for $m_\phi\gtrsim 1/T_{\rm obs}$.

In Fig.~\ref{fig:uldm_direct}, we show the constraints for all the $d$ parameters describing the scalar ULDM Lagrangian in Eq.~\eqref{eq:uldm_L}. Note that, for each panel, we assume that the parameter constrained is the only nonvanishing one. Overall, the limits are in rough agreement with the projections from~\cite{Kaplan2022} and competitive with laboratory constraints~\citep{Hees2016, Berge:2017ovy,KumarPoddar:2019ceq,Dror:2019uea, 2020arXiv200514694O}. The strongest constraints, relative to laboratory bounds, are for ULDM models coupled to the electron, $d_e$, and muon, $d_\mu$, mass terms. Indeed, relative shifts of the energy levels utilized in atomic clock experiments are insensitive to the $d_e$ coupling, since atomic energy levels in different atoms scale identically with electron mass, leading to no relative energy level shifts~\citep{VanTilburg2015, Kaplan2022}. Such an insensitivity is not a problem for the PTA observable,  though, since the pulsar phase evolution is not affected in the same way as the atomic energy levels. The lack of laboratory constraints on $d_\mu$ is simply because there is not a large number of muons to study on Earth, whereas pulsars host a large number of them. As for the gravitational signal, we can extrapolate the constraints to lower masses, where we expect them to scale as $d\propto1/m_\phi$ down to $m_\phi\sim10^{-26}\,{\rm eV}$, where $1/m_\phi$ becomes comparable to the interpulsar separation.

In Fig.~\ref{fig:uldm_vec}, we show constraints on the gauge coupling of models where the ULDM is the gauge boson of either $U(1)_{ B}$ or $U(1)_{ B - L}$. Our constraints are roughly consistent with those published by the PPTA Collaboration~\citep{Porayko:2018sfa}. This result is somewhat expected: while NANOGrav observes more pulsars, the average observation time is longer in PPTA, so roughly similar bounds are expected.

The constraints presented in Figs.~\ref{fig:uldm_direct} and \ref{fig:uldm_vec} assume that ULDM makes up the entire DM content of the universe. However, if ULDM is only a subcomponent of the total DM abundance, these constraints can be easily rescaled. Indeed, from Table \ref{tab:ULDM_effect_summary}, we see that the amplitudes for the direct coupling signals scale as $d\sqrt{\rho_\phi}$ for the scalar case and $Q\sqrt{\rho_\phi}$ for the vector case. Therefore, if ULDM is only a faction $f_\phi$ of the total DM abundance, the constraints are weakened by a factor $\sqrt{f_\phi}$.

Lastly, we comment on the prior choice for $\hat{\phi}$. Previous studies~\citep{Porayko2014, Kato2020} assumed that $\hat{\phi}$ is not a random variable in the problem and placed constraints assuming that this parameter is simply $1$. However, as pointed out by~\cite{Centers:2019dyn}, this assumption is not appropriate. Since the observation time of PTAs is much smaller than the coherence time of the DM, $\tau \sim (m_\phi v^2)^{-1}$, only one instance of the field is being sampled within each correlation patch. The DM density in any correlation patch is then a random number, which follows a Rayleigh distribution with mean $\rho_\phi$, and the priors should be chosen to reflect this. We also note that while~\cite{Porayko:2018sfa} find the distribution of $\hat{\phi}$ through numerical simulation, these results are consistent with the analytic predictions by~\cite{Foster:2017hbq}.

\subsection{Dark matter substructures}
\label{subsec:dm_substructures}
\subsubsection*{Model description}
In the $\Lambda$CDM model, the structures we observe in the universe are seeded by primordial curvature fluctuations generated during inflation and then imprinted onto the DM density field. CMB observations indicate that these fluctuations have a nearly scale-invariant power spectrum on large scales (i.e., for comoving wavenumbers $k\simeq{\rm Mpc}^{-1}$). However, on smaller scales, various theories of DM leave unique fingerprints on primordial perturbations or their evolution, resulting in different predictions for the amount of subgalactic DM substructures. Consequently, measuring local DM substructures could be crucial in determining the correct model of DM. 

PBHs are perhaps the simplest example of such small-scale DM substructures. They can be formed in inflationary theories that create large density fluctuations on small scales (like the ones described in Section \ref{subsec:sigw}). Several studies investigated the possibility of identifying a galactic PBH population by analyzing the Doppler and Shapiro signals they can leave in PTA data~\citep{Seto:2007kj, Siegel:2007fz, Dror:2019twh, Ramani:2020hdo, Lee:2020wfn, Lee:2021zqw}. In this analysis, we will closely follow the method outlined by~\cite{Lee:2021zqw} to constrain the local PBH abundance.\footnote{A similar approach could be applied to set constraints on the local abundance of DM subhalos. However, we do not consider this case, since our constraints for PBHs are already quite weak. Constraints on DM subhalos would likely be even weaker, making it a less promising target for future PTAs.}

\begin{figure}
    \centering
    \includegraphics[width=0.5\textwidth]{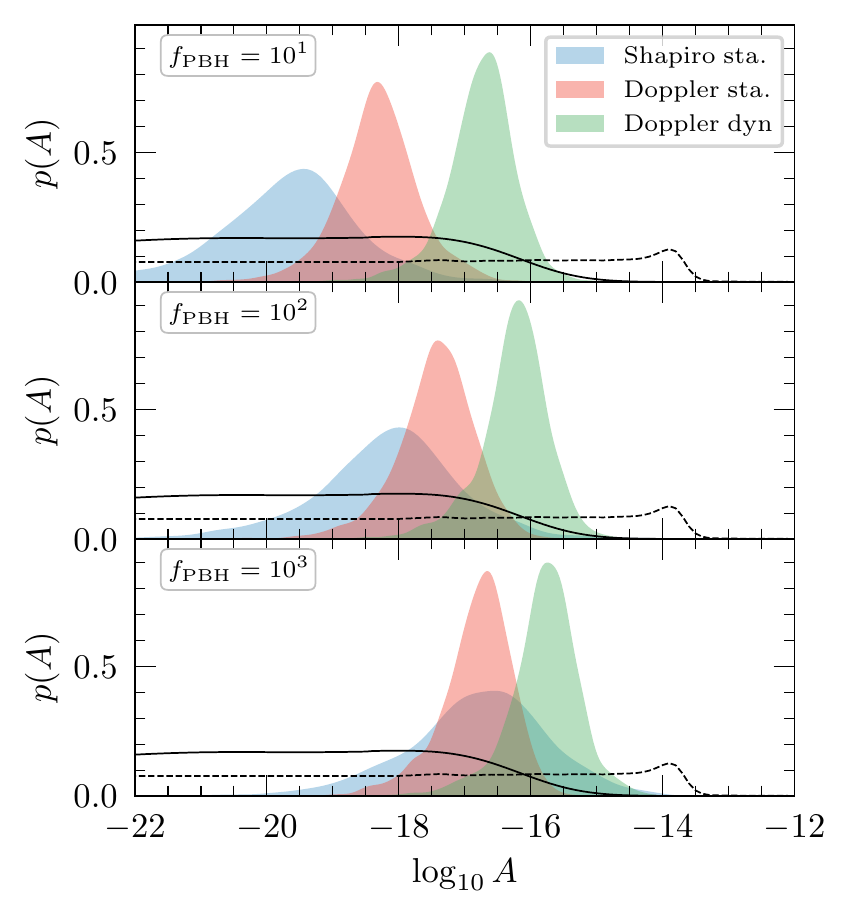}
    \caption{The black solid (dashed) lines show the posterior distributions $p(\log_{10}A_{\rm sta}|\boldsymbol{\delta t})$ ($p(\log_{10}A_{\rm dyn}|\boldsymbol{\delta t})$) for a representative pulsar (J1909-3744). The filled distributions show the conditional probability distributions $p(\log_{10}A|f_{\scriptscriptstyle \rm PBH})$ for the same pulsar and different values of $f_{\scriptscriptstyle \rm PBH}$. In this plot, $M = 10^{-6}$ ($10^{-10}$) $M_\odot$ for the Doppler static (dynamic) signal, and $M = 10 M_\odot$ for the Shapiro static signal.}
    \label{fig:pA}
\end{figure}

The Doppler signal results from the apparent shift in the pulsar spin frequency, generated by the acceleration induced by the gravitational pull of a passing PBH. According to the timescale of the transit event, $\tau$, the signal can be further classified as {\it dynamic} ({\it static}) if $\tau$ is much smaller (larger) than the observation time, $T_{\rm obs}$. In the static regime, the leading-order term of the Doppler signal that is not degenerate with the timing model is given by~\citep{Ramani:2020hdo, Lee:2020wfn, Lee:2021zqw} 
\begin{equation}\label{eq:h_D_sta}
    h_{ D, {\rm sta}}(t)=\frac{A_{ D,{\rm sta}}}{{\rm yr}^2} \, t^3 \,,
\end{equation}
where $A_{ D,{\rm sta}}$ is a dimensionless parameter that can be related to the kinematic parameters of the transiting event (see Appendix \ref{app:dm_sub} for more details). In the dynamic limit, and assuming that the signal is dominated by the closest transiting PBH, we get 
\begin{equation}\label{eq:h_D_dyn}
    h_{ D, {\rm dyn}}(t) = A_{ D, {\rm dyn}}\,(t-t_0)\,\Theta(t-t_0) \,,
\end{equation}
where $\Theta$ is the Heaviside step function, $A_{ D, {\rm dyn}}$ is a dimensionless amplitude that can be related to kinematic parameters of the transiting event, and $t_0$ is the time of closest approach (see Appendix \ref{app:dm_sub} for more details). 

The Shapiro signal refers to shifts in the TOAs caused by metric perturbations along the photon geodesic produced by PBHs transiting along the observer's line of sight. In the static limit, and after subtracting away degeneracies with timing model parameters, the leading-order term of this signal can be parameterized as
\begin{equation}\label{eq:h_S_sta}
    h_{ S,{\rm sta}}(t) = \frac{A_{ S,{\rm sta}}}{{\rm yr}^2}\,t^3 \,,
\end{equation}
where, as for the Doppler case, $A_{ S,{\rm sta}}$ is a dimensionless parameter that can be related to the kinematic parameters of the transiting event (see Appendix \ref{app:dm_sub} for more details). In the dynamic limit, there is no simple parametrization of the Shapiro signal; therefore, we do not search for this signal.

Assuming a monochromatic PBH population, our goal is to derive a posterior distribution for the PBH mass fraction, $f_{\scriptscriptstyle \rm PBH}\equiv\Omega_{\scriptscriptstyle \rm PBH}/\Omega_{\scriptscriptstyle \rm DM}$, as a function of the PBH mass, $M$: $p(f_{\scriptscriptstyle \rm PBH}|\boldsymbol{\delta t},M)$. We do this as follows:

\medskip\noindent\textbullet~For each given value of $f_{\scriptscriptstyle \rm PBH}$ and $M$, we use the Monte Carlo code developed by~\cite{Lee:2020wfn} to generate a PBH population surrounding each of the pulsars in our array. From this distribution, we derive the amplitude of the static Doppler and Shapiro signals generated by the entire PBH population and the amplitude for the dynamic Doppler signal generated by the closest transiting PBH. Finally, we repeat this procedure for $2.5\times10^3$ realizations to obtain the conditional probability distributions $p(\log_{10}A_I|f_{\scriptscriptstyle \rm PBH})$, where $I$ indexes pulsars in the array and $A$ refers to any of the PBH signal amplitudes introduced in Eqs.~\eqref{eq:h_D_sta}, \eqref{eq:h_D_dyn}, and \eqref{eq:h_S_sta}. In Fig.~\ref{fig:pA} we report some of the distributions derived in this way.\footnote{From now on, we suppress the PBH mass, $M$, in the expressions for the conditional probabilities for the sake of notation.}
    
\medskip\noindent\textbullet~One at a time, we include the PBH signals given in Eqs.~\eqref{eq:h_D_sta}, \eqref{eq:h_D_dyn}, and \eqref{eq:h_S_sta} in the timing model, and we analyze our data to derive the posterior distributions for the various PBH signal amplitudes, $p(\log_{10}\boldsymbol{A}|\boldsymbol{\delta t})$. Since the PBH signal in different pulsars is assumed to be independent, these distributions can be factorized as 
\begin{equation}
        p(\log_{10}\boldsymbol{A}|\boldsymbol{\delta t})=\prod_{ I=1}^{ N_{ P}}p(\log_{10}A_{ I}|\boldsymbol{\delta t})\,.
\end{equation}
Some of the $p(\log_{10}A_{ I}|\boldsymbol{\delta t})$ are reported in Fig.~\ref{fig:pA}. 

\medskip\noindent\textbullet~Finally, we can write
\begin{widetext}
    \begin{equation}\label{eq:dm_sub_convolution}
        p(f_{\scriptscriptstyle \rm PBH}|\boldsymbol{\delta t})=\prod_{ I=1}^{ N_{ P}}\int d \log_{10}A_{ I} \; p(f_{\scriptscriptstyle \rm PBH}|\log_{10} A_{ I})p(\log_{10} A_{ I}|\boldsymbol{\delta t}) \propto\prod_{ I=1}^{ N_{ P}}\int d \log_{10} A_{ I} \; p(\log_{10} A_{ I}|f_{\scriptscriptstyle \rm PBH})p(\log_{10} A_{ I}|\boldsymbol{\delta t})
    \end{equation}
\end{widetext}
where, in the second step, we used Bayes theorem and assumed uniform priors on $\log_{10} A_{ I}$ and $f_{\scriptscriptstyle \rm PBH}$. More details on each of these three steps can be found in Appendix~\ref{app:dm_sub} or in~\cite{Lee:2021zqw}.

\medskip
DM substructures can also possess macroscopic charges and interact with baryonic matter via long-range Yukawa interactions. These interactions can be modeled by a potential of the form
\begin{align}\label{eq:Yukawa_potential}
    V_{\mathrm{fifth}}(r) = -\tilde{\alpha}\frac{GMM_{P}}{r}e^{-r/\lambda} \, ,
\end{align}
where $M$ and $M_P$ are the masses of the DM and pulsar, respectively, $\lambda$ is the range of the interaction, and $\tilde{\alpha}$ is the effective DM-barion coupling, normalized against the  gravitational coupling (also known as the Yukawa parameter).  Here the DM can be either a particle or a macroscopic object such as a nugget of asymmetric DM~(\cite{Detmold:2014qqa,Wise:2014jva, Hardy:2014mqa,Krnjaic:2014xza,Gresham:2017zqi, Gresham:2017cvl}). 
These  Yukawa interactions can arise from an effective Lagrangian of the form $\mathcal{L}\supset g_X\phi\bar{X}X+g_n\phi\bar{n}n$, where $X$ and $n$ are the effective DM and neutron fields, and $\phi$ is a massive (but potentially light) scalar or vector field. The effective coupling is related to the coupling constants by $\tilde \alpha \approx \frac{g_n g_X}{ 4\pi G m_X m_n}$, where $m_n$ is the neutron mass. 
These interactions are constrained to be weaker than gravity for the mass range $M\lesssim 100$ GeV by the CMB, Lyman-$\alpha$ forest~\citep{Xu:2018efh}, and direct detection experiments such as X-ray Quantum Calorimeter (XQC)~\citep{Mahdawi:2018euy} and Cryogenic Rare Event Search with Superconducting Thermometers (CRESST)~\citep{CRESST:2017ues} (for a review on these constraints see \cite{Xu:2020qjk}). However, stronger-than-gravity fifth forces are allowed if $M\gg 100$ GeV, even when $X$ accounts for the entirety of the DM population.

If present, these Yukawa interactions will contribute to the pulsar's acceleration induced by a transiting DM substructure and contribute to the Doppler signal discussed before (the expression for the Yukawa contribution to the Doppler signal can be found in Appendix~\ref{app:dm_sub}). Therefore, as shown by \cite{Gresham:2022biw}, following a procedure similar to the one used to constrain the abundance of PBHs, we can constrain the value of the Yukawa parameter, $\tilde\alpha$ . Specifically, for each given value of $\tilde\alpha$ and $M$, we use the Monte Carlo code developed by~\cite{Lee:2020wfn} to generate a population of DM substructure surrounding each of the pulsars in our array. From this distribution, we derive the amplitude of the static Doppler signal generated by the closest transiting substructure by considering the acceleration induced by both the gravitational and Yukawa interaction. By repeating this procedure for multiple populations of DM substructure, we derive the distribution $p(\log_{10}A_I|\tilde\alpha)$. By plugging this quantity into an expression similar to the one given in Eq.~\eqref{eq:dm_sub_convolution}, we can derive $p(\tilde\alpha|\boldsymbol{\delta t})$ and use this quantity to constrain $\tilde\alpha$. 

\begin{figure}[t]
    \centering
    \includegraphics{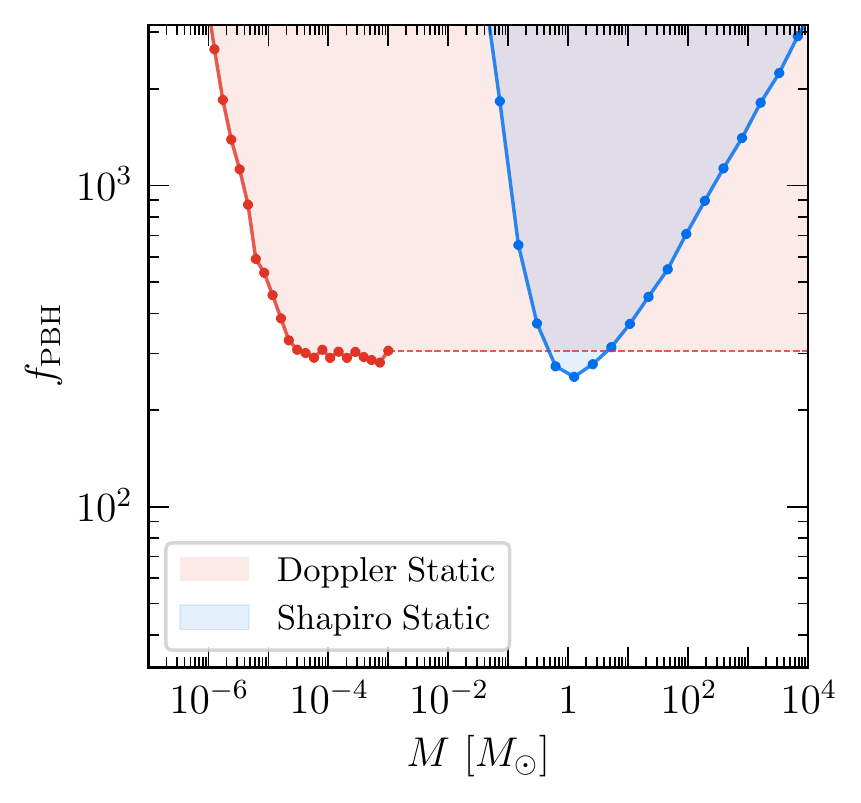}
    \caption{Constraints at the $95\%$ credible level on the local PBH abundance derived from the search for static Doppler (red shaded region) and static Shapiro signals (blue shaded region). The solid lines interpolate between the PBH masses simulated in this work, while the red dashed line shows an extrapolation of the constraints to higher masses.}
    \label{fig:pbh_limits}
\end{figure}
\subsubsection*{Results and discussion}

We start by searching for PBH signals on top of a GWB that we model as a power law with amplitude and spectral index allowed to float within the following prior ranges: $\log_{10}A_{\scriptscriptstyle\rm GWB}\in[-11,-18]$ and $\gamma_{\scriptscriptstyle\rm GWB}\in[0,7]$. We find no statistically significant evidence for any of the PBH signals described in the previous section. Therefore, we proceed to set constraints on the local PBH abundance. The prior distributions used for the PBH signal parameters are reported in Table~\ref{tab:np_priors}. 

The $95\%$ upper limits on $f_{\scriptscriptstyle \rm PBH}$ derived from the static Doppler and Shapiro signals are reported in Fig.~\ref{fig:pbh_limits}. The dynamic Doppler signal is too weak to produce any detectable signal for any of the $f_{\scriptscriptstyle \rm PBH}$ values considered.
These are the first constraints on $f_{\scriptscriptstyle \rm PBH}$ derived using real PTA data. As expected, our constraints are much weaker compared to the projections that were derived by~\cite{Lee:2021zqw} using mock data and including only white noise. Indeed, as already discussed by~\cite{Lee:2021zqw}, the presence of a common red-noise process significantly reduces the sensitivity to PBH signals. 
\begin{figure}[t]
    \centering
    \includegraphics{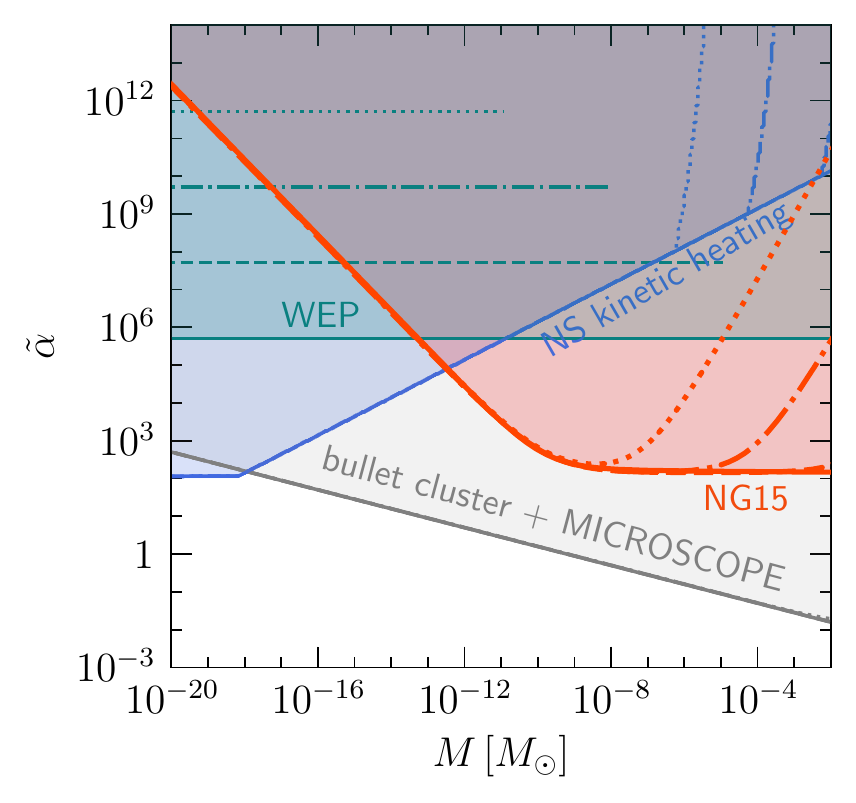}
    \caption{The 95\% credible level for the fifth-force strength $\tilde\alpha$ derived from the NG15 data (red lines) is compared with constraints from NS kinetic heating (blue lines), equivalence principle constraints (green lines), and Bullet Cluster + equivalence principle constraints (gray line). Solid (dashed) lines are deriving assuming $\lambda=1\,{\rm pc}$ ($\lambda=10^{-1}\,{\rm pc}$), while dashed-dotted (dotted) lines are derived assuming $\lambda=10^{-2}\,{\rm pc}$ ($\lambda=10^{-3}\,{\rm pc}$).}
    \label{fig:fifth_force}
\end{figure}

Finally, in Fig.~\ref{fig:fifth_force} we show the constraints on $\tilde{\alpha}$ set by NG15 data. These constraints are compared with several other constraints that can be placed on $\tilde \alpha$. Specifically, in teal we show  weak equivalence principle (WEP) constraints \citep{Wagner:2012ui, Shao:2018klg, Sun_2019} (properly rescaled to take into account the finite range of the interaction \citep{Gresham:2022biw}) derived by considering differential acceleration of baryonic test bodies toward the galactic center. In blue we report constraints from neutron star (NS) heating (assuming additional short-range DM-baryon interaction) induced by DM capture \citep{Gresham:2022biw}, derived from the coldest known NS to date - PSR J2144-3933 \citep{Guillot_2019}. And in gray we report the indirect constraints that can be derived by combining the fifth-force constraints on baryon-baryon interactions \citep{Berge:2017ovy,Fayet:2018cjy}, and Bullet Cluster constraints on DM-DM interactions \citep{Spergel:1999mh,Kahlhoefer:2013dca} (see \cite{Coskuner:2018are, Gresham:2022biw}). We find that the NG15 constraints are competitive with WEP and NS kinetic heating, especially in the $M\gtrsim 10^{-12}\,M_{\odot}$ regime. However, the combined constraint from the Bullet Cluster and MICROSCOPE dominates over all other constraints in the entire mass range of interest. We note, however, that the combined constraint is an indirect probe of the Yukawa parameter and depends on the specific form of the light mediator coupling to DM and baryons. Specifically, in deriving the combined constraints, we have assumed that the Yukawa DM-baryon interaction arises from a Lagrangian of the form $\mathcal{L}\supset g_X\phi\bar{X}X+g_n\phi\bar{n}n$. In addition, if only a subcomponent of DM interacts through the long-range fifth force, then the Bullet Cluster constraint will quickly lift, while the other constraints only deteriorate linearly with the DM fraction.

\section{Discussion}
\label{sec:results}

The analysis of the NANOGrav 15-year data set has produced the first convincing evidence for a stochastic background of GWs in the nanohertz frequency range. The origin of this background is still unknown. In this work, we considered various cosmological sources and compared them to the commonly studied astrophysical signal produced by a population of inspiraling supermassive black-hole binaries. Specifically, we considered the signals produced by nonminimal inflation scenarios,  scalar-induced GWs, cosmological phase transitions, several cosmic-string models, and domain walls.

For each of these models, we identified regions of the parameter space that are compatible with the observed signal. We find that, with the exception of stable cosmic strings of field theory origin, all new-physics models considered in this analysis are capable of reproducing the GWB signal. Many models allow us in fact to fit the signal better than the SMBHB reference model, which is reflected in Bayes factors ranging from 10 to 100 (see Fig.~\ref{fig:bf_table}). When the new-physics signals are combined with the astrophysical one, we obtain comparable results. More precisely, in several models, the addition of the SMBHB signal leads to a slight decrease of the Bayes factor, which indicates that the SMBHB contribution does not help to improve the quality of the fit but merely increases the prior volume. In other models, on the other hand, such as \textsc{pt-sound} and \textsc{dw-dr}, adding the SMBHB signal on top of the new-physics signal can lead to a slight increase of the Bayes factor, indicating that the SMBHB signal can in fact play a dominant role in the total GW spectrum. For all four stable-string models, we find Bayes factors between $0.1$ and $1$. Cosmic superstrings, on the other hand, which are also stable but not of field theory origin, can explain the data at a level comparable to other new-physics sources.

Despite the fact that some of the Bayes factors derived in this paper might suggest that a purely astrophysical interpretation of the signal is disfavored by the data, we caution against this interpretation. The Bayes factors do not account for the full range of uncertainties in both the cosmological and astrophysical signals and are prior dependent. It is conceivable that, as our understanding of SMBHB signals and our noise models improve, the tension between observations and astrophysical predictions will decrease, potentially weakening the evidence in favor of cosmological signals.

Future data sets will improve the spectral characterization of the signal and improve our ability to discriminate cosmological sources from the SMBHB signal. Unfortunately, similarities in the spectral shape and theoretical uncertainties will make it challenging to definitively determine the origin of the background using power spectrum characterization alone. However, the observation of anisotropies could eventually resolve this debate, as the expected anisotropies generated by black hole binaries are significantly larger than those produced by most cosmological sources. Similarly, the detection of a continuous wave from a single binary would provide convincing evidence in favor of an astrophysical origin of the signal. Ultimately, measurements of the GWB spectral shape and angular power spectrum may be complemented by observations of its polarization content and possible deviations from Gaussian statistics, which can again help to discriminate between an astrophysical and a cosmological origin of the signal. 

It is worth emphasizing that in all parts of our analysis  we described cosmological signals using effective parameters, e.g., $T_*$, $\alpha_*$, and $H_*R_*$, for the phase transition models. Moving forward, it will be crucial to identify microscopic models that can reproduce the values of these parameters that we found to best fit the GWB signal. That is, in order to shed more light on the various cosmological interpretations of the signal, we require a better understanding of how the NANOGrav signal could possibly result from the fundamental parameters of a particle physics Lagrangian that describes the generation of GWs in the early universe.

Along with searching for a cosmological GWB, we also analyze our data to search for deterministic signals generated by models of ULDM and DM substructures. We do not find significant evidence for either of these signals. Nonetheless, we are able to place constraints on the parameters space of these models. For a wide range of ULDM models, our constraints compete or outperform laboratory constraints in the $10^{-23}\,{\rm eV}\lesssim m_\phi\lesssim10^{-20}\,{\rm eV}$ mass window. The signal from DM substructures is harder to detect; as a consequence, we are able to set very weak constraints on the local abundance of these objects. Future data sets will improve our reach, but a better characterization of the GWB will be needed to probe realistic models of DM substructures. 

The discovery of a GWB will lead to significant breakthroughs in our understanding of cosmology and particle physics. As future PTA data sets become available, we will establish the origin of the GWB. Regardless of whether the signal is of cosmological origin, we have shown how PTAs will undoubtedly contribute to exploring new physics, either as a discovery tool or as a new way to constrain the parameter space of BSM models.

\section*{Acknowledgments}
{\it Author contributions.} An alphabetical-order author list was used for this paper in recognition of the fact that a large, decade timescale project such as NANOGrav is necessarily the result of the work of many people. All authors contributed to the activities of the NANOGrav Collaboration leading to the work presented here, and reviewed the manuscript, text, and figures prior to the paper’s submission. Additional specific contributions to this paper are as follows.

A.Mi.\ proposed and initiated the project. A.Mi.\ and K.Sc.\ coordinated and led the search. The enterprise wrapper, \texttt{PTArcade}~\citep{NG15-ptarcade}, used in this analysis was mainly developed by A.Mi., with help from D.W., K.D.O., J.N., R.v.E., T.S., and T.T. R.v.E.\ and T.S.\ developed the statistical tools needed to derive the $K$-ratio bounds. L.Z.K.\ derived the distributions of power-law fit parameters for characteristic strain spectra from SMBHB simulations. A.Mi.\ prepared all figures except those in Section \ref{subsec:uldm}, which were prepared by T.T.; Fig.~\ref{fig:fifth_force}, which was prepared by V.L.; and Fig.~\ref{fig:nrhmax}, which was prepared by R.R.L.d.S. A.Mi.\ and K.Sc.\ wrote the paper with help from K.B., K.D.O., S.Ve., and T.T.

Contributions to specific analyses are as follows. K.Sc.\ led and discussed the results of the SIGW and IGW analyses, which were performed by D.W.\ and R.R.L.d.S. A.Af.\ and R.R.L.d.S.\ derived the constraints on the parameter space of the SIGW model. D.W.\ and R.R.L.d.S.\ derived the LVK and $N_{\rm eff}$ bounds on the IGW model. A.Mi.\ led the PT analysis, which was performed by both A.Mi.\ and R.v.E.; A.Mi.\ and K.Sc.\ interpreted and discussed the results. The string analyses were led by K.D.O.\ and K.Sc., with K.D.O., K.Sc., and T.S.\ conducting the analyses; K.Sc.\ interpreted and discussed the results with help from K.D.O.\ and J.J.B.P. A.Mi.\ led the domain wall analysis, which was performed by both A.Mi.\ and D.W.; A.Mi.\ and K.Sc.\ interpreted and discussed the results. The ULDM analysis was coordinated by A.Mi., K.B., and T.T.; J.N.\ performed the analysis; and A.Mi., C.U., K.B., J.N., and T.T.\ interpreted and discussed the results. Finally, the substructure search was led by A.Mi.\ and T.T.; A.Mi., S.Ve., and V.L.\ performed the analysis; and A.Mi., S.Ve., V.L., and T.T.\ interpreted and discussed the results.

{\it Acknowledgments.}
The NANOGrav Collaboration receives support from National Science Foundation (NSF) Physics Frontiers Center award Nos. 1430284 and 2020265, the Gordon and Betty Moore Foundation, NSF AccelNet award No. 2114721, an NSERC Discovery Grant, and CIFAR. The Arecibo Observatory is a facility of the NSF operated under cooperative agreement (AST-1744119) by the University of Central Florida (UCF) in alliance with Universidad Ana G. M{\'e}ndez (UAGM) and Yang Enterprises (YEI), Inc. The Green Bank Observatory is a facility of the NSF operated under cooperative agreement by Associated Universities, Inc. The National Radio Astronomy Observatory is a facility of the NSF operated under cooperative agreement by Associated Universities, Inc. 
NANOGrav is part of the International Pulsar Timing Array (IPTA); we would like to thank our IPTA colleagues for their help with this paper.

Part of this work was conducted using the High Performance Computing Cluster PALMA II at the University of M\"unster~(\url{https://www.uni-muenster.de/IT/HPC}). This work used the Maxwell computational resources operated at Deutsches Elektronen-Synchrotron DESY, Hamburg (Germany). This work was conducted in part using the HPC resources of the Texas Advanced Computing Center (TACC) at the University of Texas at Austin. The Tufts University High Performance Computing Cluster (\url{https://it.tufts.edu/high-performance-computing}) was utilized for some of the research reported in this paper. This research used the computational resources provided by the University of Central Florida's Advanced Research Computing Center.

J.J.B.P.\ acknowledges the support by the PID2021-123703NB-C21 grant funded by MCIN/AEI /10.13039/501100011033/ and by ERDF; ”A way of making Europe,” the Basque Government grant (IT-1628-22); and the Basque Foundation for Science (IKERBASQUE). 
L.B.\ acknowledges support from the National Science Foundation under award AST-1909933 and from the Research Corporation for Science Advancement under Cottrell Scholar Award No. 27553.
P.R.B.\ is supported by the Science and Technology Facilities Council, grant No. ST/W000946/1.
S.B.\ gratefully acknowledges the support of a Sloan Fellowship and the support of NSF under award No. \#1815664.
M.C.\ and S.R.T.\ acknowledge support from NSF AST-2007993.
M.C.\ and N.S.P.\ were supported by the Vanderbilt Initiative in Data Intensive Astrophysics (VIDA) Fellowship.
Support for this work was provided by the NSF through the Grote Reber Fellowship Program administered by Associated Universities, Inc./National Radio Astronomy Observatory.
Support for H.T.C.\ is provided by NASA through the NASA Hubble Fellowship Program grant No. \#HST-HF2-51453.001 awarded by the Space Telescope Science Institute, which is operated by the Association of Universities for Research in Astronomy, Inc., for NASA, under contract NAS5-26555.
K.C.\ is supported by a UBC Four Year Fellowship (6456).
M.E.D.\ acknowledges support from the Naval Research Laboratory by NASA under contract S-15633Y.
T.D.\ and M.T.L. are supported by an NSF Astronomy and Astrophysics Grant (AAG), award No. 2009468.
The work of R.v.E.\, K.Sc., and T.S.\ is supported by the Deutsche Forschungsgemeinschaft (DFG) through the Research Training Group, GRK 2149: Strong and Weak Interactions – from Hadrons to Dark Matter. 
E.C.F.\ is supported by NASA under award No. 80GSFC21M0002.
G.E.F., S.C.S., and S.J.V.\ are supported by NSF award PHY-2011772.
The Flatiron Institute is supported by the Simons Foundation.
A.D.J.\ and M.V.\ acknowledge support from the Caltech and Jet Propulsion Laboratory President's and Director's Research and Development Fund.
A.D.J.\ acknowledges support from the Sloan Foundation.
The work of N.La.\ and X.S.\ is partly supported by the George and Hannah Bolinger Memorial Fund in the College of Science at Oregon State University.
N.La. acknowledges the support from Larry W. Martin and Joyce B. O'Neill Endowed Fellowship in the College of Science at Oregon State University.
Part of this research was carried out at the Jet Propulsion Laboratory, California Institute of Technology, under a contract with the National Aeronautics and Space Administration (80NM0018D0004).
V.S.H.L.\ is supported by the DoE under contract DE-SC0011632.
R.R.L.d.S.\ is supported by a research grant (29405) from VILLUM FONDEN. 
D.R.L.\ and M.A.M.\ are supported by NSF No. \#1458952.
M.A.M.\ is supported by NSF \#2009425.
C.M.F.M.\ was supported in part by the National Science Foundation under Grants Nos. NSF PHY-1748958 and AST-2106552.
A.Mi.\ is supported by the Deutsche Forschungsgemeinschaft under Germany's Excellence Strategy - EXC 2121 Quantum Universe - 390833306.
The Dunlap Institute is funded by an endowment established by the David Dunlap family and the University of Toronto.
K.D.O.\ was supported in part by NSF grants Nos.\ 2111738 and 2207267.
T.T.P.\ acknowledges support from the Extragalactic Astrophysics Research Group at E\"{o}tv\"{o}s Lor\'{a}nd University, funded by the E\"{o}tv\"{o}s Lor\'{a}nd Research Network (ELKH), which was used during the development of this research.
S.M.R.\ and I.H.S.\ are CIFAR Fellows.
Portions of this work performed at NRL were supported by ONR 6.1 basic research funding.
J.D.R.\ also acknowledges support from start-up funds from Texas Tech University.
J.S.\ is supported by an NSF Astronomy and Astrophysics Postdoctoral Fellowship under award AST-2202388 and acknowledges previous support by the NSF under award 1847938.
S.R.T.\ acknowledges support from an NSF CAREER award No. \#2146016.
T.T.\ contribution to this work is supported by the Fermi Research Alliance, LLC, under contract No. DE-AC02-07CH11359 with the U.S. Department of Energy, Office of Science, Office of High Energy Physics.
C.U.\ acknowledges support from BGU (Kreitman fellowship) and the Council for Higher Education and Israel Academy of Sciences and Humanities (Excellence fellowship).
C.A.W.\ acknowledges support from CIERA, the Adler Planetarium, and the Brinson Foundation through a CIERA-Adler postdoctoral fellowship.
O.Y.\ is supported by the National Science Foundation Graduate Research Fellowship under grant No. DGE-2139292.
K.Z.\ is supported by the DoE under contract DE-SC0011632, and by a Simons Investigator award.
\begin{widetext}
\appendix 

\section{Parameter ranges and limits}
\label{app:parameters}

In this appendix, we specify the prior assumptions for all model parameters used in our analyses and report characteristic values for these parameters that we extract from the corresponding reconstructed 1D marginalized posterior distributions. In Table~\ref{tab:base_priors}, we list our prior assumptions for the pulsar-intrinsic red-noise parameters $A_{\rm red}$ and $\gamma_{\rm red}$ ($A_a$ and $\gamma_a$ in Eq.~\eqref{eq:kappaa}) as well as for the SMBHB parameters $\abhb$ and $\gbhb$. In the latter case, we work with a bivariate normal distribution for $(\log_{10}\abhb,\gbhb)$ whose mean and covariance matrix are given by
\begin{equation}
\label{eq:musigma}
\boldsymbol{\mu}_{\scriptscriptstyle\rm BHB} = \begin{pmatrix} -15.6 \\ 4.7 \end{pmatrix} \,, \qquad \mat{\sigma}_{\scriptscriptstyle\rm BHB} = 10^{-1}\times\begin{pmatrix} 2.8 & -0.026 \\ -0.026 & 1.2 \end{pmatrix} \,,
\end{equation}
which we obtain by fitting the $\log_{10}\abhb$ and $\gbhb$ distributions extracted from the SMBHB simulations in the \texttt{GWOnly-Ext} library in~\citetalias{aaa+23_smbh} (see Section~\ref{sec:astro_signal}). In Table~\ref{tab:np_priors}, we list our prior assumptions for the model parameters of all new-physics models considered in this work, and in Table~\ref{tab:list}, we summarize various key values extracted from the corresponding 1D marginalized posterior distributions. Specifically, we report the Bayes estimator, including its standard deviation; the maximum posterior estimator, including the $68\%$ credible interval; and (when applicable) the upper bound based on the requirement that the $K$ ratio in Eq.~\eqref{eq:Kratio} should not be smaller than $\sfrac{1}{10}$. 

The Bayes estimator $\langle\theta\rangle$ of a parameter $\theta$ with marginalized 1D posterior probability distribution $P(\theta|\mathcal{D},\mathcal{H})$ corresponds to the expectation value with respect to the distribution $P(\theta|\mathcal{D},\mathcal{H})$, while the standard deviation $\sigma_\theta$ of the Bayes estimator corresponds to the positive square root of the associated variance $\sigma_\theta^2$,
\begin{equation}
\langle\theta\rangle = \int d\theta\:\theta\,P(\theta|\mathcal{D},\mathcal{H}) \,, \qquad \sigma_\theta^2 = \langle\theta^2\rangle - \langle\theta\rangle^2= \left[\int d\theta\:\theta^2\,P(\theta|\mathcal{D},\mathcal{H}) \right] - \left[\int d\theta\:\theta\,P(\theta|\mathcal{D},\mathcal{H}) \right]^2 \,.
\end{equation}
In practice, in a given analysis and for a given chain of MCMC samples, we compute the Bayes estimator and its standard deviation in terms of the corresponding sample mean and sample variance. The maximum posterior estimator $\theta_{\rm max}$ of a parameter $\theta$ with marginalized 1D posterior probability distribution $P(\theta|\mathcal{D},\mathcal{H})$ corresponds to the $\theta$ value where $P(\theta|\mathcal{D},\mathcal{H})$ reaches its global maximum across the predefined prior range,
\begin{equation}
P(\theta_{\rm max}|\mathcal{D},\mathcal{H}) = \underset{\theta}{\textrm{max}} \,P(\theta|\mathcal{D},\mathcal{H}) \,,
\end{equation}
and the $68\%$ Bayesian credible interval $\left[\theta_{68}^{\rm min},\theta_{68}^{\rm max}\right]$ follows from integrating the posterior distribution $P(\theta|\mathcal{D},\mathcal{H})$ over the regions of highest posterior density such that the integral returns an integrated probability of $68\%$, 
\begin{equation}
\int_{\theta_{68}^{\rm min}}^{\theta_{68}^{\rm max}} d\theta \:P(\theta|\mathcal{D},\mathcal{H}) = 0.68 \,,
\end{equation}
where $P(\theta|\mathcal{D},\mathcal{H}) > P_{68}$ for all $\theta \in \left[\theta_{68}^{\rm min},\theta_{68}^{\rm max}\right]$ and some appropriate threshold $P_{68}$. Similarly, we can also construct $95\%$ Bayesian credible intervals. Finally, we mention again that the $K$-ratio bound in the last column of Table~\ref{tab:list} indicates the value $\theta_{ K}$ of the parameter $\theta$ that returns $K = \sfrac{1}{10}$ (see Section~\ref{subsec:bayes_analysis}),
\begin{equation}
K\left(\theta_{ K}\right) = \frac{P(\mathcal{D}|\theta_{ K},\mathcal{H})}{P(\mathcal{D}|\theta_0,\mathcal{H})} = \frac{1}{10} \,.
\end{equation}
Note that, unlike all other quantities discussed in this section, the $K$ ratio is not defined in terms of a posterior probability distribution, but rather in terms of a likelihood ratio, which makes it robust against our prior assumptions.

\newpage
\renewcommand{\arraystretch}{1.5}
\setlength\LTleft{0pt}
\setlength\LTright{0pt} 
\begin{longtable*}{p{2.5cm}p{5cm}p{4.5cm}p{4cm}}
\caption{Prior distributions for the pulsar-intrinsic red-noise parameters and the parameters of the astrophysical SMBHB signal. The mean and covariance matrix of the Gaussian prior distribution for $(\log_{10}\abhb,\gbhb)$ are given in Eq.~\eqref{eq:musigma}.}
\label{tab:base_priors}\\
\toprule
\multicolumn{1}{c}{{\bf Parameter}}  & \multicolumn{1}{c}{{\bf Description}} & \multicolumn{1}{c}{{\bf Prior}} & \multicolumn{1}{c}{{\bf Comments}} \\
\hline

\multicolumn{4}{c}{{\bf Pulsar-intrinsic red noise}} \\[1pt]
$A_{\rm red}$ & red noise power-law amplitude & log-uniform $[-20, -11]$ & one parameter per pulsar  \\
$\gamma_{\rm red}$ & red noise power-law spectral index & uniform $[0, 7]$ & one parameter per pulsar \\
\hline

\multicolumn{4}{c}{\textbf{Supermassive black-hole binaries} (\textsc{smbhb})} \\[1pt]
$(\log_{10}\abhb,\gbhb)$ & SMBHB signal amplitude and slope & ${\rm normal}(\boldsymbol{\mu}_{\rm \scriptscriptstyle BHB},\boldsymbol{\sigma}_{\scriptscriptstyle \rm BHB})$ & one parameter for PTA \\
\bottomrule
\end{longtable*}

\newpage
\renewcommand{\arraystretch}{1.25}
\setlength\LTleft{0pt}
\setlength\LTright{0pt} 
\begin{longtable*}[p]{p{2.5cm}@{\hspace{0.5\tabcolsep}}p{5cm}p{4.5cm}@{\hspace{0.5\tabcolsep}}p{4cm}}
\caption{Priors distributions for the parameters of the new-physics models considered in this work. The $^*$ indicates parameters that are present only in the uncorrelated ULDM analyses.}
\label{tab:np_priors}\\
\toprule
\multicolumn{1}{c}{{\bf Parameter}}  & \multicolumn{1}{c}{{\bf Description}} & \multicolumn{1}{c}{{\bf Prior}} & \multicolumn{1}{c}{{\bf Comments}} \\
\hline

\multicolumn{4}{c}{{\bf Inflationary gravitational waves} (\textsc{igw})} \\[1pt]
$T_{\rm rh}\,[{\rm GeV}]$ & reheating temperature & log-uniform $[-3,3]$& one parameter per PTA \\
$r$ & tensor-to-scalar ratio & log-uniform $[-40,0]$& one parameter per PTA \\
$n_t$  & tensor spectral index & uniform $[0, 6]$ & one parameter per PTA \\
\hline

\multicolumn{4}{c}{{\bf Scalar-induced gravitational waves} (\textsc{sigw-delta})} \\[1pt]
$A$ & scalar amplitude & log-uniform $[-3,1]$& one parameter per PTA \\
$f_*\,[{\rm Hz}]$  & peak frequency & log-uniform $[-11, -5]$ & one parameter per PTA \\
\hline

\multicolumn{4}{c}{{\bf Scalar-induced gravitational waves} (\textsc{sigw-gauss})} \\[1pt]
$A$ & scalar amplitude & log-uniform $[-3,1]$& one parameter per PTA \\
$f_*\,[{\rm Hz}]$ & peak frequency & log-uniform $[-11, -5]$ & one parameter per PTA \\
$\Delta$ & width & uniform $[0.1,3]$& one parameter per PTA \\
\hline

\multicolumn{4}{c}{{\bf Scalar-induced gravitational waves} (\textsc{sigw-box})} \\[1pt]
$A$ & scalar amplitude & log-uniform $[-3,1]$& one parameter per PTA \\
$f_{\rm min}\,[{\rm Hz}]$ & lower frequency & log-uniform $[-11, -5]$ & one parameter per PTA \\
$f_{\rm max}\,[{\rm Hz}]$ & upper frequency & log-uniform $[-11, -5]$ & one parameter per PTA \\
\hline

\multicolumn{4}{c}{{\bf Cosmological phase transition} (\textsc{pt})} \\[1pt]
$T_*\,[{\rm GeV}]$ & transition temperature & log-uniform $[-4, 4]$ & one parameter per PTA \\
$\alpha_*$ & transition strength & log-uniform $[-2,1]$& one parameter per PTA \\
$H_*R_*$ & bubble separation & log-uniform $[-3,0]$& one parameter per PTA \\
\multirow{2}{=}{$a$ } & low-frequency slope (\textsc{pt-bubbles})& uniform $[1,3]$& one parameter per PTA \\
& low-frequency slope (\textsc{pt-sound}) & uniform $[3,5]$& one parameter per PTA \\
\multirow{2}{=}{$b$ }& high-frequency slope (\textsc{pt-bubbles}) & uniform $[1,3]$& one parameter per PTA \\
& high-frequency slope (\textsc{pt-sound}) & uniform $[2,4]$& one parameter per PTA \\
\multirow{2}{=}{$c$ }& spectral-shape width (\textsc{pt-bubbles}) & uniform $[1,3]$& one parameter per PTA \\
& spectral-shape width (\textsc{pt-sound}) & uniform $[3,5]$& one parameter per PTA \\
\hline

\multicolumn{4}{c}{{\bf Cosmic strings} (\textsc{stable})} \\[1pt]
$G\mu$  & string tension & log-uniform $[-14, -6]$ & one parameter per PTA \\
\hline

\multicolumn{4}{c}{{\bf Metastable cosmic strings} (\textsc{meta})} \\[1pt]
$G\mu$  & string tension & log-uniform $[-14, -1.5]$ & one parameter per PTA \\
$\sqrt{\kappa}$  & decay parameter & uniform $[7,9.5]$& one parameter per PTA\\
\hline

\multicolumn{4}{c}{{\bf Cosmic superstrings} (\textsc{super})} \\[1pt]
$G\mu$  & string tension & log-uniform $[-14, -6]$ & one parameter per PTA \\
$P$  & intercommutation probability & log-uniform $[-3,0]$& one parameter per PTA\\
\hline

\multicolumn{4}{c}{{\bf Domain walls} (\textsc{dw-sm})} \\[1pt]
$T_*\,[{\rm GeV}]$  & transition temperature & log-uniform $[-4, 4]$ & one parameter per PTA \\
$\alpha_*$ & energy fraction in DWs & log-uniform $[-3,0]$& one parameter per PTA \\
$b$  & high-frequency slope & uniform $[0.5,1]$& one parameter per PTA \\
$c$  & spectral-shape width & uniform $[0.3,3]$& one parameter per PTA \\
\hline

\multicolumn{4}{c}{{\bf Domain walls} (\textsc{dw-ds})} \\[1pt]
$T_*\,[{\rm GeV}]$  & transition temperature & log-uniform $[-4, 4]$ & one parameter per PTA \\
$\Delta N_{\rm eff}$ & amount of dark radiation & log-uniform $[-3, -0.41]$& one parameter per PTA \\
$b$  & high-frequency slope & uniform $[0.5,1]$& one parameter per PTA \\
$c$  & spectral-shape width & uniform $[0.3,3]$& one parameter per PTA \\
\hline

\multicolumn{4}{c}{{\bf Ultralight dark matter} (\textsc{ULDM})} \\[1pt]
$A_{i}\;[\textrm{s}]$  & ULDM signal amplitude & log-uniform $[-9, -4]$ & one parameter per PTA \\
$m_\phi\,[{\rm eV}]$  & ULDM mass & log-uniform $[-24, -19]$ & one parameter per PTA \\
$\hat\phi_{E}^2$  & Earth normalized signal amplitude & $e^{-x}$ & one parameter per PTA \\
$\hat\phi_{P}^2$  & pulsar normalized signal amplitude & $e^{-x}$ & one parameter per pulsar$^*$ \\
$\gamma_{E}$  & Earth signal phase & uniform $[0, 2\pi]$ & one parameter per PTA \\
$\gamma_{P}$  & pulsar signal phase & uniform $[0, 2\pi]$ & one parameter per pulsar\\
\hline
\multicolumn{4}{c}{{\bf Primordial black holes} (\textsc{pbh-dynamic})} \\[1pt]
$A$  & signal amplitude & log-uniform $[-20, -12]$ & one parameter per PTA \\
$T_0/T_{\rm obs}$  & normalized time of closest approach & uniform $[0,1]$& one parameter per PTA \\
\hline
\multicolumn{4}{c}{{\bf Primordial black holes} (\textsc{pbh-static})} \\[1pt]
$A$  & signal amplitude & log-uniform $[-21, -13]$ & one parameter per PTA \\
\bottomrule
\end{longtable*}

\newpage
\begin{center}
\begin{small}
\renewcommand{\arraystretch}{1.25}
\setlength\LTleft{0pt}
\setlength\LTright{0pt}
\setlength\tabcolsep{10.5pt}
\begin{longtable*}{l@{\hspace{1.25\tabcolsep}}c@{\hspace{0.7\tabcolsep}}cc@{\hspace{0.7\tabcolsep}}cc@{\hspace{0.7\tabcolsep}}c@{\hspace{1\tabcolsep}}c}
\caption{Bayesian Estimators, Maximum Posterior Values, and $68\%$ Credible Intervals for the Parameters of the New-physics Models. Values annotated with  $^*$ are at the boundary of the prior range used in the analysis.\label{tab:list}}\\
\toprule 
Parameter                                       & \multicolumn{2}{c}{Bayes Estimator} & \multicolumn{2}{c}{Maximum Posterior} & \multicolumn{2}{c}{$68\%$ Credible Interval}            & $K$ Bound   \\[-3.5pt]
                                       &{\scriptsize NP}& {\scriptsize NP+SMBHB}& {\scriptsize NP}&{\scriptsize NP+SMBHB} & {\scriptsize NP}&{\scriptsize NP+SMBHB}            &   \\ [2pt] 
\toprule
\multicolumn{8}{c}{\textbf{Inflationary Gravitational Waves} (\textsc{igw})}\\ [1pt]
$\log_{10}T_{\rm rh}/\textrm{GeV}$ & $0.02 \pm 1.60$     & $-0.07 \pm 1.61$           & $-0.53$    & $-0.60$         & $\left[-1.51,2.53\right]$ & $\left[-1.89,2.11\right]$      & $...$ \\ 
$\log_{10}r$                                    & $-14.06 \pm 5.82$   & $-15.97 \pm 7.27$          &  $-10.14$   & $-10.59$         & $\left[-22.16,-6.58\right]$   & $\left[-23.03,-7.21\right]$         & $...$ \\ 
$n_t$                                           & $2.61 \pm 0.85$   &$2.68 \pm 0.97$             & $2.02$      & $2.08$           & $\left[1.53,3.92\right]$    & $\left[1.56,4.03\right]$        & $5.72$ \\
$\log_{10}A_{\scriptscriptstyle\rm BHB}$        & $...$             & $-15.60\pm 0.56$         & $...$     & $-15.64$       & $...$                       & $\left[-16.20,-15.14\right]$  & $...$\\
$\gamma_{\scriptscriptstyle\rm BHB}$            & $...$             & $4.61\pm 0.37$           & $...$     & $4.64$         & $...$                       & $\left[4.26,5.00\right]$      & $...$\\ [2pt] \cmidrule(lr){1-8}
\multicolumn{8}{c}{\textbf{Scalar-induced Gravitational Waves} \textsc{(sigw-delta)}}\\ [1pt]
$\log_{10}A$                                    & $-0.69 \pm 0.47$  & $-0.71 \pm 0.49$         & $-0.14$     & $-0.17$          & $\left[-1.00,-0.01\right]$   & $\left[-1.03,-0.02\right]$      & $...$ \\ 
$\log_{10}f_*/{\rm Hz}$                       & $-5.90 \pm 0.60$  & $-5.93 \pm 0.67$         & $-5^*$     & $-5^*$          & $\left[-6.17,-5^*\right]$ & $\left[-6.19,-5^*\right]$   & $...$ \\
$\log_{10}A_{\scriptscriptstyle\rm BHB}$        & $...$             & $-15.77\pm 0.46$         & $...$     & $-15.71$       & $...$                       & $\left[-16.18,-15.29\right]$  & $...$\\
$\gamma_{\scriptscriptstyle\rm BHB}$            & $...$             & $4.65\pm 0.35$           & $...$     & $4.65$         & $...$                       & $\left[4.31,4.99\right]$      & $...$\\ [2pt] \cmidrule(lr){1-8}
\multicolumn{8}{c}{\textbf{Scalar-induced Gravitational Waves} \textsc{(sigw-gauss)}}\\ [1pt]
$\log_{10}A$                                    & $-0.38 \pm 0.58$  & $-0.36 \pm 0.61$         & $-0.34$    & $-0.29$         & $\left[-1.03,0.20\right]$ & $\left[-1.04,0.24\right]$     & $...$ \\ 
$\log_{10}f_*/{\rm Hz}$                       & $-6.32 \pm 0.71$  & $-6.30 \pm 0.73$         & $-7.03$    & $-6.85$         & $\left[-7.25,-5.65\right]$& $\left[-7.17,-5.57\right]$    & $...$ \\ 
$\Delta$                                        & $1.35 \pm 0.70$   & $1.30  \pm 0.70$         & $1.60$     & $1.54$          & $\left[0.51,2.07\right]$  & $\left[0.37,1.92\right]$    & $...$ \\
$\log_{10}A_{\scriptscriptstyle\rm BHB}$        & $...$             & $-15.72\pm 0.46$         & $...$     & $-15.65$       & $...$                       & $\left[-16.14,-15.22\right]$  & $...$\\
$\gamma_{\scriptscriptstyle\rm BHB}$            & $...$             & $4.65\pm 0.34$           & $...$     & $4.65$         & $...$                       & $\left[4.32,5.00\right]$      & $...$\\ [2pt] \cmidrule(lr){1-8}
\multicolumn{8}{c}{\textbf{Scalar-induced Gravitational Waves} \textsc{(sigw-box)}}\\ [1pt]
$\log_{10}A$                                    & $-1.06 \pm 0.52$  & $-1.02 \pm 0.57$         & $-1.26$    & $-1.25$         & $\left[-1.72,-0.82\right]$& $\left[-1.68,-0.74\right]$    & $...$ \\ 
$\log_{10}f_{\rm min}/{\rm Hz}$               & $-7.34 \pm 0.48$  & $-7.29 \pm 0.55$         & $-7.50$    & $-7.48$         & $\left[-8.01,-6.97\right]$& $\left[-7.97,-6.84\right]$    & $...$ \\ 
$\log_{10}f_{\rm max}/{\rm Hz}$               & $-6.06 \pm 0.65$  & $-6.12 \pm 0.81$         & $-5.40$     & $-5.36$         & $\left[-6.42,-5^*\right]$ & $\left[-6.45,-5^*\right]$     & $...$ \\
$\log_{10}A_{\scriptscriptstyle\rm BHB}$        & $...$             & $-15.70\pm 0.49$         & $...$     & $-15.64$       & $...$                       & $\left[-16.15,-15.21\right]$  & $...$\\
$\gamma_{\scriptscriptstyle\rm BHB}$            & $...$             & $4.65\pm 0.35$           & $...$     & $4.66$         & $...$                       & $\left[4.31,5.00\right]$      & $...$\\ [2pt] \cmidrule(lr){1-8}
\multicolumn{8}{c}{\textbf{Cosmological Phase Transition} \textsc{(pt-bubble)}}\\ [1pt]
$\log_{10}T_{\rm *}/\textrm{GeV}$               & $-0.76 \pm 0.49$  & $-0.71 \pm 0.70$         & $-0.90$   & $-0.89$         & $\left[-1.33,-0.39\right]$  & $\left[-1.34,-0.34\right]$    & $...$ \\ 
$\log_{10}\alpha_*$                             & $-0.26 \pm 0.47$  & $-0.23 \pm 0.52$         & $1^*$     & $0.74$          & $\left[0.03,1^*\right]$     & $\left[0.01,1^*\right]$       & $...$ \\ 
$\log_{10}H_*R_*$                               & $-0.42 \pm 0.26$  & $-0.47 \pm 0.39$         & $0^*$     & $-0.06$         & $\left[-0.56, 0^*\right]$   & $\left[-0.58,0^*\right]$      & $...$ \\ 
$a$                                             & $2.04 \pm 0.48$   & $2.07 \pm 0.49$          & $1.97$    & $2.01$          & $\left[1.49,2.54\right]$    & $\left[1.54,2.63\right]$      & $...$\\ 
$b$                                             & $1.97 \pm 0.58$   & $1.98 \pm 0.58$          & $1^*$     & $1^*$           & $\left[1^*,2.32\right]$     & $\left[1^*,2.33\right]$       & $...$\\ 
$c$                                             & $2.03 \pm 0.57$   & $2.03\pm 0.57$           & $3^*$     & $2.93$          & $\left[1.69,3^*\right]$     & $\left[1.69,3^*\right]$       & $...$\\
$\log_{10}A_{\scriptscriptstyle\rm BHB}$        & $...$             & $-15.68\pm 0.51$         & $...$     & $-15.65$        & $...$                       & $\left[-16.17,-15.21\right]$  & $...$\\
$\gamma_{\scriptscriptstyle\rm BHB}$            & $...$             & $4.64\pm 0.35$           & $...$     & $4.65$          & $...$                       & $\left[4.30,5.00\right]$      & $...$\\ [2pt] \cmidrule(lr){1-8}
\newpage\multicolumn{8}{c}{\textbf{Cosmological Phase Transition} \textsc{(pt-sound)}}\\ [1pt]
$\log_{10}T_{\rm *}/\textrm{GeV}$               & $-1.84 \pm 0.41$  & $-1.56 \pm 1.06$         & $-2.00$   & $-1.95$        & $\left[-2.33,-1.48\right]$  & $\left[-2.31,-1.30\right]$    & $...$ \\ 
$\log_{10}\alpha_*$                             & $-0.22 \pm 0.44$  & $0.14 \pm 0.56$          & $-0.21$   & $-0.15$        & $\left[-0.37, 1^*\right]$   & $\left[-0.34,0.73\right]$     & $...$ \\ 
$\log_{10}H_*R_*$                               & $-0.81 \pm 0.36$  & $-0.87 \pm 0.51$         & $-1.05$   & $-1.01$        & $\left[-1.28,-0.57\right]$  & $\left[-1.26,-0.45\right]$    & $...$ \\ 
$a$                                             & $3.58 \pm 0.47$   & $3.74\pm 0.54$           & $3^*$     & $3^*$          & $\left[3^*,3.72\right]$     & $\left[3^*,3.98\right]$       & $...$\\ 
$b$                                             & $2.87 \pm 0.57$   & $2.92\pm 0.57$           & $2^*$     & $2^*$          & $\left[2^*,3.17\right]$     & $\left[2^*,3.25\right]$       & $...$\\ 
$c$                                             & $4.16 \pm 0.56$   & $4.09\pm 0.57$           & $5^*$     & $5^*$          & $\left[3.87,5^*\right]$     & $\left[3.77,5^*\right]$       & $...$\\
$\log_{10}A_{\scriptscriptstyle\rm BHB}$        & $...$             & $-15.45\pm 0.55$         & $...$     & $-15.39$       & $...$                       & $\left[-16.04,-14.94\right]$  & $...$\\
$\gamma_{\scriptscriptstyle\rm BHB}$            & $...$             & $4.63\pm 0.38$           & $...$     & $4.67$         & $...$                       & $\left[4.27,5.03\right]$      & $...$\\ [2pt] \cmidrule(lr){1-8}
\multicolumn{8}{c}{\textbf{Cosmic Strings: Cusp-only Spectrum} \textsc{(stable-c)}}\\ [1pt]
$\log_{10}G\mu$                               & $ -10.15\pm 0.16$ & $-11.41 \pm 1.25$        & $-10.18$   & $-10.22$        & $\left[-10.33,-10.01\right]$& $\left[-12.13,-9.88\right]$& $-9.67$ \\ 
$\log_{10}A_{\scriptscriptstyle\rm BHB}$        & $...$             & $-14.95\pm 0.58$         & $...$     & $-14.56$       & $...$                       & $\left[-15.58,-14.31\right]$  & $...$\\
$\gamma_{\scriptscriptstyle\rm BHB}$            & $...$             & $4.34\pm 0.38$           & $...$     & $4.24$         & $...$                       & $\left[3.91,4.66\right]$      & $...$\\ [2pt] \cmidrule(lr){1-8}
\multicolumn{8}{c}{\textbf{Cosmic Strings: Kink-only Spectrum} \textsc{(stable-k)}}\\ [1pt]
$\log_{10}G\mu$                               & $-10.33 \pm 0.15$ & $-11.34 \pm 1.17$        & $-10.36$   & $-10.38$        & $\left[-10.50,-10.21\right]$& $\left[-11.84,-10.04\right]$& $-9.87$ \\  
$\log_{10}A_{\scriptscriptstyle\rm BHB}$        & $...$             & $-15.04\pm 0.61$         & $...$     & $-14.56$       & $...$                       & $\left[-15.68,-14.34\right]$  & $...$\\
$\gamma_{\scriptscriptstyle\rm BHB}$            & $...$             & $4.39\pm 0.38$           & $...$     & $4.28$         & $...$                       & $\left[3.95,4.72\right]$      & $...$\\ [2pt] \cmidrule(lr){1-8}
\multicolumn{8}{c}{\textbf{Cosmic Strings: Monochromatic Emission} \textsc{(stable-m)}}\\ [1pt]
$\log_{10}G\mu$                               & $-10.53 \pm 0.14$ & $-11.47 \pm 1.09$        & $-10.56$   & $-10.60$        & $\left[-10.68,-10.42\right]$& $\left[-11.91,-10.27\right]$& $-10.10$ \\ 
$\log_{10}A_{\scriptscriptstyle\rm BHB}$        & $...$             & $-15.05\pm 0.61$         & $...$     & $-14.58$       & $...$                       & $\left[-15.67,-14.34\right]$  & $...$\\
$\gamma_{\scriptscriptstyle\rm BHB}$            & $...$             & $4.39\pm 0.38$           & $...$     & $4.28$         & $...$                       & $\left[3.96,4.73\right]$      & $...$\\ [2pt] \cmidrule(lr){1-8}
\multicolumn{8}{c}{\textbf{Cosmic Strings: Numerical Spectrum} \textsc{(stable-n)}}\\ [1pt]
$\log_{10}G\mu$                               & $-10.18 \pm 0.15$ & $-11.34 \pm 1.23$        & $-10.21$   & $-10.25$        & $\left[-10.35,-10.05\right]$& $\left[-11.99,-9.90\right]$ & $-9.71$ \\ 
$\log_{10}A_{\scriptscriptstyle\rm BHB}$        & $...$             & $-14.99\pm 0.59$         & $...$     & $-14.56$       & $...$                       & $\left[-15.61,-14.32\right]$  & $...$\\
$\gamma_{\scriptscriptstyle\rm BHB}$            & $...$             & $4.37\pm 0.38$           & $...$     & $4.26$         & $...$                       & $\left[3.93,4.69\right]$      & $...$\\ [2pt] \cmidrule(lr){1-8}
\multicolumn{8}{c}{\textbf{Metastable Cosmic strings: Loops Only} \textsc{(meta-l)}}\\ [1pt]
$\log_{10}G\mu$                               & $-5.80 \pm 0.78$  & $-5.9 \pm 1.2$           & $-5.04$    & $-5.05$         & $\left[-6.14,-4.84\right]$& $\left[-6.19,-4.83\right]$    & $...$ \\ 
$\sqrt{\kappa}$                                      & $7.95  \pm 0.13$  & $7.95 \pm 0.18$          & $7.85$     & $7.84$          & $\left[7.81,8.00\right]$  & $\left[7.80,8.00\right]$      & $...$ \\ 
$\log_{10}A_{\scriptscriptstyle\rm BHB}$        & $...$             & $-15.73\pm 0.48$         & $...$     & $-15.67$       & $...$                       & $\left[-16.18,-15.25\right]$  & $...$\\
$\gamma_{\scriptscriptstyle\rm BHB}$            & $...$             & $4.64\pm 0.35$           & $...$     & $4.66$         & $...$                       & $\left[4.30,5.00\right]$      & $...$\\ [2pt] \cmidrule(lr){1-8}
\multicolumn{8}{c}{\textbf{Metastable Cosmic Strings: Loops and Segments} \textsc{(meta-ls)}}\\ [1pt]
$\log_{10}G\mu$                               & $-5.62 \pm 1.01$    & $-5.70 \pm 1.40$           & $-4.46$    & $-4.44$         & $\left[-6.26,-4.24\right]$& $\left[-6.33,-4.15\right]$    & $...$ \\ 
$\sqrt{\kappa}$                                      & $7.83  \pm 0.18$  & $7.82 \pm 0.23$          & $7.61$     & $7.61$          & $\left[7.59,7.92\right]$  & $\left[7.57,7.93\right]$      & $...$ \\ 
$\log_{10}A_{\scriptscriptstyle\rm BHB}$        & $...$             & $-15.71\pm 0.49$         & $...$     & $-15.67$       & $...$                       & $\left[-16.17,-15.23\right]$  & $...$\\
$\gamma_{\scriptscriptstyle\rm BHB}$            & $...$             & $4.64\pm 0.35$           & $...$     & $4.65$         & $...$                       & $\left[4.30,5.00\right]$      & $...$\\ [2pt] \cmidrule(lr){1-8}
\multicolumn{8}{c}{\textbf{Cosmic Superstrings} \textsc{(super)}}\\ [1pt]
$\log_{10}G\mu$                               & $-11.67 \pm 0.32$ & $-11.68 \pm 0.35$        & $-11.94$   & $-11.96$        & $\left[-12.08,-11.50\right]$& $\left[-12.09,-11.49\right]$& $-9.97$ \\ 
$\log_{10}P$                                    & $-2.23  \pm 0.55$ & $-2.21 \pm 0.57$          & $-3^*$     & $-3^*$          & $\left[-3^*,-2.01\right]$ & $\left[-3^*,-1.99\right]$      & $---$ \\  
$\log_{10}A_{\scriptscriptstyle\rm BHB}$        & $...$             & $-15.76\pm 0.46$         & $...$     & $-15.70$       & $...$                       & $\left[-16.18,-15.28\right]$  & $...$\\
$\gamma_{\scriptscriptstyle\rm BHB}$            & $...$             & $4.64\pm 0.35$           & $...$     & $4.65$         & $...$                       & $\left[4.30,4.99\right]$      & $...$\\ [2pt] \cmidrule(lr){1-8}
\multicolumn{8}{c}{\textbf{Domain Walls} \textsc{(dw-sm)}}\\ [1pt]
$\log_{10}T_{\rm *}/\textrm{GeV}$               & $-0.73 \pm 0.21$  & $-0.65 \pm 0.49$         & $-0.79$    & $-0.78$         & $\left[-0.96,-0.56\right]$  & $\left[-0.97,-0.51\right]$    & $...$ \\ 
$\log_{10}\alpha_*$                             & $-0.88 \pm 0.21$  & $-0.87 \pm 0.32$         & $-0.92$    & $-0.90$         & $\left[-1.10,-0.71\right]$  & $\left[-1.10,-0.66\right]$    & $...$ \\ 
$b$                                             & $0.74 \pm 0.14$   & $0.74\pm 0.14$           & $0.5^*$    & $0.5^*$         & $\left[0.5^*,0.83\right]$   & $\left[0.5^*,0.83\right]$     & $...$ \\ 
$c$                                             & $2.01 \pm 0.70$   & $1.92\pm 0.74$           & $3^*$      & $3^*$           & $\left[1.72,3^*\right]$     & $\left[1.57,3^*\right]$       & $...$ \\
$\log_{10}A_{\scriptscriptstyle\rm BHB}$        & $...$             & $-15.60\pm 0.50$         & $...$      & $-15.49$        & $...$                       & $\left[-16.06,-15.08\right]$  & $---$ \\
$\gamma_{\scriptscriptstyle\rm BHB}$            & $...$             & $4.66\pm 0.35$           & $...$      & $4.67$          & $...$                       & $\left[4.32,5.02\right]$      & $...$ \\ [2pt] \cmidrule(lr){1-8}
\multicolumn{8}{c}{\textbf{Domain Walls} \textsc{(dw-dr)}}\\ [1pt]
$\log_{10}T_{\rm *}/\textrm{GeV}$               & $-0.98 \pm 0.15$  & $-0.62 \pm 1.37$         & $-0.94$    & $-0.94$         & $\left[-1.10,-0.82\right]$  & $\left[-1.27,-0.67\right]$    & $...$ \\ 
$\log_{10} \Delta N_{\rm eff}$                  & $-0.48 \pm 0.06$  & $-0.87 \pm 0.71$         & $-0.41^*$   & $-0.41^*$       & $\left[-0.49,-0.41^*\right]$& $\left[-0.63,0.41^*\right]$   & $...$ \\ 
$b$                                             & $0.74 \pm 0.14$   & $0.74 \pm 0.14$          & $0.5^*$    & $0.5^*$         & $\left[0.5^*,0.97\right]$   & $\left[0.5^*,0.83\right]$     & $...$\\ 
$c$                                             & $1.95 \pm 0.68$   & $1.83 \pm 0.73$          & $3^*$      & $3^*$           & $\left[1.62,3^*\right]$     & $\left[1.44,3^*\right]$       & $...$\\
$\log_{10}A_{\scriptscriptstyle\rm BHB}$        & $...$             & $-15.33\pm 0.65$         & $...$      & $-15.59$        & $...$                       & $\left[-16.03,-14.40\right]$  & $...$ \\
$\gamma_{\scriptscriptstyle\rm BHB}$            & $...$             & $4.51\pm 0.39$           & $...$      & $4.52$          & $...$                       & $\left[4.10,4.90\right]$      & $...$ \\ [2pt] \cmidrule(lr){1-8}

\bottomrule
\end{longtable*}
\end{small}
\end{center}

\begin{figure}
	\centering
        \includegraphics[width=0.825\textwidth]{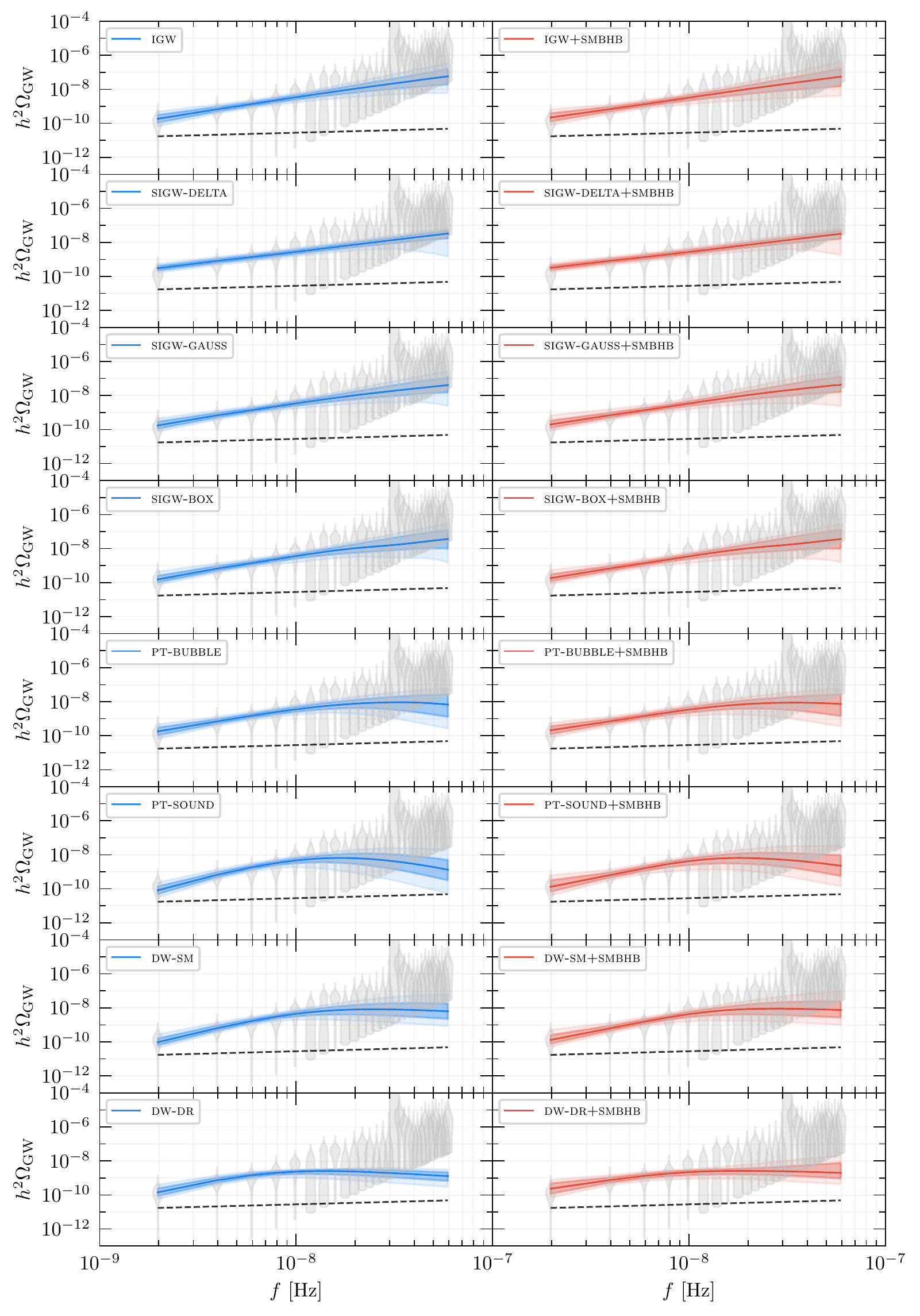}
	\caption{Median GWB spectra (solid lines) for all new-physics models considered in this work except for the cosmic-string models, together with their $68\%$ and $95\%$ posterior envelopes. Median GWB spectra for the cosmic-string models are shown in Fig.~\ref{fig:mean_spectra_all_2}. In the left column (blue shading), we show the median GWB spectra for the new-physics models alone; in the right column (red shading), we combine the new-physics signals with the signal from SMBHBs. The gray violins are symmetrical representations of the 1D marginalized posterior probability density distributions of the GW energy density at each sampling frequency of the data. The dashed black lines show the GWB spectrum produced by inspiraling SMBHBs (see caption of Fig.~\ref{fig:mean_spectra}).}
        \label{fig:mean_spectra_all_1}
\end{figure}
\begin{figure*}
	\centering
        \includegraphics[width=0.825\textwidth]{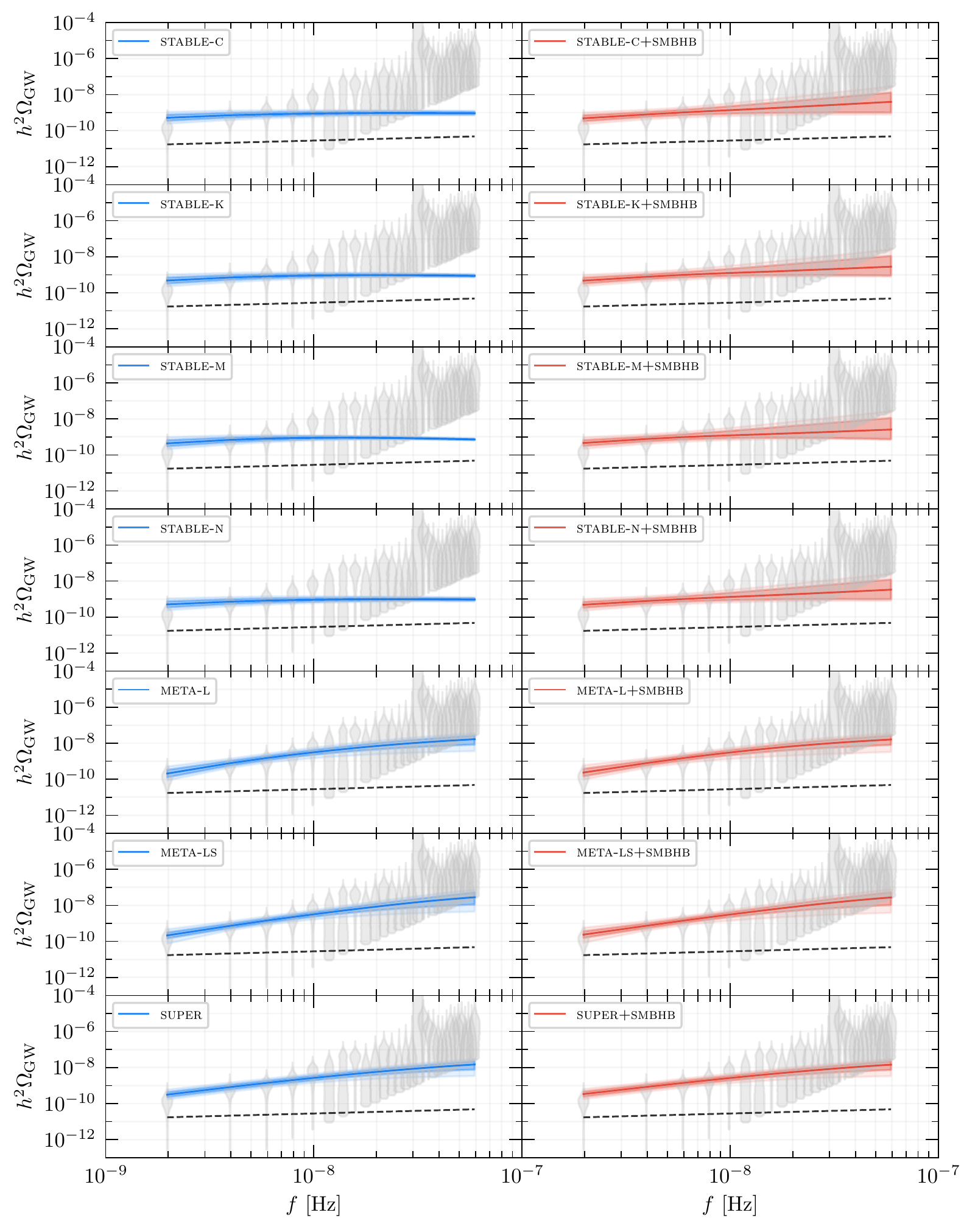}
	\caption{
        Same as Fig.~\ref{fig:mean_spectra_all_1} but for the cosmic-string models considered in this work. See Appendix~\ref{app:mediangw} for details.}
        \label{fig:mean_spectra_all_2}
\end{figure*}
\section{Median GW power spectra}
\label{app:mediangw}

In Figs.~\ref{fig:mean_spectra} and \ref{fig:landscape_spectra} in the main text, as well as in Figs.~\ref{fig:mean_spectra_all_1} and \ref{fig:mean_spectra_all_2} in this appendix, we show median GW power spectra for all of the new-physics models considered in this work. The purpose of this appendix is to explain how these spectra are constructed. The main idea is to take the output of our Bayesian analysis, i.e., the reconstructed posterior distributions in the parameter space of the respective models, and map these distributions onto distributions of $h^2\Omega_{\scriptscriptstyle \rm GW}$ values at each frequency in the NANOGrav frequency band. In practice, this means that we sample model parameter values from our posterior distributions and use these parameter values to evaluate the GW power spectrum at one frequency after another in small steps in the range from $f = 1/T_{\rm obs}$ to $f = 30/T_{\rm obs}$. At each fixed frequency, we thus obtain an induced posterior distribution of $h^2\Omega_{\scriptscriptstyle \rm GW}$ values, from which we can derive point values and uncertainty estimates. The median $h^2\Omega_{\scriptscriptstyle \rm GW}$ value of a distribution at fixed $f$ defines the value of the median GW spectrum at this frequency. Similarly, the equal-tailed $68\%$ and $95\%$ intervals around the median values provide an uncertainty estimate for the median GW spectrum; these intervals are shown in terms of the blue and red bands in Figs.~\ref{fig:mean_spectra_all_1} and \ref{fig:mean_spectra_all_2}.

We stress that the median GW spectra in Figs.~\ref{fig:mean_spectra}, \ref{fig:landscape_spectra}, \ref{fig:mean_spectra_all_1}, and \ref{fig:mean_spectra_all_2} represent effective power spectra that encapsulate global properties of the underlying distributions of actual GW spectra. In general, no individual GW spectrum at a given point in parameter space will exactly coincide with the median GW spectrum. For most models, the difference between the individual GW spectra and the median GW spectrum is rather mild. However, for some models, there can be noticeable differences, such as in the case of the SIGW models. Note, e.g., that the median GW spectrum of the \textsc{sigw-delta} model in Fig.~\ref{fig:landscape_spectra} does not have the characteristic double-peak structure that each individual GW spectrum in this model has. The reason for this is clear: the median GW spectrum follows from a distribution of many individual GW spectra whose peaks are located at different frequencies. The double-peak structure of the GW signal in the \textsc{sigw-delta} model is therefore washed out owing to the averaging over many individual GW spectra. 

Another caveat is that median GW spectra violating an experimental bound (e.g., the LVK bound) do not automatically indicate that the corresponding model is ruled out. Again, as the median GW spectrum is constructed from a distribution of individual GW spectra, the violation of an experimental bound merely signals that some (maybe most) individual GW spectra are in conflict with the experimental data. It is, however, well possible that some fraction of the underlying distribution of individual GW spectra is not ruled out and is perfectly consistent with all experimental constraints. This is, e.g., true for the GW spectra from cosmic superstrings. The median GW spectrum of the \textsc{super} model violates the LVK bound (see Fig.~\ref{fig:landscape_spectra}). However, at the level of the model parameter space, this merely means that some parameter regions are experimentally ruled out, while other regions remain viable (see Fig.~\ref{fig:string_super_corner}). Another example is the median GWB spectrum of the \textsc{igw} model, which appears to violate the $N_{\rm eff}$ bound if it is extended to high frequencies beyond the NANOGrav band (see Fig.~\ref{fig:landscape_spectra}). However, the \textsc{igw} model as a whole is not ruled out, as one can still choose the number of $e$-folds during reheating, $N_{\rm rh}$, for each individual GW spectrum. In some regions of parameter space, the $N_{\rm eff}$ bound then persists to pose a problem, despite this additional parametric freedom and even in the limit $N_{\rm rh} \rightarrow 0$; other regions, however, remain phenomenologically viable (see Appendix~\ref{app:inflation} and Fig.~\ref{fig:nrhmax}). 

\section{Supplementary material}
\label{app:supplementary}

\subsection{Cosmic inflation}
\label{app:inflation}

In Fig.~\ref{fig:nrhmax}, we present constraints on the parameter space of inflationary GWs, i.e., the \textsc{igw} model discussed in Section~\ref{subsec:inflation}. This parameter space is spanned by four quantities: the tensor-to-scalar ratio $r$, the spectral index of the primordial tensor power spectrum $n_t$, the reheating temperature $T_{\rm rh}$, and the number of $e$-folds during reheating $N_{\rm rh}$. However, because of the strong covariance between $n_t$ and $r$ in Fig.~\ref{fig:igw_corner}, we are able to eliminate $r$ and work on the 3D hypersurface in parameter space where $r$ is given by the linear fit in Eq.~\eqref{eq:ntr}. The effective parameter space spanned by $T_{\rm rh}$, $n_t$, and $N_{\rm rh}$ is then subject to two constraints: (i) the upper limit on the allowed amount of extra relativistic radiation, $\Delta N_{\rm eff}$, at the time of BBN and CMB decoupling (see Eq.~\eqref{eq:igwneff}) and (ii) the upper limit on the amplitude of the SGWB set by the LVK Collaboration based on their first three observing runs (see Eq.~\eqref{eq:LVKbound}). The $N_{\rm eff}$ bound can be saturated for a critical value of the cutoff frequency $f_{\rm end}$ in the GW spectrum, which, for given $T_{\rm rh}$, is solely controlled by the Hubble rate at the end of inflation, $H_{\rm end}$. This means that the $N_{\rm eff}$ bound can be translated to a maximally allowed cutoff frequency $f_{\rm end}^{\rm max}$, which can then be converted to a maximally allowed Hubble rate $H_{\rm end}^{\rm max}$ and ultimately an upper limit $N_{\rm rh}^{\rm max}$ on the allowed number of $e$-folds during reheating. Similarly, if the predicted GW signal at $f_{\rm lvk} = 25\,\textrm{Hz}$ should exceed the LVK bound, we can derive upper limits $H_{\rm end}^{\rm max}$ and $N_{\rm rh}^{\rm end}$ by requiring $f_{\rm end}$ not to exceed the lower end of the LVK band; for definiteness, we use $f_{\rm end}^{\rm max} = 20\,\textrm{Hz}$.

\begin{figure}
    \centering
    \includegraphics{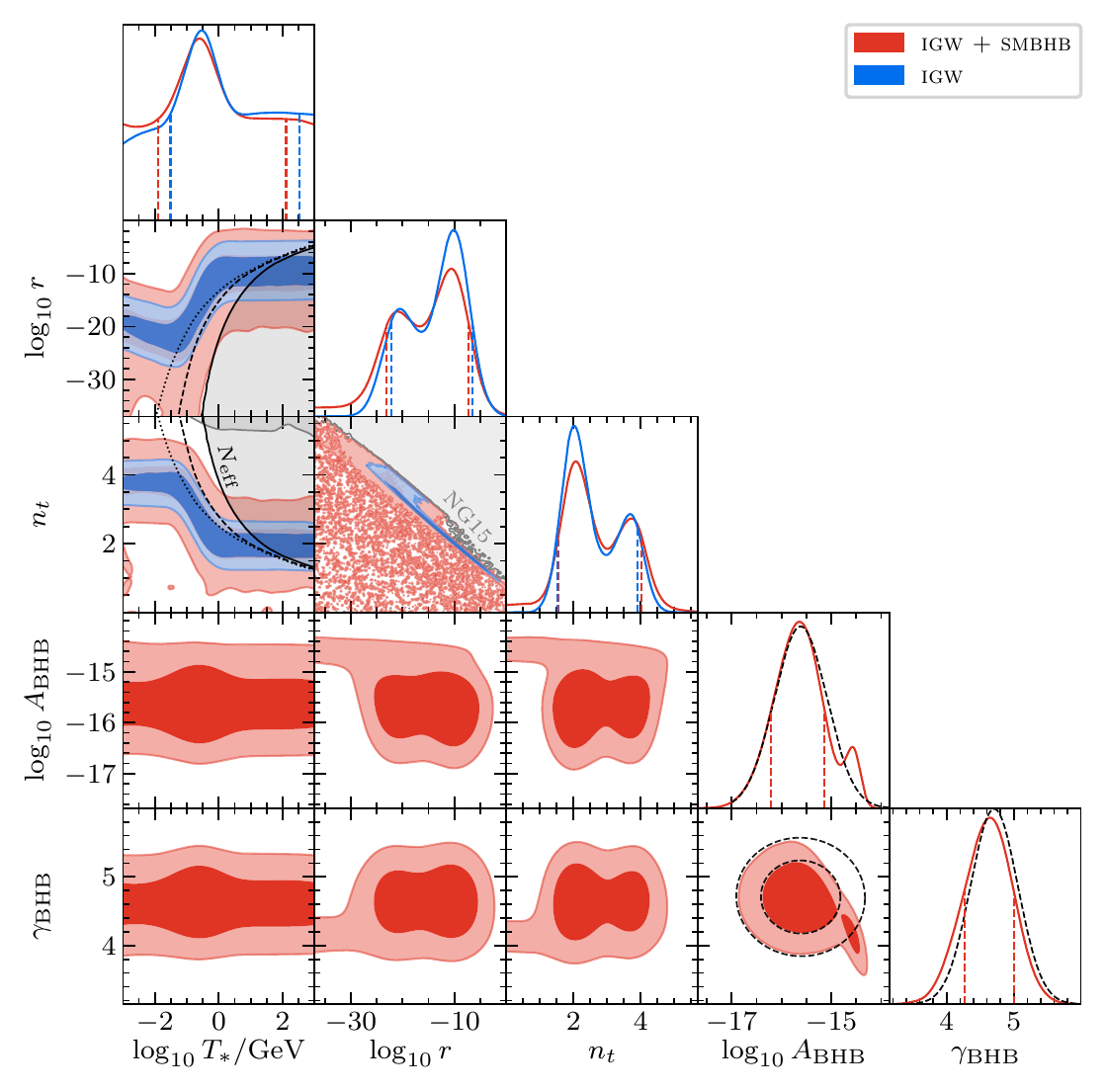}
    \caption{Same as Fig.~\ref{fig:igw_corner} but including the SMBHB parameters $\abhb$ and $\gbhb$. The black dashed lines in the three lower-right panels show the prior distributions for $\abhb$ and $\gbhb$, i.e., one 2D Gaussian and two 1D Gaussian distributions.}
    \label{fig:igw_ext_corner}
\end{figure}

\begin{figure}
	\centering
        \includegraphics[width=0.48\textwidth]{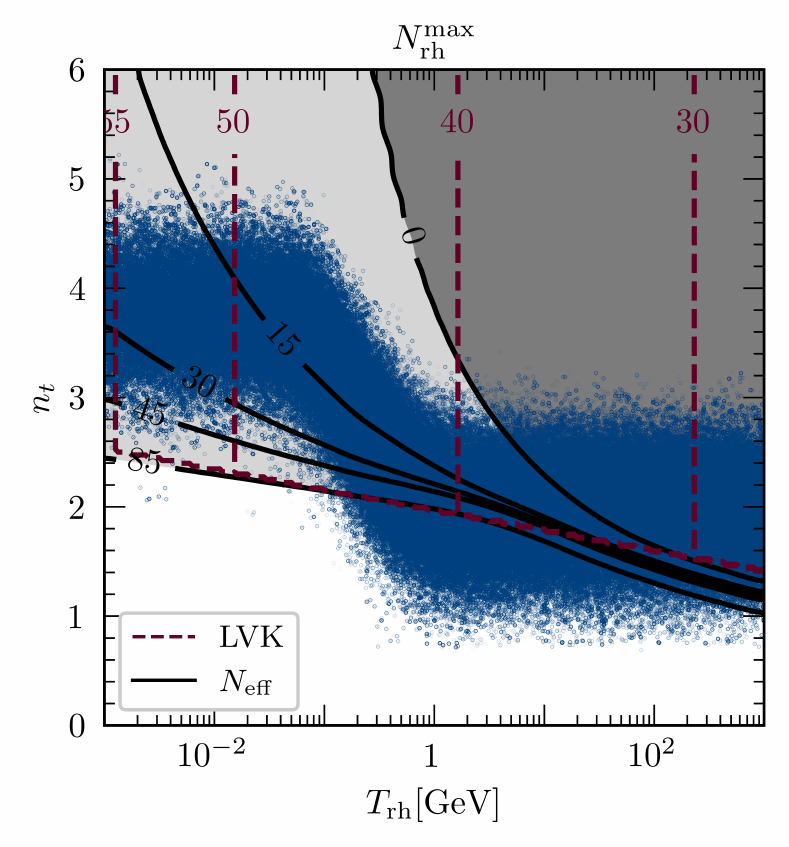}\quad
        \includegraphics[width=0.48\textwidth]{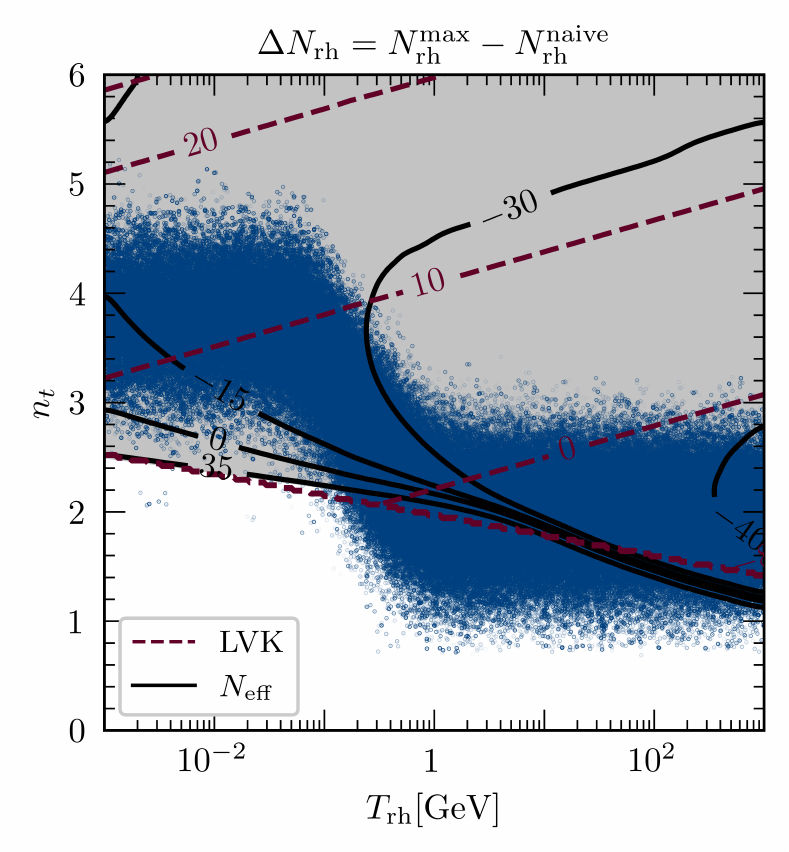}
\caption{Constraints on the parameters of inflationary GWs, i.e., the \textsc{igw} model discussed in Section~\ref{subsec:inflation}. The black solid lines refer to the $N_{\rm eff}$ bound, while the maroon dashed lines refer to the LVK bound. The left panel shows the results for $N_{\rm rh}^{\rm max}$, while the right panel compares $N_{\rm rh}^{\rm max}$ to the duration of reheating that one would naively expect in models of standard single-field slow-roll inflation. The blue points correspond to the sampled points used to obtain the 2D marginalized posterior probability density for the model parameters $T_{\rm rh}$ and $n_t$ in Section~\ref{subsec:inflation}. See Appendix.~\ref{app:inflation} for details.\label{fig:nrhmax}}
\end{figure}

Our results for $N_{\rm rh}^{\rm max}$ obtained in this analysis are presented in Fig.~\ref{fig:nrhmax} in terms of black and maroon contour lines, where the black solid lines refer to the $N_{\rm eff}$ bound, while the maroon dashed lines refer to the LVK bound. In the left panel we show our results for $N_{\rm rh}^{\rm max}$ itself, while in the right panel we compare $N_{\rm rh}^{\rm max}$ to the duration of reheating that one would naively expect in models of standard single-field slow-roll inflation (see Eqs.~\eqref{eq:Nnaive} and \eqref{eq:Hnaive}),
\begin{equation}
N_{\rm rh}^{\rm max} = \frac{2}{3}\ln\left(\frac{H_{\rm end}^{\rm max}}{H_{\rm end}^{\rm min}}\right) \,, \qquad N_{\rm rh}^{\rm naive} = \frac{2}{3}\ln\left(\frac{H_{\rm end}^{\rm naive}}{H_{\rm end}^{\rm min}}\right) \,, \qquad H_{\rm end}^{\rm naive} = \left(\frac{\pi^2}{2} r A_s\right)^{1/2} M_{\scriptscriptstyle \rm Pl} \,,
\end{equation}
where $H_{\rm end}^{\rm min}$ denotes the minimal Hubble rate that is necessary to realize the desired value of $T_{\rm rh}$ after inflation,
\begin{equation}
H_{\rm end}^{\rm min} = \left(\frac{\pi^2g_*^{\rm rh}}{90}\right)^{1/2}\frac{T_{\rm rh}^2}{M_{\scriptscriptstyle \rm Pl}} \,.
\end{equation}
In other words, for given $T_{\rm rh}$, $H_{\rm end}^{\rm min}$ represents the Hubble rate in the limit of instantaneous reheating, $N_{\rm rh} = 0$. In terms of the maximally allowed number of $e$-folds $N_{\rm rh}^{\rm max}$ and the naive expectation $N_{\rm rh}^{\rm naive}$, we can then construct
\begin{equation}
\Delta N_{\rm rh} = N_{\rm rh}^{\rm max} - N_{\rm rh}^{\rm naive} = \frac{2}{3}\ln\left(\frac{H_{\rm end}^{\rm max}}{H_{\rm end}^{\rm naive}}\right) \,,
\end{equation}
which is the quantity shown in the right panel of Fig.~\ref{fig:nrhmax}. In addition to the contour lines for $N_{\rm rh}^{\rm max}$ and $\Delta N_{\rm rh}$, we also display the samples in the $T_{\rm rh}$\,--\,$n_t$ plane that we obtain from our MCMC chains. The density of these blue points approximates the 2D marginalized posterior probability density for the model parameters $T_{\rm rh}$ and $n_t$.

In view of the results in Fig.~\ref{fig:nrhmax}, we can draw several conclusions: (i) Both the $N_{\rm eff}$ bound and the LVK bound are only relevant at large values of the spectral index. At $T_{\rm rh} \sim 10^{-3}\,\textrm{GeV}$, the number of $e$-folds during reheating is only constrained as long as $n_t \gtrsim 2.5$, while at $T_{\rm rh} \sim 10^3\,\textrm{GeV}$, the bounds on $N_{\rm rh}$ only appear at $n_t\gtrsim  1$. At $f \gg f_{\rm rh}$, these $n_t$ values translate to a slope of the GW spectrum $\alpha$ in the range from $\alpha \sim -1$ to $\alpha \sim 0.5$, which is in the regime where we expect the approximate expression in Eq.~\eqref{eq:LVKboundalpha} to be valid (see also the discussion in~\cite{Kuroyanagi:2014nba}). (ii) The LVK bound only depends on $T_{\rm rh}$. This follows from the fact that it is derived from the requirement $f_{\rm end}^{\rm max} = 20\,\textrm{Hz}$, which is independent of $r$ and $n_t$. Typically, the LVK bound on $N_{\rm rh}$ is much weaker than the $N_{\rm eff}$ bound on $N_{\rm rh}$; only in the parameter region where the $N_{\rm eff}$ bound begins to disappear, $N_{\rm rh}^{\rm max} \rightarrow \infty$, does the LVK bound becomes competitive. (iii) The $N_{\rm eff}$ bound is particularly strong at large reheating temperatures and large values of the spectral index, where $N_{\rm rh}^{\rm max}$ can become even negative. This region, which we indicate by a dark-gray shading, is phenomenologically not viable and hence ruled out (see also the regions labeled $N_{\rm eff}$ in Fig.~\ref{fig:igw_corner}). In the excluded region, we also find $\Delta N_{\rm rh} \ll 0$, which indicates that reheating lasting for the naive number of $e$-folds $N_{\rm rh}^{\rm naive}$ will definitely result in a violation of the $N_{\rm eff}$ bound. (iv) Farther away from the excluded region, the upper limits on $N_{\rm rh}$ become weak very fast, $N_{\rm rh}^{\rm max} \gg 1$. In realistic models of inflation and reheating, where we typically expect $N_{\rm rh}\sim \mathcal{O}\left(1\cdots10\right)$, these bounds should be easy to satisfy. We therefore, conclude that most of the region shaded in light gray as well as the entire white region in Fig.~\ref{fig:nrhmax} can host viable realizations of the \textsc{igw} model. This includes, in particular, scenarios with a low reheating temperature, $T_{\rm rh} \sim 10^{-(3\cdots0)}\,\textrm{GeV}$, and a large spectral index of the primordial tensor power spectrum, $n_t \sim 3\cdots 4$. 

\begin{figure}
    \centering
    \includegraphics{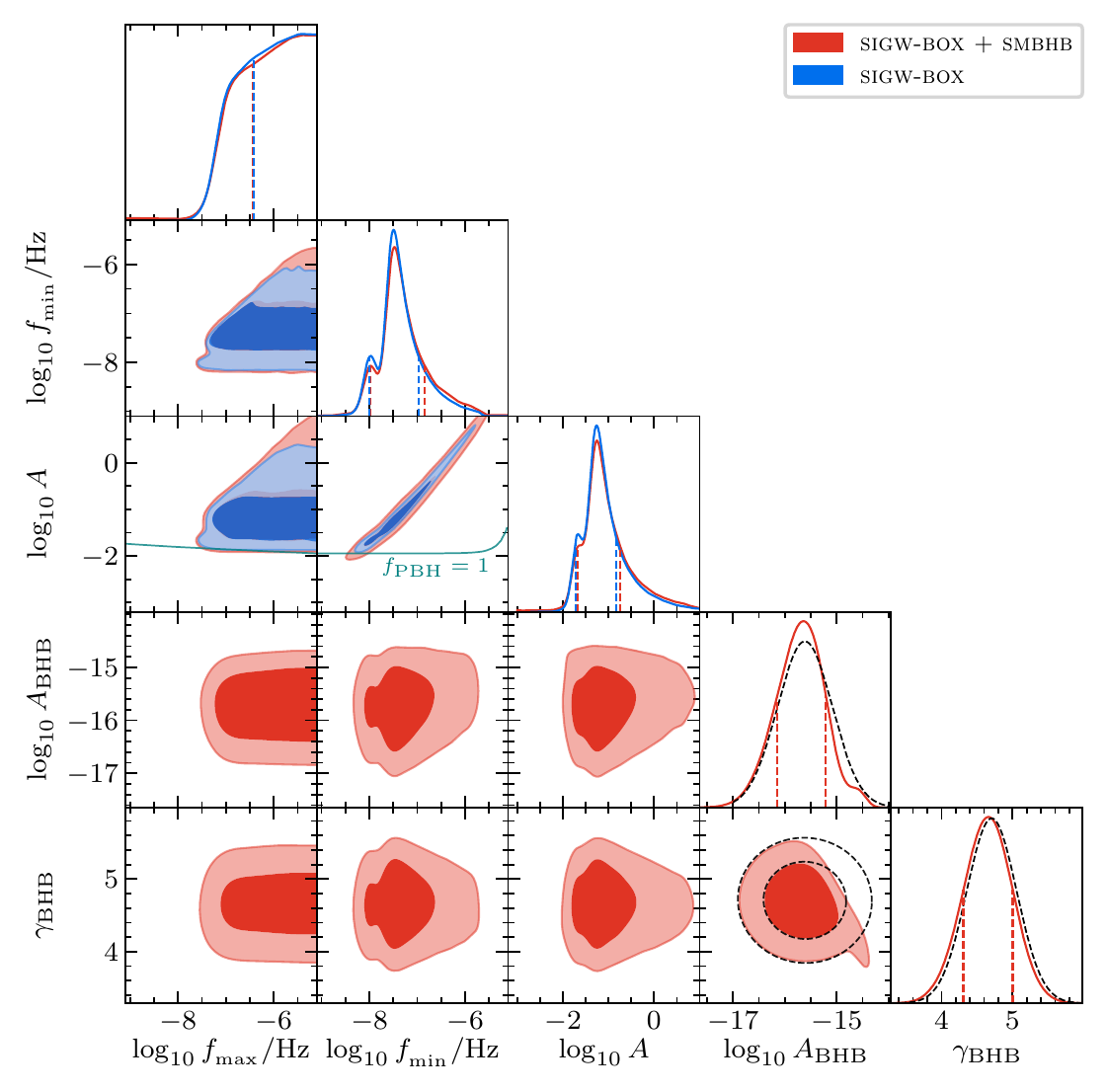}
    \caption{{Same as Fig.~\ref{fig:sigw_box_corner} but including the SMBHB parameters $\abhb$ and $\gbhb$.}
    \label{fig:sigw_box_ext_corner}}
\end{figure}

\begin{figure}
    \centering
    \includegraphics{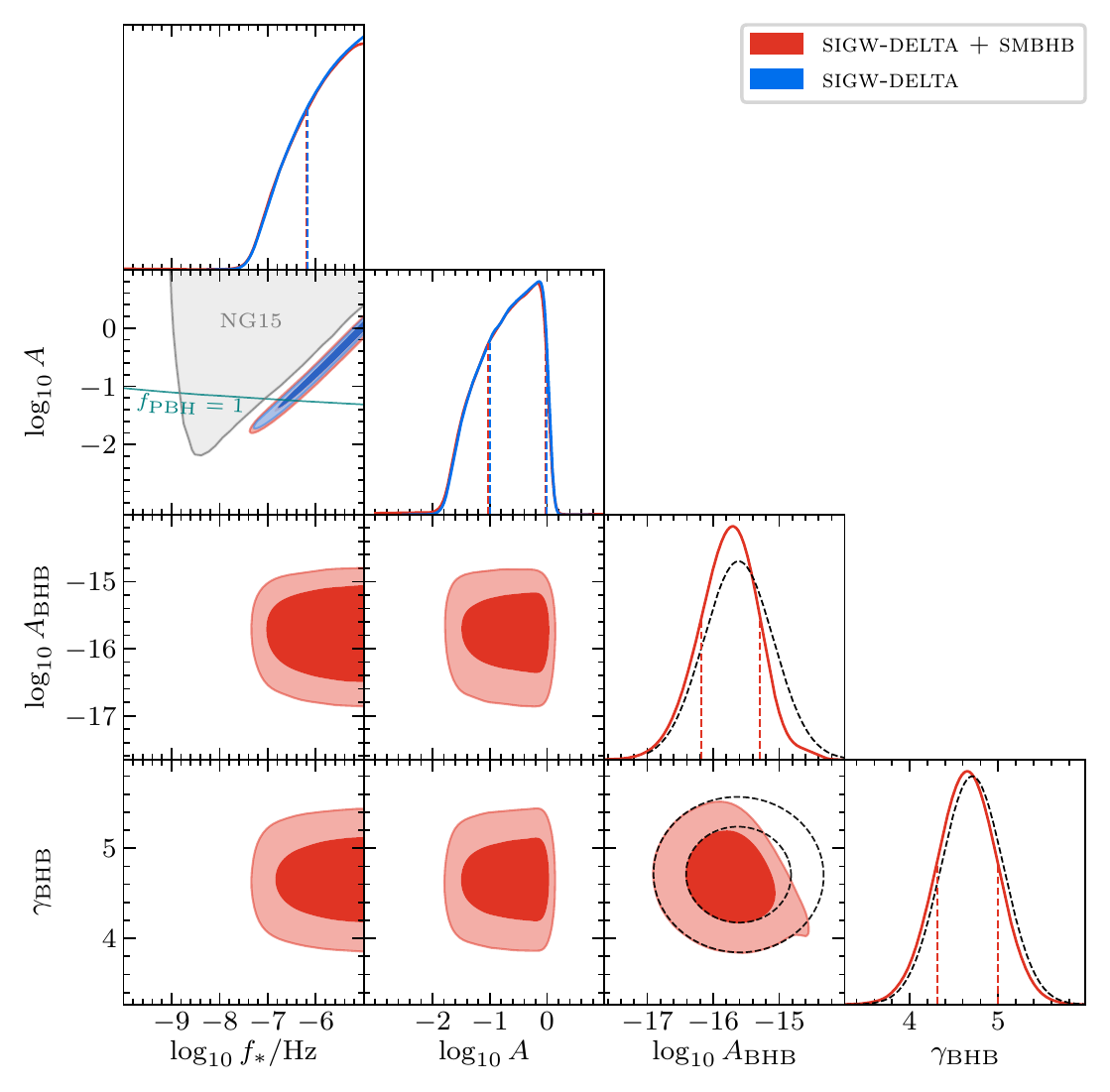}\\
    \includegraphics{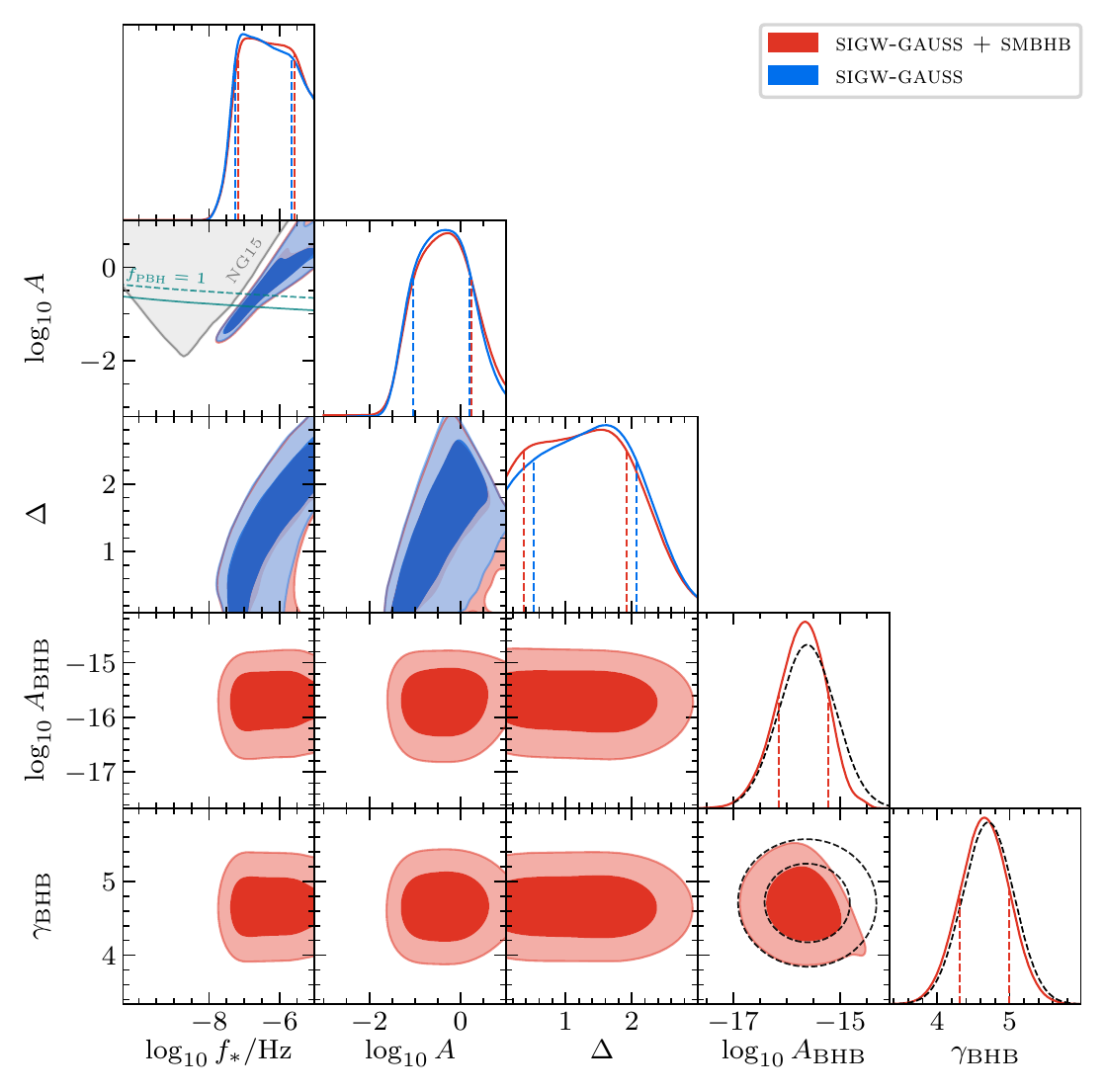}
    \caption{Same as Fig.~\ref{fig:sigw_corner} but including the SMBHB parameters $\abhb$ and $\gbhb$.}
    \label{fig:sigw_ext_corner}
\end{figure}

\subsection{Scalar-induced gravitational waves}
\label{app:sigw}

The generation of SIGWs relies on enhanced scalar perturbations on small scales, which can simultaneously lead to the production of PBHs. The parameter space of the SIGW models considered in this work is therefore subject to the constraint that the PBH mass density must not exceed the relic density of DM, $f_{\scriptscriptstyle \rm PBH} \leq 1$ (see Eq.~\eqref{eq:fpbh}). This appendix explains how we evaluate this bound on the SIGW model parameters. We use the Press--Schechter formalism for spherical collapse~\citep{Press:1973iz, Carr:1975qj} and mainly follow~\cite{Inomata:2017okj} and \cite{ Ando:2018qdb}. The main quantity of interest is $f_{\scriptscriptstyle \rm PBH}$, which denotes the total fraction of PBH DM,
\begin{equation}
f_{\scriptscriptstyle \text{PBH}} =  \int \frac{dM}{M}\, f(M) \,.
\end{equation}
with PBH mass function $f$ given as \citep{Ando:2018qdb}
\begin{equation}
    f\left(M\right)\simeq\gamma^{3/2}\left(\dfrac{\beta(M)}{1.6\times10^{-9}}\right)\left(\dfrac{10.75}{g_*(T)}\right)^{1/4}\left(\dfrac{0.12}{\Omega_{\scriptscriptstyle\text{DM}}h^2}\right)\left(\dfrac{M_{\odot}}{M}\right)^{1/2}.
\end{equation}
Here $M$ is the mass of the PBH, $T$ is the temperature at the time of PBH formation, $\gamma$ is the ratio between the PBH mass and the horizon mass, and $\beta(M)$ is the PBH production rate. In this work, we only focus on PBHs produced during the radiation-dominated era. We choose $\gamma=0.356$, following~\cite{Choptuik:1992jv},\cite{Koike:1995jm} and \cite{Niemeyer:1997mt} and assume that the mass of the PBH is proportional to the horizon mass when $k = aH$,
 \begin{equation}
     M = \gamma\, M_H \simeq \gamma\, M_{\odot} \left(\dfrac{g_*(T)}{10.75}\right)^{-1/6}\left(\dfrac{4.2\times 10^{6} \,\text{Mpc}^{-1}}{k}\right)^{2}.
 \end{equation}
 We can then express the PBH mass as a function of the temperature using the standard temperature--wavenumber relation in $\Lambda$CDM. Assuming that the density perturbation follows Gaussian statistics, $\beta(M)$ is given by
\begin{equation}
     \beta(M)=\int_{\delta_c}\dfrac{d\delta}{\sqrt{2\pi}\,\sigma(M)}\exp\left(-\dfrac{\delta^2}{2\,\sigma^2(M)}\right)\simeq \dfrac{\sigma(M)}{{\sqrt{2\pi}}\,\,\delta_c}\exp\left(-\dfrac{\delta_c^2}{2\,\sigma^2(M)}\right),
\end{equation}
 where $\delta_c$ is the critical density perturbation for the PBH formation, whose exact value depends on the shape of the curvature power spectrum and satisfies $0.4\lesssim \delta_c\lesssim 0.6$~\citep{Musco:2004ak,Musco:2020jjb}, and  $\sigma^2(M)$ is the variance of the coarse-grained density perturbation, which in the radiation domination era is given by
 \begin{equation}
      \sigma^2(k)= \dfrac{16}{81}\int \frac{dq}{q}\:\left(\dfrac{q}{k}\right)^4 \, W^2\left(\dfrac{q}{k}\right) \, \mathcal{T}^2\left(q,k^{-1}\right) \, \mathcal{P}_{\scriptscriptstyle\mathcal{R}}(q) \,.
 \end{equation}
 Here $W(x)$ is some  window function smearing over $k^{-1}$ and
 \begin{equation}
 \mathcal{T}\left(q,k^{-1}\right)= 3 \left[\sin\left(\frac{q}{\sqrt{3}k}\right)-\left(\frac{q}{\sqrt{3}k}\right) \cos\left(\frac{q}{\sqrt{3}k}\right)\right]\big/\left(\frac{q}{\sqrt{3}k}\right)^3 \,.
 \end{equation}
 is the transfer function for the radiation-dominated era. In our work, we choose $\delta_c=0.45$ as a fiducial value~\citep{Ando:2018qdb,Wang:2019kaf} and a Gaussian window function $W(x)=e^{-x^2/2}$. 
 
Our analysis is subject to a few caveats. The correct value of the critical density $\delta_c$ depends on the choice of the power spectrum and the window function~\citep{Ando:2018qdb,Musco:2018rwt,Young:2019osy,Escriva:2019nsa,Escriva:2019phb,Gow:2020bzo,Musco:2020jjb,Dandoy:2023jot}, and the nonlinear relation between density perturbations and density contrast~\citep{DeLuca:2019qsy,Young:2019yug,Escriva:2019nsa}. In this work, we use a single value of $\delta_c$ for three different power spectra for computational reasons. We also disregard any corrections from the QCD equation of state, which are expected to be important at the frequency range probed by PTA observations~\citep{Abe:2020sqb,Escriva:2022bwe,Juan:2022mir,Dandoy:2023jot,Musco:2023dak}.

 The statistics of the primordial scalar perturbations also strongly affect the abundance of PBHs since, in general, enhanced scalar perturbations come with different levels of non-Gaussianity. In this work, we focus on Gaussian primordial scalar perturbations. With non-Gaussianities, the same PBH abundance can be produced with scalar perturbations whose amplitudes are orders of magnitude smaller than those produced with Gaussian scalar perturbations. Therefore, primordial non-Gaussianity can dramatically affect the PBH bounds~\citep{Young:2013oia,Nakama:2016gzw,Pattison:2017mbe,Garcia-Bellido:2017aan,Franciolini:2018vbk,Cai:2018dig,DeLuca:2019qsy,Young:2019yug,Iacconi:2023slv}. In the parameter region where SIGWs manage to explain the NANOGrav signal, this would notably aggravate the PBH overproduction problem, which means that sizable non-Gaussianities could completely rule out the SIGW interpretation of the signal. In this work, we also neglect evolutionary effects on the PBH mass function, namely accretion and mergers, where accretion effects are expected to be small for sub-solar-mass PBHs~\citep{Ali-Haimoud:2016mbv,Raidal:2018bbj,Vaskonen:2019jpv,DeLuca:2020fpg}.

In summary, the final PBH DM fraction is quite sensitive to the assumptions discussed above. The bounds presented in the main text should be taken with a grain of salt owing to the uncertainties in the computation of $f_{\scriptscriptstyle \rm PBH}$. 

\begin{figure*}[h!]
	\centering
        \includegraphics[width=0.48\textwidth]{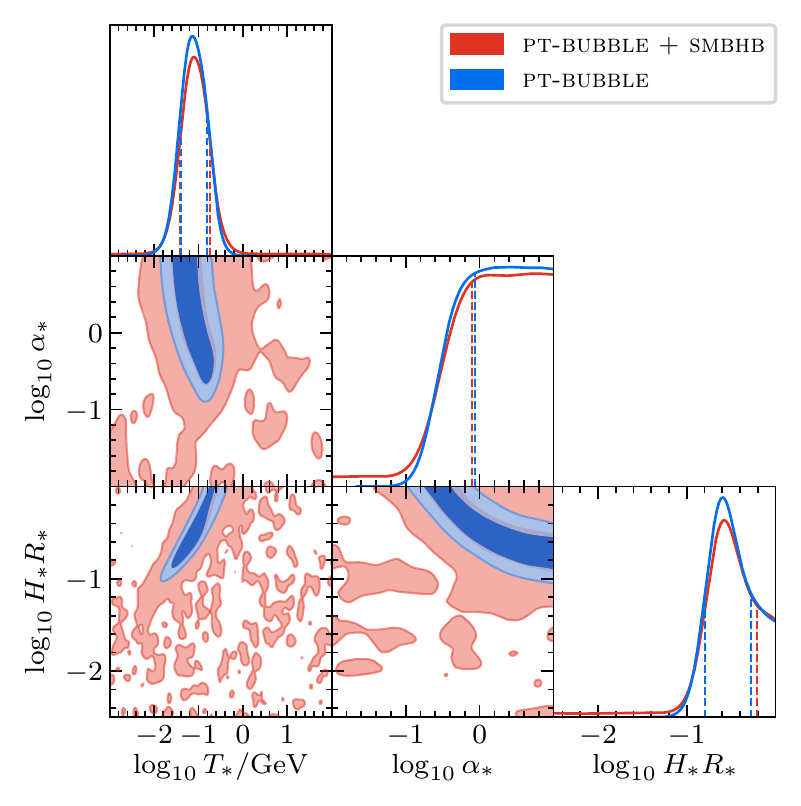}
        \includegraphics[width=0.48\textwidth]{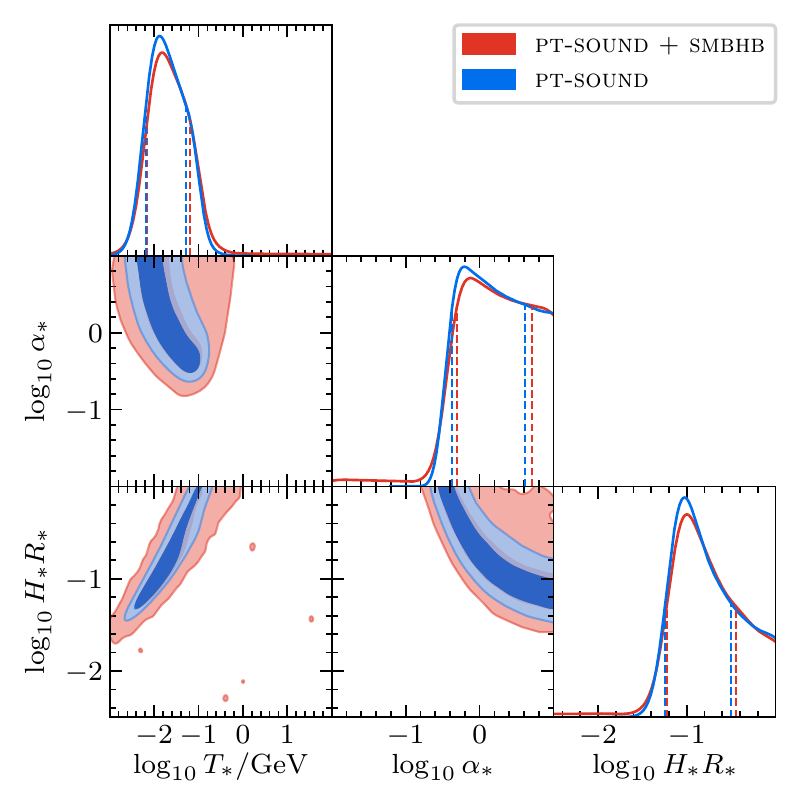}
	\caption{\label{fig:pt_corner_fix}
        Same as in Fig.~\ref{fig:igw_corner} but for the \textsc{pt-sound} (left panel) and \textsc{pt-bubble} models with a low-frequency slope fixed to the value predicted by causality, i.e., $a=3$.}
\end{figure*}

\subsection{Cosmological phase transitions}
\label{app:phase_transitions}
In the phase transition analysis discussed in the main text, we allow the low-frequency spectral index to float despite that causality predicts $a=3$. We do this to capture possible double-peak spectral features with our simple power-law parametrization. However, it is not clear whether or not a strong and fast phase transition like the one needed to explain the observed signal could produce such a double-peak structure~\citep{Hindmarsh:2017gnf, Hindmarsh:2019phv}. Therefore, in Fig.~\ref{fig:pt_corner_fix} we report the results of a phase transition analysis where we assume $a=3$. Figure~\ref{fig:pt_corner_fix} shows the reconstructed posterior distributions for the parameters $\alpha_*$, $T_*$, and $H_*R_*$ of the \textsc{pt-sound} and \textsc{pt-bubble} models, both for the case where the PT is assumed to be the only source of GWs (blue contours) and for the scenario where we consider the superposition of the PT and SMBHBs signal (red contours). For the analyses where the SMBHB signal is included, we also report the posterior distributions for $\abhb$ and $\gbhb$. For the \textsc{pt-sound} model we notice very minor differences compared to the analysis in the main text. However, for the \textsc{pt-bubble} model we notice how the posterior for $T_*$ is peaked to slightly smaller values for the reasons explained in the main text. 

In Figs.~\ref{fig:pt_bubble_corner_full} and \ref{fig:pt_sound_corner_full}, we report the posterior distributions for all the parameters of the phase transition models, including the spectral shape parameters $a$, $b$, and $c$ that were excluded from Fig.~\ref{fig:pt_corner}. As expected for the \textsc{pt-bubble} model, the low-frequency slope is peaked around $a\sim2$, which is the reconstructed slope of the GWB signal, while the posteriors for $b$ and $c$ are approximately flat. For the \textsc{pt-sound} model, the posterior for $a$ is peaked around the lower limit of the prior range $a=3$, and there is also a mild preference for larger values of the width parameter, as this would flatten the spectrum close to the peak. 
\begin{figure*}
	\centering
        \includegraphics[width=0.98\textwidth]{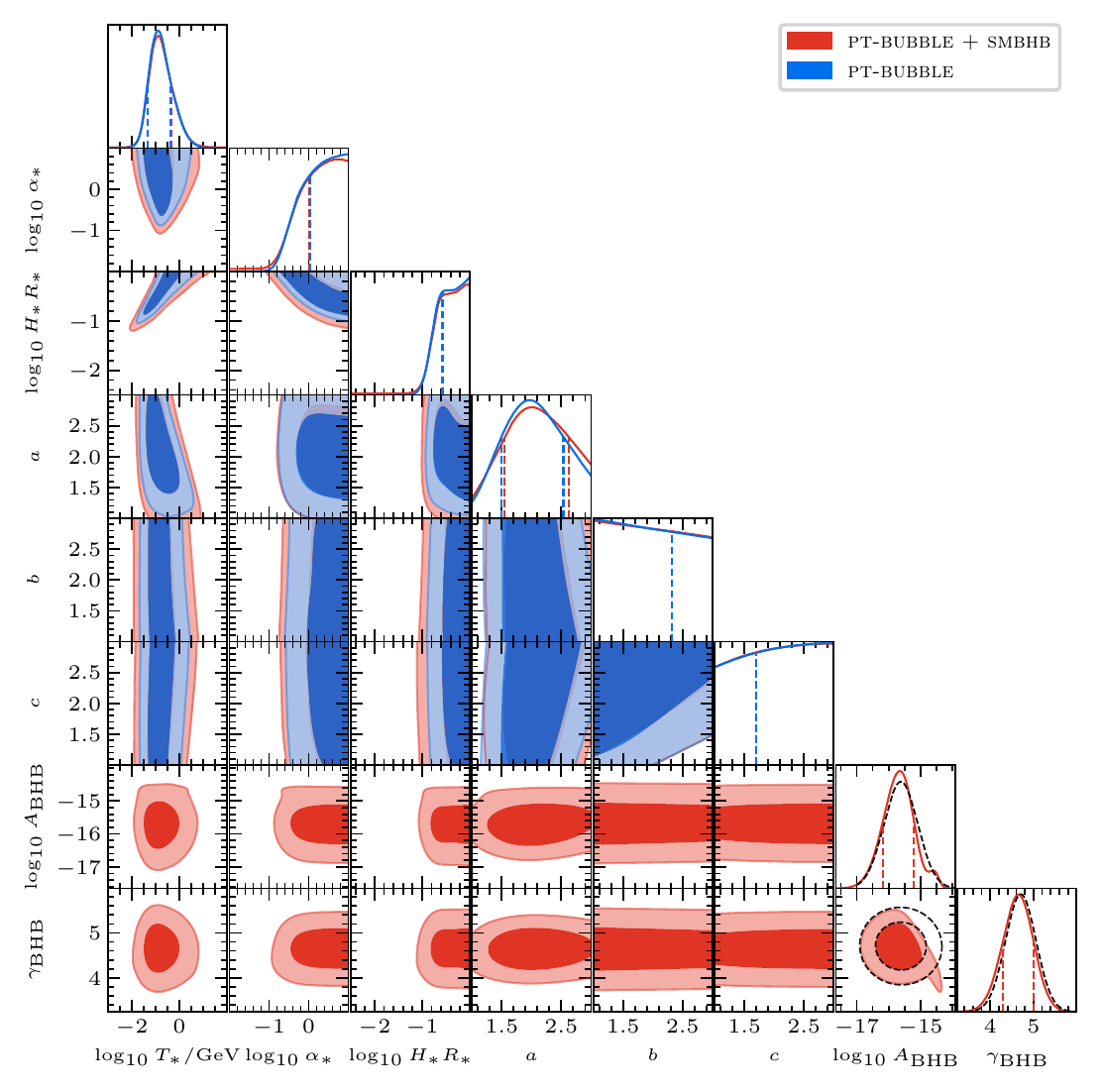}
	\caption{Same as Fig.~\ref{fig:pt_corner} but including the spectral shape parameters $a$, $b$, $c$ and SMBHB parameters $\abhb$ and $\gbhb$.}
        \label{fig:pt_bubble_corner_full}
\end{figure*}
\begin{figure*}
	\centering
        \includegraphics[width=0.98\textwidth]{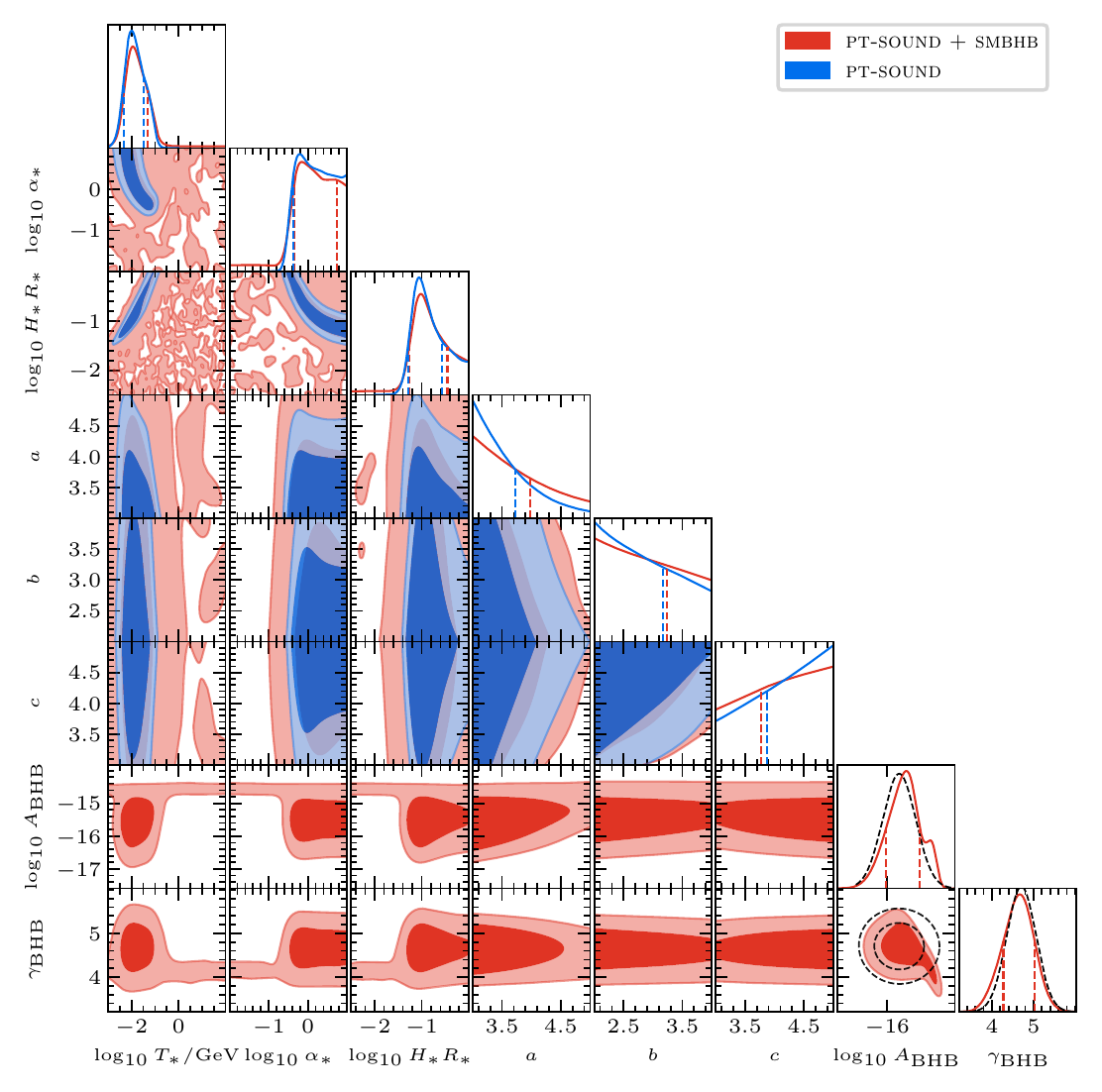}
	\caption{Same as Fig.~\ref{fig:pt_bubble_corner_full} but for the \textsc{pt-sound} model. }
        \label{fig:pt_sound_corner_full}
\end{figure*}

\subsection{Cosmic strings}
\label{app:strings}

\begin{figure}
\centering
\includegraphics{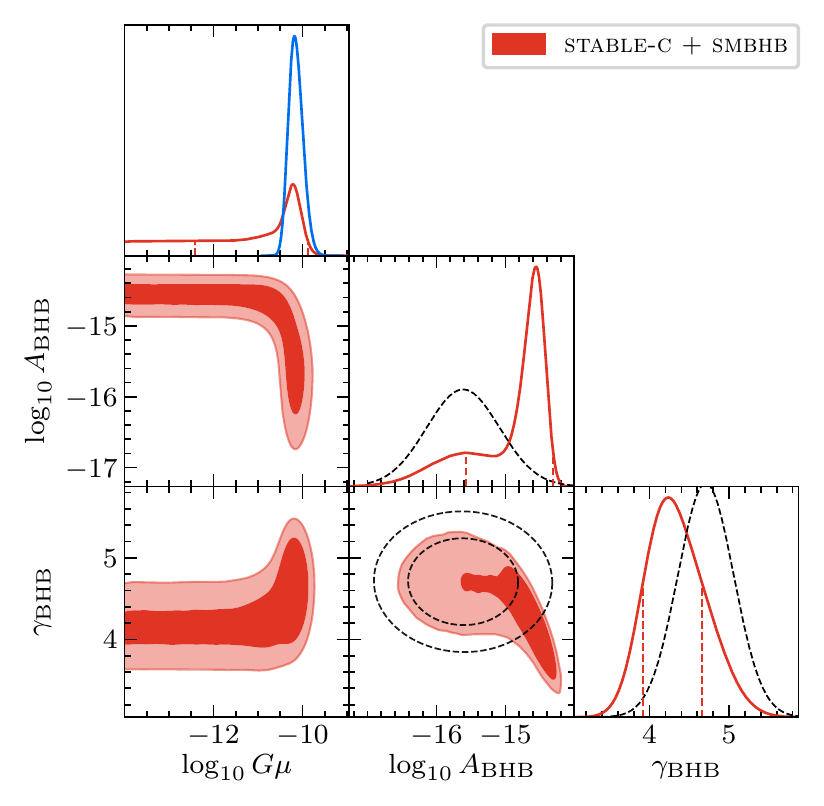}\qquad
\includegraphics{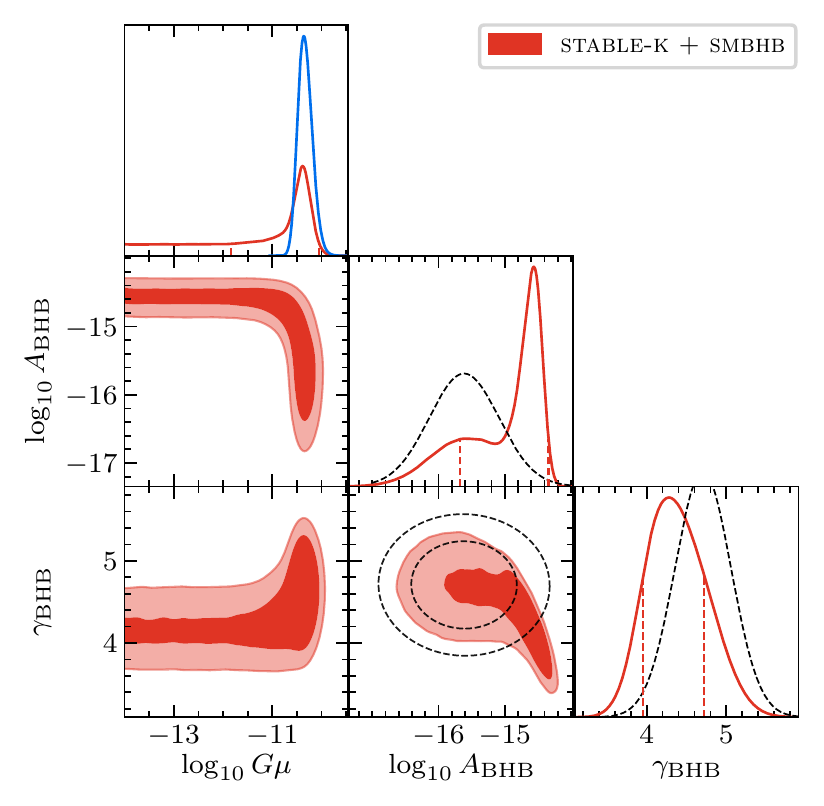}\\
\includegraphics{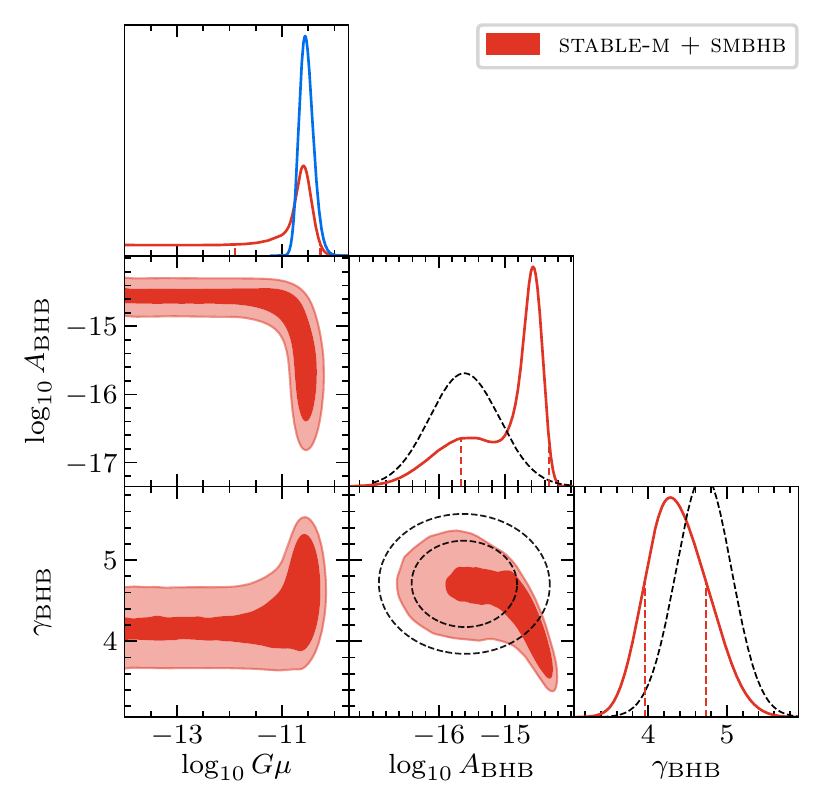}\qquad
\includegraphics{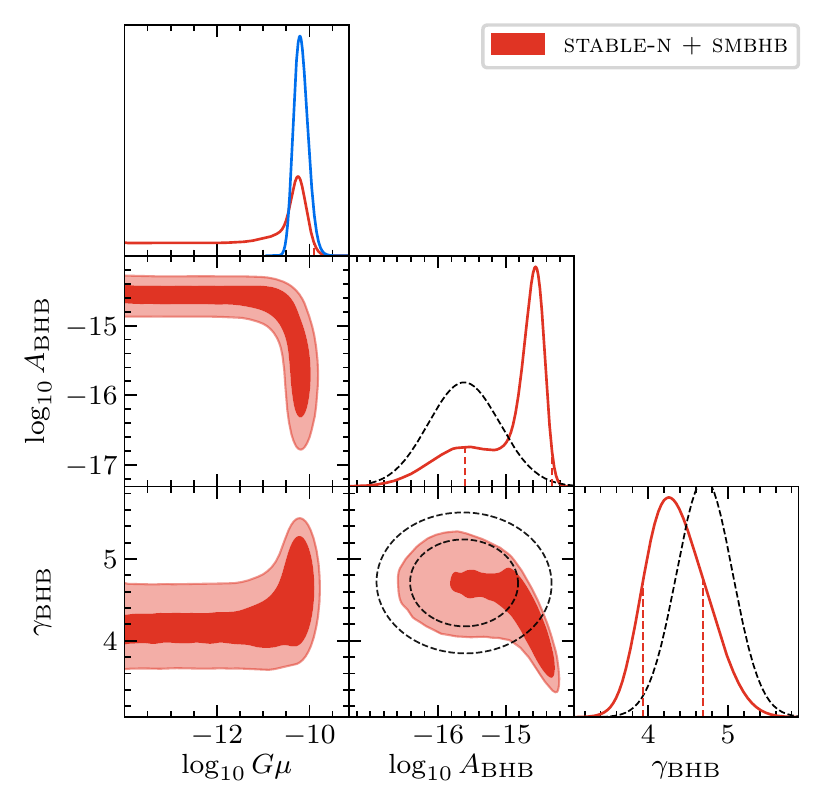}
\caption{Posterior distributions for the \textsc{stable+smbhb} strings models including the SMBHB parameters $\abhb$ and $\gbhb$. We also report the marginalized $\log_{10}G\mu$ posterior distributions for the \textsc{stable} string models (blue lines).}
\label{fig:string_stable_corner}
\end{figure}

\begin{figure}
    \centering
    \includegraphics{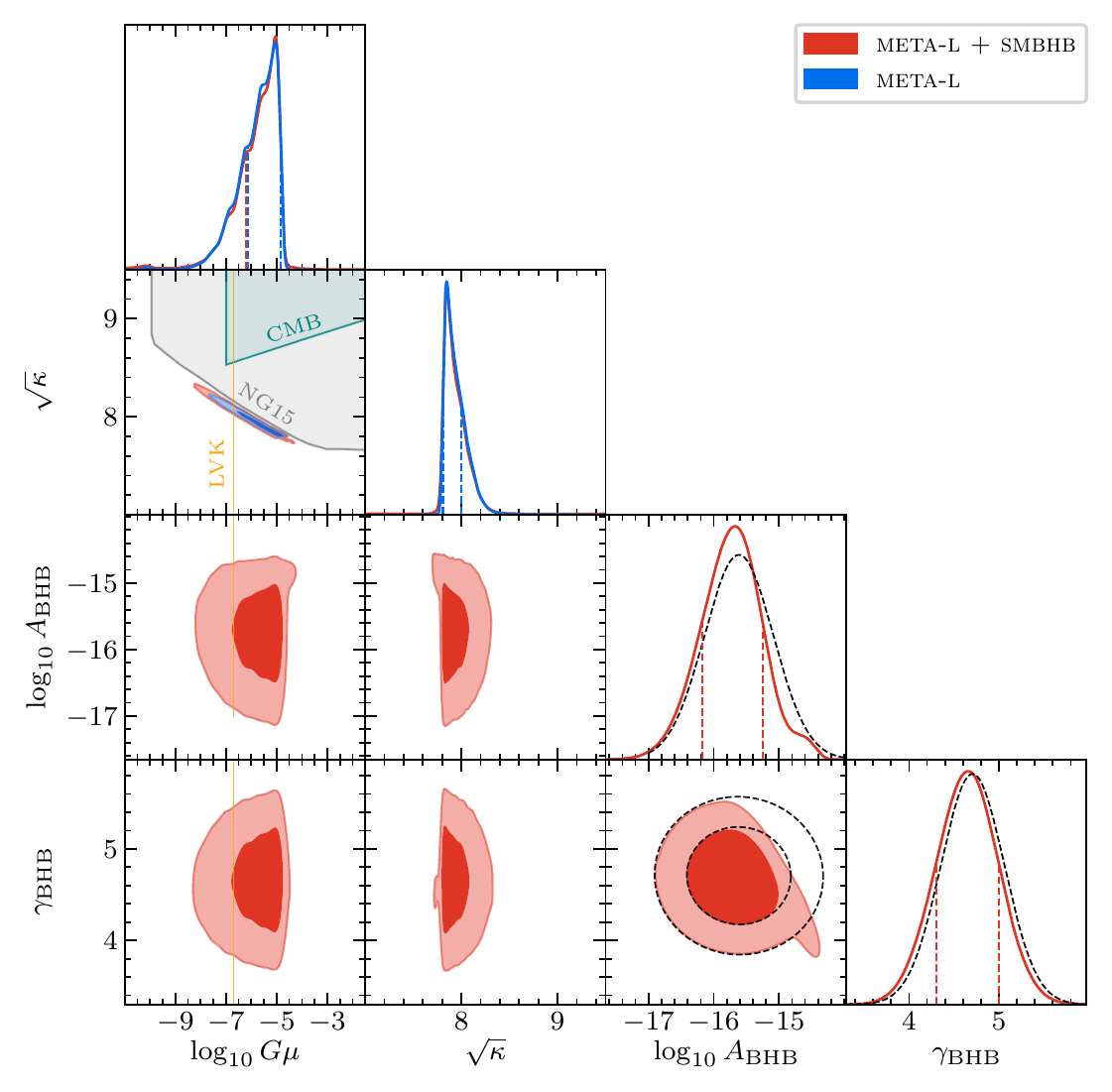}\\
    \includegraphics{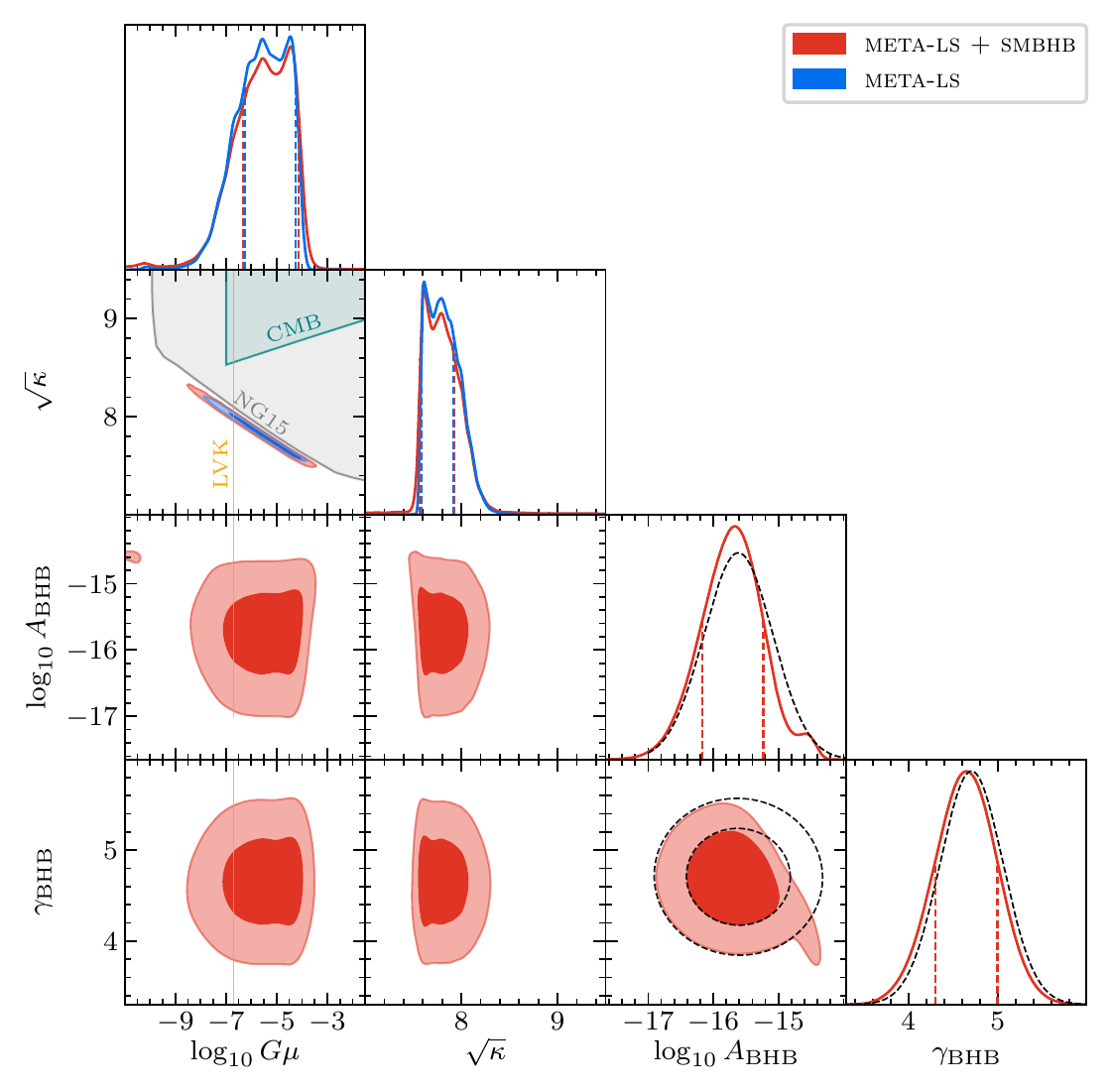}
    \caption{Same as Fig.~\ref{fig:string_meta_corner} but including the SMBHB parameters $\abhb$ and $\gbhb$.}
    \label{fig:string_meta_ext_corner}
\end{figure}

\begin{figure}
    \centering
    \includegraphics{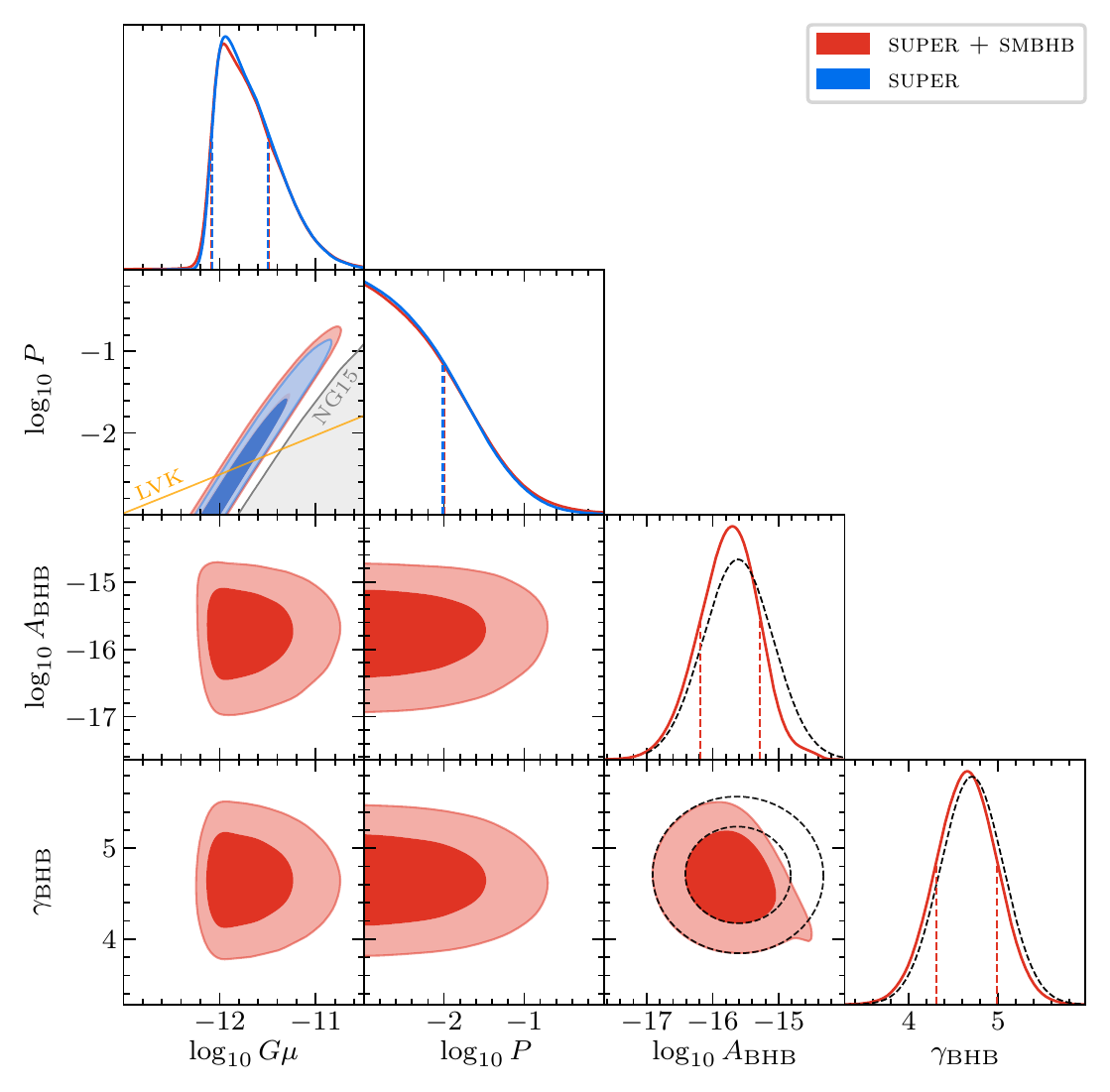}
    \caption{Same as Fig.~\ref{fig:string_super_corner} but including the SMBHB parameters $\abhb$ and $\gbhb$.}
    \label{fig:superstrings_ext_corner}
\end{figure}

In our discussion of stable cosmic strings in the main text, we only present the reconstructed marginalized 1D posterior distributions for the dimensionless cosmic-string tension, $G\mu$ (see Fig.~\ref{fig:string_stable_post}). The four strings-plus-SMBHBs models \textsc{stable-c$+$smbhb}, \textsc{stable-k$+$smbhb}, \textsc{stable-m$+$smbhb}, and \textsc{stable-n$+$smbhb}, however, feature three model parameters in total: $G\mu$, $\abhb$, and $\gbhb$. In this appendix, we complement the discussion in Section~\ref{subsec:strings} and show the corner plots for these parameters (see Fig.~\ref{fig:string_stable_corner}). A notable feature of these corner plots is that the plateau region at small values of $G\mu$, which we had already observed in Fig.~\ref{fig:string_stable_post}, now also appears in the form of flat directions in the 2D posterior distributions for $G\mu$ and $\abhb$ as well as for $G\mu$ and $\gbhb$. Meanwhile, the 2D posterior distributions for $\abhb$ and $\gbhb$ represent distorted versions of the 2D Gaussian prior distribution that we employ in our analysis, peaking at large $\abhb$ and small $\gbhb$, where SMBHBs yield the dominant contribution to the signal. All four corner plots in Fig.~\ref{fig:string_stable_corner} are qualitatively identical and only display slight numerical differences.

Next, let us turn to metastable strings. In the main text, we discuss the number density of closed string loops for the case of stable strings (see Eq.~\eqref{eq:nl-stable}) as well as for metastable strings (see Eq.~\eqref{eq:nl-meta}). In the \textsc{meta-ls} model, we need in addition the number density of string segments that form when long strings and closed loops break apart as a consequence of monopole nucleation. In order to compute the GW signal from segments, we use again Eqs.~\eqref{eq:Ogwcs} and \eqref{eq:Ikcs}, where we replace the loop number density $n_l$ by the segment number density $n_s$ and the GW power spectrum $P_k$ in Eq.~\eqref{eq:Pk} by $P_k = 4/k$, which was derived by Martin and Vilenkin in the approximation of a straight string segment in~\cite{Martin:1996cp}. Meanwhile, all relevant expressions for the segment number density $n_s$ were computed in~\cite{Leblond:2009fq} and \cite{Buchmuller:2021mbb}, which we shall summarize in this appendix. For more details, we refer to~\citep{Leblond:2009fq} and \citep{Buchmuller:2021mbb}.

First, we consider segments that form when long strings break apart. We denote their number density by $\bar{n}_s$, which we decompose into three different contributions that are relevant at different times and in different parameter regimes,
\begin{equation}
\bar{n}_s\left(\ell,t\right) = \bar{n}_s^{\rm rr}\left(\ell,t\right) + \bar{n}_s^{\rm rm}\left(\ell,t\right) + \bar{n}_s^{\rm mm}\left(\ell,t\right) \,.
\end{equation}
Here $\bar{n}_s^{\rm rr}\left(\ell,t\right)$ is the number density of segments that originate from long strings, assuming the scaling regime to end during radiation domination, evaluated during radiation domination and after the onset of the decay regime,
\begin{equation}
\bar{n}_s^{\rm rr}\left(\ell,t\right) = \Theta\left(t_{\rm eq}-t\right)\Theta\left(t-t_s\right) \,\frac{\Gamma_d^2}{\xi_r^2}\frac{\left(t+t_s\right)^2}{\left(t^3t_s\right)^{1/2}}\,\exp\left[-\Gamma_d\left(\ell\left(t+t_s\right) + \sfrac{1}{2}\,\Gamma G\mu\left(t-t_s\right)\left(t+3\,t_s\right)\right)\right] \,,
\end{equation}
where
$\bar{n}_s^{\rm rm}\left(\ell,t\right)$ is the number density of segments that originate  from long strings, assuming that the scaling regime ends during radiation domination, evaluated during matter domination and after the onset of the decay regime,
\begin{equation}
\bar{n}_s^{\rm rm}\left(\ell,t\right) = \Theta\left(t-t_{\rm eq}\right)\Theta\left(t_{\rm eq}-t_s\right) \,\frac{\Gamma_d^2}{\xi_r^2}\left(\frac{t_{\rm eq}}{t}\right)^2\frac{\left(t+t_s\right)^2}{\left(t_{\rm eq}^3t_s\right)^{1/2}}\,\exp\left[-\Gamma_d\left(\ell\left(t+t_s\right) + \sfrac{1}{2}\,\Gamma G\mu\left(t-t_s\right)\left(t+3\,t_s\right)\right)\right] \,,
\end{equation}
and $\bar{n}_s^{\rm mm}\left(\ell,t\right)$ is the number density of segments that originate from long strings, assuming that the scaling regime ends during matter domination, evaluated during matter domination and after the onset of the decay regime,
\begin{equation}
\bar{n}_s^{\rm mm}\left(\ell,t\right) = \Theta\left(t-t_s\right)\Theta\left(t_s-t_{\rm eq}\right) \,\frac{\Gamma_d^2}{\xi_m^2}\,\exp\left[-\Gamma_d\left(\ell\,t + \sfrac{1}{2}\,\Gamma G\mu\left(t-t_s\right)\left(t+t_s\right)\right)\right] \,.
\end{equation}
In view of these expressions, several comments are in order: (i) Throughout our analysis, we assume that GW emission by segments is as efficient as GW emission by loops, i.e., we work with $\Gamma = 50$ for both loops and segments. (ii) All three expressions depend on the dimensionless correlation length of the long-string network, for which we use the attractor values in the VOS model, $\xi_r = 0.27$ and $\xi_m = 0.63$ during radiation and matter domination, respectively. (iii) For a given choice of parameter values, $\bar{n}_s^{\rm rr}$, $\bar{n}_s^{\rm rm}$, and $\bar{n}_s^{\rm mm}$ never contribute simultaneously to the segment number density. For $t_{\rm eq} > t_s$, only $\bar{n}_s^{\rm rr}$ and $\bar{n}_s^{\rm rm}$ yield nonvanishing contributions (first $\bar{n}_s^{\rm rr}$ at $t<t_{\rm eq}$ and then $\bar{n}_s^{\rm rm}$ at $t> t_{\rm eq}$), whereas for $t_{\rm eq} < t_s$, the number density of segments from long strings is solely determined by $\bar{n}_s^{\rm mm}$ at all times $t>t_s$.

In addition to segments from long strings, there is also a population of segments that form when closed string loops begin to break apart because of monopole nucleation. We denote the number density of this population by $\tilde{n}_{s,1}$, where the index refers to the fact that $\tilde{n}_{s,1}$ only describes the first generation of segments that form when closed loops break apart for the first time. This first generation then gives rise to a second generation of segments that follow from monopole nucleation on first-generation segments, and so on and so forth. We comment on these higher generations further below. Before that, however, we discuss $\tilde{n}_{s,1}$, which we decompose again into three contributions,
\begin{equation}
\tilde{n}_{s,1}\left(\ell,t\right) = \tilde{n}_{s,1}^{\rm rr}\left(\ell,t\right) + \tilde{n}_{s,1}^{\rm rm}\left(\ell,t\right) + \tilde{n}_{s,1}^{\rm mm}\left(\ell,t\right) \,.
\end{equation}
Here $\tilde{n}_{s,1}^{\rm rr}\left(\ell,t\right)$ is the number density of first-generation segments that originate from closed loops that formed during radiation domination, evaluated during radiation domination and after the onset of the decay regime,
\begin{equation}
\label{eq:nss1rr}
\tilde{n}_{s,1}^{\rm rr}\left(\ell,t\right) = \Theta\left(t-t_s\right)\Theta\left(t_{\rm eq}-t\right)\Theta\left(t_{\rm eq}-t_*\right) \,\Gamma_d\left[\ell\left(t-t_s\right) + \sfrac{1}{2}\,\Gamma G\mu\left(t-t_s\right)^2\right] n_l^{\rm meta}\left(\ell,t\right)\,,
\end{equation}
where
$\tilde{n}_{s,1}^{\rm rm}\left(\ell,t\right)$ is the number density of first-generation segments that originate from closed loops that formed during radiation domination, evaluated during matter domination and after the onset of the decay regime,
\begin{equation}
\label{eq:nss1rm}
\tilde{n}_{s,1}^{\rm rm}\left(\ell,t\right) = \Theta\left(t-t_s\right)\Theta\left(t-t_{\rm eq}\right)\Theta\left(t_{\rm eq}-t_*\right) \,\Gamma_d\left[\ell\left(t-t_s\right) + \sfrac{1}{2}\,\Gamma G\mu\left(t-t_s\right)^2\right] n_l^{\rm meta}\left(\ell,t\right) \,,
\end{equation}
and $\tilde{n}_{s,1}^{\rm mm}\left(\ell,t\right)$ is the number density of first-generation segments that originate from closed loops that formed during matter domination, evaluated during matter domination and after the onset of the decay regime,
\begin{align}
\label{eq:nss1mm}
& \tilde{n}_{s,1}^{\rm mm}\left(\ell,t\right) = \Theta\left(t-t_s\right)\Theta\left(t_s-t_*\right)\Theta\left(t_*-t_{\rm eq}\right)\, \frac{\Gamma_d\,e^{-\Gamma_d\left[\ell\left(t-t_*\right) + \sfrac{1}{2}\,\Gamma G\mu\left(t-t_*\right)^2\right]}}{t^2\left(\ell + \Gamma G\mu\,t\right)^2} \\
\label{eq:nss1mm2}
& \times \Big\{0.27\left[\ell\left(t-t_s\right) + \sfrac{1}{2}\,\Gamma G\mu\left(t-t_s\right)^2\right] + 0.45 \left(\ell + \Gamma G\mu\,t\right)^{1 + 0.31}\left[F_2\left(t\right)-F_1\left(t\right)-F_2\left(t_s\right)+F_1\left(t_s\right)\right] \Big\} \,,
\end{align}
where $F_2$ and $F_1$ are shorthand notations for the following expressions involving the hypergeometric function $_2F_1$,
\begin{equation}
F_n\left(x\right) = {}_2F_1\left(n-\beta,-\beta;n+1-\beta;\frac{\Gamma G\mu}{\ell + \Gamma G\mu\,t}\,x\right)\left(\frac{\Gamma G\mu}{\ell + \Gamma G\mu\,t}\right)^{n-1} \frac{x^{n-\beta}}{n-\beta} \,, \qquad n = 1,2 \,, \qquad \beta = 0.31 \,.
\end{equation}
These results were derived in~\cite{Buchmuller:2021mbb} based on the loop number densities in the so-called BOS model~\citep{Blanco-Pillado:2013qja}, which explains the occurrence of factors such as $0.27$, $0.45$, and $0.31$ in Eqs.~\eqref{eq:nss1mm} and \eqref{eq:nss1mm2} (see~\cite{Buchmuller:2021mbb} for more details). 

Finally, the expressions in Eqs.~\eqref{eq:nss1rr}, \eqref{eq:nss1rm}, and \eqref{eq:nss1mm} enable us to estimate the total number densities of \textit{all} generations of segments that descend from closed loops breaking apart because of monopole nucleation. In principle, these number densities are described by a partial integro-differential equation where the abundance of the $n$th segment generation acts as a source term for the $(n+1)$th generation. Formally, this partial integro-differential equation can be solved analytically in terms of an infinite recursive series~\citep{Buchmuller:2021mbb}. The numerical evaluation of this series is, however, technically demanding, which is why we choose to follow a different approach for the purposes of this paper. As shown in~\cite{Buchmuller:2021mbb}, it turns out that the total number densities for segments from closed loops result in predictions for the GW spectrum that are very similar to those obtained from the corresponding first-generation number densities in Eqs.~\eqref{eq:nss1rr}, \eqref{eq:nss1rm}, and \eqref{eq:nss1mm}\,---\,up to a numerical fudge factor of $\mathcal{O}\left(10\right)$. At the level of the number densities, it therefore suffices to rescale all first-generation number densities by a constant factor in order to obtain effective number densities for all generations of segments that descend from closed loops breaking apart,
\begin{equation}
\label{eq:fudge}
\tilde{n}_s^{\rm rr}\left(\ell,t\right) \rightarrow \textrm{fudge}  \times \tilde{n}_{s,1}^{\rm rr}\left(\ell,t\right) \,, \qquad \tilde{n}_s^{\rm rm}\left(\ell,t\right) \rightarrow \textrm{fudge}  \times \tilde{n}_{s,1}^{\rm rm}\left(\ell,t\right) \,, \qquad \tilde{n}_s^{\rm mm}\left(\ell,t\right) \rightarrow \textrm{fudge}  \times \tilde{n}_{s,1}^{\rm mm}\left(\ell,t\right) \,.
\end{equation}
We stress that the functional dependence on $\ell$ and $t$ is typically not the same for the first-generation and total number densities; the rescaling in Eq.~\eqref{eq:fudge}, however, achieves comparable results at the level of the GW spectrum. In our analysis, we consistently use a fudge factor of $5$, which is a characteristic value across large regions of parameter space.

\subsection{Domain walls}
\label{app:dw}

In Fig.~\ref{fig:dw_corner_full}, we report the posterior distribution for all the parameters of the domain wall models, including the spectral shape parameters $b$ and $c$ that were excluded from Fig.~\ref{fig:dw_corner}. For both the \textsc{dw-sm} and \textsc{dw-dr} models we find a flat posterior for the high-frequency slope, $b$. This result is expected since most of the low frequency bins are fit by the low-frequency tail of the spectrum. For the width parameter, $c$, instead, both models prefer values near the upper range of our prior range. This preference for wider spectra derives from a mismatch between the reconstructed slope of the GWB, $a\sim2$ and the one predicted by causality, $a\sim3$. Larger width parameters make the spectrum from domain walls flatter near the peak and allow for a better fit to the data. 

\begin{figure*}
	\centering
        \includegraphics[width=0.65\textwidth]{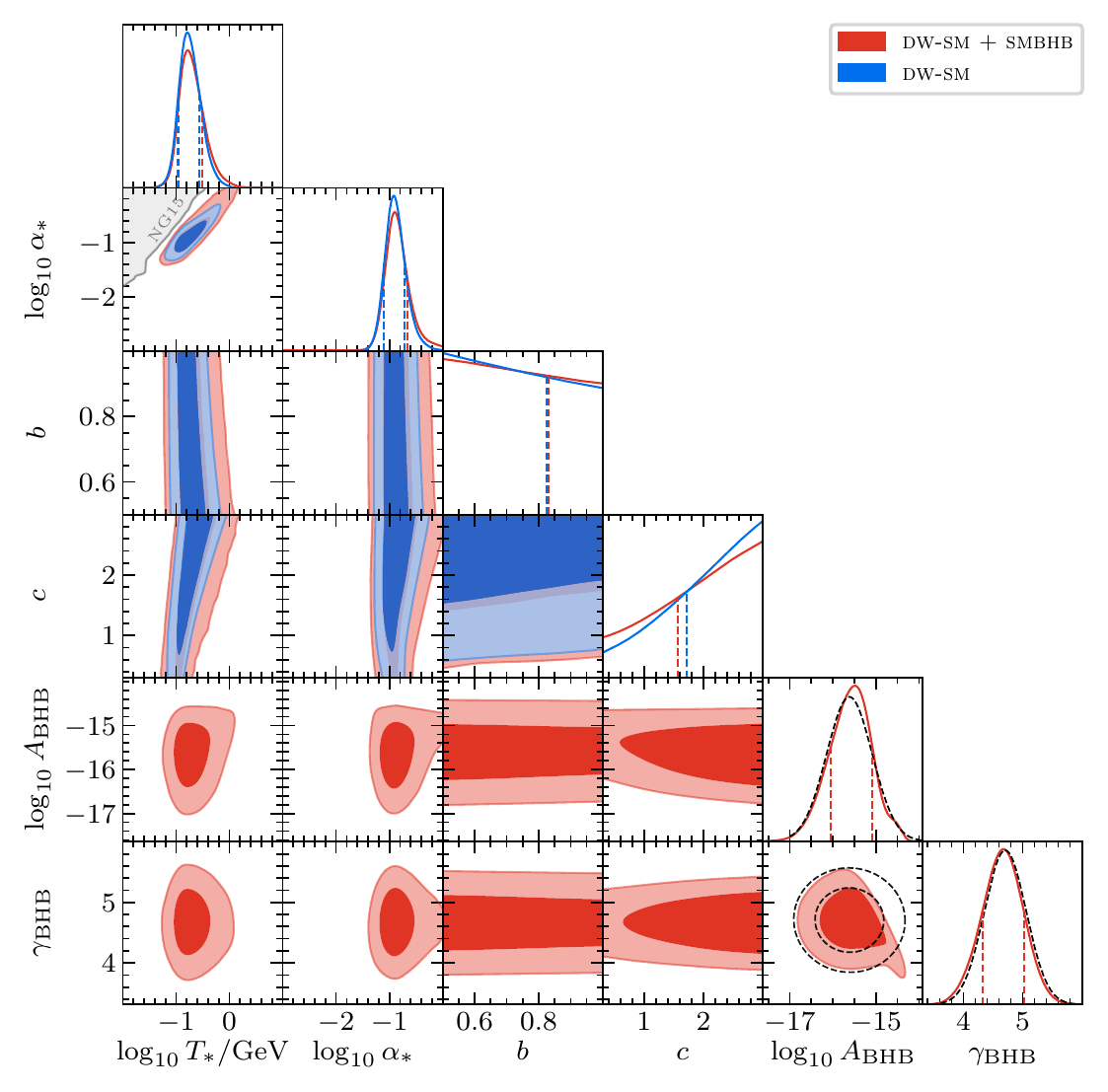}\\
        \includegraphics[width=0.65\textwidth]{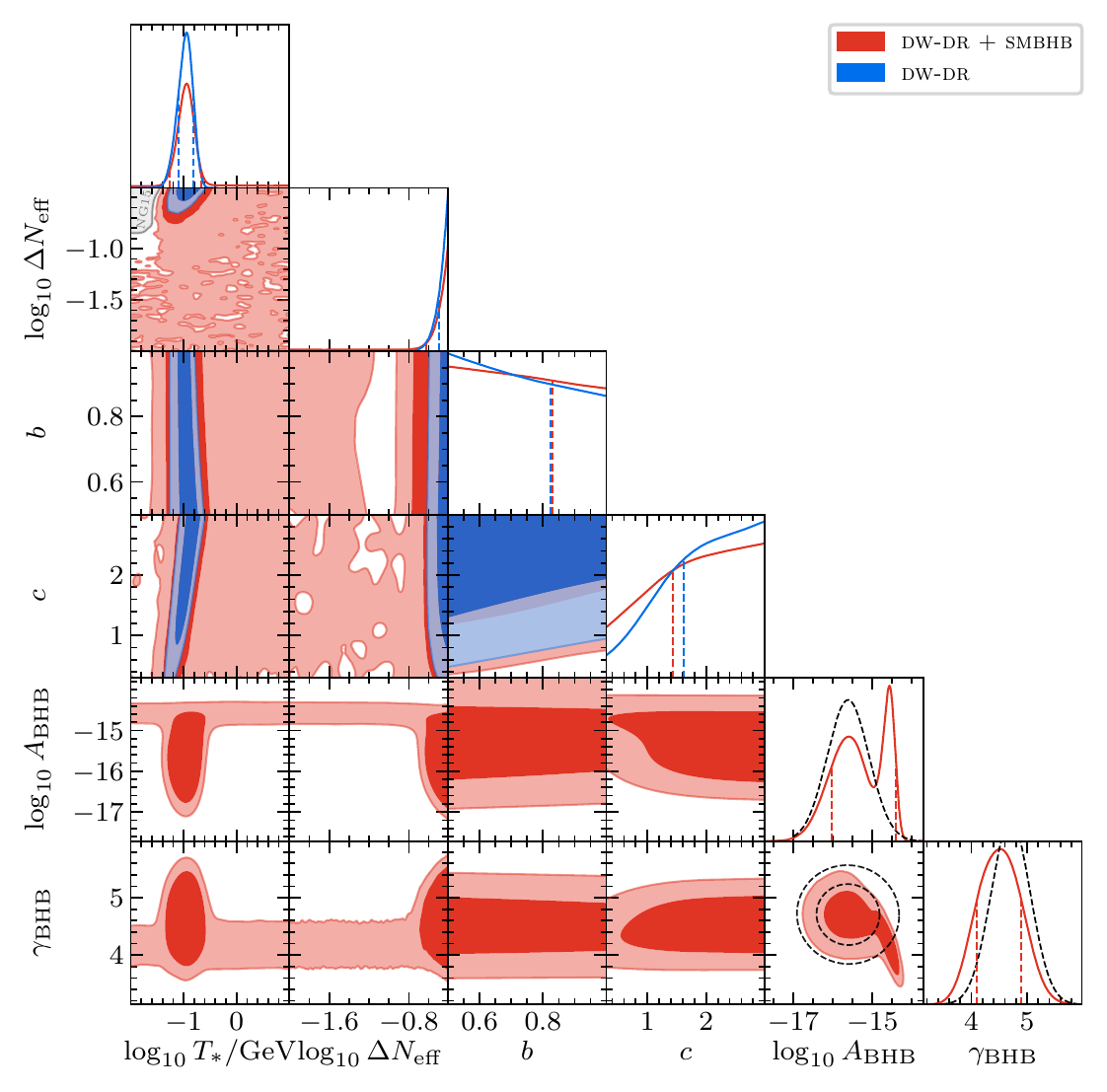}
	\caption{Same as Fig.~\ref{fig:dw_corner} but including the spectral shape parameters $b$, $c$ and the SMBHB parameters $\abhb$ and $\gbhb$.}
        \label{fig:dw_corner_full}
\end{figure*}

\subsection{Dark matter substructures}
\label{app:dm_sub}
In this appendix we provide more details on the procedure that we followed to derive the constraints on $f_{\scriptscriptstyle \rm PBH}$ reported in Section~\ref{subsec:dm_substructures}. 

Given a PBH with position relative to the pulsar given by $\bf{r}(t)=\bf{r}_0+\bf{v}t$, where $\bf{r}_0$ and $\bf{v}$ are the initial PBH position and velocity, respectively, we can write the Doppler and Shapiro signals as 
\begin{align}
    h_{ D}(t)&= \frac{G M}{v^2} \hat{\boldsymbol{d}} \cdot \big( \sqrt{1 + x_{ D}^2 }\hat{\boldsymbol{b}}_{ D} - \sinh^{-1}(x) \hat{\boldsymbol{v}} \big)\label{eq:h_D}\\
    h_{ S}(t)&= 2 G M \log(1 + x_{ S}^2)\,,\label{eq:h_S}
\end{align}
where $x_{ \text{D}} \equiv (t - t_{\text{D,0}})/\tau_{ \text{D}}$ and $x_{ \text{S}} \equiv (t - t_{ \text{S,0}})/\tau_{\text{ S}}$. These expressions only include cubic terms in time $t$ and have been previously derived in~\citep{Dror:2019twh, Lee:2021zqw}. For the static limit in which $\tau \gg T_{\text{obs}}$, these expressions can be expanded in series of $t_0/\tau$. Since the $\mathcal{O}(t^2)$ terms would be degenerate with the timing model, the measurable new-physics signal can then be parametrized as a term $\propto t^3$ as
\begin{equation}
    h(t)_{ \text{D(S)}} = \frac{A_{ \text{D (S), sta}}}{\mathrm{yr^2}} t^3\,,
\end{equation}
where $A_{ \text{D(S), sta}}$ for both the Doppler and Shapiro static signal cases are dimensionless amplitudes given by

\begin{align}
    A_{ \text{D, stat}} &= \mathrm{yr}^2 
    \frac{G M}{2 v^2} 
    \hat{\boldsymbol{d}} \cdot \bigg[ \frac{t_{ D,0}}{\tau_{ D}^4} \frac{1}{(1 + t_{ D,0}^2/\tau_{ D}^2)^{5/2}} \hat{\boldsymbol{b}}_{ D} + \frac{1}{3 \tau_{ D}^3} \frac{1 - 2 t_{ D,0}^2/\tau_{ D}^2}{(1 + t_{ D,0}^2/\tau_{ D}^2)^{5/2}} \hat{\boldsymbol{v}} \bigg] \label{eq:ADstat} \\
    A_{ \text{S, stat}} &= - \mathrm{yr}^2 \frac{4 GM}{3} \frac{t_{ S,0}}{\tau_{ S}^4} \frac{3 - t_{ S,0}^2/\tau_{ S}^2}{(1 + t_{ S,0}^2/\tau_{ S}^2)^3}\,. \label{eq:ASstat}
\end{align}

For the Doppler case, in the dynamic limit when $\tau \ll T_{\rm obs}$, the dominant contribution would come from the first term in Eq.~\eqref{eq:h_D} where $\sqrt{1 + x_{ D}^2} \propto |t - t_0|$. Upto linear order in $x_{ D}$, we can write
\begin{equation}
     h_{ D, {\rm dyn}}(t) = A_{ D, {\rm dyn}}(t-t_0)\Theta(t-t_0)\,,
\end{equation}
where $A_{\text{ D}, {\rm dyn}}$ is the dimensionless amplitude given by
\begin{equation}
    A_{ D, {\rm dyn}} = \frac{2 GM}{\tau v^2} \hat{\boldsymbol{d}} \cdot \hat{\boldsymbol{b}}_{ D}\,. \label{eq:Adyn}
\end{equation}

Given the expressions in Eqs.~\eqref{eq:h_D} and \eqref{eq:h_S} for the Doppler and Shapiro signals, we use the MC developed in~\cite{Lee:2020wfn} to derive the distributions $p(\log_{10}A_{ I}|f_{\scriptscriptstyle \rm PBH})$. Specifically, we proceed as follows:

\medskip\noindent\textbullet~For each pulsar we generate a population of $N_{\scriptscriptstyle \rm PBH}$ PBHs, where $N_{\scriptscriptstyle \rm PBH}$ is implicitly defined by the relation 
    \begin{equation}
        N_{\scriptscriptstyle \rm PBH}=f_{\scriptscriptstyle \rm PBH}\frac{\rho_{\scriptscriptstyle\rm DM} V}{M_{\scriptscriptstyle\rm PBH}}\,,
    \end{equation}
    where the simulation volume, $V$, is a sphere of radius $R=\bar v T_{{\rm obs}}$ centered around the pulsar position for the Doppler signal and a cylinder with the same radius and height given by the Earth--pulsar distance  for the Shapiro signal. Here $\bar v\simeq340\,{\rm km}/{\rm s}$ is the average PBH velocity, and $T_{{\rm obs},I}$ is the observation time of the $I$th pulsar.
    
\medskip\noindent\textbullet~For each of these PBHs we generate a random initial position and 
    velocity.  Since PTA searches are only sensitive to DM subhalos in the neighborhood of the solar system, we expect the position distribution to be uniform. Therefore, we use the probability density function $f(\bf{r})=1/V$ to sample initial positions. To sample PBHs' velocity, we use a Maxwell--Boltzmann distribution with $v_0 = 325\,{\rm km}{\rm s^{-1}}$, $v_{\rm esc} = 600\,{\rm km}{\rm s^{-1}}$, and the angular dependence assumed to be isotropic.
    
\medskip\noindent\textbullet~The PBHs' signals are then classified as dynamic if they satisfy the condition $T_{{\rm obs},I}-\tau>t_0>\tau$, and static otherwise. 

\medskip\noindent\textbullet~To evaluate $A_{\rm stat}$, we sum the static signals of all PBHs computed by using Eqs.~\eqref{eq:h_D} and \eqref{eq:h_S}, and we fit the resulting signal to a cubic polynomial in time to extract the $t^3$ term. To compute $A_{ \rm D, dyn}$, we take the closest transiting object and compute the signal amplitudes using 
    Eq.~\eqref{eq:Adyn}.

\medskip
All the previous points are repeated for $2.5\times10^3$ realizations to obtain the distributions $p(\log_{10} A_{ I}|f_{\scriptscriptstyle \rm PBH})$. Given the conditional distributions $p(\log_{10} A_{ I}|f_{\scriptscriptstyle \rm PBH})$ and the posterior distribution $p(\log_{10}A_{ I}|\boldsymbol{\delta t})$ derived by analyzing our data, we can write
\begin{equation}
    p\left(f_{\scriptscriptstyle \rm PBH}|\boldsymbol{\delta t}\right)= \prod_{ I=1}^{ N_P}\int p(f_{\scriptscriptstyle \rm PBH}|\log_{10}A_{ I})p\left(\log_{10}A_{ I}|\boldsymbol{\delta t}\right)d\log_{10}A_{ I}\,. \label{eq:f_posterior}
\end{equation}
Then, using Bayes' theorem, we can rewrite
\begin{equation}
    p\left(f_{\scriptscriptstyle \rm PBH}|\log_{10}{A}_{  I}\right) = \frac{p\left(\log_{10}A_{ I} | f_{\scriptscriptstyle \rm PBH}\right) p\left(f_{\scriptscriptstyle\rm PBH} \right)}{p \left(\log_{10}A_{ I} \right)}\,.
\end{equation}
Our priors on $p\left(f_{\scriptscriptstyle \rm PBH}\right)$ and $p\left(\log_{10}A_{ I}\right)$ are uniform, which allows us to eventually write Eq.~\eqref{eq:f_posterior} as
\begin{equation}\label{eq:PBH_posterior}
p\left(f_{\scriptscriptstyle \rm PBH}|\boldsymbol{\delta t}\right) \propto\prod_{ I=1}^{ N_{ P}}\int p(\log_{10} A_{ I}|f_{\scriptscriptstyle \rm PBH}) p(\log_{10} A_{ I}|\boldsymbol{\delta t})\;d \log_{10} A_{ I}\,,
\end{equation}
where the $\propto$ implies that the $p(f_{\scriptscriptstyle \rm PBH}|\bf{\delta t})$ would be subject to the normalization condition, $\int_0^{\infty} p(f_{\scriptscriptstyle \rm PBH}|\boldsymbol{\delta t}) \ d f_{\scriptscriptstyle \rm PBH} = 1$.

In the presence of a DM-baryon fifth force in the form of a Yukawa potential in Eq.~\eqref{eq:Yukawa_potential}, assuming point-mass DM, the pulsar frequency shift due to Doppler effect is given by \citep{Gresham:2022biw}
\begin{equation}\label{eq:Yukawa_frequency}
    \left(\frac{\delta \nu}{\nu}\right)_{\mathrm{fifth}} = \frac{\tilde{\alpha} G M}{\tau_D^2v} \hat{d} \cdot\int \frac{1}{(1+x_D^2)^{3/2}} \left(1+\frac{b}{\lambda}\sqrt{1+x_D^2}\right)e^{-(b/\lambda)\sqrt{1+x_D^2}}(\hat{b}+x_D\hat{v})dx_D \, .
\end{equation}
The integral in Eq.~\eqref{eq:Yukawa_frequency} has to be computed numerically, and the signal due to the fifth force can be computed by performing an additional integration over time and subtracting away degeneracies with timing model parameters. The total signal is the sum of the gravitational and the fifth-force contribution, $h_{D,\,\mathrm{total}}(t)=h_{D,\,\mathrm{fifth}}(t)+h_{D(S)}(t)$. In this analysis of the fifth-foce constraint, we only consider the scenario where the DM substructure makes up the entirety of the DM local density, which is equivalent to taking $f_{\mathrm{PBH}}=1$ for the gravitational contribution. Parameterizing the signal as $h_{\text{D},\,\mathrm{total}}(t)=\frac{A_{D,\,\text{total}}}{\mathrm{yr}^2}t^3$ similar to the PBH case, the amplitude $A_{D,\,\mathrm{total}}$ is a random variable dependent on $\lambda$ and $\tilde{\alpha}$. The probability distribution function $P(\log_{10}A_{D,\,\mathrm{total}}|\lambda,\tilde{\alpha})$ can be extracted again by Monte Carlo simulations and Bayes' theorem. Finally, the posterior distribution of $\tilde{\alpha}$ and $\lambda$, $P(\tilde{\alpha},\lambda|\boldsymbol{\delta t})$, is given by the analog of Eq.~\eqref{eq:PBH_posterior}
\begin{equation}\label{eq:fifth_posterior}
p\left(\tilde{\alpha},\lambda|\boldsymbol{\delta t}\right) \propto\prod_{ I=1}^{ N_{ P}}\int p(\log_{10} A_{ I}|\tilde{\alpha},\lambda) p(\log_{10} A_{ I}|\boldsymbol{\delta t})\;d \log_{10} A_{ I}\,.
\end{equation}

\clearpage
\FloatBarrier
\end{widetext}
\nocite{*}
\bibliography{bibliography.bib}{}
\bibliographystyle{aasjournal}
\end{document}